\documentclass[10pt,titlepage,onehalf,pdftex]{book} 
\title{On Charge 3 Cyclic Monopoles}
\author{Antonella D'Avanzo}
\date{2009}
\usepackage[usenames]{color} 
\definecolor{lightgr}{rgb}{.95,.95,0.95}
\usepackage{pifont}
\usepackage{lpic}
\usepackage{xypic}
\usepackage{AAtesi}
\usepackage[mine]{newchaps_A}

\usepackage[koi8-r]{inputenc}
\usepackage[russian,english]{babel}
\newcommand{\cred}[1]{\color{red}{#1}}
\newcommand{\cblu}[1]{\color{blue}{#1}}
\def\Dirac{{D\!\!\!\!/\,}}
 \newcommand{\Shat}{\hat{\Sigma}}
 \newcommand{\ah}{\hat{\mathfrak{a}}} 
 \newcommand{\bh}{\hat{\mathfrak{b}}}
 \newcommand{\azero}{\hat{\mathfrak{a}}^{{0}}} 
 \newcommand{\bzero}{\hat{\mathfrak{b}}^{{0}}}
 \newcommand{\abe}{\mathbf{a}} 
 \newcommand{\bbe}{\mathbf{b}} 
 \newcommand{\aq}{\mathfrak{a}}
 \newcommand{\bq}{\mathfrak{b}}
 \newcommand{\de}{\mathrm{d}}
 \newcommand{\uu}[1]{\hat{\mathbf{u}}_{#1}}
 \newcommand{\uzero}[1]{\hat{\mathbf{u}}^{0}_{#1}}
 \newcommand{\rot}{\mathcal{R}} 
 \newcommand{\Xhat}{\hat{X}}
  \newcommand{\Xzero}{\hat{X}_{0}}
\newcommand{\UU}{\mathbf{u}_{1}} 
\newcommand{\UT}{\mathbf{u}_{2}}
\newcommand{\UUU}{\mathbf{u}}
 
 \newcommand{\puno}{\mathbb{P}^{1}}
\newcommand{\pol}{\mathcal{P}}
\newcommand{\vrc}{\mathcal{K}}  
\newcommand{\Jac}{\mathcal{J}}  
\newcommand{\abelmap}{\mathcal{A}}  
\newcommand{\halfp}{\mathbf{e}}  
\newcommand{\divs}{\mathrm{div}}
\newcommand{\pic}{\mathrm{Pic}}  
\newcommand{\sferar}{\mathbb{P}^{1}}
\newcommand{\BH}{\widehat{{B}}}

\newcommand{\magn}{\mathbf{B}}
\newcommand{\el}{\mathbf{E}}
\newcommand{\ef}{\mathbf{F}}

\newcommand{\A}{\mathbf{A}}
\newcommand{\covde}{\boldsymbol{\mathcal{D}}}
\newcommand{\higgs}{\phi}
\newcommand{\tang}{\mathrm{T}}
\newcommand{\OO}{\mathcal{O}}
\newcommand{\e}{\mathrm{e}} 
\newcommand{\nuno}{\textbf{N1}}
\newcommand{\ndue}{\textbf{N2}}
\newcommand{\ntre}{\textbf{N3}}
\newcommand{\huno}{\textbf{H1}}
\newcommand{\hdue}{\textbf{H2}}
\newcommand{\htre}{\textbf{H3}}
\newcommand{\abel}{{\mathcal{A}}}
\newcommand{\zed}{\boldsymbol{\mathcal{Z}}}
\newcommand{\tvrc}{\widetilde{\mathcal{K}}}  
\newcommand{\shift}{s}
\newcommand{\Lpar}{A_{\lambda}}
\newcommand{\Mpar}{B_{\lambda}}
\newcommand{\spcurve}{\mathcal{C}}
\newcommand{\ie}{\textit{i.e.~}}  
\newcommand{\eg}{\textit{e.g.~}}  
\newtheorem{theorem}{Theorem}[chapter]
\newtheorem{lemma}[theorem]{Lemma}
\newtheorem{corollary}[theorem]{Corollary}
\newtheorem{proposition}[theorem]{Proposition}
\newtheorem{definition}[theorem]{Definition}
\newtheorem{fayacc}{Fay-Accola Theorem}[chapter]
\newtheorem{agm}{Extended AGM method}
\theoremstyle{remark}
\newtheorem{remark}{Remark}[section] 
\newtheorem{example}{Example}[section] 
\definecolor{darkgr}{rgb}{0.3,0.2,0.2}
\definecolor{darkred}{rgb}{0.5,0.2,0.2}
\definecolor{darkblue}{rgb}{0.2,0.2,0.7}
\usepackage{hyperref}
\hypersetup{
    colorlinks,
   citecolor=darkred,%
    filecolor=red,%
    linkcolor=darkgr,%
    urlcolor=darkblue
}
 \usepackage[all]{hypcap}
\begin{document}
\maketitle
\quad \newpage \thispagestyle{empty}
\pagenumbering{roman}
\setcounter{page}{1}
\declaration
\newpage
\quad  \newpage
  \thispagestyle{plain}
  \vspace*{6cm}
  \begin{center}
\begin{minipage}{400pt}
\noindent Part of the work presented hereby also appears in the following paper:

\cite{BDE10} H. W. Braden, A. D'Avanzo, and V. Z. Enolski. \emph{On charge-3 cyclic monopoles}. arXiv:  \htmladdnormallink{math-ph/1006.3408}{http://arxiv.org/abs/1006.3408}. 
\\

 Some of the results obtained hereby rely on calculations performed with \enf{Maple}, in particular making use of the package \texttt{algcurves}, part of Maple, and of the code \texttt{extcurves}, written by Timothy Northover, which can be downloaded at  \url{http://gitorious.org/riemanncycles}, where  the relevant documentation is also to be found.\\
The Maple files relevant for the work presented in this thesis can be requested from the author at antonella.davanzo@gmail.com.\\

\noindent  Figures \ref{be06fig1}, \ref{be06fig}-\ref{cycg2} were realized by Daniela D'Avanzo. Figure \ref{extim} was produced by Timothy Northover.
\end{minipage}\end{center}
\newpage\quad \newpage
\cleardoublepage
\phantomsection
\addcontentsline{toc}{chapter}{Abstract}
\abstr{Monopoles are solutions of an $SU(2)$ gauge theory in $\mathbb{R}^{3}$ satisfying a lower bound for energy and certain asymptotic conditions, which translate as topological properties encoded in their charge. Using methods from integrable systems, monopoles can be described in algebraic-geometric terms via their {spectral curve}, \ie an algebraic curve, given as a polynomial P in two complex variables, satisfying certain constraints. In this thesis we focus on the Ercolani-Sinha formulation, where the coefficients of P have to satisfy the Ercolani-Sinha constraints, given as relations amongst periods. \\

In this thesis a particular class of such monopoles is studied, namely charge 3 monopoles with a symmetry by $C_{3}$, the cyclic group of order 3. This class of cyclic 3-monopoles is described by the genus 4 spectral curve $\Xhat$, subject to the Ercolani-Sinha constraints: the aim of the present work is to establish the existence of such monopoles, which translates into solving the Ercolani-Sinha constraints for $\Xhat$.\\

Exploiting the symmetry of the system, we manage to recast the problem entirely in terms of a genus 2 hyperelliptic curve $X$, the (unbranched) quotient of $\Xhat$  by $C_{3}$ . A crucial step to this aim involves finding a basis for $H_{1}(\Xhat,\mathbb{Z})$, with particular symmetry properties according to a theorem of Fay. This gives a simple form for the period matrix of $\Xhat$ ; moreover, results by Fay and Accola are used to reduce the Ercolani-Sinha constraints to hyperelliptic ones on $X$. We solve these constraints on $X$  numerically, by iteration using the tetrahedral monopole solution as starting point in the moduli space. We use the Arithmetic-Geometric Mean method to find the periods on $X$: this method is well understood for a genus 2 curve with real branchpoints; in this work we propose an extension to the situation where the branchpoints appear in complex conjugate pairs, which is the case for $X$. \\

We are hence able to establish the existence of a curve of solutions corresponding to cyclic 3-monopoles.}
\newpage\thispagestyle{empty} 
\quad \newpage 
\thankyou{\vspace{-3mm}\thispagestyle{empty}
\begin{minipage}{400pt}
First of all, many thanks are due to my supervisor, Prof.~Harry Braden, for having borne with me all these years, for the continuous encouragement and the belief in my capabilities: without his patience and perseverance this work would have never been completed. Especially in the last stages of this thesis, his support has been invaluable. \\
I will always be deeply grateful to Prof.~Victor Enolskii, for his support, the enthusiasm with which he shared a lot of his personal experience and plenty of new ideas; not to mention the exquisite hospitality: \begin{otherlanguage}{russian}
спасибо
\end{otherlanguage}!
\\
I owe a lot to Tim Northover, for  his Maple code, which he developed  in a crucial moment for my research, and for the many useful discussions. \\
Many thanks to my second supervisor, Prof.~Jos\'e Figueroa-O'Farrill, for having always been there during tough moments, for the enthusiasm he shared about maths and physics, and for being the catlyst of the EMPG.\\
An ``extra''-special thanks to ``the Italian committee'': Daniele and Patricia, for all the deep conversations, and the silly ones, for a great help in not losing my mind completely in these final months, for the cooking and the currys and the proofreading, and, really, just for being such good friends; and Moreno and Erida (honorary!) who together with them have been my family in Edinburgh, and added so much to my life here.\\
Thanks to all the people who have made my days here pleasant over the years: Enrique, Andrea, honorary student here, the mathematician ``girls'',  Fatima, Dimitra, Elena, Elizabeth, Rosemary (and Patricia again!) with whom I shared so much more than a few hours in the same department.\\
Thanks to what's left of my friends from Italy: Francesco, for having hosted me so many times lately
; and George, my dear friend, for everything we have shared and still have to share after so long.\\
I have been lucky to have a family I can always rely on: thanks mamma and pap\'a, Costantino and Daniela, and Maria zia Pia, for having supported me along all my decisions in life, for the strong bond that we still manage to keep despite the distance and my long silences, and for being always there for me.\\
And thanks to Alessandro, ``for the support, the surreality and the rebel words'' (cit.); for all the experiences we shared together, funny and sad ones, and intense, and unforgettable; for continuing to surprise me all the time; for the sharp and original thinking, the ``lively'' discussions and the sweet actions; and for so much more that wouldn't fit in this few lines... grazie!
\end{minipage}}
\newpage \quad \thispagestyle{empty} 
\tableofcontents
%
%
%
%
%
\clearpage
\cleardoublepage
\phantomsection
\addcontentsline{toc}{chapter}{Introduction}
\chapter*{Introduction}
\setcounter{page}{1}\pagenumbering{arabic}
The concept of a \enf{magnetic monopole} as a pointlike magnetic charge was introduced by Dirac almost 80 years ago in \cite{dirac}, in the context of electromagnetism. It was a ground-breaking idea, as  for the first time it put electricity and magnetism truly on the same footing in Maxwell's equations, and more importantly provided an argument for the  quantisation of electric charge. \\
With the study of nonabelian gauge theories, where the $U(1)$ gauge group of Maxwell's theory is enlarged to a nonabelian one, it was soon realised that objects like the Dirac monopole are not just peculiar of electromagnetism. In the case of $SU(2)$,  \,'t Hooft \cite{th} and Polyakov \cite{pol} numerically found that the Yang-Mills equations admit soliton solutions, namely smooth fields configurations localised in a region of space, which at large distances behave like Dirac monopoles.
Shortly afterwards, in \cite{bs} Prasad and Sommerfield  exhibited analytic solutions in a simplified case, with vanishing Higgs potential; Bogomolny studied this case further \cite{bog}, introducing the Bogomolny equations \eqref{bogomintro}. Hence this class of monopoles is referred to as \enf{BPS monopoles}: these are the monopoles we are concerned with in the present thesis.\\ 

Interestingly enough, monopoles have never been found experimentally. Nevertheless,  the physical interest in the consequences of the existence of such objects, and, more relevant to us, the beauty and richness of their mathematical description make monopoles  a rich and dynamic area of research still nowadays. \\

Consider an $SU(2)$ gauge theory on the Euclidean $\mathbb{R}^{3}$, with connection $\A$, and Higgs field $\higgs$.  Monopoles are then solutions of  the \enf{Bogomolny equations}
\begin{equation}\label{bogomintro}
 \ef=\star\covde\higgs
\end{equation}
(with certain boundary conditions); here $\ef$ is the field strength and $\covde$ the covariant derivative associated to a gauge field $\A$, while $\star$ is the Hodge star. \\
The Bogomolny equations can be obtained via a dimensional reduction of the selfduality equations in 4 dimensions,  keeping one of the coordinates constant: several remarkable properties can hence be understood using tools coming from instanton theory. In particular, in \cite{hitgeo} Hitchin, adapting the twistor approach, showed that monopoles may be identified  with certain bundles over $\puno$, subject to some nonsingularity conditions; the solution is then determined by an {algebraic curve} $\spcurve$ in $\puno$. \\
Moreover, Nahm managed to adapt  the ADHM instanton construction to the case of monopoles in \cite{nahm}. The resulting Nahm's equations can be presented in Lax form: this allows to use methods from integrable systems in this context. In particular an algebraic curve can be constructed, the \enf{spectral curve}, which is again $\spcurve$. \\
These two methods, which Hitchin showed to be equivalent in \cite{hitmon}, both yield the {spectral curve} as fundamental object. The main advantage of this description is that the spectral curve is a gauge invariant object which determines the solutions completely, in principle. In practice, however, very little is known about explicit solutions. Most of the work in this direction concerns spectral curves which are, or can be reduced to, elliptic curves. This is due to the fact that  elliptic  curves are indeed  well understood, and are subject of a vast literature spanning centuries now.
Symmetric monopoles constitute a class where some explicit solutions are known: in general, symmetry simplifies the problem, reducing the solutions to ones written again in terms of elliptic functions.\\

Ercolani and Sinha provided a powerful insight in the investigation of exact monopole solutions \cite{es}. They showed how one can solve the Nahm equations in terms of \enf{Baker-Akhiezer functions}, introducing methods from the algebraic-geometric theory of integrable systems in the monopole setting (in particular \cite{kr77}). Their procedure, although conceptually simple, proves rather challenging to implement explicitly, especially for higher charge monopoles. This is due to the fact that it requires a rather explicit knowledge of several objects on the spectral curve, which is in general a difficult task for higher genus curves. Braden and Enolskii \cite{BE06} manage to implement Ercolani-Sinha construction for a {charge 3 monopole} with a particular symmetry, hence finding explicit solutions for monopoles in this class.
In the present work of thesis, we aim to extend the results of \cite{BE06}  to a more general class of monopoles, namely charge 3 cyclic monopoles. \\

A \enf{charge 3 cyclic monopole} has a (genus 4) spectral curve $\Xhat$ of equation
\begin{equation}\label{cu}
w^{3}+\alpha w z^{2}+\beta z^{6}+\gamma z^{3}-\beta=0,
\end{equation}
with $\alpha,\beta, \gamma\in\mathbb{R}$, subject to some constraints. We focus in particular on solving one set of Ercolani-Sinha constraints, involving relations amongst the periods of $\Xhat$: the number of cases when these constraints can be solved is fairly limited. \\
 The  cyclic group $C_{3}$ acts on $\Xhat$, and we have an  unbranched  covering $\pi:\Xhat \rightarrow X$ of a genus $2$ curve $X$. Results by Fay \cite{fay} and Accola \cite{accola} allow us to express several objects on the genus 4 curve in terms of those on the genus 2 curve. For this simplification to take place, the first step is the explicit construction of a basis of $H_{1}(\Xhat, \mathbb{Z})$ with particular symmetry properties: this proves to be nontrivial, and constitutes one of the main results of this thesis.\\
 Using the theory of unbranched covers, we manage to reduce the Ercolani-Sinha constraints to hyperelliptic ones, hence making them more manageable. Moreover, the genus 4 theta function is expressed as a product of hyperelliptic ones, which means that the solutions for this class of monopoles are expressible in terms of hyperelliptic functions. \\
 
In order to solve the Ercolani-Sinha constraints, we make use of the \enf{Arithmetic-Geometric Mean (AGM)} method. This method was developed by Gauss \cite{gauss99} to find elliptic integrals in a more efficient way, and modified by Richelot and Humbert for the case of genus 2 surfaces with real branchpoints in \cite{richelot2,humbert}. We manage to extend it to genus 2 curves with complex conjugate branchpoints: we give all the details for the subclass of such curves to which the quotient monopole belongs; we remark that the methods we use make possible such an extension also to \enf{generic} genus 2 curves with complex conjugate branchpoints.\\

Using this extended AGM method, we solve the Ercolani-Sinha constraints numerically, via an iteration starting from the known solutions of  the tetrahedral monopoles. We find a \enf{one parameter family of charge 3 cyclic monopoles}, by giving explicit expressions for those parameters $\alpha$, $\beta$, $\gamma$ in \eqref{cu} satisfying the Ercolani-Sinha constrants: this provides a quantitative confirmation of results obtained by Hitchin, Manton and Murray \cite{symmmon}, and Sutcliffe \cite{sutcyclic} obtained by geodesic monopole scattering arguments.\\
Finally, a study of the Igusa invariants for the curve \eqref{cu} allows us to conclude that 4 of the above solutions correspond to curves admitting elliptic subcovers: hence, in the class of cyclic charge 3 monopoles there are \enf{4 elliptic ones}.

\subsection*{Plan of the work}
Chapter 1 gives  some background about monopoles. After a short overview of the 
original differential geometric formulation, we briefly present the Nahm and 
Hitchin descriptions; within 
this setting we introduce symmetric monopoles, which are a main theme of 
this work. We then describe the Ercolani-Sinha formulation, which, using tools 
from integrable systems, provides a suitable framework for exact monopole solutions: in this context, we examine in some detail a certain symmetric charge 
3 monopole studied by Braden and Enolskii. \\

In Chapter 2 we recall some elements from the algebraic-geometric theory of integrable systems. In particular, we focus on the Lax formulation and, within this framework, we discuss the spectral curve associated to a system in Lax form and how the solution can be expressed in terms of theta functions on this curve. This introduces some of the techniques that are at the basis of the study of monopoles introduced in Chapter 1, upon which the present work of thesis is based.\\

We study the spectral curve for the cyclic charge 3 monopole in detail  in Chapter 3. We examine its general properties, and in particular the $C_{3}$ symmetry, and its quotient with respect to $C_{3}$, making use of Fay's theory of unbranched covers. One of the main results of this work is indeed the explicit construction of a particularly symmetric basis which satisfies Fay's theorem. This allows to express several objects on $\Xhat$ in terms of objects on $X$: in particular, the Ercolani-Sinha constraints reduce to hyperelliptic constraints, for which we are able to find a solution.\\

Chapter 4 is dedicated to the study of the invariants of the curve $X$, to investigate for which values of the parameters the curve $X$ admits a reduction to an elliptic curve.\\

Finally, solutions for the Ercolani-Sinha constraints reduced to $X$ are calculated explicitly in Chapter 5, using the Arithmetic-Geometric Mean method. After examining the original method, due to Gauss, and its adaptation by Richelot and Humbert  to genus 2 surfaces with real branchpoints, we present a further extension to curves with complex conjugate branchpoints. Using this extended AGM method, we manage to solve the Ercolani-Sinha constraints iteratively, finding a curve of solutions in the moduli space.
From the analysis of the curve invariants performed in Chapter 4, we also deduce the existence of 4 elliptic monopoles within the class of those with cyclic symmetry and charge 3.
\newpage

\chapter{Monopoles}
\introcap{We give here some background about monopoles. After a short overview of the original differential geometric formulation, we briefly present the Nahm and Hitchin descriptions, which are more algebraic-geometric in flavour; within this setting we introduce symmetric monopoles, which are a main theme of this work. We then describe the Ercolani-Sinha formulation, which, using tools from integrable systems, provides a suitable framework for exact monopole solutions: in this context, we examine in some detail a certain symmetric charge 3 monopole studied by Braden and Enolskii.
}

\numberwithin{equation}{section}
\section{A very short introduction to monopoles}
The setting is a \enf{gauge theory}, the prototype of which is electromagnetism. \\
In the geometric formulation of electromagnetism, the fundamental objects are a 1-forms on $\mathbb{R}^3$, the electric field $\el$, and a 2-form on $\mathbb{R}^3$, the magnetic field $\magn$. Out of these, one can build the Maxwell field tensor $\ef=\magn+c\; \de t \wedge \el \;\in\Omega^{2}(\mathcal{M}_{4})$, where  $\mathcal{M}_{4}=\mathbb{R}^3 \times \mathbb{R}$ is the four dimensional Minkowski space-time. In these geometric terms, Maxwell equations (in absence of sources) can be written as follows:
\begin{equation}\label{maxwelleq}
 \de \ef=0,\qquad \de \star \ef=0,
\end{equation}
where $\star$ is the Hodge star. The first of eq. (\ref{maxwelleq}) implies that $\ef$ is closed on $\mathcal{M}_{4}$, hence can be expressed as $\ef=\de \A$, where $\A$ is a 1-form, the vector potential.\\
From its very definition, the vector potential is defined up to an exact form, or in other words  it is invariant under a gauge transformation $\A \to \A+\de \alpha$. Geometrically this means that  $\A$ is a connection form of a (trivial) $U(1)$ bundle over $\mathcal{M}_{4}$, and $\ef$ is its curvature. Note that the setting where the necessity of the bundle description really arises is, in fact, that of the so-called \enf{Dirac monopole} \cite{dirac}. A Dirac monopole is a solution of the Maxwell equation with a magnetic charge at the origin of $\mathbb{R}^3$. The magnetic  field $\magn$ corresponding to such  a magnetic monopole is given by
\begin{equation*}
 \magn = \frac{1}{4\pi r^3}(x_1 \de x_2 \wedge \de x_3 + cyclic).
\end{equation*}
In terms of the ``usual'' vector $\vec{B}$ this would read
\begin{equation*}
 \vec{B}(\mathbf{x})= \frac{\mathbf{x}}{4\pi r^3}
\end{equation*}
where $\mathbf{x}=(x_{1},x_{2},x_{3}),\;r=|\mathbf{x}|$.\\
With this magnetic field, there exists no globally defined form $\A$ such that   $\ef=\de \A$. However, it is possible to find such a form locally, for instance 
\begin{align*}
\A_N=\;\frac{1}{4\pi r} \frac{1-\cos\theta}{\sin\theta} \mathbf{e}_{\phi}\quad \mathrm{on}\quad U_{N}\\
\A_S=-\frac{1}{4\pi r} \frac{1+\cos\theta}{\sin\theta} \mathbf{e}_{\phi}\quad \mathrm{on} \quad U_{S},
\end{align*}
where $U_N=S_N\times\mathbb{R}$ and $U_S=S_S\times\mathbb{R}$, where $S_N$ (resp. $S_S$) is the sphere minus the north (resp. south) pole. 
Thus $\ef=\de \A$ holds locally. Note that the following relation holds:
\begin{equation*}
 \A_N=\A_S+\de\alpha, \qquad \mathrm{where} \; \alpha=-\frac{1}{2\pi}\phi.
\end{equation*}
which is precisely the Maurer-Cartan equation for a connection on a $U(1)$ bundle.\\
In general, the connection viewpoint is extremely powerful  when the space has a nontrivial topology, as in this case, where there is a singularity at the origin. However, in this work we are mostly interested in monopoles without singularities, defined in the whole of $\mathbb{R}^3$: this means that for a connection there is always a globally defined gauge potential. 
Nevertheless, these monopoles share important similarities with the Dirac monopole, including some sort of topological nature.

\subsection{$SU(2)$ monopoles}

Pure Yang-Mills theory is a gauge theory directly modelled on electromagnetism, but with gauge group a non-abelian Lie group instead of $U(1)$. The connection $\A$ is interpreted (in the quantum theory) as giving rise to a vector particle for every generator of the Lie algebra of $G$, with zero mass; in order to incorporate mass in a gauge invariant way, one has to include in the theory a new field $\higgs$, the \enf{Higgs field}.\\The action is then (proportional to)
\begin{equation}\label{actionYM}
 S=\int_{\mathcal{M}}\Big((\ef,\ef)+(\covde \higgs,\covde \higgs)+V(\higgs)\de vol \Big) 
\end{equation}
where $V$ is a gauge invariant potential, $\covde$ is the covariant derivative associated to the connection $\A$,  
and $(\ef,\ef)=\mathrm{Tr}(\ef\wedge \star \ef)$.\\
In the following, we only consider a theory on $\mathbb{R}^3$ equipped with the Euclidean metric, with gauge group $SU(2)$, a Higgs fields taking values in the adjoint bundle to the principal $SU(2)$ bundle on $\mathbb{R}^3$ and zero potential.\\
The action then reads
\begin{equation*}
 S=\int_{\mathbb{R}^3}\Big((\ef,\ef)+(\covde \higgs,\covde \higgs)\Big);
\end{equation*}
since $S$  is interpreted as the energy of the system, one has to ensure that it is finite. This is achieved by imposing the following field behaviour at infinity\footnote{
Note that this ensures the finiteness of the energy only when the fields satisfy the Bogomolny equations \ref{bogom}.
} 
(for details see \cite{jt}):%
\begin{align}\label{bc1}
\begin{split}
||\higgs||= 1+O(r^{-1})\\
 \frac{\partial ||\higgs||}{\partial \Omega}= O(r^{-2}),\\
 ||\covde \higgs||= O(r^{-2}),
 \end{split}
\end{align}
where $ \frac{\partial }{\partial \Omega}$ denotes the angular derivative (along a sphere).
\\
Integrating the action density over a ball of radius $R$ centered at the origin, one has:
\begin{align}\label{eqb1}
 \int_{B^{2}_{R}}\big(\left(\ef,\ef\right)+\left(\covde \higgs,\covde \higgs\right)\big) &=
\int_{B^{2}_{R}}\big(\left(\ef-\star\covde \higgs,\ef-\star\covde \higgs\right)+2\left(\star\covde \higgs,\ef\right)\big).
\end{align}
Upon using the Bianchi identity for $\ef$:
\begin{align*}
 \de(\higgs,\ef)&%
 =(\star\covde\higgs,\ef),
\end{align*}
which implies that 
\begin{align*}
 \int_{B^{2}_{R}}(\star\covde \higgs,\ef)\;=&\int_{S^{2}_{R}}(\higgs,\ef),
\end{align*}
where $S^{2}_R=\partial B^{2}_{R}$.  Let the connection $\A$ be defined on a rank $2$ vector bundle $E$.  For large $R$, as $||\higgs||\to 1$ by \eqref{bc1}, its eigenspaces define complex line bundles $L$ and $L^{*}$  over $S^{2}_{R}$, with Chern class $\pm k$. 
With the decay conditions $||\covde \higgs||= O(r^{-2})$, the projection of the curvature $\ef$ on these line bundles approaches the curvature of the bundles themselves:
\begin{equation*}
 \lim_{R\to\infty}\int_{S_{R}}(\higgs,\ef) =\pm 4\pi k,
\end{equation*}
hence
\begin{equation*}
 S=\int_{\mathbb{R}^3}(\ef-\star\covde \higgs,\ef-\star\covde \higgs)\;\pm4\pi k.
\end{equation*}
Therefore, if $k \geqslant 0$, the action $S$ is minimised if $(\A,\higgs)$ satisfies the \enf{Bogomolny equations}:
\begin{equation}\label{bogom}
 \ef=\star\covde\higgs.
\end{equation}
A Higgs field solving these equations can in fact be shown to have (in a suitable gauge) the following asymptotic expansion:
\begin{equation*}
\higgs \sim \begin{pmatrix}i & 0 \\ 0 & -i\end{pmatrix} -\frac{k}{2r} \begin{pmatrix}i & 0 \\ 0 & i\end{pmatrix}  +O(r^{-2}).
\end{equation*}
This adds a further constraint to the asymptotic conditions (\ref{bc1}); hence we can now give the following
\begin{definition}
 An \enf{$SU(2)$ monopole of charge $k$} is a solution of the above \enf{Bogomolny equations} with (minimal) energy $4\pi k$ and boundary conditions
 \begin{align}\label{bc}
\begin{split}
 ||\higgs||&= 1-\frac{k}{2r}+O(r^{-2})\\
 \frac{\partial ||\higgs||}{\partial \Omega}&= O(r^{-2}),\\
||\covde \higgs||&= O(r^{-2}),
 \end{split}
\end{align}
\end{definition}

\begin{remark}
We notice that monopoles  also arise as dimensional reduction of a Yang-Mills theory in the four dimensional Euclidean space. More specifically, consider an $SU(2)$ gauge theory on the Euclidean $\mathbb{R}^4$, with Higgs field taking values in the adjoint bundle and action  as in eq. (\ref{actionYM}), with a non zero potential: 
\begin{equation}\label{actionYM4}
S=\int_{\mathbb{R}^4}\left((\ef,\ef)+(\covde \higgs,\covde \higgs)+\lambda(1-|\higgs|^2)^2\de^4 x\right)  
\end{equation}
 If we  consider only  the solutions which are invariant by translations along, for instance, the fourth coordinate, these can be dimensionally reduced to $\mathbb{R}^3$, where the first three components of $\A$ can be interpreted as components of a connection on $\mathbb{R}^3$, and the fourth as the Higgs field.\\ 
Taking $\lambda=0$ in the potential, and keeping the boundary conditions given by the finiteness of the action \eqref{actionYM4} results in the dimensional reduction of the self-duality equations $\ef=\star\ef$  to the Bogomolny equations (\ref{bogom}).\\
Thus, \enf{$SU(2)$ monopoles} can be interpreted as \enf{static self-dual solutions of Yang-Mills equations on $\mathbb{R}^4$}.
 \end{remark}
 
\begin{remark}
Note that setting $\lambda=0$ in (\ref{actionYM4}) is known (mainly in the physics literature) as taking the BPS limit of the Yang-Mills action. Indeed,  \,'t Hooft and Polyakov found the first monopole solution in a non abelian gauge theory, namely a static solution of the fields equation coming from eq. (\ref{actionYM4}) with $\lambda\neq 0$; their analysis was mainly numerical. Shortly after this, Prasad and Sommerfield found an analytic form for the solution in the case $\lambda=0$, which is a spherically symmetric monopole of charge 1. Then Bogomolny studied further this $\lambda=0$ limit, also  introducing the Bogomolny equations (\ref{bogom}). Hence this class of monopoles is often referred to as BPS monopoles: in this work, we refer to them simply as monopoles.
\end{remark}

\begin{remark}
Finally, we point out that in the case of a gauge group $U(1)$, Bogomolny equations reduce to 
$$B= \mathrm{grad}~\higgs$$ where $\higgs=\frac{k}{2r}$. This corresponds indeed to the Dirac monopole.
\end{remark}

\section{Nahm construction}
The above  description of monopoles, whilst very beautiful, does not prove very useful when finding explicit monopoles solutions. Hence, several different approaches have been developed to this aim, both approximate and exact. In this work we are mainly concerned with Nahm's approach \cite{nahm}, and the subsequent formulations given by Hitchin (\cite{hitmon, hitgeo}), and Ercolani and Sinha (\cite{es}); in this section, we describe Nahm's formulation.\\

In \cite{nahm} Nahm managed to adapt the Atiyah-Hitchin-Drinfeld-Manin (ADHM, see \cite{ADHM}) construction for instantons to the case of monopoles. The equivalence between the geometric description of monopoles and Nahm's algebraic description was proven by Hitchin in \cite{hitmon}, where he also showed the equivalence of yet another formulation, that we refer to as Hitchin's formulation (explained in section \ref{hitchinform}). \\

The crucial construction needed to understand this alternative formulation is the \enf{Nahm transform}, which is a two way transform: it takes solutions of the Bogomolny equations (with boundary conditions (\ref{bc})) to solutions of the Nahm's equations, a first order differential equation between matrices satisfying certain conditions (see end of this section), and vice versa.\\
In analogy to the ADHM construction, Nahm considers a Dirac operator on $\mathbb{R}^{3}$, given by:
\begin{equation*}
\Dirac=\left(
\begin{array}{cc}
    0 & D_{z}   \\
     D^{\dagger}_{z} & 0
\end{array}
\right)=
\left(
\begin{array}{cc}
    0 &\boldsymbol{\sigma}\cdot \covde -(\higgs+iz)  \\
     \boldsymbol{\sigma}\cdot \covde +(\higgs+iz) & 0
\end{array}
\right)
\end{equation*}
where $\boldsymbol{\sigma}=(\sigma_{1},\sigma_{2},\sigma_{3})$ are the Pauli matrices,  $\Dirac$ acts on  $\mathcal{L}^{2}$-normalizable $\mathbb{C}^{2}\times \mathbb{C}^{2}$ valued\footnote{
Both $\mathbb{C}^{2}$s are representations of $\mathfrak{su}(2)$: the first copy is the spin space where the $\sigma_{i}$ act, the other is the space where the connection and the Higgs live.
} 
 functions on $\mathbb{R}^{3}$, and $D^{\dagger}$ is the adjoint of $D$.\\
After using Bogomolny equations, one finds:
\begin{equation*}
D_{z}D^{\dagger}_{z}= \covde \cdot \covde -(\higgs+iz)(\higgs+iz),
\end{equation*}
which means  $D_{z}D^{\dagger}_{z}$ is a positive operator, and hence has no ($\mathcal{L}^{2}$-normalizable) zero modes; this in particular  implies that the $\mathcal{L}^{2}$-kernel of $D^{\dagger}_{z}$ is empty. One can show\footnote{Using an $\mathcal{L}^{2}$ index theorem it follows that $D_{z}$ has an index $k$ if  $z\in(-1,1)$ , and zero otherwise.} that $D_{z}$ has $k$-dimensional $\mathcal{L}^{2}$-kernel $N_{z}$ only if $z\in(-1,1)$ . \\
Consider then an orthonormal basis for $N_{z}$,  namely $k$ functions $\psi^{i}$ such that
\begin{equation*}
D_{z}\psi^{i}=0, \qquad \int_{\mathbb{R}^{3}}(\psi^{i},\psi^{j})=\delta_{ij},
\end{equation*}
and satisfying
\begin{equation*}
\int_{\mathbb{R}^{3}}(\psi^{i},\frac{\partial \psi^{j}}{\partial z})=0.
\end{equation*}
These functions can be used to define the three \enf{Nahm matrices} as follows:
\begin{equation}\label{nahmmatrices}
T^{ij}_{a} (z)=\int_{\mathbb{R}^{3}}(\psi^{i},x_{a}\psi^{j}).
\end{equation}
These matrices satisfy a number of properties,  given below.
\begin{itemize}
\item[\nuno] \enf{Nahm equations}: 
\begin{equation}\label{nahmeq}
\frac{\de T_{i}}{\de t}=\frac{1}{2}\sum_{j,k} \varepsilon_{ijk}[T_{j},T_{k}].
\end{equation}
\item[\ndue] $T_{i}$ is analytic for $z\in(-1,1)$ and has simple poles at $z=-1$, $z=1$, where the residues form irreducible $k$-dimensional representations of $\mathfrak{su}(2)$.
\item[\ntre] $\qquad T_{i}(z)=-T_{i}^{{\dagger}}(z)  \qquad T_{i}(z)=T_{i}^{T}(-z)$.
\end{itemize}

A proof of these properties can be found either in Nahm's paper \cite{nahm}, or in the review \cite{wy}.\\

A solution of eqs. \eqref{nahmeq} satisfying conditions \nuno, \ndue , \ntre\, can be mapped back to a connection and a Higgs field via the inverse Nahm transform, which has a similar structure to the map introduced earlier, and is given schematically as follows (for details see again \cite{nahm} or  \cite{wy}).\\
Consider the differential operator
\begin{equation*}
D_{x}:=\frac{\de}{\de z} - T_{i}\sigma^{i}-x_{j}\sigma_{j}. 
\end{equation*}
It can be shown that its $\mathcal{L}^{2}$-kernel $E_{x}$ is 2-dimensional (this makes use of \ndue). Choosing an orthonormal basis\footnote{Note that the ambiguity in the choice of this basis is reflected in an ambiguity in the definition (\ref{nahmfields}) of the fields up to a gauge transformation} $(v_{1},v_{2})$ for $E_{x}$, define a connection and a Higgs field by
\begin{align}\label{nahmfields}
\begin{split}
\higgs_{ab}&=i\int_{-1}^{1} z v^{\dagger}_{a}(z)v_{b}(z) \de z,\\
\A_{ab}&=i\int_{-1}^{1} z v^{\dagger}_{a}(z)\frac{\partial}{\partial x_{i}}v_{b}(z) \de z.
\end{split}
\end{align}
$\A$ and $\higgs$ so defined are indeed smooth solutions of the Bogomolny equations with the appropriate boundary conditions for a charge $k$ monopole.\\
We can then state the following theorem.
\begin{theorem}[\enf{Nahm \cite{nahm}, Hitchin \cite{hitmon}}]
An \enf{$SU(2)$ monopole of charge $k$}, namely a solution of the Bogomolny equations with (minimal) energy $4\pi k$ and boundary conditions (\ref{bc}), is equivalent to a solution of the \enf{Nahm equations} \nuno, subject to the conditions \ndue, \ntre.
\end{theorem}\vspace{3mm}

\begin{remark}
The Nahm transform and its inverse are isometries, and the Nahm transform followed by its inverse gives back the original monopole.
\end{remark}\vspace{2mm}
\begin{remark}
Nahm's equations can be cast in \enf{Lax form}, which allows to use methods from integrable systems; we explain  this in some more detail.\\
Define
\begin{align*}
A_{-1}&=(T_{1}+iT_{2}), &  A_{0}&=-2iT_{3} , &  A_{1}=T_{1}-iT_{2},
\end{align*}
and hence, introducing a \enf{spectral parameter} $\zeta\in\puno$:
\begin{align*}
A&=A_{-1}\zeta^{-1}+A_{0}+A_{1}\zeta,    &  M&=\frac{1}{2}A_{0}+A_{1}\zeta.
\end{align*}
Then \nuno\; becomes
\begin{equation*}
\frac{\de A}{\de z}=[A,M],
\end{equation*}
\ie  a \enf{Lax form for Nahm's equations}. This permits to use integrable system methods to find a solution to Nahm's equations. Indeed, an algebraic curve appears, namely the \enf{spectral curve}, given by the equation (see Chapter \ref{intsys})
\begin{equation}\label{nahmcurve}
P(\zeta,\eta)=\mathrm{det}(\eta\mathbb{I}-A)=
\mathrm{det}\left(\eta+(T_{1}+iT_{2})-2iT_{3}\zeta+(T_{1}-iT_{2})\zeta^{2}\right)=0,
\end{equation}
and Nahm's equations describe a linear flow on the Jacobian on this curve. As described in Chapter \ref{intsys}, there are (almost) algorithmic approaches to deal with this class of systems:  in particular, in this work we make use of the finite-gap integration method, due to Krichever (see section \ref{krth} for more details), in a framework due to Ercolani and Sinha (see \cite{es} and section \ref{esform}). 
\end{remark}\vspace{2mm}
\begin{remark}
Finally, it has been possible to solve Nahm's equations in a number of nontrivial cases, and then to carry out the inverse Nahm transform, at least numerically: in this way a number of monopole solutions of different charges  have been constructed explicitly; this has proven useful especially to describe qualitative features of monopoles, such as the energy density distribution.\\
\end{remark}

\section{Hitchin data}\label{hitchinform}
Another approach to monopoles in terms of a certain class of holomorphic bundles on $\tang\puno$ is given by Hitchin; here we only sketch very briefly this construction, referring to \cite{hitgeo} for details.\\

The idea is to consider the set of oriented geodesics on $\mathbb{R}^{3}$ with the Euclidean metric, which is in fact isomorphic to $\tang\puno$; a solution to the Bogomolny equations gives rise to a holomorphic bundle on this surface.\\

To set notation, introduce homogeneous coordinates $[\zeta_{0}:\zeta_{1}]$  on $\puno = U_{0} \cup U_{1}$, with $U_{i}$ standard open sets. Let $\zeta=\zeta_{0}/\zeta_{1}$ be a coordinate on $U_{0}$; hence  standard coordinates $(\zeta,\eta)$ on $\pi^{-1}(U_{0})\subset  \tang\puno$ are defined by $(\zeta,\eta)\to \eta\dfrac{\de}{\de \zeta}$ (and similarly for $U_{1}$).\\
With respect to these coordinates and the open covering of $\tang\puno$ given by $\tilde{U}_{i}=\pi^{-1}(U_{i})$, the line bundle $L^{j}$ is defined by the transition function $\exp(-j\tfrac{\eta}{\zeta})$ in the intersection $\tilde{U}_{0}\cap\tilde{U}_{1}$.\\
Recall that in $\puno$  the transition function $g_{01}=\zeta^{n}$ on $U_{0}\cap U_{1}$ defines the line bundle $\OO(n)$ (which is the unique line bundle of degree $n$ on $\puno$); denote the pullback of this line bundle to $\tang\puno$ also by $\OO(n)$. Take then the tensor product of  this line bundle $\OO(n)$ with the line bundle $L^{j}$ defined earlier, to obtain the bundle $L^{j}(n):= L^{j}\otimes\OO(n) $.\\

Consider the differential operator:
\begin{equation*}
D_{\gamma}=\covde_{\gamma}-i\higgs,
\end{equation*}
where $\covde_{\gamma}$ is the covariant derivative along the oriented straight line $\gamma$.\\
Define a bundle $\tilde{E}$ on the space of geodesics by associating to each $\gamma$ the (2-dimensional) kernel  $E_{\gamma}$ of $D_{\gamma}$. If $\A$ and $\higgs$ satisfy the Bogomolny equations, one can endow $\tilde{E}$ with a \emph{holomorphic structure} (with some additional properties, see Theorem 4.2 in \cite{hitgeo}).\\
The bundle $\tilde{E}$ has two holomorphic sub-bundles, $E^{\pm}$, whose fibers $E^{\pm}_{\gamma}$ are solutions of $D_{\gamma}f=0$ which decay at $\pm\infty$ along the line $\gamma$. It can be seen (Theorem 6.3 in \cite{hitgeo}) that $E^{+}\simeq L(-k)$ and $E^{-}\simeq L^{*}(-k)$.%
The set of those curves $\gamma$ for which $E^{+}_{\gamma}= E^{-}_{\gamma}$ forms a curve in $\tang\puno$, called the \enf{spectral curve}  $S$ of the monopole. Since a decaying solution decays exponentially, the spectral curve is also the set of lines along which there is an $\mathcal{L}^{2}$-solution.\\
One can give a more precise characterisation of the spectral curve. That the quotients of $\tilde{E}$ by $E^{\pm}$ satisfy $$\tilde{E}/E^{+}\simeq L(k),\qquad \tilde{E}/E^{-}\simeq L^{*}(k).$$
The curve $S$ is defined by the vanishing of the map $E^{+}\to \tilde{E}/E^{-}$ and hence by a section of $(E^{+})^{*}\otimes  \tilde{E}/E^{-} \simeq \mathcal{O}(2k) $. In terms of the standard coordinates  $(\zeta,\eta)$ on $\tang\puno$, $S$ is then defined by an equation of the form:
 \begin{equation}\label{hcurve}
 P(\eta,\zeta)=\eta^{k}+a_{1}(\zeta)\eta^{k-1}+\ldots +a_{k-1}(\zeta)\eta+a_{k}(\zeta),
 \end{equation}
where $a_{j}(\zeta)$, $0\leq j\leq k$ are polynomial of degree at most $2j$ in $\zeta$.\\
This curve satisfies the following properties (\cite{hitgeo, hitmon})
\begin{itemize}

\item[\textbf{H0}]   $S$ is compact and has no multiple components.
\item[\huno] \enf{Reality condition}: the curve $S$ is invariant under the standard real structure $\tau$ on $\tang\puno$. The involution $\tau$ is defined by reversing the orientation of the lines in $\mathbb{R}^{3}$, and in coordinates takes the following form
\begin{equation}\label{realstr}
\tau:(\zeta,\eta) \rightarrow \left(-\dfrac{1}{\bar{\zeta}},-\dfrac{\bar{\eta}}{\bar{\zeta}^{2}}\right).
\end{equation}
This invariance condition translates to the following relations amongst the coefficients $a_{i}$
\begin{equation*}
a_{j}(\zeta)=(-1)^{j}\zeta^{2j}\overline{a_{j}\left(-\frac{1}{\bar{\zeta}}\right)}.
\end{equation*}
\item[\hdue] $L^{2}$ is holomorphically trivial on $S$ and $L(k-1)$ is real.\\
The first statement follows from the fact that, on $S$, $E^{+}$ and $E^{-}$ are isomorphic, which is equivalent to say that $E^{+}\otimes E^{-}\;(\simeq L^{2})$. Both these properties are consequences of the boundary conditions (\ref{bc}).
\item[\htre] $H^{0}(S,L^{s}(k-2))=0$ for $s\in(0,2)$. This is equivalent to the non-singularity of the monopole determined by $S$.
\end{itemize}
The above findings can be summarised  in the following theorem.
\begin{theorem}[\enf{Hitchin \cite{hitmon}}]\label{hitthm}
An \enf{$SU(2)$ monopole of charge $k$}, namely a solution of the Bogomolny equations with (minimal) energy $4\pi k$ and boundary conditions (\ref{bc}), is equivalent to a spectral curve of the form (\ref{hcurve}), subject to the \enf{Hitchin conditions \huno,  \hdue, \htre}.
\end{theorem}
Incidentally,  note that from the above properties one can deduce, using Riemann-Hurwitz formula, that the genus of $S$ is $g_{S}=(k-1)^{2}$.\\

If $k=1$, the spectral curve (\ref{hcurve}) satisfying \huno, \hdue, \htre\; takes the form :
\begin{equation}\label{kuno}
\eta=(x_{1}+ix_{2})-2x_{3}\zeta-(x_{1}-ix_{2})\zeta^{2},
\end{equation}
where $(x_{1},x_{2},x_{3})=\mathbf{x}\in\mathbb{R}^{3}$ and can be interpreted as the \enf{centre} of the monopole; indeed, such a curve is the set of all oriented lines through the point $\mathbf{x}$. This curve is also called a \emph{real section} as it defines a real section of the bundle $\tang\puno\to\puno$.\\
A $k$-monopole also has a well defined centre (as well as a total phase), whose coordinates $(c_{1},c_{2},c_{3})$  enter in the spectral curve as coefficients of the polynomial $a_{1}(\zeta)$ as follows:
\begin{equation}\label{a1center}
a_{1}(\zeta)=(c_{1}+ic_{2})-2c_{3}\zeta-(c_{1}-ic_{2})\zeta^{2},
\end{equation}\vspace{2mm}

\begin{center}
\begin{minipage}{350pt}
\enf{Note}:  Nahm's  and Hitchin's approaches are indeed equivalent: this is shown in the fundamental paper \cite{hitmon}, where a link among these two formulations, and the original geometric description is given. We remark that both in Hitchin's and in Nahm's data an algebraic curve appears, directly as in $\huno$, or indirectly as in eq. (\ref{nahmcurve}): this is indeed the \enf{same curve} in two different manifestations.
\end{minipage}\vspace*{5mm}
\end{center}
\section{Symmetric Monopoles}\label{symmmon}
Although the methods outlined above provide a great deal of simplification, in general it proves quite involved to find an exact solution for a monopole, and not many solutions have been found thus far: a notable exception is the case of \enf{symmetric monopoles}. Symmetric monopoles are $k$-monopole solutions which are \enf{invariant under various symmetries}:  this constrains their spectral curves, and in some cases it is quite immediate to infer the non-existence of monopoles with certain symmetries. It is more difficult to obtain results on the existence and on the exact form of the solutions, although some solutions have been found already in the early paper \cite{symmmon}. The  fundamental paper \cite{symmmon} is the basis for the exposition below: we have chosen to present this topic in Hitchin formalism, rather than in Nahm formalism because, despite using the latter in most of this thesis, we believe that symmetry is more natural and transparent in the former. \\

A general monopole curve has the form given in eq. (\ref{hcurve}), subject to conditions \huno, \hdue, \htre; here we investigate the form of these curves when the monopole is required to be invariant with respect to certain symmetry groups of $\mathbb{R}^{3}$ of \enf{rotations around the origin}. To implement this symmetry, the  monopole is taken to be centered at the origin, which means, from eq. (\ref{a1center}), that $a_{1}(\zeta)=0$. But before doing this,  a special type of symmetry, namely \enf{inversion}, is discussed.

\subsection{Inversion}
The inversion map in $\mathbb{R}^{3}$ is the reflection in the $(x_{1},x_{2})$ plane, namely:
\begin{equation*}
I:(x_{1},x_{2},x_{3})\longrightarrow (x_{1},x_{2},-x_{3}).
\end{equation*}
As this map reverses orientation, it induces an anti-holomorphic map on the space $\tang\puno$, which in standard coordinates reads
\begin{equation*}
\phi:\; (\zeta,\eta) \rightarrow \left(\dfrac{1}{{\bar{\zeta}}}, -\dfrac{\bar{\eta}}{\bar{\zeta}^{2}}\right) 
\end{equation*}
Notice that $\phi$ is very similar to the real structure $\tau$ of eq. (\ref{realstr}); in fact $$\phi\circ\tau(\zeta,\eta)= (-\zeta,\eta).$$
It follows then that if $P(\zeta,\eta)=0$ is a monopole curve, and hence invariant under $\tau$, then the inverted curve is $P(-\zeta,\eta)=0$.\\
It is interesting to study the monopoles which are invariant under inversion; in particular, their moduli space presents an interesting structure (see \cite{symmmon}, section 4). Moreover, they form a large class, and  are also easy to characterise: their spectral curve is given by polynomials $P(\zeta,\eta)=0$ which are \emph{even} in $\zeta$.

\subsection{Rotations}\label{symmrotcurv}
Let us now recall that an $SO(3)$ rotation in $\mathbb{R}^{3}$ corresponds to a M\"{o}bius transformation in $\puno$ and hence $\tang\puno$. In particular, a rotation by an angle $\theta$ around the unit vector $(n_{1},n_{2},n_{3})$ is represented in $\tang\puno$ as :
\begin{align*}
\zeta \to \zeta' =\dfrac{(d+ic)\zeta+(b-ia)}{-(b+ia)\zeta+(d-ic)}, \qquad 
\eta \to \eta'  =\frac{\eta}{\left(-(b+ia)\zeta+(d-ic)\right)^{2}}
\end{align*}
where
\begin{equation*}
n_{1}\sin\frac{\theta}{2}=a,\quad  
n_{2}\sin\frac{\theta}{2}=b,\quad
n_{3}\sin\frac{\theta}{2}=c,\quad
\cos\frac{\theta}{2}=d,\quad
\end{equation*}
Thus, a monopole is invariant under a rotation if its spectral curve $P(\zeta,\eta)=0$ is invariant under the associated M\"{o}bius transformation, namely $P(\zeta',\eta')=0$ is the same curve. One can then examine the consequences of invariance under a number of uncomplicated subgroups of $SO(3)$.\\

\begin{example}[\enf{Cyclic and dihedral monopoles}]
The first simple example is the \enf{cyclic group of rotations $C_{n}$} around the $x_{3}$ axis. It is generated by
\begin{equation*}
\zeta'=\e^{\tfrac{2\pi i}{n}}\zeta, \qquad \eta'=\e^{\tfrac{2\pi i}{n}}\eta.
\end{equation*}
Hence, for $P(\zeta,\eta)=0$ to be invariant under $C_{n}$, all terms must have the same total degree, $mod$ $n$; since the leading term is $\eta^{k}$, all terms must have degree $k$ $mod$ $n$. In particular, for $P(\zeta,\eta)=0$ to be invariant under $C_{k}$, all terms must have degree $0$, $mod$ $k$.\\

The groups $C_{n}$ are extended to \enf{dihedral groups} $D_{n}$ by adding to the generators a rotation by $\pi$ around the $x_{1}$-axis, which corresponds to the transformation:
\begin{equation}\label{extrasymm}
\zeta'=\frac{1}{\zeta},\quad \eta'=-\frac{\eta}{\zeta^{2}}.
\end{equation}

Let us consider some examples of such symmetric curves, namely charge $3$ monopoles with either $C_{3}$ or $D_{3}$ symmetry. A general charge 3 monopole curve takes the form
\begin{align*}\begin{split}
\eta^{3}+\eta&(\alpha_{4}\zeta^{4}+\alpha_{3}\zeta^{3}+\alpha_{2}\zeta^{2}+\alpha_{1}\zeta^{1}+\alpha_{0})+\\
\qquad &(\beta_{6}\zeta^{6}+\beta_{5}\zeta^{5}+\beta_{4}\zeta^{4}+\beta_{3}\zeta^{3}+\beta_{2}\zeta^{2}+\beta_{1}\zeta^{1}+\beta_{0})=0,
\end{split}\end{align*}
where the coefficients are subject to the following conditions:
\begin{align*}
\alpha_{4}&=\bar{\alpha}_{0}, &\alpha_{3}&=\bar{\alpha}_{1},& \alpha_{2}&=\bar{\alpha}_{2}; & \; &\\
\beta_{6}&=-\bar{\beta}_{0},& \beta_{5}&=\bar{\beta}_{1},& \beta_{4}&=-\bar{\beta}_{2}, &\beta_{3}&=\bar{\beta}_{3}.
\end{align*}
Imposing the $C_{3}$ symmetry, the curve reduces to
\begin{equation}\label{c3curve}
\eta^{3}+\alpha\eta\zeta^{2}+\beta\zeta^{6}+\gamma\zeta^{3}-\bar{\beta}=0,
\end{equation}
where $\alpha,\gamma\in\mathbb{R}$, and also $\beta$ can be made real by a change of coordinates (rotation about the $x_{3}$-axis).\\

Imposing on (\ref{c3curve}) the extra symmetry (\ref{extrasymm}), one obtains that for $D_{3}$ symmetric curve $\gamma=0$, hence:
\begin{equation}\label{d3curve}
\eta^{3}+\alpha\eta\zeta^{2}+\beta(\zeta^{6}-1)=0.
\end{equation}
\end{example}\vspace{2mm}
\begin{example}[\enf{Platonic monopoles}]
An interesting class is given by those monopoles which are invariant under the \enf{Platonic groups}; in these cases, Klein's theory of invariant polynomials proves very useful in implementing the symmetry requirements on the spectral curve. For instance, by symmetry considerations alone one can conclude that no \enf{octahedrically} symmetric monopole exists for $k=2$ and $k=3$; for $k=4$, it would have the following form:
\begin{equation}\label{oct}
\eta^{4}+a(\zeta^{8}+14\zeta^{4}+1)=0.
\end{equation}
Analogously, the simplest monopole curve with \enf{tetrahedral} symmetry is of the form (for $k=3$)
\begin{equation}\label{thetr}
\eta^{3}+ia\zeta(\zeta^{4}-1)=0.
\end{equation}
Notice that after a rotation, this curve becomes
\begin{equation}\label{thetr1}
\eta^{3}+a(\zeta^{6}+5\sqrt{2}\zeta^{3}-1)=0,
\end{equation}
thus exhibiting the $C_{3}$ symmetry around the $x_{3}$ axis (cf. eq. (\ref{c3curve}) ). The simplest curve with \enf{icosahedral} symmetry is (for $k=6$)
\begin{equation}\label{ico}
\eta^{6}+a\zeta(\zeta^{10}+11\zeta^{5}-1)=0.
\end{equation}
We remark that the curves (\ref{thetr}), (\ref{oct}), (\ref{ico}) only satisfy the reality conditions \huno: for them to describe monopoles, they also have to satisfy conditions \hdue\, and \htre. One can prove (see \cite{symmmon}), that the curves (\ref{thetr}), (\ref{oct}), are indeed spectral curves for monopoles of charge 3 and 4, for suitable values of $a$; on the other hand, there is no charge $6$ monopole with icosahedral symmetry.\\
Note that the quotients of the curves (\ref{thetr}), (\ref{oct}), (\ref{ico}) by the respective symmetry groups are \enf{elliptic curves}: this proved a great advantage in obtaining the above results. Indeed, the explicit solutions for the tetrahedral and icosahedral monopoles are expressed in terms of elliptic functions, and the constant $a$ in terms of elliptic periods.
\end{example}

These last examples are in fact paradigmatic of the general situation for symmetric monopoles: the symmetry, and reality conditions impose (sometimes stringent) constraints on the spectral curve, but this is not enough to conclude that the constrained curve does indeed correspond to a monopole. For this to be the case, conditions \hdue\,and \htre\,(or \ndue\, and \ntre\, in Nahm's formulation) need to be satisfied as well: this proves to be a very difficult step in general, as it involves relations between integrals on the spectral curve, which become harder to find explicitly the higher the genus of the curve. In the above cases, the spectral curve admits an elliptic quotient, so the integral reduce to hyperelliptic ones, which are manageable; we shall see that the same happens for a certain symmetric 3-monopole (cf. \cite{BE06} and section \ref{sec:be}). In the case of the cyclic 3-monopole, which is the object of the present work, the general quotient is hyperelliptic (cf. section \ref{sec:quotient}), which is a slightly more involved case.

\section{Ercolani-Sinha formulation}\label{esform}
Based on Nahm's description of monopoles,  Ercolani and Sinha developed a method for finding solutions of Nahm's equations making use of Krichever's method for solving integrable systems in terms of Baker-Akhiezer functions on the spectral curve $S$. They provided an explicit expression for these functions, and hence translated Hitchin's constraints \hdue, \htre\,to constraints on these functions, and hence to constraint on periods on the curve $S$.\\
We describe the Ercolani-Sinha method in some detail here, as this is the method that we apply in this thesis to the case of the cyclic 3-monopole: we follow the expositions given in the original paper \cite{es}, and in the more recent \cite{BE06}.\\

Ercolani and Sinha consider the differential operator
\begin{equation*}
\frac{\de}{\de z} +M(z)= \frac{\de}{\de z} +\frac{1}{2}A_{0}(z)+A_{1}(z)\zeta
\end{equation*}
 related to the Lax equations; studying its spectral theory, one sees that
 \begin{equation*}
\left(  \frac{\de}{\de z} +\frac{1}{2}A_{0}(z)\right)\varphi=-\zeta A_{1}(z)\varphi
 \end{equation*}
 does not take the usual form of an eigenvalue problem, since $A_{1}$ is dependent on $z$. One can obtain the standard eigenvalue equation
 \begin{equation}\label{stev}
\left(  \frac{\de}{\de z} +Q_{0}(z)\right){\Phi}=-\zeta Q_{1}(0)\Phi 
 \end{equation}
 after performing the following gauge transformation
 \begin{equation*}
 Q_{i}(z)=C^{-1}(z)A_{i}(z)C(z), \qquad \varphi=C(z)\Phi,
 \end{equation*}
 with
 \begin{equation}\label{QC}
 C(z)^{-1}C'(z)=\frac{1}{2}Q_{0} \quad \Leftrightarrow \quad  \left(\frac{\de}{\de z} +\frac{1}{2} Q_{0}(z) \right) C(z)^{-1}=0
 \end{equation}
 Hence the reduced Nahm equations become:
 \begin{equation}\label{rednahmeq}
 \frac{\de Q}{\de z}=[Q,Q_{+}],
 \end{equation}
 where $Q=\zeta^{-1}Q_{-1} +Q_{0}+\zeta Q_{1}$ and $Q_{+}=Q_{0}+\zeta Q_{1}$.\\
Note that $Q_{1}(z)=Q_{1}(0)=A_{1}(0)$, and since $A_{1}$ is symmetric, we can assume it is diagonal: $Q_{1}(0)=A_{1}(0)= \mathrm{diag}(\rho_{1},\ldots,\rho_{k})$.\\
Comparing with the expression (\ref{nahmcurve}) for $P(\zeta,\eta)$, we see that the $\rho_{j}$ are indeed the roots of $P(\zeta,\eta)/\zeta^{2k}$ near $\zeta=\infty$:
\begin{equation}\label{prho}
\frac{P(\zeta,\eta)}{\zeta^{2k}} \sim \prod_{j=1}^{k}\left(  \frac{\eta}{\zeta}-\rho_{j} \right).
\end{equation}

One can apply Krichever's method to the reduced Nahm equations (\ref{rednahmeq})  to find a solution for $Q_{0}$ in terms of Baker-Akhiezer functions. This relies on the following fundamental theorem by Krichever (more details can be found in section \ref{krth}).
\begin{theorem}[\enf{Krichever, 1977}]\label{krithm}
Let $S$ be a smooth algebraic curve of genus $g_{S}$ with $n\geqslant 1$ punctures $P_j $, $j
=1,\ldots, n$.  Then for each set of $g_S+n-1$ points $\delta
_{1},\ldots ,\delta _{g_S+n-1}$ in general position, there exists
a unique function $\Psi _{j }\left( t,P\right) $ and local
coordinates $w_{j }(P)$ for which $w_{j }(P_{j })=0$, such that
\begin{enumerate}
\item  the function $\Psi _{j }$ of $P\in S$ is meromorphic
outside the punctures and has at most simple poles at $\delta
_{s}$ (if all of them are distinct);
\item  in a neighbourhood of the puncture $P_{l }$ the function $\Psi _{j }$ takes the form
\begin{equation*}
\Psi _{j }\left( z, P\right) =e^{z\,{w_{l }}^{-m}}\left( \delta
_{j l }+\sum\limits_{k=1}^{\infty }\alpha\sp{k}_{j l }\left(
z\right) w_{l }^{k}\right), \qquad w_{l }=w_{l }\left(
P\right),\qquad m\in\mathbb{N}\sp{+}. 
\end{equation*}
\end{enumerate}
\end{theorem}

We can use this result to express $\Phi$ in eq. (\ref{stev})  as follows. Let $\Phi_{l}$ be the columns of $\Phi$ in eq. (\ref{stev}), normalised so that
\begin{equation}\label{normphi}
\exp(\zeta A_{1}(0)z)\Phi|_{\infty}=Id_{n}
\end{equation}
Consider the $k$ points $P_{j}$ above $\zeta=\infty$ as punctures; we denote their $\eta$ coordinate by $\infty_{j}$. Then one has the following result: 
\begin{theorem}[\enf{Ercolani-Sinha \cite{es} }]\label{esthm}
The $j$-th component of $\Phi_{l}$, normalised by (\ref{normphi}) is given by Theorem \ref{krithm} on $S$, where the punctures are the $k$ points $P_{j}$ above $\zeta=\infty$.\\
From this, the matrix $Q_{0}$ can be reconstructed from
\begin{equation*}
(Q_{0})_{jl}=-(\rho_{j}-\rho_{l})\alpha_{jl}=-(\rho_{j}-\rho_{l}) \lim_{P\to\infty_{l}} \zeta \exp(\zeta\rho_{l}z) \Psi_{j}(z,P).
\end{equation*}
\end{theorem}

To summarise the complete Ercolani-Sinha method, one can express the components of $\Phi$  in terms of these Baker-Akhiezer functions, and hence reconstruct $Q_{0}$; then, using \ref{QC}, one can determine $C(z)$, and therefore $A_{-1},A_{0}, A_{1}$, from which reconstruct the matrices $T_{i}$. \\

We explain in detail how to obtain $\Phi$ and $Q_{0}$ in the next subsections.

\subsection{Solving Nahm's equations via the spectral transform}
\subsubsection{Baker-Akhiezer functions}
The functions $\Psi_{l}$ of  section  \ref{krithm}  can be expressed explicitly in terms of theta functions on $S$; in order to do so, we introduce the following basic ingredients:
\begin{itemize}
\item[1.] \enf{The differential $\gamma_{\infty}$}. From eq. (\ref{prho}) we see that the rational function $\eta/\zeta$ has the following asymptotic behaviour:
\begin{equation*}
\frac{\eta}{\zeta} \sim \rho_{j}\zeta \quad \mathrm{near }\;\infty_{j}.
\end{equation*}
Hence its differential satisfies
\begin{equation}\label{etazetainf}
\de\left( \frac{\eta}{\zeta} \right) \sim \left(-\frac{\rho_{j}}{t^{2}} +O(1)\right)\de t \quad \mathrm{near }\;\infty_{j},
\end{equation}
where $t=1/\zeta$ is a choice of local coordinate.
Using the above as motivation, one can define a second kind differential $\gamma_{\infty}$ such that
\begin{align}
\gamma_{\infty}& \sim \left(-\frac{\rho_{j}}{t^{2}} +O(1)\right)\de t \quad \mathrm{near }\; \infty_{j};\label{asym}\\
\oint_{\mathfrak{a}_{i}}&\gamma_{\infty}=0; \label{ginfnorm}\\
\gamma_{\infty}&\;\mathrm{holomorphic\; away\; from}\;\infty.
\end{align}
\item[2.]\enf{The vector $U$}. 
After Theorem \ref{krithm} and the subsequent observation, one can introduce, for  a particular flow,  the vector with components:
\begin{equation}\label{vecU}
U_{j}=\frac{1}{2\pi i} \oint_{\mathfrak{b}_{j}}\gamma_{\infty}.
\end{equation}
This differential encodes the flow on the Jacobian corresponding to the time evolution of  the system.

\item[3.] \enf{Normalisation factors $\nu_{j}$}. To obtain the normalisation required in Theorem \ref{krithm} we introduce the following normalisation factors
\begin{equation*}
	\nu_{j}=\lim_{P\to\infty_{j}}\left[  \left( \int_{P_{0}}^{P}\gamma_{\infty} \right) +\frac{\eta}{\zeta}\right].
\end{equation*}
Note that these in fact depend on the basepoint $P_{0}$, and are finite quantities, because of the asymptotic property (\ref{asym}).
\end{itemize}
We can now give an explicit expression for the non normalised function $\tilde{\Psi}_{i}$ in terms of theta functions; we introduce first the following functions:
\begin{equation}\label{phithetam1}
{F_{i}}=\frac{
\theta \left(  \abel(P) - \zed_{j}+z U  \right) }
{\theta\left(\abelmap(P)-\sum_{j=1}^{g}\abelmap(\delta_{j})-\vrc\right)}
\frac{\prod_{j\neq i}\theta(\abelmap(P)-\boldsymbol{R}_j)}
{\prod_{j= i}^{N-1}\theta(\abelmap(P)-\boldsymbol{S}_j)},
\end{equation}
where $\abelmap$ denotes the Abel map, $\vrc$ the vector of Riemann constants based at $P_{0}$ (see Appendix \ref{apprs} for some properties of these objects). Here we have set
\begin{align*}
\boldsymbol{\mathcal{Z}}_{j } &=\boldsymbol{\mathcal{Z}}_{T}+\abel \left( P_{j }\right)\equiv \abel(\Delta_j)+\vrc ,&
\boldsymbol{\mathcal{Z}}_{T}&=\sum\limits_{s=1}^{g_S+n-1}\abel \left( \delta_{s}\right) -\sum\limits_{j =1}^{n}\abel \left( P_{j}\right)+\vrc,\\
\boldsymbol{R}_j&= \sum\limits_{s=1}^{g_S-1}\abel \left( \delta_{s}\right) +\abel \left( P_{j }\right)+\vrc, &
\boldsymbol{S}_j&=\sum\limits_{s=1}^{g_S-1}\abel \left(\delta_{s}\right) +\abel \left( \delta_{g_S-1+j}\right)+\vrc.
\end{align*}
Also, by Abel's theorem, $\boldsymbol{\mathcal{Z}}_{j }$  is equivalent to an \enf{effective divisor $\Delta_{j}$ of degree $g_{s}$}.\\ 
From eq. \eqref{phithetam1}, we obtain the following expression for the normalised functions $\Psi_{j}$
\begin{equation}\label{phitheta}
\Psi_{j}=\frac{F_{i}(P)}{F_{i}(P_{i})} e^ {z\,\left(\int\limits_{P_0}^{P}\gamma_{\infty}- \nu_{j} \right)}.
\end{equation}

The functions $\Psi_{j}$ in eq. (\ref{phitheta}) have all the required properties of Theorem \ref{krithm}: for a further discussion on this point we refer to section \ref{krth}, where we examine in more detail how to build Baker-Akhiezer functions on a spectral curve in a more general context. \\

Using the expression (\ref{phitheta}) for the $\Phi_{j}$ in Theorem \ref{esthm}, we obtain the following expression for $Q_{0}$ in terms of Baker-Akhiezer functions:
\begin{equation}\label{q0es}
\left(Q_0\right)_{jl}=-(\rho_j-\rho_l)\,c_{jl}\,e\sp{z[\nu_l-\nu_j]}\,
\frac{ \theta \left( \abel (P_l)-\zed_{j }
+z\,U\right)
\theta \left(\abel \left( P_{j }\right)
-\zed_{j } \right) }
{ \theta \left(\abel \left( P_{j }\right)
-\abel_{j }+
z\,U\right) \theta \left( \abel (P_l)
-\zed_{j }\right) },
\end{equation}
where
\begin{equation}\label{cijdef}
c_{jl}=\lim_{P\rightarrow\infty_l}\zeta\, g_j(P),\qquad
P=(\zeta,\eta)\in S.
\end{equation}
\subsection{The Ercolani-Sinha constraints}
As we have now expressed $Q_{0}$ concretely in terms of Baker-Akhiezer functions, we are able to impose Hitchin's constraints $\hdue,\,\htre$ explicitly on these Baker-Akhiezer functions: these are implemented as constraints on the periods of the curve $S$. We examine this in some detail in this section, since in the applications of the Ercolani-Sinha method we are most interested in, namely the 3-monopole of \cite{BE06}, and the cyclic 3-monopole studied later in this thesis, a large part of the analysis has been devoted in fact to the implementation of these constraints.
\\

Let us consider then Hitchin's constraints in detail.
\begin{itemize}
\item[i] \enf{$L(k-1)$ is real.} \cite{har}.

\item[ii] \enf{Triviality of $L^{2}$}. $L^{2}$ is trivial if and only if it admits  a (nowhere vanishing) holomorphic section $f$, namely, in local coordinates, if and only if one can find holomorphic functions $f_{0}$ on $U_{0}\cap S$ and $f_{1}$ on $U_{1}\cap S$, such that, on $U_{0}\cap U_{1}\cap S$ one has
$$  f_{0}=\exp\left(-2\frac{\eta}{\zeta}\right)f_{1} .$$
Taking the logarithmic derivative we get
\begin{equation}\label{sectf}
\de\log f_{0}= - \de\left( 2\frac{\eta}{\zeta} \right)+\de\log f_{1}.
\end{equation}
In order to avoid any essential singularity of $f_{0}$ on $U_{0}$, then  $\de\log f_{1}$  has to cancel those of  $- \de\left( 2{\eta}/{\zeta} \right)$ at $\infty$; this means, using eq. (\ref{etazetainf}), that
\begin{equation}\label{delogf0inf}
\de\log f_{1} \sim  2 \left(-\frac{\rho_{j}}{t^{2}} +O(1)\right) \quad \mathrm{near }\;\infty_{j}.
\end{equation}
Then, using eq. (\ref{sectf}) we can define
\begin{align*}
m_{j}&=-\frac{1}{2\pi i} \oint_{\mathfrak{a}_{j}}\de\log f_{0}=
-\frac{1}{2\pi i} \oint_{\mathfrak{a}_{j}}\de\log f_{1},\\
n_{j}&=\;\;\,\frac{1}{2\pi i} \oint_{\mathfrak{b}_{j}}\de\log f_{0}=\;\;\,
\frac{1}{2\pi i} \oint_{\mathfrak{b}_{j}}\de\log f_{1}.
\end{align*}
Comparing eqs. (\ref{etazetainf})  and  (\ref{delogf0inf}), we can write:
\begin{equation*}
\gamma_{\infty}=\frac{1}{2}\de\log f_{1}+i\pi\sum_{l=1}^{g}m_{j},
\end{equation*}
where the $\omega_{j}$ are the canonically $\mathfrak{a}$-normalised holomorphic differentials on $S$, and the second addend ensures that the normalisation (\ref{ginfnorm}) is achieved.\\
Integrating $\gamma_{\infty}$ around $\mathfrak{b}$-cycles yields the \enf{Ercolani-Sinha constraints}
\begin{equation}\label{escon}
\oint_{\mathfrak{b}_{j}}\gamma_{\infty} = i\pi n_{j}+i\pi\sum_{l=1}^{g}\tau_{jl}m_{l},
\end{equation}
where $\tau$ is the period matrix of $S$; these constraints are equivalent to the triviality of $L^{2}$ on $S$. This condition implies that the vector $U$ of eq. (\ref{vecU}) (which appears in the Baker-Akhiezer functions (\ref{phitheta})) takes the following form:
\begin{equation}\label{vecUes}
U=\frac{1}{2}\mathbf{n}+\frac{1}{2} \tau \mathbf{m}
\end{equation}
or, in other words, $U$ is a half-period.\\
Note that Ercolani and Sinha in \cite{es} take $\mathbf{n}$ to be zero: Braden and Enolskii in \cite{BE06} show that this need not be the case. 

\item[iii] \htre: $H^{0}(S,L^{s}(k-2))=0$. This condition can be restated in terms of the theta divisor as  \begin{equation}\label{h3es}
\qquad L^{z} (k-2) \in \Jac ^{g_{S}-1} \backslash \Theta \qquad \mathrm{for }\; z \in(-1, 1).
\end{equation}
This constraint must be checked using explicit knowledge of the $\Theta$-divisor.
\end{itemize}

Summarising, we have the following theorem:
\begin{theorem}[\enf{Ercolani-Sinha Constraints \cite{es}}]\label{escthm} The following are equivalent:
\begin{enumerate} 
\item $L\sp2$ is trivial on $\mathcal{C}$.
\item 
\begin{minipage}{350pt}\;\vspace{-4mm}
\begin{equation}\label{EScond}
2U\in \Lambda\Longleftrightarrow
U=\frac{1}{2\pi\imath}\left(\oint_{\mathfrak{b}_1}\gamma_{\infty},
\ldots,\oint_{\mathfrak{b}_g}\gamma_{\infty}\right)\sp{T}= \frac12
\boldsymbol{n}+\frac12\tau\boldsymbol{m} .
\end{equation}
\end{minipage}
\end{enumerate}
\end{theorem}
\vspace{5mm}
We mention here another equivalent characterisation given by Houghton, Manton and Ram\~ao, namely: 
\begin{theorem}[\enf{Houghton, Manton and Rom\~ao \cite{hmr}}]\label{hmrthm} The following condition is equivalent to the conditions of Theorem \ref{escthm}: there exists a 1-cycle
$\mathfrak{c}=\boldsymbol{n}\cdot{\mathfrak{a}}+
\boldsymbol{m}\cdot{\mathfrak{b}}$ such that for every holomorphic
differential $\Omega$,
\begin{equation*}
\oint\limits_{\mathfrak{c}}\Omega=-2\beta_0,\label{HMREScond}
\end{equation*}
\end{theorem}
Moreover, Braden and Enolskii show a further property of this cycle in \cite{BE06}:
\begin{corollary}
The cycle $\mathfrak{c}$ satisfies
$$\tau_*\mathfrak{c}=-\mathfrak{c}.$$
\end{corollary}
\subsection{Braden-Enolskii extensions to the Ercolani-Sinha theory}
Braden and Enolskii in \cite{BE06} propose an extension to the Ercolani-Sinha formulation, also giving simpler expressions for $\Psi$ and $Q_{0}$ (see section 3 of \cite{BE06}).\\

We begin by providing an expression for $\Psi_{j}$ in terms of theta functions,  alternative to eq. (\ref{phitheta}):
\begin{equation}\label{newphitheta}
\Psi _{j }\left( z, P\right) =g_{j }(P)\, \frac{
\theta_{\frac{\boldsymbol{m}}{2},\frac{\boldsymbol{n}}{2}} \left(
\abel (P)-\abel(\infty_{j
})+z\,U-\tvrc\right)
\theta_{\frac{\boldsymbol{m}}{2},\frac{\boldsymbol{n}}{2}}
\left(-\tvrc \right) } {
\theta_{\frac{\boldsymbol{m}}{2},\frac{\boldsymbol{n}}{2}}
\left(\abel (P)-\abel(\infty_{j })
-\tvrc\right)
\theta_{\frac{\boldsymbol{m}}{2},\frac{\boldsymbol{n}}{2}} \left(
z\,U-\tvrc\right) }\,
e^{z\,\int\limits_{P_0}^{P}\gamma_{\infty}-z\,\nu_j}
\end{equation}
Here
$\theta_{\frac{\boldsymbol{m}}{2},\frac{\boldsymbol{n}}{2}}$ are
theta functions with characteristics, $\abel$ is the Abel map,
$z\in(-1,1)$, and $P\in S$. The vector $\tvrc$ %
is defined by
\begin{equation}\label{bevrc}{\tvrc}=
\vrc+\abel\left((n-2)
\sum_{k=1}\sp{n}\infty_k\right),
\end{equation}
where $\vrc$ is, as above, the vector of Riemann constants.\\
In \cite{BE06} it is established that
\begin{lemma}\label{beh3thm} The following equivalence holds:
\begin{equation}\label{beh3}
H^{0}(S,L^{s}(k-2))\ne0\Longleftrightarrow
\theta(sU-\tvrc)=0
\end{equation} 
for $s\in(0,2)$, with $\tvrc$ as in \eqref{bevrc}.
\end{lemma}
Thus the problem is to determine when the (real) line
$sU-\tvrc$ intersects the theta divisor $\Theta$. \\

We remark that $\tvrc$ satisfies the following properties:
\begin{enumerate}\item $\tvrc$ is independent of the choice
of base point of the Abel map; \item $
\theta(\tvrc)=0$;
\item $2\tvrc \in \Lambda$; \item for $n\ge3$ we
have $ \tvrc\in \Theta_{\rm singular}$.
\end{enumerate}
The point $\tvrc$ is the distinguished point
Hitchin uses to identify degree $g-1$ line bundles with
$\Jac(\mathcal{C})$. The proof of these properties together with
the following lemma further constraining the Ercolani-Sinha vector
may be found in \cite{BE06}:
\begin{lemma}\label{ueventheta}$U\pm
\tvrc$ is a non-singular even theta
characteristic.
\end{lemma}\vspace{3mm}

Moreover, Braden and Enolskii give another expression for  $Q_0(z)$:
\begin{theorem}[\enf{Braden and Enolskii \cite{BE06}}] The matrix $Q_0(z)$ (which has poles of first order at
$z=\pm1$) may be written
\begin{equation}\label{ourq0}
Q_{0}(z)_{jl}  = \epsilon_{jl}\,\frac{\rho_{j}-
\rho_{l}}{\mathcal{E}(\infty_j,\infty_l)}\,e\sp{i\pi\boldsymbol{\tilde
q}\cdot(\abel(\infty_l)-\abel(\infty_j))}\,
    \,\frac{\theta(\abel(\infty_{l}) -\abel(
    \infty_{j}) + [z+1]U - \tvrc)}{\theta(
    [z+1]U
    - \tvrc)}\,e^{z(\nu_{l} - \nu_{j})}.
\end{equation}
Here $E(P,Q)=\mathcal{E}(P,Q)/\sqrt{dx(P)dx(Q)}$ is the
Schottky-Klein prime form, $U -
\tvrc=\frac12\boldsymbol{\tilde
p}+\frac12\tau\boldsymbol{\tilde q}$ ($\boldsymbol{\tilde p}$,
$\boldsymbol{\tilde q}\in\mathbb{Z}\sp{g}$) is a non-singular even
theta characteristic, and $\epsilon_{jl}=\epsilon_{lj}=\pm1$ is
determined (for $j<l$)
 by
$\epsilon_{jl}=\epsilon_{jj+1}\epsilon_{j+1j+2}\dots\epsilon_{l-1l}$.
The $n-1$ signs $\epsilon_{jj+1}=\pm1$ are arbitrary.
\end{theorem}

For the proof of this theorem, we refer again to the paper \cite{BE06}.

\section{A certain symmetric charge 3 monopole}\label{sec:be}
We explore here in some detail a class of charge three monopoles with a certain symmetry, introduced  by Braden and Enolskii in \cite{BE06}: this constitutes a particularly important example, because it is one of the few cases (and the only one of charge greater than 2) where the Ercolani-Sinha method has been fully implemented. Moreover, it is particularly relevant for this thesis as we aim to generalise this to a slightly less symmetric case, the cyclic 3-monopole; the work we present in the later chapters is, in fact, partly based on this.

\subsection{The spectral curve, a homology basis and the Riemann period matrix.}\label{be1}
Braden and Enolskii in \cite{BE06} consider the following spectral curve of a charge 3 monopole 
(cf. eq. (\ref{d3curve}) )
\begin{equation}\label{BEcurv}
\eta^{3}+\beta \zeta^{6}+ \gamma \zeta^{3} - \beta =0. 
 \end{equation}
The corresponding Riemann surface, which we denote\footnote{This choice of notation is clarified in sections \ref{spcurve}, \ref{alpha0}.} by $\Xzero$, has genus 4; viewing it as a branched three sheeted cover of the Riemann sphere $\puno$, $\Xzero$ has 6 branchpoints, $\lambda_{i},\; i=1\ldots 6$:
\begin{align}\label{lambda}%
\lambda_{1}&=\frac{1}{6\beta}\left( -3\gamma + \frac{1}{3}\Delta^{1/2} \right), 
&\qquad \lambda_{4}&=\frac{1}{6\beta}\left( -3\gamma - \frac{1}{3}\Delta^{1/2} \right),\nonumber\\
\lambda_{2}&=\rho \lambda_{1}, \; \lambda_{3}=\rho^{2} \lambda_{1},
&\lambda_{5}&=\rho \lambda_{4},\; \lambda_{6}=\rho^{2} \lambda_{4},\qquad
\end{align}
where 
$\Delta= 27(12\beta^2+ \gamma^2) $. Hence the curve can be parametrised in the following way, which we use later 
\begin{equation}
w^3=\prod_{i=1}^6(z-\lambda_i), \qquad \mathrm{where}\quad w=-\beta\sp{-\frac{1}{3}}\eta,\;z=\zeta. \label{curvegena}
\end{equation}
We  also mention here the basis of the holomorphic differentials used in \cite{BE06}:
\begin{align*}
\mathbf{v}_{1}=&\frac{\de z}{w}, & 
\mathbf{v}_{2}=&\frac{\de z}{w^{2}}, &
\mathbf{v}_{3}=&\frac{z\;\de z}{ w^{2}},&
\mathbf{v}_{4}=&\frac{z^{2}\de z}{ w^{2}}.& 
\end{align*}
This curve (\ref{BEcurv}) admits the following automorphisms:
\begin{align}
\shift : \;(z,w)& \; \to \; (z,\; \rho w),   \label{shift} \\
\rot :\; (z,w)&\; \to \; (\rho z,\; w), \label{rot0} \\
 \tau:\; (z,w)&\rightarrow \left(-\dfrac{1}{\bar{z}},-\dfrac{\bar{w}}{\bar{z}^{2}}\right),\label{realinv0} \\
 \phi:\; (z,w)&\rightarrow \left(-\dfrac{1}{{z}},-\dfrac{{w}}{{z}^{2}}\right),\label{inv0} 
\end{align}
where $\rho=\exp(2 i\pi/3)$;  $\shift$ corresponds to shifting a point up  one sheet, $\rot$ rotates by $2\pi/3$ without shifting sheet, $\tau$ is the real involution of eq. (\ref{realstr}), and $\phi$ is an inversion.\\

The first task of \cite{BE06} is providing a canonical basis for $H_{1}(\Xzero,\mathbb{Z})$: this is given here in Figure \ref{be06fig1}. Also, we give an expansion for these cycles in terms of ``basic arcs'' as follows. Denote by $\gamma_k(i,j)$ the arc going from branchpoint $\lambda_{i}$ to branchpoint $\lambda_{j}$ on sheet $k$:
\begin{equation*}
\gamma_k(i,j)=\mathrm{arc}_k ( \lambda_{i},\lambda_{j}  ),\quad i\neq j=1,\ldots,6\quad \text{on $k$-th sheet} 
\end{equation*}
Then:
\begin{align}\begin{split}
\abe_1&=\gamma_1(1,2)+\gamma_{2}(2,1),\quad 
\bbe_1=\gamma_{1}(2,1)+\gamma_3(1,2),\\
\abe_2&=\gamma_{2}(3,4)+\gamma_3(4,3),\quad 
\bbe_2=\gamma_2(4,3)+\gamma_1(3,4),\\
\abe_3&=\gamma_{3}(5,6)+\gamma_1(6,5),\quad 
\bbe_3=\gamma_3(6,5)+\gamma_2(5,6), \label{arcexpzeroBE} \\
\abe_0&=\gamma_{3}(1,2)+\gamma_1(2,6)+\gamma_{3}(6,5)+\gamma_2(5,1),\\
\bbe_0&=\gamma_{3}(1,2)+\gamma_{1}(2,5)+\gamma_2(5,6)+\gamma_3(6,3)+\gamma_1(3,4)+\gamma_2(4,1)
\end{split}
\end{align}

  We remark that here the monodromy (calculated with respect to infinity) is $[1,2,3]$ (more details on monodromy and how to calculate it are given in section \ref{remarks}).\\ 
This basis has the following property under the symmetries $\shift$ and $\rot$:
\begin{align}\label{BEbsymm}
\rot_{*}^{k}(\abe_{i})&=\abe_{i+k},&\rot_{*}^{k}(\bbe_{i})&=\bbe_{i+k},;\\
\quad \shift_{*}(\bbe_{0})&=\abe_{0}, & \quad \shift_{*}(\bbe_{k})&=\abe_{k},  n=1,2,3 \nonumber
\end{align}
where the last map $ \shift_{*}$ has the effect of just shifting up (in our conventions) one sheet.\\

\begin{figure}
\begin{minipage}{100pt}%
\begin{lpic}[draft,clean]{BE06_123(7cm,)}
   \lbl[t]{135,70;$\color{red}{\abe_1}$}
   \lbl[t]{128,90;${\cblu{\bbe_1}}$}
   \lbl[t]{45,90;$\cred{\abe_2}$}
   \lbl[t]{25,70;${\cblu{\bbe_2}}$}
   \lbl[t]{60,20;$\color{red}{\abe_3}$}
   \lbl[t]{100,28;${\cblu{\bbe_3}}$}
   \end{lpic}
\end{minipage}\hspace*{40mm}
\begin{minipage}{100pt}\hspace*{15mm}
   \begin{lpic}[draft,clean]{BE06_0(7cm,)}
   \lbl[t]{85,90;$\color{red}{\abe_0}$}
   \lbl[t]{72,70;${\cblu{\bbe_0}}$}
   \end{lpic}
\end{minipage}
\caption{Symmetric homology basis of \cite{BE06}}\label{be06fig1}
\end{figure}
The choice of such a symmetric homology basis allows us to relate the matrices of $\abe$ and $\bbe$-periods, and hence  simplifies greatly the Riemann period matrix.\\
Introducing vectors
\begin{align*}
\boldsymbol{x}&=(x_1,x_2,x_3,x_4)^T
=\left(
\oint_{\abe_1}\mathbf{v}_{1},\ldots, \oint_{\abe_4}\mathbf{v}_{1} \right)^T ,
\\
\boldsymbol{b}&=(b_1,b_2,b_3,b_4)^T
=\left(
\oint_{\abe_1}\mathbf{v}_{2},\ldots,\oint_{\abe_4}\mathbf{v}_{2} \right)^T ,\\
\boldsymbol{c}&=(c_1,c_2,c_3,c_4)^T
=\left(
\oint_{\abe_1}\mathbf{v}_{3},\ldots,\oint_{\abe_4}\mathbf{v}_{3} \right)^T ,\\
\boldsymbol{d}&=(d_1,d_2,d_3,d_4)^T =\left(
\oint_{\abe_1}\mathbf{v}_{4},\ldots,\oint_{\abe_4} \mathbf{v}_{4}
\right)^T ,
\end{align*}
one then finds the following result:
\begin{proposition}[\enf{Wellstein \cite{well99}; Matsumoto \cite{matsum}; Braden-Enolskii, \cite{BE06}}]
\label{matsumoto1}For the curve $\Xzero$ of equation (\ref{BEcurv}), one has
\begin{align}\begin{split}
\mathcal{A}&=\left(
  \oint\limits_{\abe_k} \mathbf{v}_i\right)_{i,k=1,\ldots,4}
=(\boldsymbol{x},\boldsymbol{b},\boldsymbol{c},\boldsymbol{d})
\\
 \mathcal{B}&=\left(
  \oint\limits_{\bbe_k} \mathbf{v}_i\right)_{i,k=1,\ldots,4}
= (\rho H\boldsymbol{x},\rho^2 H\boldsymbol{b},\rho^2
H\boldsymbol{c}, \rho^2
H\boldsymbol{d})=H\mathcal{A}\Lambda,\end{split}\label{calba}
\end{align}
where $H=\mathrm{diag}(1,1,1,-1)$ and
$\Lambda=\mathrm{diag}(\rho,\rho\sp2,\rho\sp2,\rho\sp2)$.
Then the Riemann period matrix is of the form
\begin{align}
\tau_{{b}}&=\mathcal{A}\mathcal{B}^{-1}=\rho\left(
H-(1-\rho)\frac{\boldsymbol{x}\boldsymbol{x}^T} {\boldsymbol{x}^T
H\boldsymbol{x}}   \right),\label{taumat}
\end{align}
and $\tau_{\mathfrak{b}}$ is
positive definite if and only if
$ \bar{\boldsymbol{x}}^T H \boldsymbol{x} <0$.
\end{proposition}
\subsection{Solving the Ercolani-Sinha constraints}\label{sectesbe06}
For convenience we rewrite the curve of eq. (\ref{BEcurv}) as follows:
\begin{align}\label{BEcurv1}
w^{3}=z^{6}+bz^{3}-1=(z^{3}-a^{3})\left( z^{3}-\frac{1}{a^{3}}\right)
\end{align}
where $b=-\tfrac{\gamma}{\beta}$ and $a^{-3}=\tfrac{1}{2}(b+\sqrt{b^{2}+4})$.\\
The Ercolani-Sinha constraints here read:
\begin{equation}
\boldsymbol{n}^T \mathcal{A}+\boldsymbol{m}^T\mathcal{B}
=\nu(1,0,0,0).\label{esconda}
 \end{equation}
where\footnote{
To see this observe that (\ref{HMREScond}) requires that
$-2\delta_{1k} = \oint_{\boldsymbol{n}\cdot\mathfrak{a}+ \boldsymbol{m}\cdot\mathfrak{b}} \Omega^{(k)}$ 
for the
differentials 
$ \Omega^{(1)} = \dfrac{\eta^{n-2} \de \zeta} {\frac{\partial P}{\partial \eta}} = \dfrac{\de \zeta}{n \eta}$,\\
$\Omega^{(2)} = \dfrac{\eta^{n-3} \de \zeta} {\frac{\partial P}
{\partial \eta}} ,\dots $.
In the parametrisation (\ref{curvegena}) we are using we have that
    $$x_{i} = {\oint_{\mathfrak{a}_i}}\dfrac{\de z}{w} = {\oint_{\mathfrak{a}_i}} \dfrac{\de\zeta}{-\beta^{-\frac{1}{3}}\eta} = -3\beta^{\frac{1}{3}}{\oint_{\mathfrak{a}_i}} \Omega^{(1)}.$$
Imposing 
    $-2 = \oint_{\boldsymbol{n}\cdot\mathfrak{a}+
\boldsymbol{m}\cdot\mathfrak{b}} \Omega^{(1)}  =
    -\frac{1}{3}\beta^{-\frac{1}{3}}(\boldsymbol{n}.\boldsymbol{ x} + \rho
    \boldsymbol{m}.H.\boldsymbol{x}) $ and so $
    \boldsymbol{n}.\boldsymbol{ x} + \rho
    \boldsymbol{m}.H.\boldsymbol{x}= \nu $,
with the value of $\nu$ stated.
}
 $\nu=6\beta^{\frac{1}{3}}$ (in general $\nu$ depends on normalizations).\\
These constraints may be solved using the following result.
\begin{proposition}\label{beh2}
The Ercolani-Sinha constraints (\ref{esconda}) are satisfied for
the curve (\ref{BEcurv}) if and only if\vspace*{-3mm}
\begin{align}\label{expxx}
    \boldsymbol{x} & = \xi(H\boldsymbol{n} + \rho^{2}\boldsymbol{m}),
\end{align}\vspace*{-3mm}
where \vspace*{-3mm}
\begin{align}
\xi &   = \frac{\nu}{[\boldsymbol{n}.H
    \boldsymbol{n}-\boldsymbol{m}.\boldsymbol{n}+\boldsymbol{m}.H
    \boldsymbol{m}]} = \frac{6
\beta^{\frac{1}{3}}}{[\boldsymbol{n}.H
    \boldsymbol{n}-\boldsymbol{m}.\boldsymbol{n}+\boldsymbol{m}.H
    \boldsymbol{m}]}.\label{expxxx}
\end{align}
\end{proposition}\vspace*{3mm}
The proof, which follows from Proposition \ref{matsumoto1}, can be found in \cite{BE06}.\\
Thus, solving the Ercolani-Sinha constraints amounts to imposing the four constraints (\ref{expxx}) on the periods $x_k$. 
For the curve (\ref{BEcurv}) one can actually calculate these integrals in terms of special functions (Gauss hypergeometric functions) and the constraints in fact reduce to a number theoretic one.\\

We remark that in this case all the integrals reduce to just 8, namely:
\begin{align*}
\int\limits_0^{\tilde{\alpha}}{d}u_i=\mathcal{I}_i,\quad
\int\limits_0^{\tilde{\beta}}{d}u_i=\mathcal{J}_i,\quad
i=1,\ldots,4,
\end{align*}
where $\lambda_{1}=(\tilde{\alpha}, 0), \lambda_{4}=(\tilde{\beta}, 0)$ and $\tilde{\beta}=\frac{1}{\tilde{\alpha}}$, and it is intended that these integrals are computed on the first sheet. This follows from the symmetries of the curve, for details see  \cite{BE06}, p. 50 (also cf. section \ref{altpm}). In particular, using $\mathcal{I}_1$ and $\mathcal{J}_1$, which in terms of hyperelliptic function read
\begin{align}\begin{split}
\mathcal{I}_1(\tilde{\alpha})&=\int\limits_{0}^{\tilde{\alpha}}\frac{{d}z}{w}
=-\frac{2\pi\sqrt{3}\tilde{\alpha}}{9} {_2F_1}\left(\frac13,\frac13;1;-\tilde{\alpha}^6\right),\\
\mathcal{J}_1(\tilde{\alpha})&=\int\limits_{0}^{-\frac{1}{a}}\frac{{d}z}{w} =
\frac{2\pi\sqrt{3}}{9\tilde{\alpha}} {_2F_1}\left(\frac13,\frac13;1;
-\tilde{\alpha}^{-6}\right),\end{split} \label{integralsij}\end{align}
the $x$-periods are
\begin{equation}\label{xper}
\begin{array}{rlrl}
x_{1}&=-(2\mathcal{J}_1+\mathcal{I}_1)\rho -2\mathcal{I}_1
-\mathcal{J}_1, &x_{2}&=(\mathcal{J}_1- \mathcal{I}_1 )\rho+
\mathcal{I}_1+2\mathcal{J}_1,
\\
x_{3}&=(\mathcal{J}_1+2\mathcal{I}_1)\rho-\mathcal{J}_1+\mathcal{I}_1,
&x_{4}&=3(\mathcal{J}_1-\mathcal{I}_1)\rho+3\mathcal{J}_1,
\end{array}
\end{equation}
where we remark that $x_{2}=\rho x_{1}$ and $x_{3}=\rho^{2}x_{1}$ by symmetry.\\

Hence, using  (\ref{expxx}) and (\ref{xper})  the Ercolani-Sinha constraints can be rewritten as
\begin{equation}\label{esred}
    x_{i} = \xi( \epsilon_{i}n_{i} + \rho^{2}m_{i}) = (\tilde{\alpha}_{i}\mathcal{I}_1 +
    \tilde{\beta}_{i} \mathcal{J}_1) + (\gamma_{i}\mathcal{I}_1 + \delta_{i}
    \mathcal{J}_1)\rho, 
\end{equation}
where $\epsilon_{i}=1$ for $i-1,2,3$ and $\epsilon_{4}=-1$.
Now this expression can be solved for $n_{i}, m_{i}$, as in the following
\begin{proposition}\label{propbe09}
For each pair of relatively prime integers $m,n$ such that
$$(m + n)(m -2n)<0$$
we obtain a solution $\boldsymbol{n}=\begin{pmatrix} n& m-n&-m&2n-m \end{pmatrix}$, $\boldsymbol{m}=\begin{pmatrix} m& -n&n-m&-3n \end{pmatrix}$ to the Ercolani-Sinha constraints for the curve (\ref{BEcurv1}) as follows. First we solve the equation
\begin{equation}\label{treq}
\dfrac{2n-m}{m + n}=\frac{{_2F_1}(\frac{1}{3},
\frac{2}{3}; 1,t)}{{_2F_1}(\frac{1}{3}, \frac{2}{3}; 1,1-t)}
\end{equation}
for $t$; we have then
\begin{equation*}
b=\frac{1-2t}{\sqrt{t(1-t)}},\qquad t=
\frac{-b+\sqrt{b^2+4}}{2\sqrt{b^2+4}},
\end{equation*}
and we obtain $\beta$ from
\begin{equation}\label{ESbeta}
\beta^{\frac{1}{3}} = -(n + m )\, \frac{2 \pi}{3
    \sqrt{3}}\ \frac{a}{(1+a\sp6)\sp\frac13}\ {_2F_1}(\frac{1}{3}, \frac{2}{3}; 1, t)
\end{equation}
with $a\sp6=t/(1-t)$.
\end{proposition}

We remark that the constraint  $\gcd(m,n)=1$ ensures that we have indeed a primitive vector in the lattice $\mathbb{Z}^{8}$. We also point out that eq. (\ref{treq}) can be rewritten as follows:
\begin{equation}\label{ar}
\mathcal{R}=\dfrac{2n-m}{m + n}=-\dfrac{\mathcal{I}_{1}}{\mathcal{J}_{1}}
\end{equation}
The quantity $\mathcal{R}$ plays an important role in simplifying the period matrix of a curve whose parameters satisfy eq. (\ref{treq}) (see eq. (\ref{ABsym})).\\

Hence, the problem of finding solutions to the Ercolani-Sinha constraints has been reduced to that of finding solutions to the transcendental equation (\ref{treq}): remarkably, this can be done. One finds in Ramanujan's second notebook several results of generalised modular equations, and various theta functions identity, some of them recently proven (see \cite{BBG95}, \cite{ber98}). We report here some of the solutions to the Ercolani-Sinha constraints found using these methods by Braden and Enolskii, referring to \cite{BE06} for details.
\begin{equation}\label{tableBE}
\begin{array}{|c|c|c|c|c|} \hline
n&m&(2n-m)/(m+n)&t&b\\
\hline 2&1&1&\frac{1}{2}&0\\
\hline
1&0&{2}&\frac{1}{2}+\frac{5\sqrt{3}}{18}&-5\sqrt{2}\\
\hline1&1&\frac{1}{2}&\frac{1}{2}-\frac{5\sqrt{3}}{18}&5\sqrt{2}\\
 \hline 4&-1&3&(63+171\sqrt[3]{2}-18\sqrt[3]{4})/250&
(44+38\sqrt[3]{2}+26\sqrt[3]{4})/3\\
\hline 5&-2&4&\frac{1}{2}+\frac{153\sqrt{3}-99\sqrt{2}}{250}&
9\sqrt{458+187\sqrt{6}}\\
\hline
\end{array}
\end{equation}
Hence, we can see that only a discrete (countable) number of solutions appear, meaning that there is at most a countable number of monopoles with spectral curve of the form (\ref{BEcurv}).
In particular, we point out from the table above that the tetrahedral monopole of eq. (\ref{thetr})  is one of them.

\subsection{Covers and reduction}\label{coversbe06}
We briefly describe how the curve under consideration covers elliptic curves, which provides a noteworthy simplification of the period matrix (as well as a better understanding of some of the results based on Ramanujan's identities, see \cite{BE06}). This is particularly relevant in view of our subsequent study of the cyclic 3-monopole, where again a quotient appears.\\

The results about the covers are summarised in the following
\begin{lemma}\label{coverlem}%
The curve with equation $ y^3=x^6+bx^3-1$, with arbitrary values of the parameter $b$, is a simultaneous covering of the four elliptic curves $\mathcal{E}_{\pm}$, $\mathcal{E}_{1,2}$ as indicated in the diagram, where $\mathcal{C}\sp*$ is an
intermediate genus 2 curve

\vskip 0.5cm
\begin{center}
\unitlength=1mm
\begin{picture}(60,30)(0,0)
\put(9,23){\makebox(0,0){${\Xzero}=(x,y)$}}
\put(6,20){\vector(-4,-1){20}} \put(-8,20){\makebox(0,0){$\pi^*$}}
\put(-18,15){\makebox(0,0){$X_{0}$}}
\put(-26,0){\makebox(0,0){$\mathcal{E}_+=(z_+,w_+)$}}
\put(-20,12){\vector(-1,-1){9}} \put(-16,12){\vector(1,-1){9}}
\put(0,0){\makebox(0,0){$\mathcal{E}_-=(z_-,w_-)$}}
\put(-9,10){\makebox(0,0){$\pi_-$}}
\put(-26,10){\makebox(0,0){$\pi_+$}}
\put(10,19){\vector(1,-1){16}} \put(16,10){\makebox(0,0){$\pi_1$}}
\put(28,0){\makebox(0,0){$\mathcal{E}_1=(z_1,w_1)$}}
\put(12,20){\vector(2,-1){34}} \put(38,10){\makebox(0,0){$\pi_2$}}
\put(55,0){\makebox(0,0){$\mathcal{E}_2=(z_2,w_2)$}}
\end{picture}
\end{center}
\vskip 5pt
The equation for $X_{0}$ is
\begin{equation*}
\nu^{2}=(\mu^{2}+b)^{2}+4
\end{equation*}
The equations of the elliptic curves are
\begin{align}
\mathcal{E}_{\pm}:&\qquad%
w_{\pm}^2=z_{\pm}(1-z_{\pm})
(1-k_{\pm}^2z_{\pm})   ,\label{curvepm}\\
\mathcal{E}_1:&\qquad
z_1^3+{w_1}^3+3z_1+b=0,\label{curve1}\\
\mathcal{E}_2:&\qquad {w_2}^3+{z_2}^2+bz_2-1 =0 ,
\label{curve2}
\end{align}
where the Jacobi moduli $k_{\pm}$ are given by
\begin{equation*}
k_{\pm}^2=-\frac{\rho(\rho M \pm 1)(\rho M \mp 1)^3  } {(M\pm
1)(M\mp 1 )^3}
\end{equation*}
with
$M=\tfrac{K}{L},\; K=(2\imath -b)^{\frac13},
L=(b^2+4)^{\frac16}$.\\
\end{lemma}

When a curve covers one of lower genus, the period matrix admits a reduction, \ie can be expressed in terms of a lower genus period matrix, and similarly, the associated theta functions can be expressed in terms of lower dimensional theta functions.\\
As the curve under consideration has many covers, it admits many reductions too;  the case where a true simplification occurs is when the Ercolani-Sinha vector reduces as well: this is the object of the rest of this section.\\

Before doing this, we give an expression for the period matrix of a curve of the form (\ref{BEcurv}), satisfying Ercolani-Sinha constraints. 
Combining equations (\ref{calba}), (\ref{taumat}) and (\ref{xper}), one obtains:

\begin{align}\begin{split}
\mathcal{A}&= \left(\begin{array}{cccc}
-1-2\rho-(2+\rho)\mathcal{R}&1+2\rho&1+2\rho+(1-\rho)\mathcal{R}&-1+\rho\\
2+\rho+(1-\rho)\mathcal{R}&-2-\rho&1-\rho-(2+\rho)\mathcal{R}&-1+\rho\\
-1+\rho+(1+2\rho)\mathcal{R}&1-\rho&-2-\rho+(1+2\rho)\mathcal{R}&-1+\rho\\
3+3\rho-3\rho\mathcal{R}&0&-3\rho-3(1+\rho)\mathcal{R}&0
\end{array}\right)\left(\begin{array}{cccc} \mathcal{J}_1&&&\\
                           &\mathcal{I}_2\\
                           &&\mathcal{J}_3\\
                           &&&\mathcal{I}_2
\end{array}\right),\\
\mathcal{B}&=\left(\begin{array}{cccc}
2+\rho+(1-\rho)\mathcal{R}&1-\rho&1-\rho-(2+\rho)\mathcal{R}&2+\rho\\
-1+\rho+(1+2\rho)\mathcal{R}&1+2\rho&-2-\rho+(1+2\rho)\mathcal{R}&2+\rho\\
-1-2\rho-(2+\rho)\mathcal{R}&-2-\rho&1+2\rho+(1-\rho)\mathcal{R}&2+\rho\\
3-3(1+\rho)\mathcal{R}&0&3-3\rho\mathcal{R}&0
\end{array}\right)\left(\begin{array}{cccc} \mathcal{J}_1&&&\\
                           &\mathcal{I}_2\\
                           &&\mathcal{J}_3\\
                           &&&\mathcal{I}_2
\end{array}\right), \end{split}\label{ABsym}
\end{align}
where $\mathcal{R}$ is a defined in eq. (\ref{ar}).\\
The Riemann matrix $\Pi= (\mathcal{A},\mathcal{B})^{T}$ admits a reduction with respect to each one of its columns: we consider here a reduction with respect to the first column.\\
Using eq. (\ref{expxx}) it follows that
\begin{align}
\Pi\lambda=\left(%
\begin{array}{c}
  \mathcal{A} \\
  \mathcal{B} 
\end{array}%
\right) \,\left(%
\begin{array}{c}
  1 \\
  0 \\
  0 \\
  0 \\
\end{array}%
\right)=\left(\begin{array}{c}\boldsymbol{x}\\\boldsymbol{y}
\end{array}\right)=
\left(\begin{array}{c}\oint_{\mathfrak{a}_i}{d} u_1
\\ \oint_{\mathfrak{b}_i}{d} u_1
\end{array}\right)=
\left(\begin{array}{c}\xi(H\boldsymbol{n}+\rho^2\boldsymbol{m})\\
\xi(\rho \boldsymbol{n}+H\boldsymbol{m})
\end{array}\right)=\xi\,M\left(\begin{array}{c}1\\\rho
\end{array}\right),\label{escond2}
\end{align}
where $M$ is the $2g\times2$ integral matrix
\begin{equation*} 
M^T=\left(\begin{array}{cccccccc}n_1-m_1&n_2-m_2&n_3-m_3&-n_4-m_4
&m_1&m_2&m_3&-m_4      \\ -m_1&-m_2&-m_3&-m_4&n_1&n_2&n_3&n_4
\end{array}\right).
\end{equation*}
For every two Ercolani-Sinha vectors $\boldsymbol{n}$,
$\boldsymbol{m}$ one has
\begin{equation*}
M^TJM=d\left(\begin{array}{cc}0&1\\-1&0\end{array}\right),\quad
d=\boldsymbol{n}.H\boldsymbol{n}- \boldsymbol{m}. \boldsymbol{n}
+\boldsymbol{m}.H\boldsymbol{m}= \sum_{j=1}^4(\varepsilon_j
n_j^2-n_jm_j+\varepsilon_j m_j^2).
\end{equation*}

\begin{theorem}\enf{(Braden and Enolskii \cite{BE06})}\label{symwp} For the
symmetric monopole one can reduce by the first column using the
vector (\ref{escond2}) whose elements are related as in Proposition \ref{propbe09}, with $\gcd(n_1,m_1)=1$. Then
$$d=2(n_1+m_1)(m_1-2n_1)$$ and for $d\ne0$ there exists an element $\sigma$ of the
symplectic group $\mathrm{Sp}_{2g}(\mathbb{Z})$ such that
\begin{align*}
\tau'_{\mathfrak{b}}=\sigma\circ\tau_{\mathfrak{b}}=
\left(\begin{array}{ccccc}
(\rho+2)/d&{\alpha}/{d}&0&\ldots&0\\
{\alpha}/{d}&&&&\\
0&&{\tau}^{\#}&&\\
\vdots&&&&\\ 0&&&&
\end{array}\right).
\end{align*}
Letting \, $p\, m_1+q\, n_1=1$ then
\begin{equation*}
\alpha=\gcd(m_1+4n_1-q\,[m_1-2n_1],n_1-2m_1-p\,[m_1-2n_1]).
\end{equation*}
When $\alpha=1$ a further symplectic transformation allows the
simplification $\tau'_{11}=\rho/d$.

Under $\sigma$ the Ercolani-Sinha vector becomes
\begin{equation*}
  \sigma\circ{\boldsymbol U}=  \sigma\circ({\boldsymbol m}\sp{T}+{\boldsymbol
    n}\sp{T}\tau_{\mathfrak{b}})=(1/2,0,0,0).
\end{equation*}
\end{theorem}
The proof of Theorem \ref{symwp} is constructive: in particular there is an algorithm by Martens \cite{ma92,ma92a} for constructing $\sigma$ which has been implemented in  $Maple$ by Braden and Enolskii \cite{BE06}.\\

The theory of Weiestrass reduction implies that the theta functions reduce correspondingly, taking the form\footnote{When $\gcd(\alpha,d)\ne1$ a
smaller multiple than $d_1=d$ would suffice here with
correspondingly fewer terms in the sums $0\le m\le d_1-1$.}:
\begin{align*}\theta((z,w);\tau'_\mathfrak{b})&=\sum_{m=0}\sp{d-1}
\theta(z+\frac{m\alpha}{d};\frac{\rho+2}{d})\,\theta\left[%
\begin{array}{ccc}
  \frac{m}{d} & 0 & 0 \\
  0 & 0 & 0 \\
\end{array}%
\right](Dw;D{\tau}^{\#}D)\\
&=\sum_{m=0}\sp{d-1} \theta \left[%
\begin{array}{c}\frac{m}{d}\\0\end{array}%
\right](dz;d(\rho+2))\,\theta((w_1+\frac{m\alpha}{d},w_2,w_3);{\tau}^{\#}),
\end{align*}
where $D=\mathrm{diag}(d,1,1)$.\\

Using the above expression for the theta function in eq. (\ref{ourq0}), we conclude that one can express $Q_{0}$ in terms of genus 1 theta functions, or
\begin{proposition}[\enf{Braden and Enolskii \cite{BE06}}]\label{symel} For symmetric monopoles the theta function $z$-dependence of
$Q_0(z)$ is expressible in terms of elliptic functions.
\end{proposition}

 This means in particular that the last constraint \enf{H3} in Ercolani-Sinha formulation \eqref{h3es} can be reduced to the problem of finding zeros of elliptic functions. Using this, Braden and Enolskii manage to limit the countable number of possible solutions of table \eqref{tableBE}, establishing a stronger result:
 \begin{theorem}[\enf{Braden and Enolskii \cite{BE09}}]\label{mainthmBE}
The only curves of the family \eqref{BEcurv} that yield BPS monopoles correspond to tetrahedrally symmetric monopoles. These have $b=\pm5\sqrt{2}$ and $(m,n)=(1,1)$ or $(0,1)$.
\end{theorem}
%
%

%
%
%
%
%
%
%
%
%
%
%
%
\chapter{Algebraic-geometric aspects of integrable systems: some basics results}\label{intsys}

\introcap{Integrable systems are a vast area of research, and we do not attempt to give even just a short summary here. We only  recall some elements on the Lax formulation, which leads to a description of such systems in terms of Riemann surfaces; this is the point of view we take in this work of thesis. We illustrate how to associate a \enf{spectral curve} to a Lax pair with a parameter, and how this allows to realise the dynamical evolution of the system as a linear flow on the Jacobian of this curve.}
\ding{125}%
{{Integrable systems, what are they? It's not easy to answer precisely. The question can occupy a whole book,  or be dismissed as Louis Armstrong is reputed to have done once when asked what jazz was - `If you gotta ask, you'll never know!'}}\ding{126}
\vspace{-3mm}\begin{flushright}\small{N. Hitchin \cite{IShitch}} \end{flushright}
\vspace{5mm}
The study of integrable systems is a very old subject, which has been of interest to both mathematician and physicists for centuries. Despite this, or perhaps because of this, it is difficult to give a general definition that encompasses all the different ways in which integrability manifests itself: the main feature of such systems is considered to be, depending on the situation, the existence of ``many'' conserved quantities, or the ability to give explicit solutions, or the appearance of algebraic geometry, or the presence of certain topological or algebraic structures, etc. . \\
Hence, much work has been done in the last century on the subject on integrable systems, in both finite and infinite dimensions, using and indeed developing many sophisticated mathematical techniques, from PDEs to symplectic geometry, from differential topology to representation theory. Here we are mainly concerned with algebraic-geometric methods in integrable systems, involving, for instance, algebraic curves, theta functions, Abelian varieties etc. .\\

We already see the appearance of such objects  in several classical dynamical systems: it was known since the nineteenth century that solutions for many classical tops, or already the simple pendulum, were expressible in terms of elliptic functions. {A posteriori}, we see that there is an algebraic curve in the background. Indeed, modern techniques make this explicit: we can build this curve from the beginning, use it to obtain constants of motion, and eventually to represent the solutions. This is achieved through the \enf{Lax formulation}, which is the main subject of this chapter. \\

Several classical integrable systems admit a Lax description: for instance, in finite dimensions, all the completely integrable Hamiltonian systems, which constitute a large and important class; 
moreover, several nonlinear equations, \eg the Korteveg-de Vries equation, the Kadomtsev-Petviashvili equation, the nonlinear Schr\"odinger equation, etc. . In particular, the theory of the so called finite gap solutions of KdV type equations led to the more general results for solutions to Lax equations in terms of Baker-Akhiezer functions. The main result that we discuss in this chapter is in fact how to express solutions of a Lax equation in terms of Baker-Akhiezer functions, using a result due to Krichever \cite{kr77}. We remark, though, that the introduction of algebraic-geometric methods in integrable systems has many contributors: we refer to \cite{BBEIM}  and references therein for more details, and to \cite{introis} for an introduction.\\ 
This chapter is based on \cite{introis,hurt}.

\section{The Lax formulation}
Within the realm of integrable systems, what we are concerned with in this thesis are systems admitting a \enf{Lax formulation}:
\begin{definition}
A dynamical system admits a \enf{Lax formulation} if its time evolution can be described by an equation of the form
\begin{equation}\label{laxdef}
\frac{\de}{\de t} A=\dot{A}= [A,B],
\end{equation}
where $A$ and $B$ are $N\times N$ matrices, and $[A,B]=AB-BA$. The pair $(A,B)$ is called a \enf{Lax pair}.
\end{definition}
It can be seen that the entries of $B$ are functions of those of $A$, hence the equation of the motion \eqref{laxdef} is \enf{nonlinear}.\\

In general, the Lax formulation for a given system is not unique: from the form of eq. \eqref{laxdef} it is immediate to see that if $A$, $B$ are a Lax pair for a certain system, then so are $A_{1}=MAM^{-1} $ and $B_{1}=MBM^{-1}  +\dot{M}M$, for any $N\times N$ matrix $M$.\\

An important feature of such systems is the immediate appearance of conserved quantities. Indeed, the solution of eq. \eqref{laxdef} is of the form
\begin{equation*}
A(t)=g(t)A(0)g(t)^{-1}\qquad \mathrm{with}\;g(t)\; \mathrm{s.t.} \quad  B=\frac{\de g}{\de t} g^{-1}
\end{equation*}

It follows that all the quantities of the form
\begin{equation}\label{tracescc}
I^{(k)}=\mathrm{Tr} A^{k}
\end{equation}
are constants of motion. The presence of ``many'' constants of motion is  in general a fundamental feature of integrable systems; note, though, that from eq. \eqref{tracescc} nothing can be inferred about their independence.\\

Since the components of the characteristic polynomial for $L$ are functions of these traces, the whole spectrum of $L$ is conserved: the time evolution of a system in Lax form is \enf{isospectral}.
\section{Lax pairs with a parameter}
In many cases, the Lax pair depends on a complex parameter $\lambda$:
\begin{equation}\label{laxpar}
\dot{A}_{\lambda}= [\Lpar,\Mpar].
\end{equation}
 Lax pairs of this form arise quite naturally for many systems, and the study of the analytical properties of the Lax equation depending on a parameter gives considerable insight into the understanding of the dynamical evolution of the system. A fundamental object  for this aim is the {spectral curve}.
 \begin{definition}%
 Given a Lax pair $\Lpar$, $\Mpar$ depending on a parameter , the characteristic equation for $\Lpar$
 \begin{equation}
 P(\mu,\lambda)=\det(\Lpar- \mu\mathbb{I})=0
 \end{equation}
 defines an algebraic curve $\spcurve$, called the \enf{spectral curve}.
 \end{definition}
 In the following we assume it is irreducible, and that it can be completed to obtain a \enf{Riemann surface}.\\
 
 Notice that, since the coefficients of the characteristic polynomials are functions of the $I^{(k)}$ of eq.\eqref{tracescc}, the spectral curve is \enf{preserved by the flow}.\\
 To understand equations of the type \eqref{laxpar}, one has to study the spectral curve and its structure. In particular, the dynamical information about the system is encoded in a line bundle on $\spcurve$, which describes a linear flow on the Jacobian $\Jac(\spcurve)$ of $\spcurve$. Under these circumstances, together with a restriction on the form of $\Mpar$, the solution can be written in terms of theta functions. The rest of this chapter explains this in some more detail.\\
 
Firstly, we recall that the genus of the spectral curve can be calculated to be
\begin{equation*}
g=\frac{N(N-1)}{2}\sum_{k}n_{k}-N+1,
\end{equation*}
where $n_{k}$ is the order of $\Lpar$ at a pole $\lambda_{k}$, and the sum runs over all poles.\\

 The first important object on $\spcurve$ is the eigenvector bundle, a line bundle on $\spcurve$ built as follows. For every values of the pair $(\mu,\lambda)=P\in\spcurve$ we have an eigenvector $v(P,t)\in V$, hence an eigenspace: assuming the eigenvalues all have multiplicity $1$, this eigenspace has dimension $1$ at every point on $\spcurve$; this defines the bundle of eigenvectors $L^{*}_{t}$. \\
 Introducing the eigenvector map
 \begin{equation*}
 f_{t}: \spcurve \to \mathbb{P} V; \quad P\to \mathbb{C} v(P,t)\subset V,
 \end{equation*}
 we define the \enf{eigenvector bundle}\footnote{Note that this is the dual of the bundle of eigenvectors.} $L_{t}$ as 
 \begin{equation*}
 L_{t}=f^{*}_{t}(\mathcal{O}_{\mathbb{P} V}(-1)) \in \pic^{d}(\spcurve),
 \end{equation*}
 where %
  $\pic^{d}(\spcurve),$ is the group of line bundles of degree d. This degree $d$ of  $L_{t}$ is  computed to be ( \eg in \cite{introis})
  \begin{equation*} d= g+N-1. \end{equation*}

Note that the eigenvector bundle does depend on $t$, since, although the eigenvalues are constant with time, the corresponding eigenvectors are not. Recalling that  $ \pic^{d}(\spcurve)\simeq \Jac(\spcurve)$, we have that the time evolution of the dynamical system in Lax form \eqref{laxpar} can be translated into time evolution on the Jacobian of its spectral curve:
 \begin{equation}\label{linfl}
 \phi: \mathbb{R} \to  \Jac(\spcurve),\quad t \to L_{t}.
 \end{equation}
 It is possible to show the following result (see \eg \cite{linflow}).
 \begin{theorem}\label{linthm}
 Under certain conditions on $\Mpar$, the flow \eqref{linfl} is linear on $\Jac(\spcurve)$.
 \end{theorem}
 
 We recall that the Jacobian of a Riemann surface is a complex torus. Hence, the above  result means that, for an integrable system in the form \eqref{laxpar}, it is possible to realise the linear flow on the torus $\Jac(\spcurve)$. Note the relation with completely integrable Hamiltonian systems here: for  such systems, the Arnold-Liouville theorem states that the phase space admits a (local) fibration by tori, built such that the dynamical flow is linear on them.%
 The interesting thing is the appearance of a torus in both descriptions: the advantage is that the Jacobian has a given canonical linear structure, while the Arnold-Liouville tori have a linear structure defined by the flow itself.\\
 
 The conditions on $\Mpar$ in Theorem \ref{linthm},  for the flow defined by \eqref{laxpar} to be linear, are that
 \begin{equation}\label{condmlin}
 \Mpar=(P(\Lpar, \lambda^{-1}))_{+}
 \end{equation}
 where $P(z, w)$ is an arbitrary complex polynomial of two variables,  and $(\;,\;)_{+}$ denotes taking the regular part of $P$, with respect to its Laurent series expansion in $\lambda$.\\
 
 Note that here, given a matrix equation of the form \eqref{laxpar}, we have built a linear flow of line bundles  $L_{t}$. It is possible to do the converse, namely, given a linear flow of line bundles on $\spcurve$,  of degree $g+N-1$ to construct  a solution $\Lpar(t)$ to eq. \eqref{laxpar} explicitly in terms of theta functions on $\Jac(\spcurve)$.  This is done as follows. \\
 
 Take a linear flow $L_{t}$ of line bundles of degree $g+N-1$ on $\spcurve$. Take an ``initial value'' line bundle $L_{0}$, represented by a positive  divisor $\delta$, or  $L_{0}=L(\delta)$. \\
 Denote by $\psi^{1},\ldots,\psi^{N}$ a basis  for $H^{0}(\spcurve, L_{0}) $. We remark that $L_{0}$ has precisely $N$ sections when\footnote{This follows from Riemann-Roch theorem, upon using that $\mathrm{deg}L_{0}=g+N-1$:
 $$ \mathrm{dim}H^{0}(\spcurve,\mathcal{O}( L_{0})) = \mathrm{deg} L_{0}+1-g+\mathrm{dim}H^{1}(\spcurve, \mathcal{O}(L_{0})) = N+\mathrm{dim}H^{1}(\spcurve, \mathcal{O}(L_{0})).$$}
 \begin{equation}\label{nonspecialdiv}
 \mathrm{dim} H^{1}(\spcurve, \mathcal{O}(L_{0})) =0,
 \end{equation}
 \ie if the divisor $\delta$ is nonspecial: we assume that this is the case throughout.\\ 

  Choose a $\lambda_{0}\in\mathbb{C}$ over which there are $N$ distinct points on $\spcurve$, ordered as $P_{1}=(\lambda_{0},\mu_{1}),\ldots,P_{N}=(\lambda_{0},\mu_{N})$, and define the $N\times N$ matrix $\hat{\Psi}$ as follows:
 \begin{equation*}
(\hat{\Psi}(\lambda_{0}))_{ij}=\psi^{j}(P_{i}).
 \end{equation*}
\begin{proposition}
The Lax matrix $\Lpar$ can be expressed as 
\begin{equation}\label{Lparpsi}
A_{\lambda_{0}}=\Psi(\lambda_{0})\cdot \boldsymbol{\mu}\cdot(\Psi(\lambda_{0}))^{-1},
\end{equation}
where $\boldsymbol{\mu}=\mathrm{diag}(\mu_{1},\ldots,\mu_{N})$.
\end{proposition}
 Note that eq. \eqref{Lparpsi} can be extended by continuity to give $\Lpar$ in a neighbourhood of $\lambda_{0}$, and hence globally, since $\Lpar$ is a polynomial.\\
 
 In order to have an explicit expression for $\Lpar$, we need an explicit representation for the basis $\psi^{k}$ of $L_{0}$ ; we choose one as follos. As the``initial value'' line bundle $L_{0}$ is represented by the divisor $\delta$, sections of $L_{0}$ are then meromorphic function of $\spcurve$ with poles at $\delta$, \ie
 $$ \divs(\psi)\geq-\delta.$$
Finding a  basis for such sections amounts to a gauge choice for $\Lpar$. We can fix the gauge by imposing the $\Lpar$ is diagonal for a fixed value of $\lambda$: we choose the value $\lambda=\infty$. This results in the eigenvectors being proportional to the canonical vectors $e_{i}$ at $\lambda=\infty$.\\
Let $D^{{\infty}}=P^{{\infty}}_{{1}}+\ldots P^{{\infty}}_{{N}}$, where the $P^{{\infty}}_{{1}}$ are the $N$ points above $\lambda=\infty$, and let $\Delta^{{i}}=\delta-D^{{\infty}}+P^{\infty}_{i}$, $D^{i}=D^{{\infty}}-P^{\infty}_{i}$. It follows from the assumption \eqref{nonspecialdiv} that\footnote{For a thorough discussion of this point we refer to \cite{BE06},  sect. 2.3.2.}
\begin{equation*}
 \mathrm{dim} H^{0}(\spcurve, \mathcal{O}(L_{\Delta_{i}})) =1.
\end{equation*}
This  implies that sections of $L_{\Delta_{i}}$  also represents sections of $L_{0}=L(\delta)$; in particular, for each $i$ there is one meromorphic section $\psi^{i}$ such that
 \begin{equation}\label{divpsi}
  \divs(\psi^{i})\geq-\delta+D^{{i}}.
  \end{equation}
 The $\psi^{i}$ may be normalised by\footnote{This is possible since $\psi^{i}(P^{{\infty}}_{i}) \neq 0$, as the eigenvectors are proportional to the canonical vectors $e_{i}$ at $\lambda=\infty$.}
 \begin{equation}\label{normpsi}
 \psi^{i}(P^{{\infty}}_{i})=1.
 \end{equation}
 These last two conditions uniquely determine a basis $(\psi_{i})$ for $ H^{0}(\spcurve, \mathcal{O}(L_{0}))$.\\
 
 \section{Theta functions}
 The final step is writing these sections in terms of theta functions; the Riemann theorem on the zeros of the theta function is fundamental in this construction and is recalled in Appendix \ref{apprs}, together with the basic properties of some objects used below, such as the Abel map $\abelmap$, the vector of Riemann constants $\vrc$, and the theta function $\theta$.\\
 
 Write the divisors $\Delta^{k}$ and $\delta$ as  $\Delta^{k}=\sum_{i=1}^{g}\Delta^{k}_{i}$ and $\delta=\sum_{i=1}^{g+N-1}\delta_{i}$.\\
 One can find $e\in\Theta$ such that
$\theta(e+\abelmap(\Delta^{k}_{i}))\neq0,\;
\theta(e+\abelmap(P^{\infty}_{i}))\neq0, \;
\theta(e+\abelmap(\delta_{i}))\neq0$.\\
Then a function satisfying \eqref{divpsi} is given by
\begin{equation}\label{psithetamer}
\tilde{\psi^{i}}=\frac{\theta\left(\abelmap(P)-\abelmap(\Delta^{i})-\vrc\right)}
{\theta\left(\abelmap(P)-\sum_{j=1}^{g}\abelmap(\delta_{j})-\vrc\right)}
\frac{\prod_{j\neq i}\theta(\abelmap(P)-\abelmap(P^{\infty}_{j})+e)}
{\prod_{j= i}^{N-1}\theta(\abelmap(P)-\abelmap(\delta_{g+j})+e)}.
\end{equation}

First, we notice that the theta functions in the expression \eqref{psithetamer} have been assembled so that their automorphy factors cancel, i.e. the function $\tilde{\psi_{i}}$ is indeed well defined.\\

By the Riemann theorem, the zeros of $\theta\left(\abelmap(P)-\sum_{j=1}^{g}\abelmap(\delta_{j})-\vrc\right)$ are the $g$ points $\delta_{1},\ldots, \delta_{g}$, hence $\tilde{\psi_{i}}$ has poles at these points.\\

As for the rest, we notice that the second fraction in \eqref{psithetamer} can be expressed as a product of $N-1$ quotients of the form $\theta(\abelmap(P)-\abelmap(P^{\infty}_{k})+e)/\theta(\abelmap(P)-\abelmap(\delta_{l})+e)$, for appropriate $k$ and $l$. 
Let us  analyse the properties of $\theta(\abelmap(P)-\abelmap(Q)+e)$ first. From the Riemann theorem, we know it has $g$ zeros: one of them is at
$P_{1}=Q$, and the others, say $P_{2},\ldots ,P_{g}$, are independent of it. Indeed, it follows from the Riemann theorem that $\abelmap(P_{1})+\ldots +\abelmap(P_{g})=\abelmap(Q)+e-\vrc$. Since $P_{1}=Q$, the remaining $g-1$ zeros are determined by an equation that is independent of $Q$. In particular, this implies that the function $\theta(\abelmap(P)-\abelmap(P^{\infty}_{k})+e)/\theta(\abelmap(P)-\abelmap(\delta_{l})+e)$
has only one zero at $P^{\infty}_{k}$ and one pole at $\delta_{l}$; the other
$g-1$ zeros of the theta functions at numerator and denominator cancel. So overall the product contributes $N-1$ poles at  $\delta_{1},\ldots, \delta_{g}$ and$N-1$ zeros at $P^{\infty}_{j}$, with $j\neq i$.\\

Hence, a  section normalised as in eq. \eqref{normpsi} is given by
\begin{equation}
{\psi^{i}(P)}=\frac{\tilde{\psi^{i}}(P)}{\tilde{\psi^{i}}(P^{\infty}_{i})}
\end{equation}
\subsection{Time evolution and Baker-Akhiezer functions}\label{krth}
In the previous section we have reconstructed the Lax matrix $A_{\lambda}(0)$ for a specific ``initial value'' line bundle $L_{0}$ at $t=0$. To describe the flow of matricial polynomials $\Lpar(t)$  corresponding to a linear flow of line bundles $L_{t}$ we can repeat the above construction for every time $t$ to obtain a time dependant basis $\psi^{i}_{t}$ for $H^{0}(\spcurve,\mathcal{O}(L_{t}))$, and hence a time dependent Lax matrix. Instead of doing this, though, the time dependance can be made more explicit using Baker-Akhiezer functions rather than meromorphic functions.\\

The sections $\psi^{i}_{t}$ of $L_{t}$ may be represented as functions with zeros at $\Delta_{i}$, poles at $\delta$ and exponential singularities on a divisor $D$.\\
\begin{theorem}[\enf{Krichever, 1977}]\label{krithm1}
Given a nonspecial divisor $\delta$ of degree $g+n-1$, there exists
a unique function $\psi _{j }\left( t,P\right) $ and local
coordinates $\mu_{j }(P)$ for which $\mu_{j }(D_{j })=0$, such that
\begin{enumerate}
\item  the function $\psi _{j }$ of $P\in S$ is meromorphic
outside $D$ and has at most simple poles at $\delta$ (if all of them are distinct);
\item  in a neighbourhood of the puncture $D_{l }$ the function $\psi _{j }$ takes the form
\begin{equation*}
\psi _{j }\left( z, P\right) =e^{z\,{w_{l }}^{-m}}\left( \delta
_{j l }+\sum\limits_{k=1}^{\infty }\alpha\sp{k}_{j l }\left(
z\right) w_{l }^{k}\right), \qquad w_{l }=w_{l }\left(
P\right),\qquad m\in\mathbb{N}\sp{+}. \label{baexp}
\end{equation*}

\end{enumerate}
\end{theorem}
\noindent The integer $m$ is arbitrary, but in applications it is determined by the given flow; in the case of monopoles it is $1$. \\

Introduce the second kind differential $\Omega^{[m]}_{D}$ such that\footnote{Note  that in the monopole case there we have used a different second kind differential $\gamma_{\infty}$, with different normalisation properties \eqref{asym}. The only difference lies in a different normalization for the exponential $\exp(\int_{P_{0}}^{P}t\Omega^{[m]}_{D})$ in \eqref{htkr}.}
\begin{align}
\Omega^{[m]}_{D}& \sim \de \left(\mu_{j}^{-m}+O(\mu_{j}) \right)   %
\quad \mathrm{near }\; D_{j};\label{asym1}\\
\oint_{\mathfrak{a}_{i}}&\Omega^{[m]}_{D}=0; \label{ginfnorm1}\\
\Omega^{[m]}_{D}&\;\mathrm{is \;holomorphic\; away\; from}\;D.
\end{align}
Introduce then the vector
 \begin{equation}\label{vecU1}
U^{[m]}_{j}=\frac{1}{2\pi i} \oint_{\mathfrak{b}_{j}}\Omega^{[m]}_{D}.
\end{equation}
The function $\psi_{j}$ of the above theorem is then given, for our case, by defining
\begin{equation}\label{htkr}
h^{i}_{t}=\frac{\theta\left( \abelmap(P)+tU^{[m]}-\abelmap(\Delta^{{i}})-\vrc \right)}
{\theta\left( \abelmap(P)-\abelmap(\Delta^{{i}})-\vrc \right)}e^{\int_{P_{0}}^{P}t\Omega^{[m]}_{D}};
\end{equation}
hence the expression for a normalised section $\psi^{i}_{t}$ of $L_{t}$ %
 reads
\begin{equation}\label{psiBA}
{\psi^{i}_{t}(P)}=\frac{h^{i}_{t}(P)\tilde{\psi^{i}_{t}}(P)}{h^{i}_{t}(P^{\infty}_{i})\tilde{\psi^{i}_{t}}(P^{\infty}_{i})}.
\end{equation}

From the considerations after eq. \eqref{psithetamer}, we conclude that this function is well defined, and  has the required poles and zeros; moreover, it has exponential singularities at $D$.\\
We point out that the multiplication by $h^{i}_{t}$ does not add any extraneous poles, as its denominator cancels with one of the factors at the numerator of eq. \eqref{psithetamer}. Thus, the poles of $\psi^{i}_{t}(P)$ only come from those of  \eqref{psithetamer}, hence are independent of time.\\
 Denote by $D_{0}(t)$ the divisor of the zeros  of $h^{i}_{t}$, which  hence contains all the time evolution information: by the Riemann theorem it satisfies 
\begin{equation}
\abelmap(D_{0}(t))-\abelmap(\Delta^{i})=-tU^{[m]}.
\end{equation}
So we recover the fact that the flow is linear on $\Jac(\spcurve)$.\\

To conclude, using this theta function representation in eq. \eqref{Lparpsi}, we obtain a solution for $\Lpar(t)$ in terms of theta functions, realised as a linear flow on $\Jac(\spcurve)$.
\section{General remarks}
This chapter has been a quick overview of algebraic-geometric methods in solving Lax systems, to give a general idea of why these methods are so powerful in this context. Firstly, algebraic geometric techniques have a wide range of applications, since many known integrable systems admit Lax pairs: from classical completely integrable Hamiltonian systems, to many nonlinear PDEs, to quantum systems as well.\\

Moreover, in this context, one manages to make precise sense of the idea that an integrable system admits a ``simple'' or closed solution, either ``by quadratures'', or expressible in terms of known functions, or linear in terms, for instance, of angle action variables. We see that for Lax systems the flow is indeed linear with respect to the given linear structure on the Jacobian of the spectral curve.\\

 The most important  outcome of these algebraic-geometric techniques is   probably the possibility of obtaining an explicit expression for the dynamical flow of a Lax system in terms of Baker-Akhiezer functions. The beauty and value of this result lie in the generality and compactness of this expression. In practice, though, the concrete realisation of this formula requires a deep understanding of the spectral curve, to obtain an explicit expression for the objects needed, \ie the vector of Riemann constants, the theta function, hence the period matrix, which in turn depends on a specific choice of homology basis. This is easier to do for the case of elliptic or hyperelliptic curves, but already for trigonal curves it becomes a more complicated task: this is, for instance, the case of the monopole curve \eqref{c3curve}, which is the object of the present thesis. \\
 It is often the case, however, that in the presence of symmetry, even for more complicated spectral curves it is possible to simplify eq. \eqref{psiBA} by reducing the solution. Introducing a particularly symmetric homology basis, the period matrix takes a simpler form and it is then possible to express the theta functions in terms of lower genus ones, often hyperelliptic or elliptic. 
 Again, this is the approach we take for the cyclic charge 3 monopole.\\

\chapter{The spectral curve}\label{curve}
\introcap{\noindent This  chapter is dedicated to the study of a (family of) curve(s) corresponding to a charge three monopole symmetric under the cyclic group $C_{3}$, introduced by Hitchin, Manton and Murray in \cite{symmmon} (see also section \ref{symmmon}).
Some general properties of the curve are examined, in particular its $C_{3}$ symmetry, and the corresponding quotient, in view of Fay's theory of unbranched covers \cite{fay}. The main result is the explicit construction of a particular symmetric basis, which satisfies Fay's theorem. This allows to express several objects on this genus 4 curve in terms of objects on the genus 2 quotient: in particular, the Ercolani-Sinha constraints can be reduced to relations between periods of a genus 2  curve, hence becoming more manageable from a computational point of view. Throughout, a constant comparison is made with the class of symmetric monopoles studied by Braden and Enolskii in \cite{BE06}, corresponding to the limit $\alpha\to 0$ of the present case, which serves both as a motivation and a consistency check.
}
\numberwithin{equation}{section}
\section{The Spectral Curve}\label{spcurve}
Some general properties of the curve introduced in eq. \eqref{c3curve} are presented here; we remark incidentally that we should more correctly refer to this as a class of curves, depending on the parameters $\alpha$, $\beta$, $\gamma$: for each value of these parameters we have a curve, which corresponds to a monopole only if it satisfies the Ercolani-Sinha constraints of Theorem \ref{escthm}.
\subsection{Properties}\label{properties}
\begin{figure} 
 \hspace*{-6mm}
\includegraphics[height=250pt]{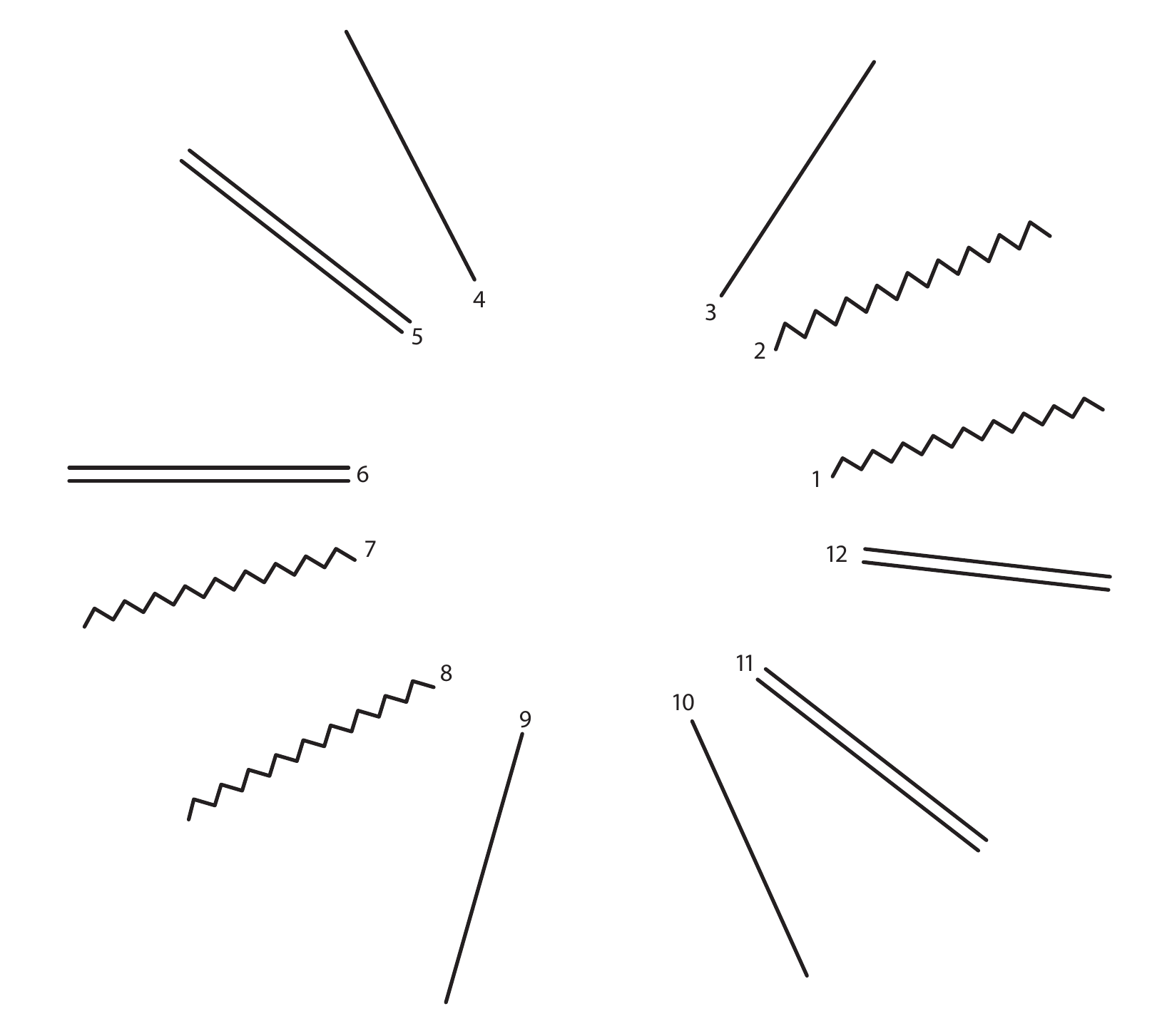}
\hspace*{-5mm}
\begin{minipage}{100pt}\vspace*{-9mm}
 \includegraphics[height=80pt]{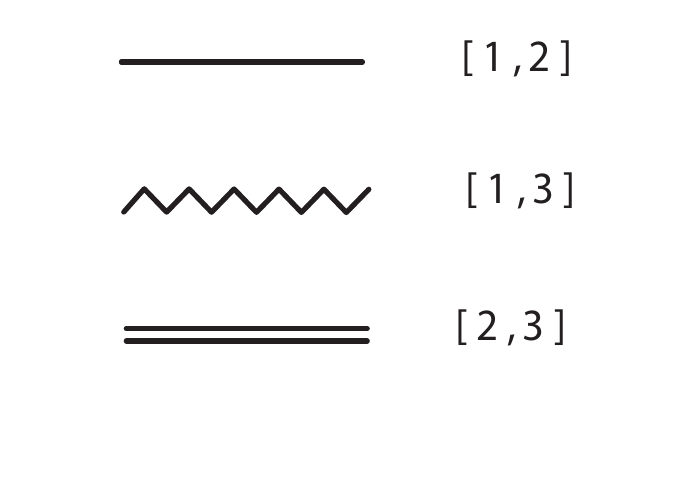}
\end{minipage}
\caption{Branchpoints and monodromy for $\Xhat$}\label{bpts4}
\end{figure}

Let us consider the Riemann surface $\hat{X}$ defined by the following equation:
 \begin{equation}\label{eq:xhat}
 w^{3}+\alpha w z^{2}+\beta z^{6}+ \gamma z^{3} - \beta =0,
 \end{equation}
 where\footnote{The most general form for this surface would be $w^{3}+\alpha w z^{2}+\beta z^{6}+ \gamma z^{3} - \beta^{*} =0$, with $\alpha, \beta, \gamma\in\mathbb{C}$. But for it to be the spectral curve of a monopole, it has to be invariant under the real involution $\tau$ (condition  \enf{H1} in section \ref{hitchinform}); this implies that $\alpha , \gamma\in\mathbb{R}$; moreover it can be easily seen that $\beta$ can be made real with a change of coordinate (a rotation of $z$ of an opportune angle)} $\alpha, \beta, \gamma\in\mathbb{R}$.\\
 The surface $\hat{X}$ is not hyperelliptic, and has genus 4, hence a basis for the holomorphic differentials consists of 4 elements, for which a standard choice is
\begin{align}\label{diffxhat}
\begin{split}
\uu{1}=\;&\frac{\de z}{3 w^{2}+\alpha z^{2}}, \qquad\qquad
\uu{2}=\frac{z\de z}{3 w^{2}+\alpha z^{2}}, \\ 
\uu{3}=\;&\frac{z^{2}\de z}{3 w^{2}+\alpha z^{2}}, \qquad\qquad
\uu{4}=\frac{w\de z}{3 w^{2}+\alpha z^{2}}. 
\end{split}
\end{align}
The above curve can be seen as a cover of the complex sphere with $3$ sheets, and 12 ramification points, whose $z$-coordinates are given by 
\begin{align}\label{bptsexp}
\BH^{z}_{1,2}&=\left[\frac{1}{18\beta}\left( -9\gamma-2\imath\sqrt{3}\alpha^{3/2}\pm \Delta_{+}^{1/2} \right)\right]^{1/3}, &\BH^{z}_{3,4}=\left[\frac{1}{18\beta}\left( -9\gamma+2\imath\sqrt{3}\alpha^{3/2}\pm \Delta_{-}^{1/2} \right)\right]^{1/3}\\
&\BH_{i+4k}^z=\rho^k (\BH^{z}_{i}),\quad& i=1,\ldots,4,\; k=0,1,2 , \quad \nonumber
\end{align}
where 
$$\Delta_{\pm}= 324\beta^2-3(2\alpha^{3/2}\pm 3\imath \sqrt{3}\gamma)^2 ,
\qquad \rho=e^{\frac{2}{3}\pi\imath}.$$
In Figure \ref{bpts4} we give a qualitative sketch of the branchpoints; we remark that, despite having chosen specific values for $\alpha,\beta,\gamma$ for Figure \ref{bpts4}, this is the picture of the curve $\Xhat$ that we keep in mind for the rest of the present work: indeed, there is no lack of generality in doing so, as the general properties of $\Xhat$ do not change with the parameters (unless $\alpha=0$, which is a degenerate case, examined in section \ref{alpha0}); this is due to the fact that the relative positions of the $z$-coordinates of the $\BH_{i}$ do not change varying $\alpha,\beta,\gamma$. %
The values for the parameters in Figure \ref{bpts4} are chosen for numerical convenience.\\ 

Moreover, we are also able to find explicitly the monodromy around each branchpoint:
\begin{align*}
\BH_{{1}},\;& \BH_{{2}}, \;\BH_{{7}}, \;\BH_{{8}}&\longrightarrow &\quad& [1,3]\\
\BH_{{3}},\; &\BH_{{4}}, \;\BH_{{9}},\; \BH_{{10}}&\longrightarrow &\quad& [1,2]\\
\BH_{{5}},\; &\BH_{{6}}, \;\BH_{{11}},\, \BH_{{12}}&\longrightarrow &\quad &[2,3]
\end{align*}
This is represented as well  in Figure \ref{bpts4}: as indicated in the figure, a continuous line means that around branchpoint $\BH_{i}$ there is monodromy $[1,2]$, and so on. More details about monodromy are given in section \ref{remarks}, here we only remark that it is calculated numerically, but it doesn't change varying the parameters (as long as $\alpha\neq 0$).\\

The curve admits the following automorphisms
\begin{align}
\qquad& \enf{inversion}:&   \phi:\; (z,w)&\rightarrow \left(-\dfrac{1}{{z}}, -\dfrac{{w}}{{z}^{2}}\right);\label{inv} \\ 
\qquad& \enf{real involution}:& \tau:\; (z,w)&\rightarrow \left(-\dfrac{1}{\bar{z}},-\dfrac{\bar{w}}{\bar{z}^{2}}\right);\label{realinv} \\
\qquad& \enf{cyclic involution}:& \qquad\sigma: \; (z,w)&\rightarrow(\rho z,\rho w). \label{cyclic}
\end{align}
Comparing the above with the symmetries of the curve  (\ref{BEcurv}) studied in \cite{BE06}, we note that the inversion and the real involution appear in that case too, the former being an inversion with respect to the origin, and the latter corresponding to the reality condition of constraint \enf{H1} in Hitchin formulation (see section \ref{hitchinform}).\\
The \enf{cyclic symmetry} is the symmetry by the action of $C_{3}$: this is again a symmetry of the curve (\ref{BEcurv}), and in the notation of section \ref{be1} it reads $\sigma=\rot \circ\shift$. This last automorphism indeed characterises this family of monopoles, as explained in section \ref{symmrotcurv}; in fact, this is the main focus of the present investigation: we are hence going to exploit this symmetry further in what follows.

\subsection{The quotient with respect to $C_{3}$}\label{sec:quotient}
First, we remark that  $\sigma$ sends the branchpoints one into the other as follows (cf. eq. \eqref{bptsexp})
\begin{align*}
\BH_{1}\;\overset{\sigma}{\longrightarrow}&\;\BH_{5}\;\overset{\sigma}{\longrightarrow}\BH_{9},\\
\BH_{2}\;\overset{\sigma}{\longrightarrow}&\;\BH_{6}\;\overset{\sigma}{\longrightarrow}\BH_{10},\\
\BH_{3}\;\overset{\sigma}{\longrightarrow}&\;\BH_{7}\;\overset{\sigma}{\longrightarrow}\BH_{11},\\
\BH_{4}\;\overset{\sigma}{\longrightarrow}&\;\BH_{8}\;\overset{\sigma}{\longrightarrow}\BH_{12}.
\end{align*}

The quotient under the cyclic involution is described by the map\footnote{
Note that in fact it is the composition of the quotient map with a map that puts the quotient curve in hyperelliptic standard form.
}
 \begin{equation}\label{quotient}
 \pi:   (z,w)\longrightarrow (x,y)=\left(\frac{w}{z},\; \beta(z^{3}+\frac{1}{z^{3}})\right). 
\end{equation}
So the quotient curve, which we denote $X$, is
\begin{equation}\label{quotientcurve}
y^{2}=(x^{3}+\alpha x +\gamma)^{2}+4\beta^{2}.
\end{equation}
It is a \emph{genus 2}, hence hyperelliptic, Riemann surface, in standard form. Seen as a 2-sheeted cover of the Riemann sphere, it has six branchpoints whose $x$-coordinates are
\begin{align}
\begin{split}\label{bpquotient}
B^{x}_{1}&=\frac{1}{6}\,\rho\,\delta_{-}^{\frac{1}{3}}-\frac {2\rho^{2}a}{\delta_{-}^{\frac{2}{3}}}, \qquad
B^{x}_{2}=\frac{1}{6}\,\rho\,\delta_{+}^{\frac{1}{3}}-\frac {2\rho^{2}a}{\delta_{+}^{\frac{2}{3}}}, \qquad
B^{x}_{3}=\frac{1}{6}\,\delta_{-}^{\frac{1}{3}}-\frac {2a}{\delta_{-}^{\frac{2}{3}}},  \\
B_{4}^{x}&=\frac{1}{6}\,\delta_{+}^{\frac{1}{3}}-\frac {2\,a}{\delta_{+}^{\frac{2}{3}}},  \qquad \quad\;\,
B_{5}^{x}=\frac{1}{6}\,\rho^{2}\,\delta_{-}^{\frac{1}{3}}-\frac {2\rho\, a}{\delta_{-}^{\frac{2}{3}}}, \qquad
B_{6}^{x}=\frac{1}{6}\,\rho^{2}\,\delta_{+}^{\frac{1}{3}}-\frac {2\rho\,a}{\delta_{+}^{\frac{2}{3}}},
\end{split}
\end{align}
where
\begin{equation*}
\delta_{\pm}= -108\,\gamma-216\beta\,i+12\,\sqrt {12\,\alpha^{3}+81\, \left( \gamma\pm2\beta\,i \right) ^{2}}.
\end{equation*}
As $\delta_{-}=\overline{\delta_{+}}$, these branchpoints can be split in complex conjugate pairs
\begin{align}\label{bpquotientcc}
B_{6}&=\overline{B_{1}}, & B_{5}&=\overline{B_{2}} ,& B_{4}&=\overline{B_{3}}.
\end{align}

We remark in particular that the quotient map $\pi:\Xhat\to X$ is an \enf{unbranched cover} of $X$ by $\Xhat$. This can be seen using the Riemann-Hurwitz formula, \ie
\begin{equation*}
2\cdot(\hat{g}-1)=2\cdot\mathrm{deg}(\pi)\cdot(g-1)+\sum_{p}\mathrm{mult}(\pi)_{P} \quad\Longrightarrow \quad \sum_{p}\mathrm{mult}(\pi)_{P}=0
\end{equation*}
This fact is very relevant, in that it allows us to use the theory of unbranched covers (see section \ref{unbrquot}) with interesting results.\\

\begin{figure} 
\begin{center}
 \hspace*{-6mm}
\includegraphics[height=250pt]{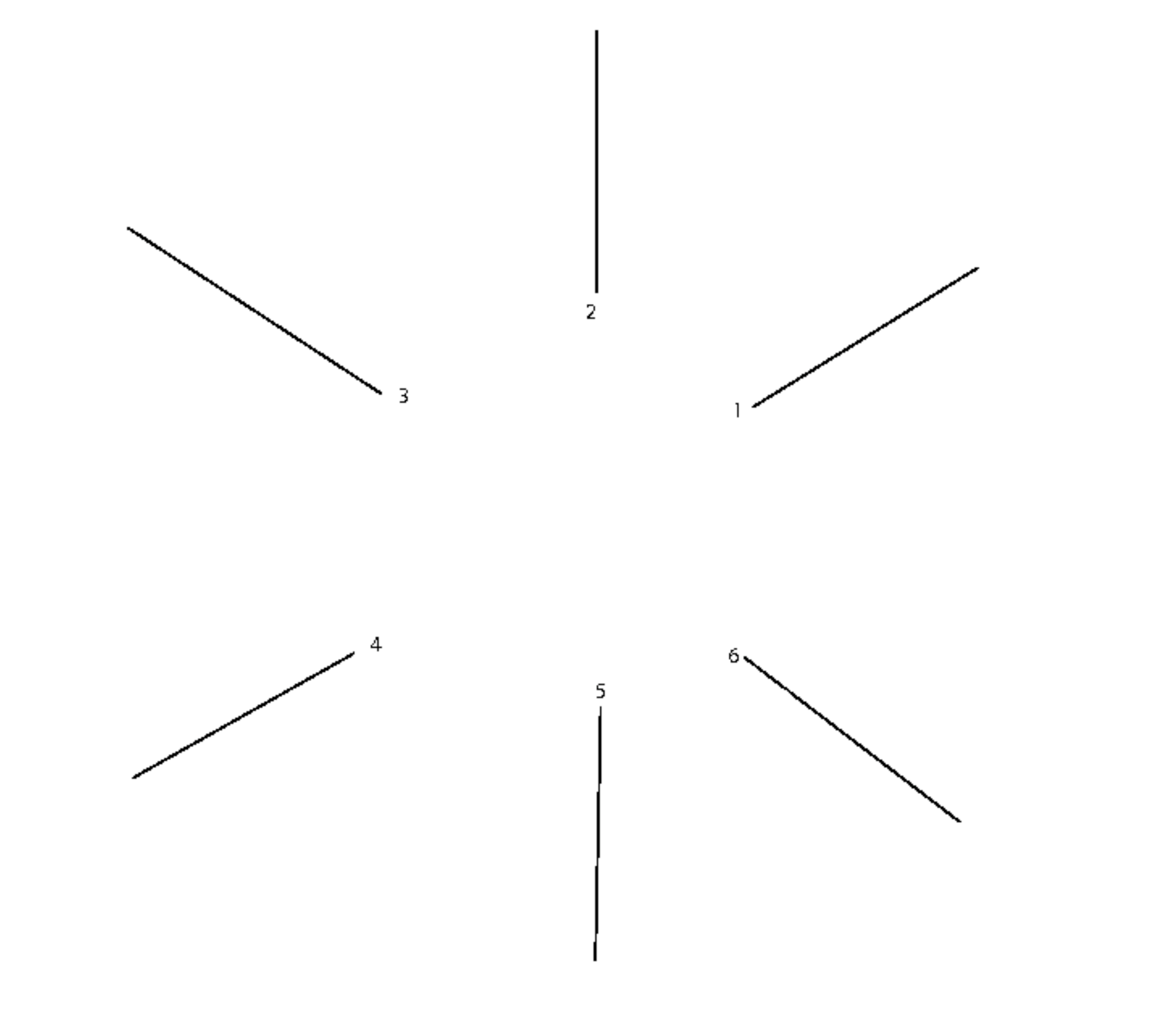}
\hspace*{-5mm}
\caption{Branchpoints and monodromy for $X$}\label{bpts2}
\end{center}\end{figure}

Note that the branchpoints $B_{i}$ \emph{are not} the images of the branchpoints of $\Xhat$ under $\pi$: this is due to the form (\ref{quotient}) of the map $\pi$, which mixes the coordinates $z$ and $w$. \\
Figure \ref{bpts2} shows again a qualitative sketch of the branchpoints for the curve $X$, and their monodromy, with the same choice of parameters of Figure \ref{bpts4}, and the same \textit{caveats} as in the previous case;  more details about monodromy are given in section \ref{remarks}.\\

A standard basis for the holomorphic differentials on $X$ is given by
\begin{equation}
\UU=\frac{\de x}{y}, \quad 
\UT=\frac{x\;\de x}{y}.
\end{equation}
 We notice that the differentials $\uu{2}$ and $\uu{4}$ on $\Xhat$ are invariant under $\sigma$ and hence descend to differentials on $X$; or better, in other terms: 
\begin{equation}
 \pi^{*}\;\UU=-3\uu{2}, \qquad \pi^{*}\;\UT=-3\uu{4}.
\end{equation}
This observation allows us to considerably simplify some integrals, and hence the period matrix: this is examined in section \ref{symmpm}.

\numberwithin{equation}{section}
\section{The case $\alpha=0$}\label{alpha0}
This section is dedicated to the analysis of the spectral curve in the case $\alpha=0$, namely the case studied by Braden and Enolskii in \cite{BE06}, and	reviewed in section \ref{sec:be}. This is indeed a genuinely  different case, in that we have truly different curves, with different properties, \eg more symmetry.  We are particularly interested in making contact with the case $\alpha=0$, as the present work aims to generalise the results in \cite{BE06}, and therefore, it is useful to have that as both a comparison and an inspiration in what follows.\\

\begin{figure}
\begin{center}
 \includegraphics[height=250pt]{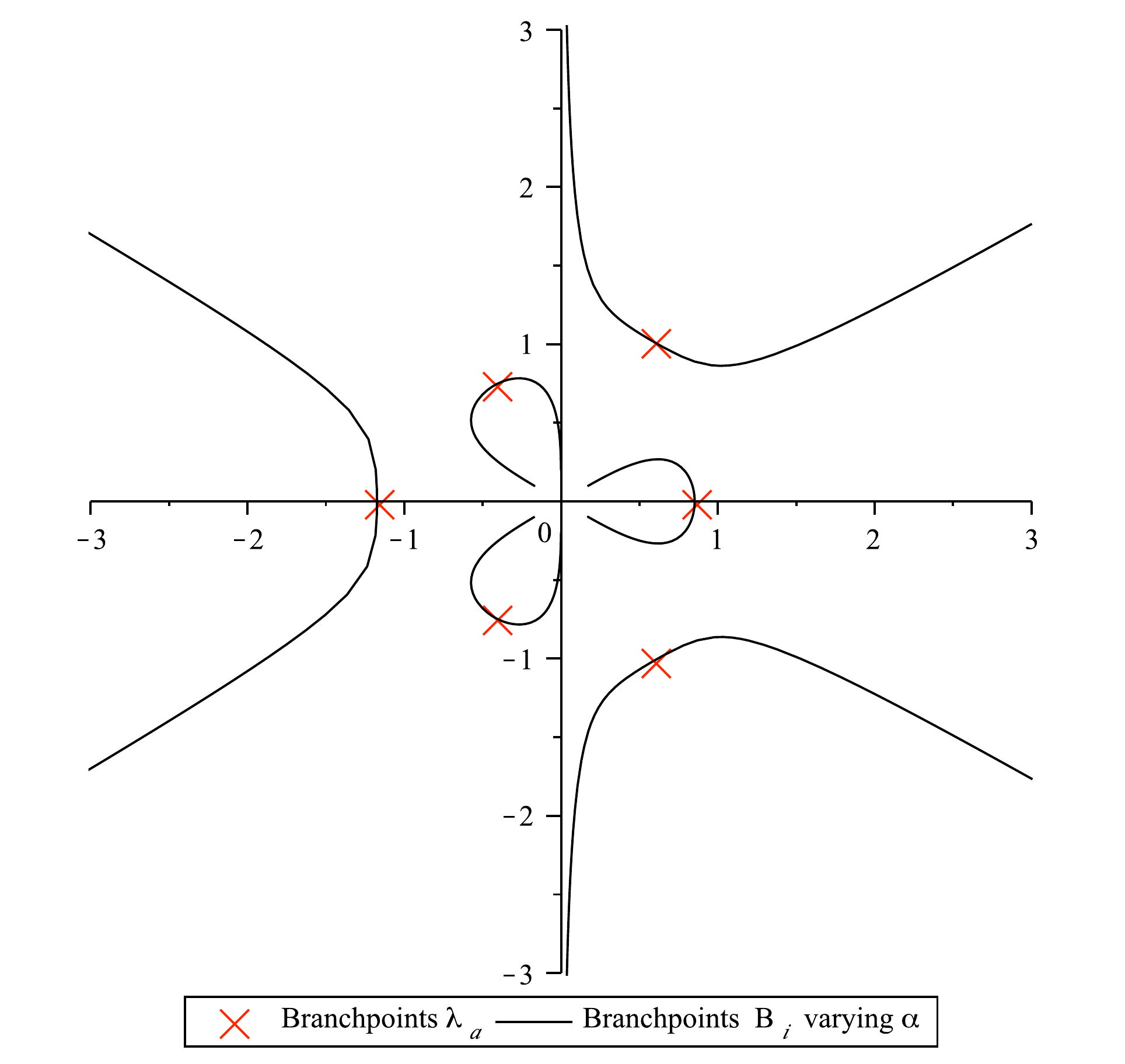}
\caption{Branchpoints for $\alpha\to 0$}\label{alphato0}
\end{center}
\end{figure}

In the case $\alpha=0$ we recover the curve of  eq. \eqref{BEcurv}, which we report here again for convenience
\begin{equation}\label{eq:xzero}
 w^{3}+\beta z^{6}+ \gamma z^{3} - \beta =0,
 \end{equation}
 together with some of its basic properties.  The corresponding Riemann surface, which we denote by $\Xzero$, also has  genus 4, but only 6 branchpoints, $\lambda_{i},\; i=1\ldots 6$, given in (\ref{lambda}). 
Indeed, letting $\alpha\to 0$ in eq. \eqref{bptsexp}, we see that the branchpoints $\BH_{i}$ collide pairwise to give the $\lambda_{i}$ (see Figure \ref{alphato0}).\\

\begin{figure}
\begin{center}
\begin{minipage}{160pt}\vspace*{-2mm}\hspace*{-7mm}
 \includegraphics[width=190pt]{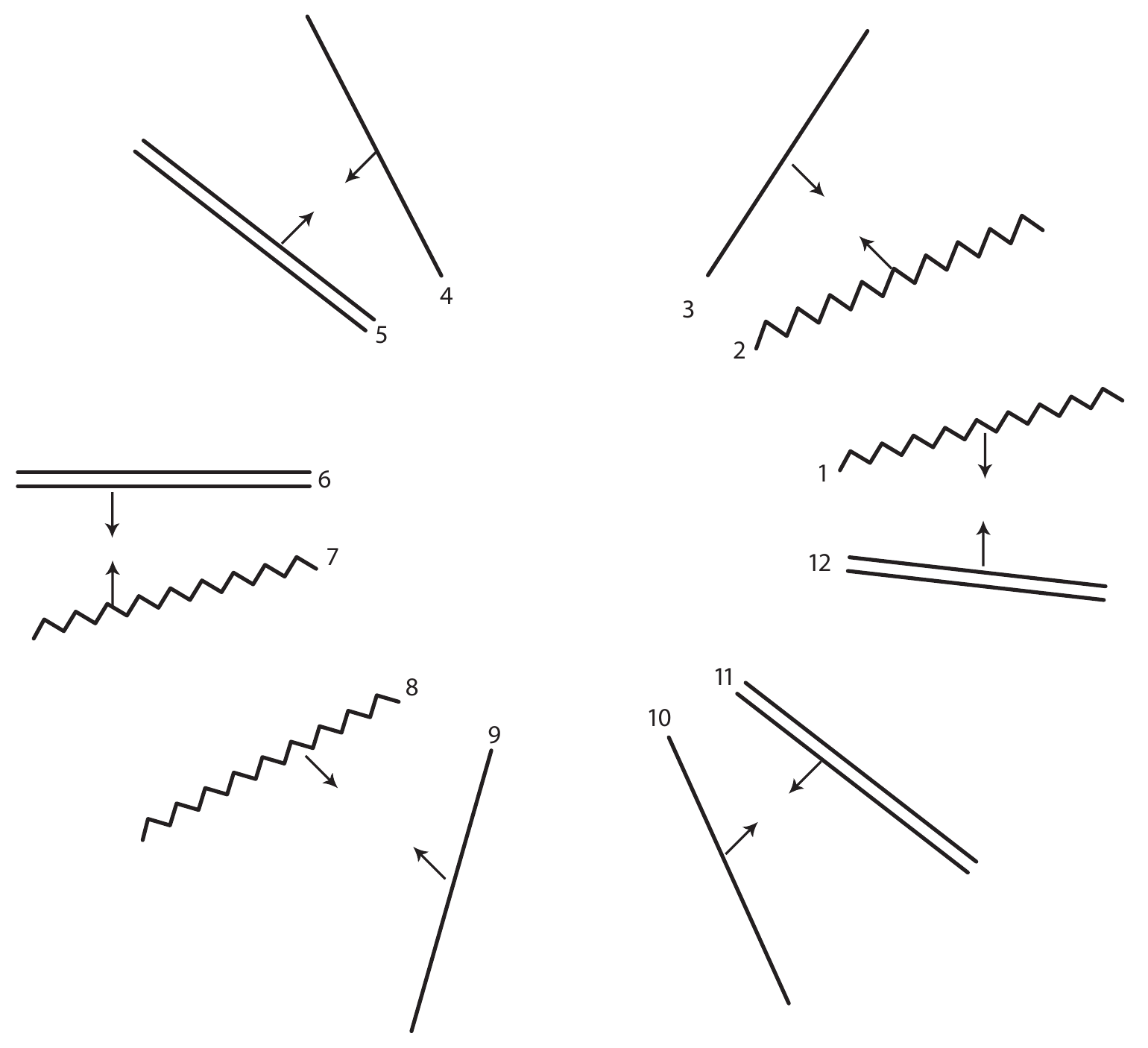}
\end{minipage} %
\begin{minipage}{160pt}\vspace*{-7mm}%
\includegraphics[width=190pt]{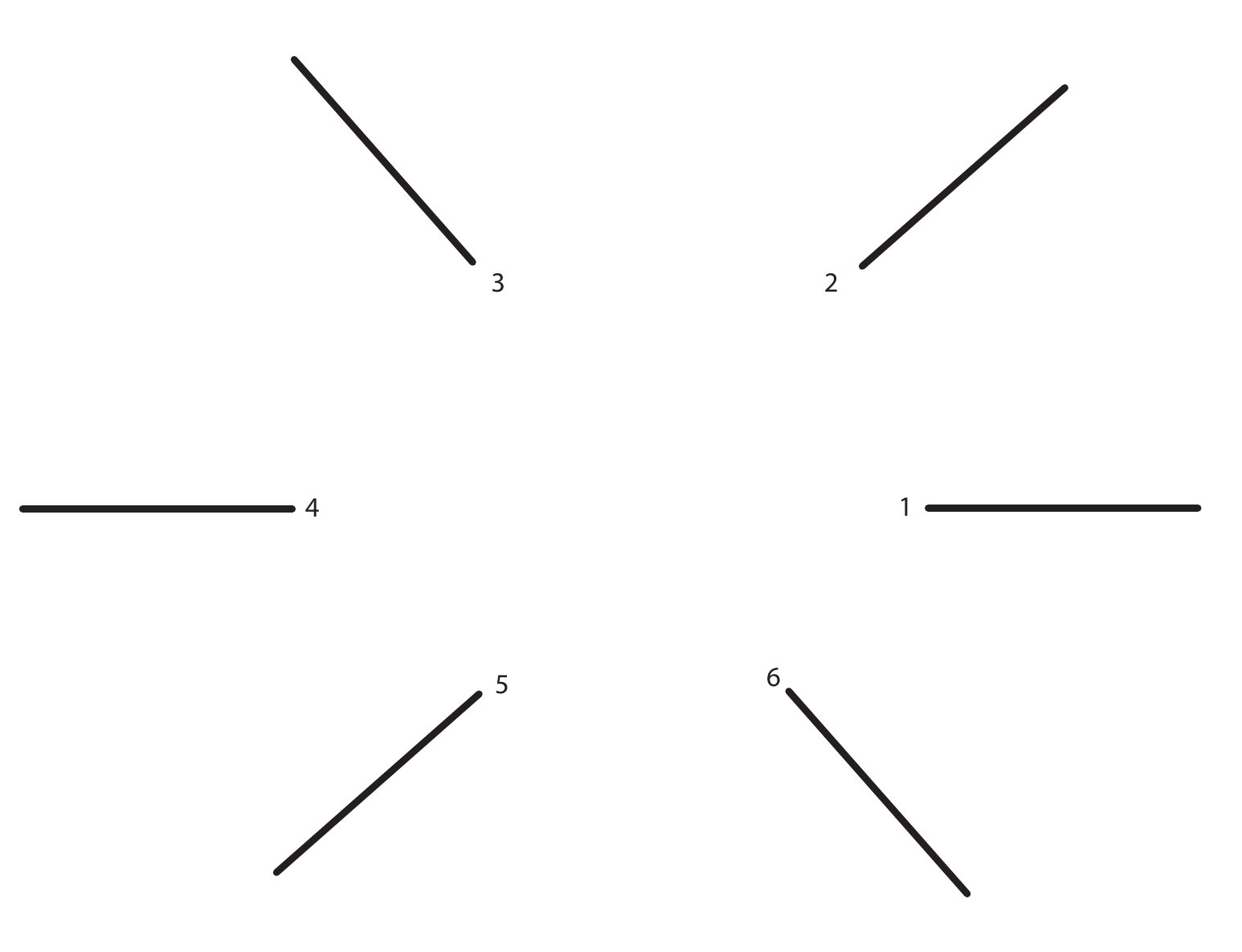} 
\end{minipage}
\caption{Branchpoints and monodromy for $\Xzero$}\label{bpts0}
\end{center}
\end{figure}

In Figure \ref{bpts0} we also give the monodromy, which is the same for every branchpoint, namely $[1,2,3]$; this can be seen from an explicit calculation, but also by appropriately taking the limit of the monodromies of  $\Xhat$. This is explained in detail in section \ref{remarks}.\\

The basis of the holomorphic differentials used in the following is
\begin{align}
\uzero{1}=&\frac{\de z}{w^{2}}, & 
\uzero{2}=&\frac{z\;\de z}{w^{2}}, &
\uzero{3}=&\frac{z^{2}\de z}{ w^{2}},&
\uzero{4}=&\frac{w\;\de z}{ w^{2}}. \nonumber
\end{align}
Note that the ordering is different from that used in \cite{BE06}, as this basis corresponds, up to a factor of $1/3$ to the one for $\Xhat$ given in eq. \eqref{diffxhat}
\begin{equation*}
\lim_{\alpha\to 0}\uu{i}=\frac{1}{3}\uzero{i}.
\end{equation*}

Taking the limit for $\alpha\to 0$ in the \emph{quotient} curve $X$, we obtain the curve $X_{0}$. We notice that the limiting operation does not change the curve dramatically as it changes the curve $\Xhat$: indeed, no degeneration appears, unlike the $\Xhat$ case, and the branchpoints just move slightly; the one remarkable effect is that this curve still has the extra symmetry observed above.\\

\introcap{\enf{Notation}
 \;We recall that all the objects with a hat are objects on the genus $4$ curve $\Xhat$ \eqref{eq:xhat}, \ie the spectral curve, while objects without a hat belong to the quotient curve $X$, obtained quotienting $\Xhat$ of \eqref{quotientcurve} with respect to the $C_{3}$ action. Also, objects with a hat and a superscript $0$ are intended to be on the limit for $\alpha\to 0$ of the curve $\Xhat$ \eqref{eq:xzero}. Summarizing this in a diagram:
$$
\xymatrix@=40pt{\Xhat  \ar[r]^-{\qquad\alpha \to 0\qquad} \ar[d]^-{\pi}& \Xhat_{0}\\
X }
$$
}
\numberwithin{equation}{section}
\section{Remarks }\label{remarks}
In this section we examine in more detail some technical aspects of the previous work.\\
Throughout this work, we view a Riemann surface $\Sigma$, given by an equation of the form $\pol(z,w)=0$, as a branched $n$-sheeted cover of the Riemann sphere, with covering map $\pi:\Sigma \to \puno$. In practice, this means that we treat it as $n$ copies of the complex plane, each corresponding to a sheet, that we glue appropriately along the branchcuts to obtain the surface. First we examine in detail how to explicitly calculate monodromy and how to order sheets in practice given a Riemann surface, emphasising the fact that these are two distinct operations. We then explain how they actually tie together, giving information on how to glue sheets to a branchcut.
All this is needed to understand the picture of the Riemann surfaces $\Xhat$, $\Xzero$ and $X$ of the last section, and their homology bases, which are introduced later in this chapter: therefore we conclude this section applying these considerations to our specific case.
\subsubsection{Monodromy}
Let us remark a very important point (which is often overlooked in the literature): \enf{the monodromy around a branchpoint depends on the {basepoint}}, the point where the loop is based. Indeed, we recall that the monodromy representation of a holomorphic map $\pi:X\to Y$ is defined as the representation of the fundamental group based at a point of the base, $\pi_{1}(Y,P)$ in terms of permutations of $n$ elements, $n$ being the number of preimages of $P$ under $\pi$ (cf. \eg \cite{miranda}, p. 84).\\

So, in our explicit calculation of the monodromy around a branchpoint $B_{i}$, we first choose a reference point $z^{P}$ on the base $\puno$, which is the same for every branchpoint, and we order its preimages  on $X$, $P_{1}=(z^{P},w^{P}_{1})$,$\ldots$, $P_{n}=(z^{P},w^{P}_{n})$; then we consider, on the base space $\mathbb{P}^{1}$, a loop based at $z_{0}$ around $B_{i}$. We lift this closed path to the Riemann surface (using the \texttt{Acontinuation} command in Maple); if the lifted path is an open path, its initial point is going to be $P_{i}$ and its final point $P_{j}$:  then we say that the monodromy around the branchpoint has cycle structure $[i,j,\ldots]$. In other words, we start from an ordering of the $w$-values above $z^{P}$, \ie $(w^{P}_{1},w^{P}_{2},\ldots , w^{P}_{n})$, and analytically continuing these values along  path enrcircling the ramification point, we return to the same set of $w$-values, but permuted $(w^{P}_{\sigma_{i}(1)},w^{P}_{\sigma_{i}(2)},\ldots , w^{P}_{\sigma_{i}(n)})$. This defines a permutation $\sigma_{i}$, the monodromy data for the branchpoint $B_{i}$;  the permutations $\sigma_{i}$  generate the monodromy group based at $z^{P}$. A different choice of ordering the $w$-values at $z^{P}$ leads to  a conjugate set of permutations and isomorphic monodromy group. \\

\begin{figure}
\begin{center} 
\includegraphics*[scale=0.75]{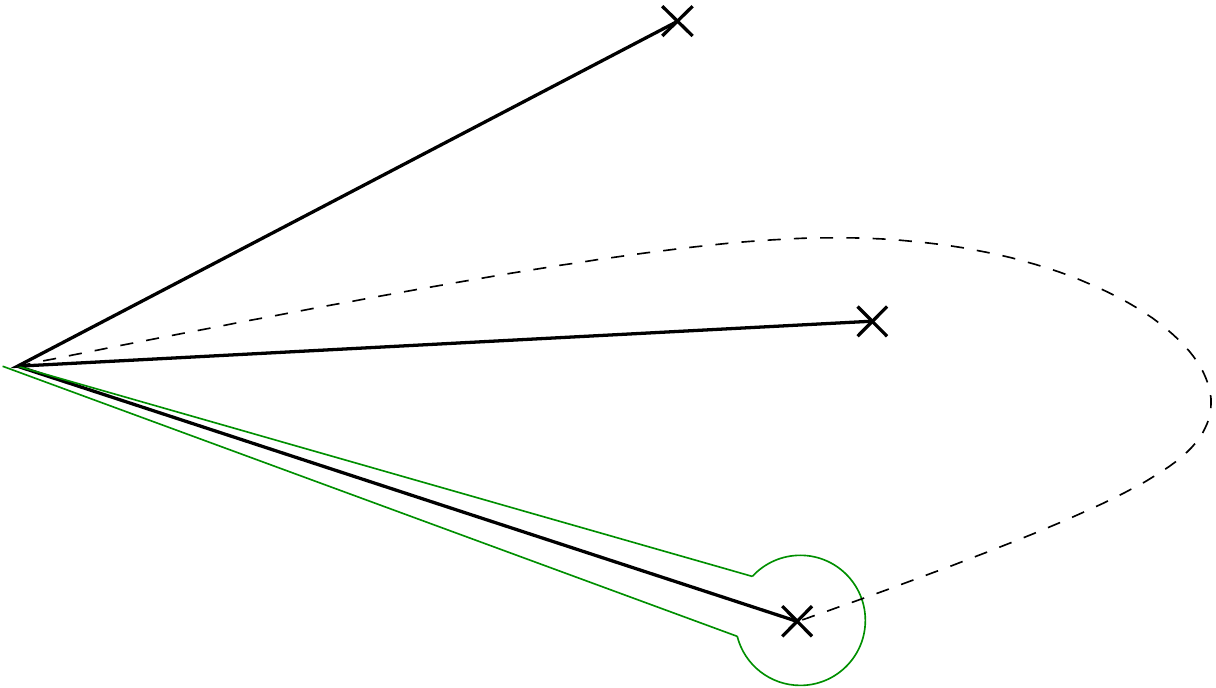}
\caption{Cuts and paths for monodromy calculations}\label{mon1}
\end{center}
\end{figure}

In this picture, the concept of \enf{branchcut} is introduced, namely, using the classical terminology, we ``cut'' the complex sphere by eliminating, for every branchpoint $B_{i}$, a curve (usually a straight line) connecting $B_{i}$ to the basepoint $P_{0}$, hence obtaining the sheet, which is a plane. In practice, to calculate the monodromy the closed path we use is the one as in Figure \ref{mon1}: qualitatively, the path follows the cut on one side, then encircles the branchpoint along a small circle of radius $\epsilon$, then follows the cut on the other side. Note that conventionally this path is oriented such that it goes around the branchpoint \enf{anticlockwise}. Notice also that the monodromy depends on how we choose the cut: for instance, if in Figure \ref{mon1} we consider the dashed cut and follow the path around it as just described, the monodromy around $B_{i}$ is different, and in fact conjugate to that obtained following the other cut. Thus, when we talk about monodromy around a branchpoint we always mean monodromy calculated with respect to a specific choice both of basepoint, and of cut; in particular, in section \ref{spcurve}, the monodromy is always understood to be calculated along the path explained above, using the point as infinity as basepoint, and the cuts as in Figures \ref{bpts4}, \ref {bpts0}. Details about this are given at the end of this section.
\subsubsection{Sheet ordering}
\begin{figure}
\begin{center} 
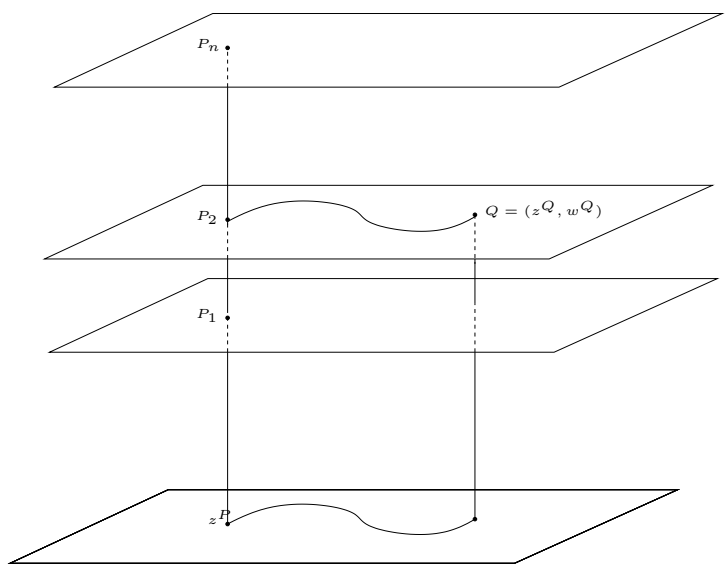
\caption{Sheet labeling}\label{sheets}
\end{center}
\end{figure}

In the picture above, we wish to think of the $n$ cut planes as sheets.
Firstly, we need a specific way of identifying, or labelling, sheets. The notion of sheet ordering is distinct from our ordering at $z_{0}$. In practice we achieve this as follows: we fix another (regular) point on $\puno$, the base of the covering, say $z^{{S}}$. Corresponding to it we have $n$ points on the Riemann surface $S_{i}=(z^{{S}},{w}^{S}_{i})$, where ${w}^{S}_{i}$ are solutions of $\pol(z^{S},w)=0$; now, we order the $w^{S}_{i}$, and we say that $w^{S}_{1}$ labels sheet $1$, etc. . The points $S_{i}:=(z^{{S}},w^{S}_{i})$ are then our reference points for each sheet: to understand on which sheet a given point $Q=(z^{{Q}},w^{{Q}})$ is, we have to compare its $w$-value with that of one of the reference points $S_{i}$. \\
In order to do so, choose a path $\gamma$ on $\puno$ starting from $z^{Q}$ and ending on $z^{S}$, with the only constraint  that it does not have to cross any cuts. Then lift it to a path $\tilde{\gamma}\in \Sigma$ (for instance, using the \texttt{Acontinuation} command in Maple) with the condition that $\tilde{\gamma}(0)={Q}$: the endpoint of $\tilde{\gamma}$ on $\Sigma$ is one of the $S_{i}$, which tells us then on which sheet $Q$ is. This is pictured in Figure \ref{sheets}.\\

In this work we have always chosen $z=0$ as reference point for sheet ordering (or sheet base, in short), and have ordered the sheets as follows:
\begin{equation}\label{shord}
\text{sheet base} \; z=0 \to 
\begin{cases}
\text{sheet 1}& w=a\in \mathbb{R} \\
\text{sheet 2}& w=\rho^{2} \;a\\
\text{sheet 3}& w=\rho \;a .
\end{cases}
\end{equation}
This ordering in the limit $\alpha\to 0$ yields the sheets ordering in \cite{BE06}, thus making the comparison with the Braden-Enolskii case much easier.

\subsubsection{Monodromy and gluing sheets}
Before explaining in detail how to calculate monodromy for the curve $\Xhat$ (where we choose the basepoint to be at infinity), we clarify one further very important point concerning the very meaning of monodromy. \\

The notion of sheet ordering at $z^{S}$ is distinct from the ordering at $z^{P}$. Indeed, depending on the path chosen to link $z^{S}$ with $z^{P}$ we could reach a different $w_{i}^{P}$, so the $w$-values at $z^{P}$ do not distinguish sheets alone. In practice which  $w_{i}^{P}$ one reaches depends on the ``sector'' between cuts where the path approaches  $z^{P}$.
Sheets and  monodromy then appear as distinct constructs\footnote{We thank Tim Northover for many useful discussions concerning this point.}.\\
 In particular one cannot say just from the monodromy information at a branchpoint which sheets come together, or vice versa. In fact,  the monodromy data around a branchpoint, together with the sheet ordering, gives information about how to attach the $n$ copies of $\mathbb{C}$ to a given cut, rather than about which sheets are connected at a certain cut. Thus, if a branchpoint has monodromy $[i,j,\ldots]$, it does not necessarily mean  that crossing the corresponding cut one goes from sheet $i$ to sheet $j$. We make this more clear from an example, as it is a fundamental point in understanding the homology basis of  section \ref{hombasis}.
 
 \begin{example}Consider the curve of equation
 $$w^{3}+z^{2}w+z^{3}+1=0,$$
 with $6$ branchpoints
\begin{align*}
\mathrm{b}^{z}_{1}=d_{+}^{-\frac{1}{3}}, \quad \mathrm{b}^{z}_{2}=\rho\,d_{-}^{-\frac{1}{3}}, \quad  \mathrm{b}^{z}_{3}=\rho\, d_{+}^{-\frac{1}{3}},  \quad \mathrm{b}^{z}_{4}=\rho^{2}d_{-}^{-\frac{1}{3}}, \quad \mathrm{b}^{z}_{5}=\rho^{2}\mathrm{b}^{z}_{1}, \quad  \mathrm{b}^{z}_{6}=d_{-}^{-\frac{1}{3}},
\end{align*}
 with $d_{\pm}=1\pm 2\,i\sqrt{3}/9$ and $\rho=\exp(2i\pi/3)$; they are represented in Figure \ref{extim} as grey spheres.  \\

\begin{figure}[t!]
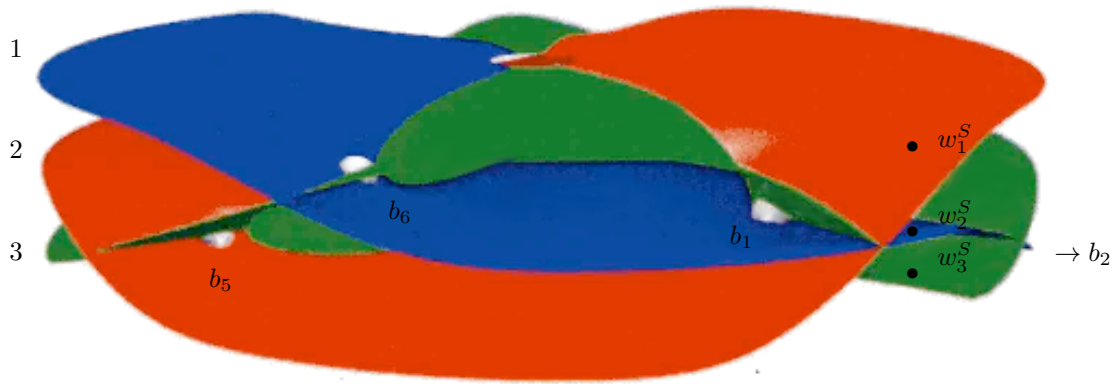

\begin{center}
\makebox[320pt]{
\begin{lpic}[draft,clean]{tim-curve3(17cm,)}
    \lbl[t]{140,33;${\to b_2}$}
   \lbl[t]{100,35;$b_{1}$}
   \lbl[t]{60,38;$b_{6}$}
   \lbl[t]{39,30;$b_{5}$}
   \lbl[t]{120,45;$\bullet$}   \lbl[t]{125,47;$w^{S}_{1}$}
   \lbl[t]{120,35;$\bullet$}   \lbl[t]{125,38;$w^{S}_{2}$}
   \lbl[t]{120,30;$\bullet$}   \lbl[t]{125,33;$w^{S}_{3}$}
\lbl[t]{15,57;${1}$} 
\lbl[t]{15,45;${2}$}
\lbl[t]{15,33;${3}$}
\end{lpic}}
\end{center}
\caption{Example I}\label{extim}
\end{figure}

Take the point $z^{S}=1/2\exp(i\pi/3)$ as basepoint for the sheet ordering, which lies in the ``sector'' between $\mathrm{b}_{1}$ and $\mathrm{b}_{2}$, and choose an ordering for the sheets $(w_{1}^{S},w_{2}^{S},w_{3}^{S})$ (the black dots in Figure \ref{extim}). This means, looking at Figure \ref{extim}, that sheet 1 corresponds to the colour red, sheets 2 to blue and sheet 3 to green.  \\
Take the origin to be $z^{P}$, the reference point for monodromy, and impose an ordering  of the $w_{i}^{P}$ out of that at $z^{S}$, by analytic continuation along a straight line from $z^{S}$ to $z^{P}$: $(w_{3}^{P},w_{2}^{P},w_{1}^{P})=(-1,-\rho, -\rho^{2})$. In the figure this is represented by  the \emph{height}. Then the monodromy around each branchpoint is computed to be
 $$
{\mathrm{b}_{1}}, \mathrm{b}_{4}  \to [-1,-\rho^{2}]\equiv [1,3],  \qquad \mathrm{b}_{2}, \mathrm{b}_{5}  \to  [-\rho,-\rho^{2}]\equiv [2,3],  \qquad \mathrm{b}_{3}, \mathrm{b}_{6}  \to  [-1,-\rho]\equiv  [1,2],
 $$
and in the figure is represented as a change in height.\\

 \begin{figure}
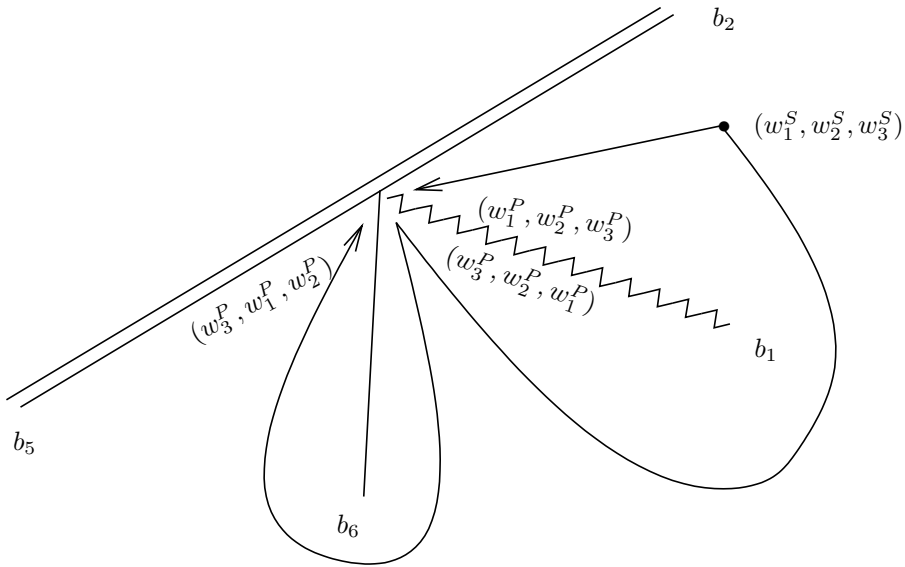

 \begin{center}
\begin{lpic}[draft,clean]{tim-graph3(11cm,)}
    \lbl[t]{69,43;${\bullet}$}     \lbl[t]{79,44;$\small{(w_{1}^{S},w_{2}^{S},w_{3}^{S})}$}
       \lbl[t]{53,35,-8;$\small{(w_{1}^{P},w_{2}^{P},w_{3}^{P})}$}
          \lbl[t]{50,29,-19;$\small{(w_{3}^{P},w_{2}^{P},w_{1}^{P})}$}
             \lbl[t]{24,27,28;$\small{(w_{3}^{P},w_{1}^{P},w_{2}^{P})}$}
\lbl[t]{69,54;$b_{2}$}
\lbl[t]{73,22;$b_{1}$}
\lbl[t]{33,5;$b_{6}$}
\lbl[t]{2,13;$b_{5}$}
  \end{lpic}
  \end{center}
\caption{Example II}\label{extim1}
\end{figure}

This pictorial representation clarifies the fundamental differences between sheets and monodromy, together with their link: the sheet ordering at $z^{S}$  imposes an ordering of the preimages $w_{i}^{P}$ of $z^{P}$; the monodromy at a branchpoint is a permutation of this $w_{i}^{P}$ ordering, after going around a branchpoint (represented in Figure \ref{extim} by a change in height); this tells us how to attach a sheet at each cut.\\ %
For instance, starting in the sector between $\mathrm{b}_{1}$ and $\mathrm{b}_{2}$, the preimages of $z^{P}$ are ordered $(w_{1}^{P},w_{2}^{P},w_{3}^{P})$%
; going around $\mathrm{b}_{1}$, the monodromy is $\sigma_{1}=[1,3]$, hence the order is now $(w_{\sigma_{1}(1)}^{P},w_{\sigma_{1}(2)}^{P},w_{\sigma_{1}(3)}^{P})=(w_{3}^{P},w_{2}^{P},w_{1}^{P})$: this means that sheets $1$ and $3$ come together at this cut. Going from there around $\mathrm{b}_{6}$, the monodromy is $\sigma_{6}=[1,2]$, hence the order is now $(w_{\sigma_{6}(3)}^{P},w_{\sigma_{6}(2)}^{P},w_{\sigma_{6}(1)}^{P})=(w_{3}^{P},w_{1}^{P},w_{2}^{P})$: comparing with the ordering we had in the previous ``sector'', this means that sheets $2$ and $3$ come together at this cut. This is sketched in Figure \ref{extim1}.\\
From this we see that starting at the point $(z^{P},w_{1}^{P})$ on sheet 1, the red one, after crossing the cut at $\mathrm{b}_{1}$ we end up on the green sheet, namely sheet 3;  after crossing one more cut at $\mathrm{b}_{6}$ we end up on the blue sheet, namely sheet 2; and so on.\\
One can continue in this fashion for every branchpoint: the case of the cuts at $\mathrm{b}_{1}$ and $\mathrm{b}_{6}$ is already explicative of the general situation.
\end{example}

\subsubsection{Monodromy calculation for $\Xhat$ and $\Xzero$}
We can now clarify the technical details about how we calculate monodromy in section {\ref{spcurve}, where we consider, for both curves, the point at infinity as the basepoint for monodromy, and the cuts are straight lines from the branchpoints to infinity. The way to deal with this, numerically, is to use the \emph{inversion} (\ref{inv}) to bring the point at infinity to the origin, hence making all the paths finite. Note also that the branchpoints get mapped one into the other as follows:
\begin{align*}
\phi(\BH_{1})&=\BH_{6}, & \phi(\BH_{2})=&\BH_{5}, & \phi(\BH_{3})=&\BH_{4},\\ 
\phi(\BH_{7})&=\BH_{12}, & \phi(\BH_{8})=&\BH_{11}, & \phi(\BH_{9})=&\BH_{10},
\end{align*}
(and the rest are obtained recalling that $\phi^{2}=\mathbb{Id}$):
so, for instance, the monodromy around $B_{1}$ with cuts going to infinity is obtained from the monodromy around $B_{7}$ with cuts going to zero. This is represented in Figure \ref{invbp}.\\

\begin{figure}[t!]
\begin{center} 
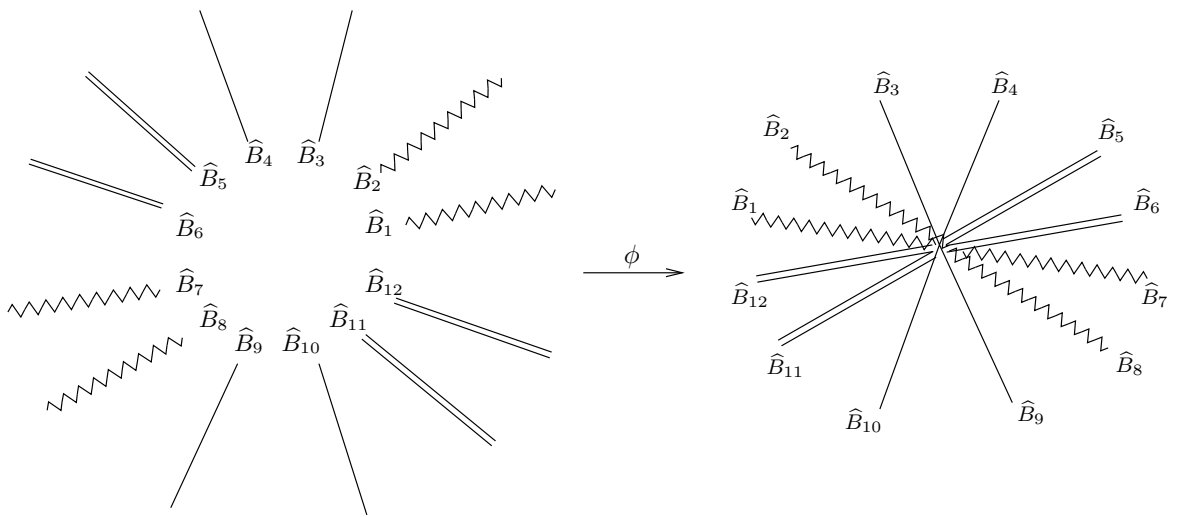
\caption{Branchpoints and cuts under inversion}\label{invbp}
\end{center}
\end{figure}

There is just a small subtlety, namely that the sheet ordering of (\ref{shord}), \ie using the origin $z=0$ as the sheet base\footnote{
Since the origin is already used as monodromy base, we cannot take it as sheet base as well; this is solved by slightly moving the sheet base away from the origin (we choose the point $z=0.1+0.1i$), this also has the advantage that after inversion it corresponds to a finite point, which makes it easier to compare the respective ordering of the $w$-values by analytic continuation to the $S_{i}$. %
}, is not preserved by $\phi$: hence, to obtain the sheet ordering of eq. (\ref{shord}) in the case of outwards cuts to infinity, we have to choose a different ordering in the case with cuts to zero, induced from the previous one via $\phi$.\\

\begin{figure}
\begin{center} 
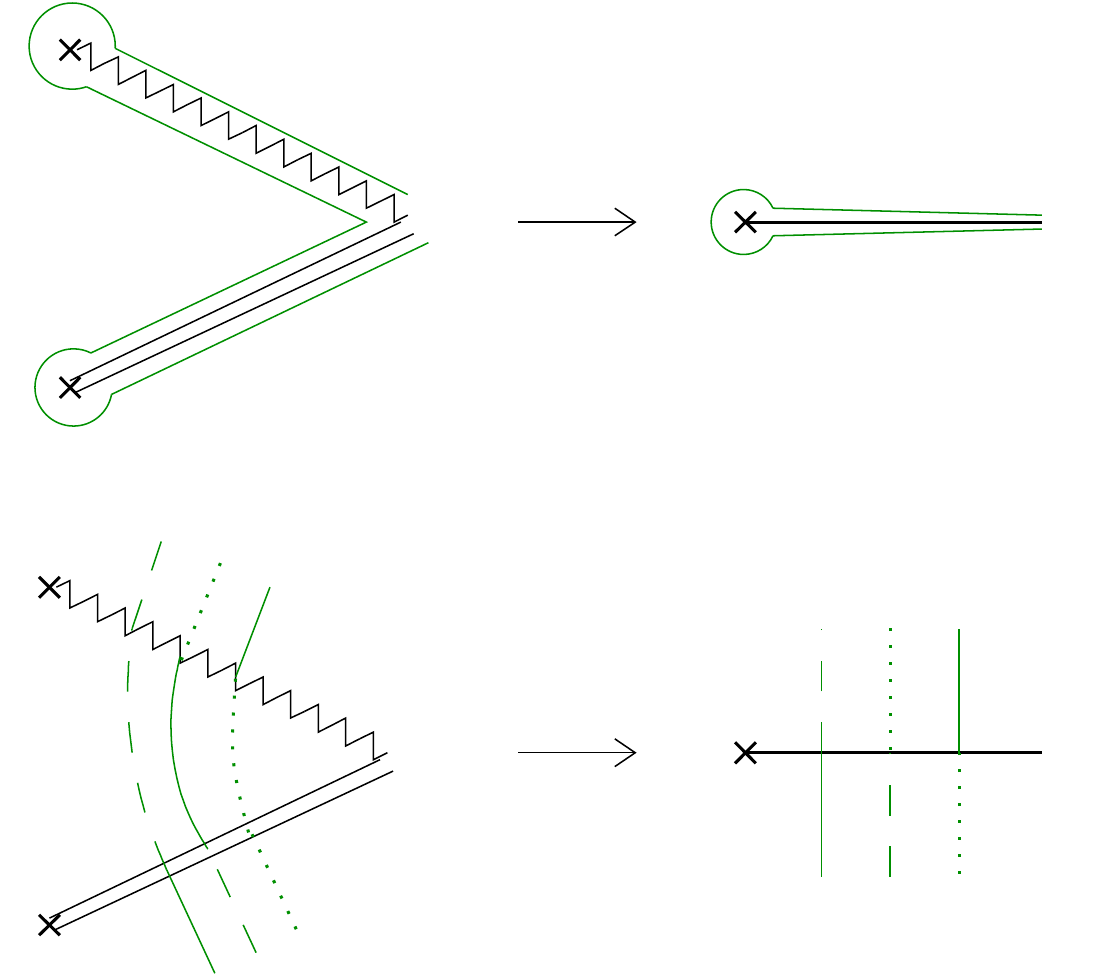
\caption{Cuts for $\alpha\to 0$}\label{joint}
\end{center}
\end{figure}
This last  remark allows us to clarify a final point, namely what happens in the limit when $\alpha\to 0$. In this limit the branchpoints in $\Xhat$ collide pairwise, and so do the corresponding cuts: from this one can deduce the monodromy around the resulting cut out of that of the starting two cuts. Let us focus for simplicity on the colliding branchpoints $\BH_{12}$ and $\BH_{1}$: as said earlier, in practice we calculate the monodromy after inversion, so the cuts we consider are those of Figure \ref{joint}, with basepoint $z^{P}$. Lifting a closed loop based at $z^{P}$, if it starts, say, at $P_{1}$, going around $\BH_{1}$ it ends  at $P_{3}$, then going around $\BH_{12}$ it ends at $P_{1}$; now, in the limit, the cuts collide, and hence only the initial and final part of the path remain, meaning that, in the limit $\alpha\to 0$, if the lifted path starts at $P_{1}$, going around $\lambda_{1}$, it ends at $P_{2}$ (as in the upper part of Figure \ref{joint}).\\
This can be seen in an equivalent way, considering paths crossing the cuts, as shown in the bottom part of Figure \ref{joint}.\\

 Doing this again for the other cases one obtains that the monodromy around $\lambda_{1}$ is $[1,2,3]$ along the cut going inwards; as explained earlier, the ordering of the sheets has been chosen  to match before and after inversion, so we can conclude that the monodromy around $\lambda_{1}$ along the cut going outwards is indeed $[1,2,3]$.\\
One proceeds in the same fashion the other branchpoints (and corresponding cuts), to obtain the monodromy  in Figure \ref{bpts0}.
\section{Interlude: unbranched quotients}\label{unbrquot}
In this section we describe some results in the theory of \enf{unbranched covers}, which plays an important role in the study of the curve $\Xhat$, and simplifies the problem  significantly.\\
The general setting is that of a Riemann surface with a cyclic group acting on it, such that the quotient induced by this action is unbranched.\\
In the first subsection we present a theorem of Fay \cite{fay} about the existence, under these circumstances, of a basis for the first homology group with certain symmetry properties. This proves to be crucial in expressing several objects on our curve $\Xhat$ in a much simpler way \cite{fay,accola}; moreover, we remark that the theorem is non constructive, and %
there are few examples in the literature of such a basis: %
here we provide a new one.
\subsection{Fay's  symmetric homology basis}
Consider a Riemann surface $\Shat$ of genus $\hat{g}$ which admits an (effective holomorphic) action of a \enf{finite cyclic group} $G$ of order $p$
\begin{equation*}
\Phi:\Shat \longrightarrow \Shat.
\end{equation*}
This action induces a quotient
\begin{equation*}
\pi :\Shat \longrightarrow \Sigma,
\end{equation*}
 where $\Sigma$ is another Riemann surface of genus $g$. We consider the case in which the covering $\pi$ has \enf{no branchpoints}.\\

It is then possible to find a canonical basis (see Appendix \ref{apprs}) of $H_{1}(\Shat,\mathbb{Z})$, whose  $2\hat{g}$ cycles we denote by\footnote{Throughout the following discussion quantities with a hat are understood to be  on the ``original'' curve, quantities without a hat are on the quotient curve.}
\begin{equation}\label{eq:basesopra}
\ah_{0}, \ah_{1},\ldots ,\ah_{\hat{g}-1}; \quad \bh_{0}, \bh_{1},\ldots ,\bh_{\hat{g}-1};
\end{equation}
and a canonical basis of $H_{1}(\Sigma,\mathbb{Z})$, whose  $2g$ cycles we  denote by
\begin{equation}\label{eq:basesotto}
\aq_{0},\aq_{1}\cdots \aq_{g-1};\quad \bq_{0},\bq_{1},\cdots \bq_{g-1};
\end{equation}
with the following properties with respect to the lift of the covering $\pi_{*}$
\begin{align}\label{eq:symmpi}
\pi_{*}(\ah_{k+n(g-1)})&=\aq_{k}, & \pi_{*}(\ah_{0})&=\aq_{0},\\
 \pi_{*}(\bh_{k+n(g-1)})&=\bq_{k}, &  \pi_{*}(\bh_{0})&=p,\; \bq_{0};\nonumber \\ \quad& \;& \quad
  & \footnotesize{n =0,\ldots ,p-1; \; k=1,\ldots,g}. \nonumber
\end{align}
In particular, this implies  that the cycles on $\Shat$ have the following behaviour under $\Phi_{*}$ 
\begin{align}\label{eq:symmphi}
\Phi_{*}^{n}(\ah_{k})=&\ah_{k+n(g-1)},&  \quad \\ 
 \Phi^{n}_{*}(\bh_{k})=&\bh_{k+n(g-1)}. &\Phi_{*}^{n}(\bh_{0})=&\bh_{0}. \nonumber
\end{align}
Here, and throughout in the following, the equality signs mean equals in $H_{1}(\Shat,\mathbb{Z})$, so when representing the cycles as closed paths on the surfaces we have to replace ``equals'' with ``homologous''.\\

The above properties can be stated at the level of the holomorphic differentials, \ie for $H^{1}(\Shat,\mathbb{R})$. If $\hat{\boldsymbol{\omega}}_{0},\ldots, \hat{\boldsymbol{\omega}}_{\hat{g}-1}$ is a basis for $H^{1}(\Shat,\mathbb{R})$ dual to that in eq. \eqref{eq:basesopra}, then eqs. \eqref{eq:symmphi} become
\begin{equation}
\Phi^{*n}( \hat{\boldsymbol{\omega}}_{0})= \hat{\boldsymbol{\omega}}_{0} , \quad
\Phi^{*n}( \hat{\boldsymbol{\omega}}_{k})= \hat{\boldsymbol{\omega}}_{k+n(g-1)}.
\end{equation}

\subsection{Simplification in the period matrix, theta functions and vector of Riemann Constants}\label{fayaccolavrc}
Using a basis satisfying the symmetry properties above, one manages to obtain a particularly symmetric expressions for the Riemann period matrix, and a consequent simplification in the associated theta function.

\begin{fayacc} \label{fayaccolathm1}
With respect to the ordered canonical homology bases
$\ah_{0},\ldots ,\ah_{\hat{g}-1}$, $\bh_{0}, \bh_{1},\ldots ,\bh_{\hat{g}-1}$ and
$\aq_{0},\cdots \aq_{g-1}$, $\bq_{0},\bq_{1},\cdots \bq_{g-1}$
satisfying \eqref{eq:symmphi}, the $\mathfrak{a}$-normalised Riemann period
matrices of~   $\hat{\Sigma}$ and $\Sigma$ take the
respective forms
\begin{equation}\hat{\tau}%
=\left( \begin{array}{ccccc} 
\hat{\tau}_{00}&\hat{\tau}_{0}&\hat{\tau}_{0}& \cdots &\hat{\tau}_{0}\\
\hat{\tau}_{0}^{\mathsf{T}}&   M   &   M' &   \cdots   &   M^{(p-1)}\\
\vdots  & \ddots & \;& \;& \vdots \\
\hat{\tau}_{0}^{\mathsf{T}}&M'&M'' &  \cdots & M
\end{array}\right),
\qquad {\tau}
=\left( \begin{array}{cc} 
\dfrac{1}{p}\hat{\tau}_{00}&\hat{\tau}_{0}\\
\hat{\tau}_{0}^{\mathsf{T}}&   N   \\
\end{array}\right),
\end{equation}
where
\begin{align*}
 \hat{\tau}_{0i}&=\oint_{\phi^{n}\bh_{0}}\hat{\boldsymbol{\omega}}_{i},\qquad 
 M^{(n)}_{ij}=\oint_{\phi^{n}\bh_{j}}\hat{\boldsymbol{\omega}}_{i} ,\\
 N_{ij}&=\hat{\tau}_{ij}+\hat{\tau}_{i , j+g-1}+\ldots +\hat{\tau}_{i, \,j(p-1)(g-1)}, \qquad i,j =1,\ldots g-1
\end{align*}
\end{fayacc}
This particularly symmetric form for the period matrices allows to simplify both the theta function and the vector of Riemann constants for $\Shat$,  that can be expressed in terms their correspondents on $\Sigma$.\\
Recall that the map $\pi:\Shat  \to \Sigma$ can be lifted to a map  $\pi^{*}:\Jac(\Sigma) \to \Jac(\Shat) $, and hence to a map from $\mathbb{C}^{g}$ to $\mathbb{C}^{\hat{g}}$, which we still denote $\pi^{*}$: with the above choices \eqref{eq:basesopra},\eqref{eq:basesotto} of homology bases, we have, for $\boldsymbol{ z}\in\mathbb{C}^{g}$
$$
\pi^{*}\boldsymbol{ z}=\pi^{*}(z_{0},z_{1},\ldots,z_{g-1})=(pz_{0},z_{1},\ldots,z_{g-1},\ldots,z_{1},\ldots,z_{g-1}),
$$
or in characteristic notation
$$
\pi^{*}\left( \begin{array}{cccc}
\alpha_{0} &  \alpha_{1}  & \ldots & \alpha_{g-1}\\
\beta_{0} &  \beta_{1}  & \ldots & \beta_{g-1}
\end{array}\right)=
\left( \begin{array}{cccccccc}
\alpha_{0} &  \alpha_{1}  & \ldots & \alpha_{g-1} &\ldots & \alpha_{1}  & \ldots & \alpha_{g-1}\\
p\beta_{0} &  \beta_{1}  & \ldots & \beta_{g-1} & \ldots &  \beta_{1}  & \ldots & \beta_{g-1}
\end{array}\right).
$$

We can then state the following theorem, proven by Accola\footnote{Note that in Accola's proof the covering need not be induced by the action of a cyclic group, but here we give the result in this particular case as it is the case under consideration. Also, in this more general setting, the form of $\halfp$ is not known explicitly.} \cite{accola} and Fay \cite{fay} independently.
\begin{fayacc} \phantomsection\label{fayaccolathm2}
For $\boldsymbol{ z}\in \mathbb{C}^g$ one has
\begin{equation}
\dfrac{\hat{\theta}[\hat{\halfp}](\pi^{*}\boldsymbol{ z};\hat{\tau})}
{\prod_{k=0}^{p-1}\theta [\halfp] \left(\boldsymbol{ z};\tau
\right)}
=c_0(\widehat{\tau}%
). \label{fafactora}
\end{equation}
where $c_0(\hat{\tau})$  is a non-zero constant independent of $\boldsymbol{ z}$ and\footnote{Note that this differs from proposition 4.1 in \cite{fay} by a factor of 2 in the denominator in the first entry: this is due to a different choice of conventions in the basis we use for the Jacobian, which differs by a factor $2\pi$ from what Fay considers.}
\begin{equation}\label{halfpfay}
\halfp=\left ( \frac{p-1}{2\,p}, 0, \ldots , 0\right), \qquad \hat{\halfp}=\pi^{*}\halfp.
\end{equation}
Moreover, the following relation between the two vectors of Riemann constants holds
\begin{equation}\label{vrcaccola}
\pi^{*}\vrc_{Q}=\hat{\vrc}_{\hat{Q}}+(g-1)\sum\limits_{n=0}^{p-1} \abelmap( \Phi^{n}(\hat{Q}))+ \hat{\halfp}.
\end{equation}
\end{fayacc}

\section{A symmetric homology basis: preliminaries}\label{basispre}
\subsection{Fay's symmetric basis for the curve $\Xhat$}
As the curve $\Xhat$ satisfies the hypotheses  above, we can now restate the properties that Fay's basis has to satisfy in this particular case. 
Using the same notation as above (cf. eqs. \eqref{eq:basesopra}, \eqref{eq:basesotto}), eqs. \eqref{eq:symmpi} become 
\begin{align}\label{eq:symmpiX}
\pi_{*}(\ah_{i})&=\aq_{1},& \pi_{*}(\ah_{0})&=\aq_{0}, \\
\pi_{*}(\bh_{i})&=\bq_{1},& \pi_{*}(\bh_{0})&=3 \;\bq_{0}, \qquad i=1,2,3, \nonumber
\end{align}
and eqs. \eqref{eq:symmphi} become
\begin{align}\label{eq:symmrot}
\sigma_{*}^{n}(\ah_{k})&=\ah_{k+n},&\quad& \\
\quad \sigma^{n}_{*}(\bh_{k})&=\bh_{k+n}, & \sigma_{*}^{n}(\bh_{0})&=\bh_{0}, \qquad n=1,2,3 .\nonumber
\end{align}
It turns out that in this case it is most effective to find a basis on $\Xhat$ satisfying \eqref{eq:symmrot}, and then project it to obtain a basis  on $X$ satisfying \eqref{eq:symmpi}.
In $H^{1}(\Shat,\mathbb{Z})$, eqs. \eqref{eq:symmrot} mean that one cycle stays invariant under the group action, and three pairs are mapped one into the other by this action.
\subsection{$\ah_{0}$ is invariant under $\sigma$}\label{a0invariant}
We remark that Fay's conditions in the case of a finite cyclic group of order 3 require  $\ah_{0}$ to be invariant as well.\\
 Indeed, from equations  \eqref{eq:symmpi}  it follows that the matrix representing the involution $\sigma$ on a basis with such symmetry properties is of the following form
 \begin{equation*}%
 M_{\sigma}=
\left(
\begin{array}{cccccccc}
a_{1}  & a_{2}  & a_{3} & a_{4}  & a_{5} & a_{6}  & a_{7} & a_{8}    \\
 0 & 0  & 1 & 0 & 0  & 0 & 0  & 0   \\
 0 & 0  & 0 & 1 & 0  & 0 & 0  & 0   \\   
 0 & 1 & 0 & 0 & 0  & 0 & 0  & 0   \\
 0 & 0  & 0 & 0 & 1  & 0 & 0  & 0   \\
 0 & 0  & 0 & 0 & 0  & 0 & 1  & 0   \\
 0 & 0  & 0 & 0 & 0  & 0 & 0  & 1   \\
 0 & 0  & 0 & 0 & 0  & 1 & 0  & 0   
\end{array}
\right)
\end{equation*}
Only the first row  is arbitrary, as there are no conditions on how $\ah_{0}$ behaves under $\sigma_{*}$. But imposing that that $\sigma$ is a symplectic transformation and that  $\sigma_{*}^{3}$ is the identity is enough to fix all the $a_{i}$ to vanish except for $a_{1}=1$. Hence the cycle $\ah_{0}$ is indeed invariant. \\

Therefore, we can rewrite Fay's conditions \eqref{eq:symmrot} as follows
\begin{align}
\sigma_{*}^{n}(\ah_{k})&=\ah_{k+n},&\sigma_{*}^{n}(\ah_{0})&=\ah_{0}, \label{fayfinal}\\
\quad \sigma^{n}_{*}(\bh_{k})&=\bh_{k+n}, & \sigma_{*}^{n}(\bh_{0})&=\bh_{0}. \qquad n=1,2,3 \nonumber.
\end{align}
These are the conditions we focus on in the following.\\

As remarked above, here the equalities are to be intended in $H_{{1}}(\Xhat)$; so, when we represent the cycles by closed paths in $\Xhat$, the equal signs should actually mean  ``homologous''. Indeed, it turns out that even at the level of paths ``true'' equalities hold: the only exception is $\ah_{0}$, whose images under $\sigma^{k}$ are just homologous. %
\section{The relation to the case $\alpha=0$}
It turns out that the above requirements  \eqref{fayfinal} do not determine a homology basis uniquely: indeed, we found many. An obvious ambiguity can be seen already at this stage, for instance, given a basis satisfying \eqref{fayfinal}, cyclically interchanging the cycles $\ah_{i},\bh_{i},i=1,2,3$  yields a basis with the same symmetry; there is in fact even more freedom in the choice of the $\ah_{i},\bh_{i},i=1,2,3$. Therefore we can impose some extra conditions on this basis, in particular that in the limit $\alpha\to 0$ some of the cycles are mapped to those of the homology basis in \cite{BE06}. This would be a desirable feature as it allows us to relate the present work  with  \cite{BE06}, whose results we aim to generalise.\\
 We will be more precise on this in section \ref{hombasis}, and we refer to section \ref{be1} for more details on the basis of \cite{BE06}: here we only  give again this basis in Figure \ref{be06fig}. We remark again that here the monodromy (calculated again with respect to infinity) is the same for every branchpoints, namely $[1,2,3]$. The notation for the cycles is explained in the figure, and is used throughout this work.
We  also recall  that this basis satisfies different symmetry properties than those we are interested in, see eqs. \eqref{BEbsymm}.

\begin{figure}
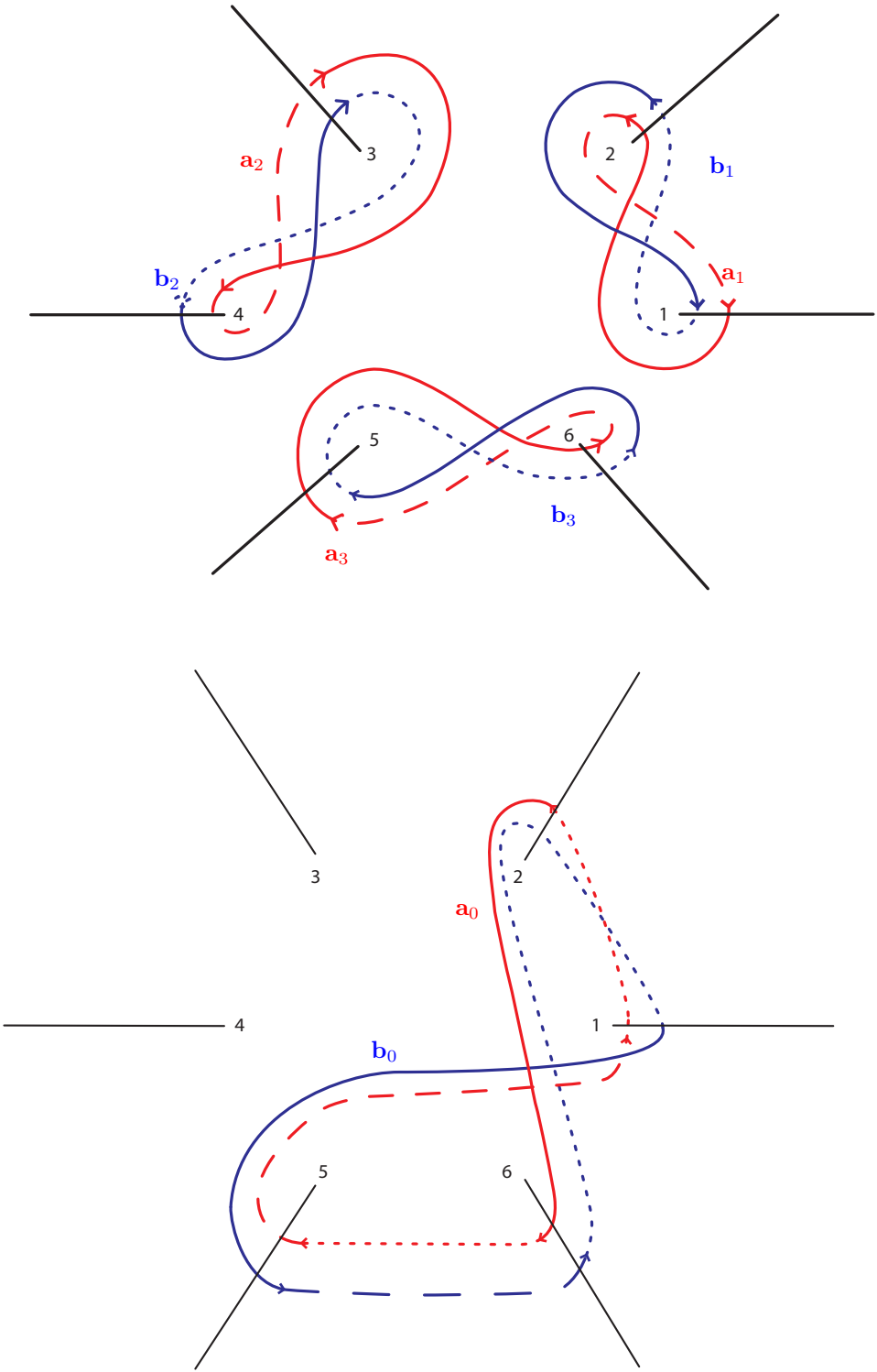

\begin{lpic}[draft,clean]{BE06_123(13cm,)}
   \lbl[t]{130,70;$\color{red}{\abe_1}$}
   \lbl[t]{128,90;${\cblu{\bbe_1}}$}
   \lbl[t]{45,90;$\cred{\abe_2}$}
   \lbl[t]{30,70;${\cblu{\bbe_2}}$}
   \lbl[t]{60,20;$\color{red}{\abe_3}$}
   \lbl[t]{100,28;${\cblu{\bbe_3}}$}
   \end{lpic}
   \begin{lpic}[draft,clean]{BE06_0(13cm,)}
   \lbl[t]{85,90;$\color{red}{\abe_0}$}
   \lbl[t]{70,65;${\cblu{\bbe_0}}$}
   \end{lpic}
\caption{Symmetric homology basis of \cite{BE06}}\label{be06fig}
\end{figure}
\section{A symmetric homology basis}\label{hombasis}
In this section we find explicitly a basis for $H_{1}(\Xhat)$ satisfying Fay's symmetry conditions, as in eqs. \eqref{fayfinal}.\\
We remark again that, as Fay's result is non-constructive, providing an example is not a trivial task. Here we do not provide an algorithm to implement this;  we think that indeed this would be possible, in principle. In fact, there are already algorithmic ways to construct a homology basis: for instance, Tretkoff and Tretkoff algorithm (see \cite{TT} and Appendix \ref{apptt}), which is algebraic in nature. Once the Tretkoff and Tretkoff's basis is found, one could impose on it the required symmetry conditions, and get a symmetric basis in terms of the original Tretkoff and Tretkoff's basis. The problem with this is that usually one does not get a ``simple'' result: for instance, the Tretkoff and Tretkoff's basis is already quite complex (in the sense that the cycles are very involved, cf. Appendix \ref{apptt}), and a linear combination of its elements is even more so.  This defeats the whole purpose of having a basis with a certain symmetry, which is indeed to simplify the problem. In particular, in the calculation of the periods, having ``simpler'' cycles leads to a much simpler and compact result, see section \ref{symmpm}.\\

Hence, the way we follow is not algorithmic;  nevertheless, the methods used provide some insight about how to tackle this sort of problems, and the same logic can be used in similar cases.\\
Moreover, we emphasise that many results described below were greatly simplified by Tim Northover's {package} for Maple described in Appendix \ref{timscode}.
\subsection{The cycles  $\ah_{i},\bh_{i}$}
First, we focus on the search of the three pairs $\ah_{i},\bh_{i}$, with the condition that they are mapped to one another by $\sigma$. \\
As mentioned earlier, we want to relate out constructions as much as possible with those $\alpha=0$ as in \cite{BE06}, hence initially we work in this limit, and then we try to extend the cycles to the case $\alpha\neq 0$. 
We have remarked earlier that the cycles $\abe_{i},\bbe_{i}, i=1\ldots 3$ in Figure \ref{be06fig}  do not have Fay's symmetry properties. Nevertheless, we can choose one pair of them, say $\abe_{i},\bbe_{i}$, to be $\azero_{1},\bzero_{1}$, and then obtain the others applying $\sigma$:
\begin{align}
\azero_{1}&=\abe_{1}, & \bzero_{1}&=\bbe_{1},\\
\azero_{k+1}&=\sigma_{*}^{k}\abe_{1}, & \bzero_{k+1}&=\sigma_{*}^{k}\bbe_{1},\qquad k=1,2.
\end{align}
They are shown in the upper part of  Figure \ref{cyc123} .\\
Note that the cycles thus chosen still satisfy the second equality of \eqref{BEbsymm}, namely
\begin{equation*}
 \shift_{*}(\bzero_{k}) =\rot_{*}^{2}\circ\sigma_{*}(\bzero_{k})=\azero_{k},  \qquad n=1,2,3. 
\end{equation*}

To obtain the corresponding cycles on $\Xhat$, we exploit some properties of the limit ${\alpha\to 0}$. We recall that in this limit the branchpoints in $\Xhat$ and the corresponding cuts collide pairwise; this means that if a cycle crosses both cuts, its arc between the cuts becomes degenerate. This is illustrated in the example in the bottom part of Figure \ref{joint}, where we examine the cuts corresponding to $\BH_{12}$ and $\BH_{1}$. In that picture consider, for instance, the path on $X$ starting on sheet 2, crossing the cut and hence continuing on sheet 3: we see immediately that there is a unique path on $\Xhat$ that gives the previous path; the same happens for the other possible paths crossing this cut.\\ 
 From this example it is then clear how to carry out the inverse process, \ie starting from a cycle in the case $\alpha=0$ how to reconstruct its ``missing'' arc between the collided cuts, unambiguously. Hence, we obtain the cycles $\ah_{i},\bh_{i}$, as in Figure \ref{cyc123} with the required properties as in eqs. \eqref{fayfinal}:
\begin{align}
\sigma^{n}_{*}(\bh_{k})&=\bh_{k+n}, & \sigma_{*}^{n}(\bh_{0})&=\bh_{0}, \qquad n=1,2,3 .\nonumber,
\end{align}
and also
\begin{align}
\lim_{\alpha\to0}\ah_{1}&=\abe_{1}, & \lim_{\alpha\to0}\bh_{1}&=\bbe_{1}.
\end{align}
\begin{figure}
\begin{center}
\begin{lpic}[draft,clean]{cycles0_123(13cm,)}
   \lbl[t]{135,75;$\color{red}{\azero_1}$}
   \lbl[t]{127,98;${\cblu{\bzero_1}}$}
   \lbl[t]{47,97;$\cred{\azero_2}$}
   \lbl[t]{30,77;${\cblu{\bzero_2}}$}
   \lbl[t]{63,29;$\color{red}{\azero_3}$}
   \lbl[t]{102,36;${\cblu{\bzero_3}}$}
   \end{lpic}
   \begin{lpic}[draft,clean]{cycles_123bis(13cm,)}
  \lbl[t]{131,98;$\color{red}{\ah_1}$}
   \lbl[t]{124,108;${\cblu{\bh_1}}$}
   \lbl[t]{43,109;$\cred{\ah_2}$}
   \lbl[t]{30,98;${\cblu{\bh_2}}$}
   \lbl[t]{75,39;$\color{red}{\ah_3}$}
   \lbl[t]{102,39;${\cblu{\bh_3}}$}   \end{lpic}
\caption{Cyclic homology basis}\label{cyc123}
\end{center}
\end{figure}
\;
\begin{figure}
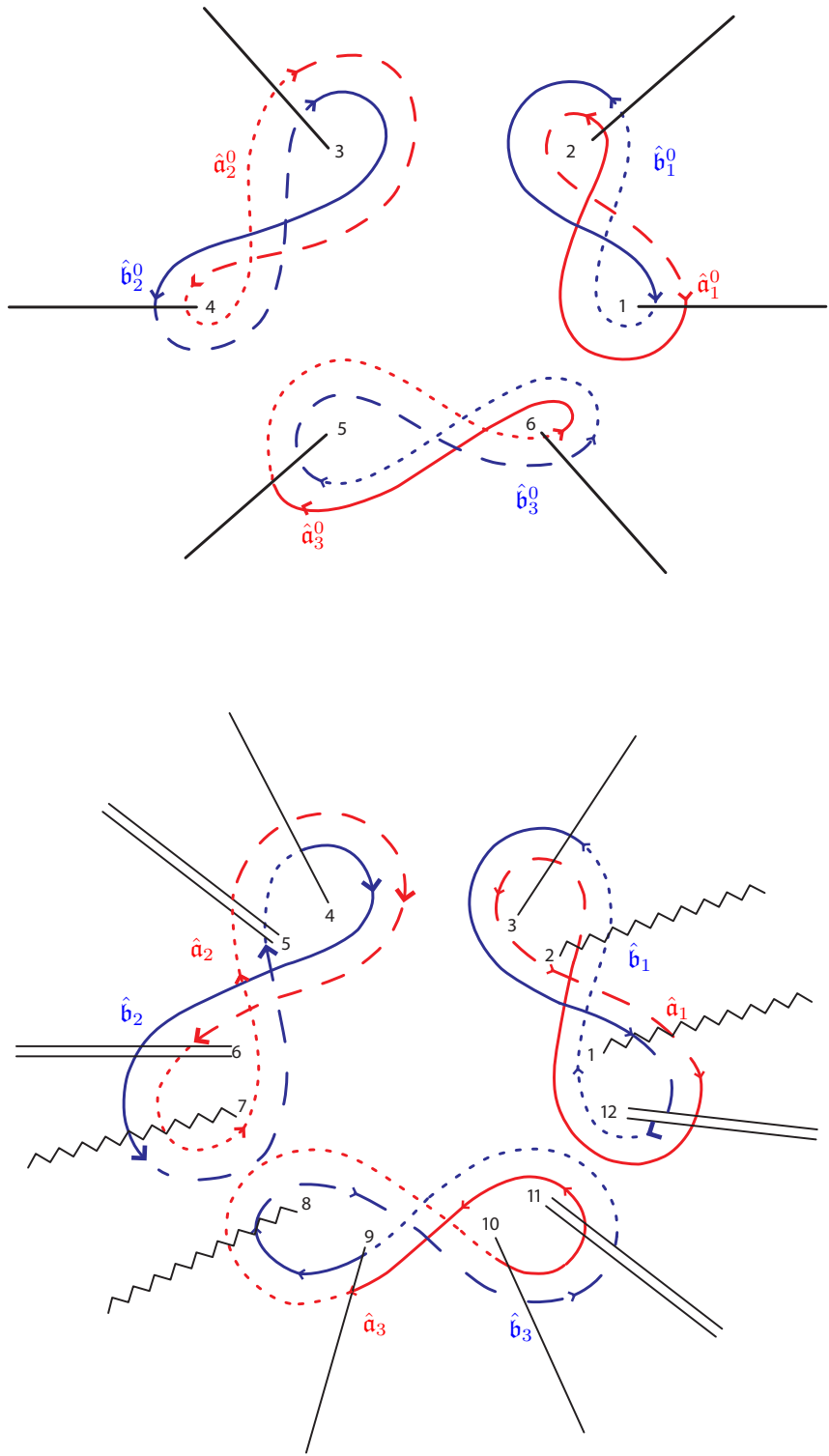

\begin{center}
   \begin{lpic}[draft,clean]{cycles0_0(12cm,)}
   \lbl[t]{107,100;$\color{red}{\azero_0}$}
   \lbl[t]{125,100;${\cblu{\bzero_0}}$}
   \end{lpic}
   \begin{lpic}[draft,clean]{cycles_0(12cm,)}
      \lbl[t]{117,100;$\color{red}{\ah_0}$}
   \lbl[t]{134,104;${\cblu{\bh_0}}$}
   \end{lpic}
\caption{Cyclic homology basis}\label{cyc0}
\end{center}
\end{figure}

\subsection{The cycles $\ah_{0}, \bh_{0}$}\label{a0b0}
As for the remaining two cycles, we focus first on $\bh_{0}$.\\
It can be found using similar considerations: $\bh_{0}$ has to be invariant under $\sigma$ and is a three-fold covering under $\pi$ of a cycle on $X$, and needs to have intersection zero with all the other cycles.
Implementing first the symmetry requirements, we split the $z$ and $w$ planes in three sectors, and as above we look for a cycle which has a component of similar shape in each sector. From these considerations, we choose a cycle $\bh_{0}$ as in the lower part of Figure \ref{cyc0}.\\
One can check that its intersection with the $\ah_{i},\bh_{i}$ is zero, either by inspection, or using Northover's \texttt{extcurves} (see Appendix \ref{appmaple}).\\

As for $\ah_{0}$, we focus first on the condition of its intersection with $\bh_{0}$ to be $-1$, and used the invariance under $\sigma$ (as shown in section \ref{a0invariant})  as a consistency check. Since we want the maximal possible contact with \cite{BE06} in the case $\alpha=0$, one can check whether one of the cycles $\abe_{0},\bbe_{0}$ has the correct intersections with the other cycles in the limit: it turns out that $\abe_{0}$ does. As explained in the previous section, we can lift this cycle to the case $\alpha$ nonzero, to obtain the cycle in Figure \ref{cyc0}.\\
One can then check, for instance with Northover's code, that the basis so obtained is indeed canonical, and moreover that it satisfies the required symmetry properties. In particular, we notice that while the cycle $\bh_{0}$ is manifestly invariant (just looking at the figure), the cycle $\ah_{0}$ is invariant up to homology.
\subsection{Summary and arc expansion}\label{secarcex}
We have found a canonical basis for $H_{{1}}(\hat{X})$ with the symmetry properties as in \eqref{eq:symmrot}, namely
\begin{align}
\sigma_{*}^{n}(\ah_{k})&=\ah_{k+n},&\quad& \\
\quad \sigma^{n}_{*}(\bh_{k})&=\bh_{k+n}, & \sigma_{*}^{n}(\bh_{0})&=\bh_{0}; \qquad n=1,2,3 \nonumber.
\end{align}
Moreover, its limit for $\alpha \rightarrow 0$ satisfies
$$\rot_{*}^{2}\circ\sigma_{*}\left( \lim_{\alpha\rightarrow0}\bh_{i}  \right)=
 \lim_{\alpha\rightarrow0}\ah_{i},\qquad i=1,2,3,
 $$
 and also
$$\azero_{1}=\lim_{\alpha\rightarrow0}\ah_1=\abe_1,\qquad
\bzero_{1}=\lim_{\alpha\rightarrow0}\bh_1=\bbe_1,\qquad
\azero_{0}=\lim_{\alpha\rightarrow0}\ah_0=\abe_4.$$\vspace*{3mm}
These cycles can be expanded in terms of ``basic arcs'' as follows.\\
 Denote by $\gamma_k(i,j)$ the arc going from branchpoint $ \widehat{B}_{i}$ to branchpoint $ \widehat{B}_{j}$ on sheet $k$:
\begin{equation*}
\gamma_k(i,j)=\mathrm{arc}_k ( \widehat{B}_{i},\widehat{B}_{j}  ),\quad i\neq j=1,\ldots,12\quad \text{on sheet } k.  
\end{equation*}
Then we have the following
\begin{align}\label{arcexpsymmbasis}\begin{split}
\ah_1&=\gamma_1(1,2)+\gamma_2(2,1),  \quad\;\quad
\bh_1=\gamma_1(3,1)+\gamma_2(1,12)+\gamma_3(12,3),  \\
\ah_2&= \gamma_2(5,6)+\gamma_3(6,5), \quad\quad\;
\bh_2=  \gamma_2(7,5)+\gamma_3(5,4)+\gamma_1(4,7),    \\
\ah_3&= \gamma_3(9,10)+\gamma_1(10,9),   \quad\;
\bh_3=  \gamma_3(11,9)+\gamma_1(9,8)+\gamma_2(8,11),
\\
\ah_0&=\gamma_1(3,10)+\gamma_3(10,9)+\gamma_1(9,8)+\gamma_2(8,12)+\gamma_3(12,3)
, \\
\bh_0&= \gamma_1(2,8)+\gamma_{2}(8,11)+\gamma_{3}(11,4)+\gamma_{1}(4,7)+
\gamma_2(7,12)+\gamma_3(12,2)
.
\end{split}
\end{align}
and, as $\alpha\to 0$,
\begin{align}\begin{split}
\azero_1&=\gamma_1(1,2)+\gamma_{2}(2,1),\quad 
\bzero_1=\gamma_{1}(2,1)+\gamma_3(1,2),\\
\azero_2&=\gamma_{2}(3,4)+\gamma_3(4,3),\quad 
\bzero_2=\gamma_2(4,3)+\gamma_1(3,4),\\
\azero_3&=\gamma_{3}(5,6)+\gamma_1(6,5),\quad 
\bzero_3=\gamma_3(6,5)+\gamma_2(5,6), \label{arcexpzero} \\
\azero_0&=\gamma_{3}(1,2)+\gamma_1(2,6)+\gamma_{3}(6,5)+\gamma_2(5,1),\\
\bzero_0&=\gamma_{3}(1,2)+\gamma_{1}(2,5)+\gamma_2(5,6)+\gamma_3(6,3)+\gamma_1(3,4)+\gamma_2(4,1).
\end{split}
\end{align}
We point out explicitly that both these basis are indeed canonical, \ie that the intersection matrix for both is\footnote{We have chosen the minus sign to agree with the conventions of \cite{BE06}.} $-J$, where
$J=\left(
\begin{smallmatrix}
 \;\;\mathbf{O}_{4}  &   \mathbf{I}_{4}   \\
- \mathbf{I}_{4}  &   \mathbf{O}_{4} 
\end{smallmatrix}
\right).$
This can be checked by manually, computing the intersections of each pair of cycles in Figures \ref{cyc123},  \ref{cyc0}; here it has also been verified using Northover's code \texttt{extcurves}, which provides a function to find the intersection matrix of a given homology basis (see \ref{timscode}).\\

For completeness we also report here, for the case $\alpha=0$, the matrix of change of basis between the \enf{cyclic} basis, described above (cf. Figure \ref{cyc0},\ref{cyc123}), and the \enf{symmetric} basis of \cite{BE06} (cf. Figure \ref{be06fig}):
\begin{equation}\label{Mchangeb}
M:=\left[ \begin {array}{cccccccc} 1&0&0&0&0&0&0&0\\0&
1&0&0&0&0&0&0\\0&0&0&0&0&0&1&0\\0&0
&0&-1&0&0&0&-1\\-1&0&0&0&1&0&0&0\\0
&0&0&0&0&1&0&0\\0&0&-1&0&0&0&-1&0
\\0&0&0&1&0&0&0&0\end {array} \right] 
\end{equation}
This matrix has been obtained numerically using Northover's code \texttt{extcurves} for Maple (see Appendix \ref{appmaple} and the Maple file \texttt{change\_basis\_alpha0.mw}).\\
\quad
\section{The projection of the symmetric basis on the quotient curve}\label{secquotbasis}
Fay's basis projects to a basis for $H_{{1}}(X)$ as follows (see \cite{fay})
\begin{align}\label{eq:symmpiX1}
\pi_{*}(\ah_{i})&=\aq_{1},& \pi_{*}(\ah_{0})&=\aq_{0} \\
\pi_{*}(\bh_{i})&=\bq_{1},& \pi_{*}(\bh_{0})&=3 \bq_{0}; \qquad i=1,2,3. 
\end{align}
Thus to obtain such a basis one just projects the cycles of Figures \ref{cyc123}, \ref{cyc0}: the fact that the branchpoints of $\hat{X}$ do not get mapped by $\pi$ to branchpoints of $X$ (cf. section \ref{sec:quotient}) makes the projection less straightforward. This has been  implemented in Maple with Northover's \texttt{extcurves} (see Appendix \ref{appmaple} and Maple file \texttt{projected\_basis.mw}), and the result is given in Figure \ref{cycg2}.\newpage

\begin{figure}[h!]
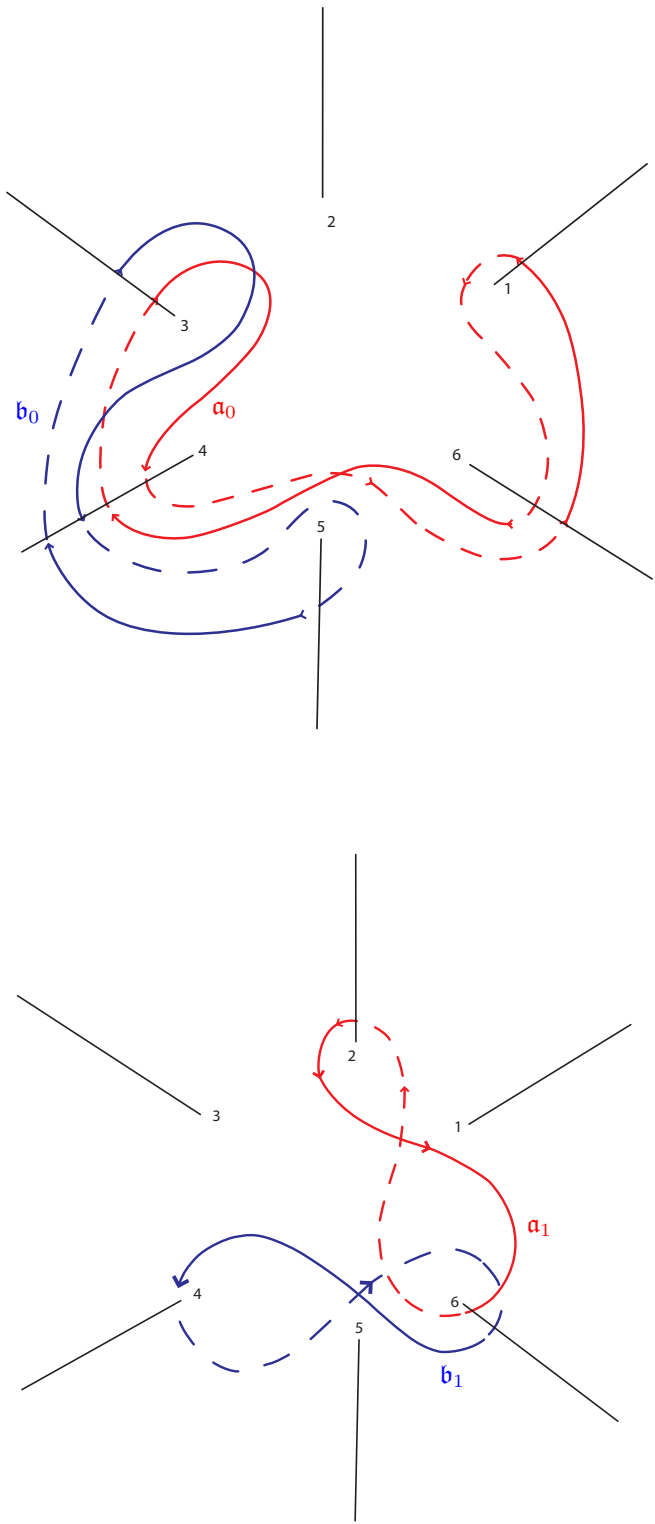

\begin{center}
\begin{lpic}[draft,clean]{cycles_g2_0(11cm,)}
    \lbl[t]{63,90;$\cred{\aq_0}$}
   \lbl[t]{22,90;${\cblu{\bq_0}}$}
   \end{lpic}
   \begin{lpic}[draft,clean]{cycles_g2_1(11cm,)}
  \lbl[t]{134,78;$\color{red}{\aq_1}$}
      \lbl[t]{115,46;${\cblu{\bq_1}}$}   \end{lpic}
\caption{Cyclic homology basis for the quotient curve $X$}\label{cycg2}
\end{center}
\end{figure}
   \;   \newpage
With the same notation as in the previous section, the arc expansion for these cycles reads:
\begin{align}\label{basisqexp} \begin{split}
\aq_{1}&=\gamma_{1}(2,6)+\gamma_{2}(6,2),\\
\bq_{1}&=\gamma_{1}(6,4)+\gamma_{2}(4,6),\\
\aq_{0}&=\gamma_{1}(3,4)+\gamma_{2}(4,6)+\gamma_{1}(6,1)+\gamma_{2}(1,6)+\gamma_{1}(6,4)+\gamma_{2}(4,3),\\
\bq_{0}&=\gamma_{1}(3,4)+\gamma_{2}(4,5)+\gamma_{1}(5,4)+\gamma_{2}(4,3).
\end{split}\end{align}
We therefore have a homology basis for the hyperelliptic curve $X$ which is different from the usual one for an hyperelliptic curve in standard form. Since it is the projection of the basis for the homology of the genus $4$ curve $\hat{X}$, it turns out that many objects on $\hat{X}$ which depend on the basis used (\eg the period matrix), can be ``reduced'', namely expressed in terms of objects on $X$.
As the curve $X$ is hyperelliptic, and in standard form, calculations are much easier in this case, thus the problem simplifies greatly.
\section{Symmetries and the period matrix}\label{symmpm}
We are now in a position to use the basis found in section \ref{hombasis} to obtain a very simple form for the period matrix of $\Xhat$.
Consider all the symmetries of the monopole curve, the real involution $\tau$ (\ref{realinv}), the inversion $\phi$ (\ref{inv}), and the cyclic symmetry $\sigma$ (\ref{cyclic}).
These maps lift as maps from $H_{1}(\hat{X},\mathbb{Z})$ to $H_{1}(\hat{X},\mathbb{Z})$, and their matrix form in the basis of Figures \ref{cyc123},  \ref{cyc0} is
\begin{small}
\begin{multline}\label{tauM}
M_{\tau}:=\left( \begin {array}{rrrrrrrrr} 
2&0&0&0&0&-1&-1&-1\\
1&0&1&1&-1&0&1&1\\
1&1&0&1&-1&1&0&1\\
1&1&1&0&-1&1&1&0\\
6&1&1&1&-2&-1&-1&-1\\
1&0&0&0&0&0&-1&-1\\
1&0&0&0&0&-1&0&-1\\
1&0&0&0&0&-1&-1&0
\end {array} 
 \right);\quad M_{\sigma}:=
\left(
\begin{array}{cccccccc}
 1 &  0 & 0 & 0  & 0 & 0  & 0 &0  \\
 0 & 0  & 1 & 0 & 0  & 0 & 0  & 0   \\
 0 & 0  & 0 & 1 & 0  & 0 & 0  & 0   \\   
 0 & 1 & 0 & 0 & 0  & 0 & 0  & 0   \\
 0 & 0  & 0 & 0 & 1  & 0 & 0  & 0   \\
 0 & 0  & 0 & 0 & 0  & 0 & 1  & 0   \\
 0 & 0  & 0 & 0 & 0  & 0 & 0  & 1   \\
 0 & 0  & 0 & 0 & 0  & 1 & 0  & 0   
\end{array}
\right);\\
M_{\varphi}:=\left( \begin{array}{rrrrrrrr} 
-1&0&0 &0&0&0&0&0\\
0&0&-1&0&0&0&0&0\\
0&-1&0 &0&0&0&0&0\\
0&0&0&-1&0&0&0&0\\
0&0&0&0&-1&0&0&0\\
0&0&0&0&0&0&-1&0\\
0&0&0&0&0&-1&0&0\\
0&0&0&0&0&0&0&-1
\end {array} \right).\qquad\quad
\end{multline}
\end{small}
The first two are found using Northover's \texttt{extcurves} for Maple  , while the last follows from the very defining properties of the symmetric basis (cf. section \ref{a0invariant}) and has indeed been checked with Northover's code.\\ 

These symmetries hugely simplify the period matrix as follows.\\
Firstly, we exploit the cyclic the symmetry $\sigma$; it acts on the differentials as follows
\begin{align}
\sigma^{*}\uu{1}&=\rho^{2}\uu{1}, & \sigma^{*}\uu{2}&=\uu{2},\\
\sigma^{*}\uu{3}&=\rho\;\uu{3}, &\sigma^{*}\uu{4}&=\uu{4}.
\end{align}
Hence, the matrices of periods take the following form
\begin{equation}\label{pihatsigma}
\hat{\mathcal{A}}=\left( \begin{array}{cccc} 
0&x_{{0}}&0&y_{{0}}\\
z_{{1}}&x_{{1}}&w_{{1}}&y_{{1}}\\
{\rho}^{2}z_{{1}}&x_{
{1}}&\rho\,w_{{1}}&y_{{1}}\\
\rho\,z_{{1}}&x_{{1}}&{\rho}^{2}w_{{1}}&y_{{1}}\end {array}\right), \quad
\hat{\mathcal{B}}= \left(\begin{array}{cccc}
0&X_{{0}}&0&Y_{{0}}\\
Z_{{1}}&X_{{1}}&W_{{1}}&Y_{{1}}\\
{\rho}^{2}Z_{{1}}&X_{{1}}&\rho\,W_{{1}}&Y_{{1}}\\
\rho\,Z_{{1}}&X_{{1}}&{\rho}^{2}W_{{1}}&Y_{{1}}
\end {array} \right). 
\end{equation}
For instance, we get
\begin{align*}
z_{0}&=\oint_{\ah_{0}}\uu{1}=\oint_{\sigma_\ast\mathfrak{a}_0}\uu{1}=\oint_{\mathfrak{a}_0}\sigma\sp\ast \uu{1}=\oint_{\mathfrak{a}_0}\rho\sp2  \uu{1}=\rho\sp2 z_0\qquad \Rightarrow z_{0}=0,\\
z_2&=\oint_{\mathfrak{a}_2}\uu{1}=\oint_{\sigma_\ast\mathfrak{a}_1}\uu{1}=
\oint_{\mathfrak{a}_1}\sigma\sp\ast
\uu{1}=\oint_{\mathfrak{a}_1}\rho\sp2  \uu{1}=\rho\sp2\,z_1,
\end{align*}
and similarly with the others, to obtain \eqref{pihatsigma}.\\

We now use the other symmetry $\varphi$, acting on the holomorphic differentials as follows
\begin{equation}\label{phi13}
\varphi^{*}\uu{1}=\uu{3}, \quad \varphi^{*}\uu{3}=\uu{1}.
\end{equation}
With considerations similar to those reported above for $\sigma$,  we have the following identities
\begin{align}
w_{2}&=-z_{1}, & w_{1}&=-z_{2}, & w_{3}&=-z_{3},\label{phiida}\\
W_{2}&=-Z_{1},  &W_{1}&=-Z_{2}, & W_{3}&=-Z_{3}.\label{phiidb}
\end{align}

Hence the matrices of periods simplify to
\begin{equation}\label{pihatsigma1}
\hat{\mathcal{A}}=\left( \begin{array}{cccc} 
0&x_{{0}}&0&y_{{0}}\\
z_{{1}} & x_{{1}}& -\rho^{2}z_{1} & y_{{1}}\\
{\rho}^{2}z_{{1}}&x_{{1}}&-z_{1}&y_{{1}}\\
\rho\,z_{{1}}&x_{{1}}&-\rho z_{1}&y_{{1}}\end {array}\right),\quad
\hat{\mathcal{B}}= \left(\begin{array}{cccc}
0&X_{{0}}&0&Y_{{0}}\\
Z_{{1}}&X_{{1}}&-\rho^{2}Z_{1}&Y_{{1}}\\
{\rho}^{2}Z_{{1}}&X_{{1}}&-Z_{1}&Y_{{1}}\\
\rho\,Z_{{1}}&X_{{1}}&-\rho Z_{1}&Y_{{1}}
\end {array} \right). 
\end{equation}

As for the real involution, one gets:
\begin{align}
\tau^{*}\uu{1}&=\;\;\overline{\uu{3}}, & \tau^{*}\uu{3}&=\overline{\uu{1}}, \label{tau13}\\
\tau^{*}\uu{2}&=-\overline{\uu{2}}, & \tau^{*}\uu{4}&=-\overline{\uu{4}}. \label{tau24}
\end{align}
These relations, together with the above expression \eqref{tauM} for $M_{\tau}$, allow us to simplify further the structure of the period matrix. We have then that
\begin{align}
Z_{1}&=-(\bar{w}_{1}+z_{1})=\rho\bar{z}_{1}-z_{1}, &W_{1}&=-(\bar{z}_{1}+w_{1})=\rho^{2}z_{1}-\bar{z}_{1}, \label{ZWtau}\\
3X_{1}&=2x_{0}+\bar{x}_{0},  & 3Y_{1}&=2y_{0}+\bar{y}_{0}, \\
X_{0}&=2x_{1}+x_{0}+\bar{x}_{1}+2X_{1}, &
Y_{0}&=2y_{1}+y_{0}+\bar{y}_{1}+2Y_{1}. 
\end{align}
We prove for instance the first one:
\begin{align*}
\overline{w_{1}}&=\int_{\ah_{1}}\overline{\uu{3}}=\int_{\ah_{1}}\tau^{*}\uu{1}=\int_{\tau_{*} \bh_{1}}\uu{1} && \text{by \eqref{tau13}}\\
\quad&=\left(\int_{\ah_{0}}+\int_{\ah_{2}} +\int_{\ah_{3}}-\int_{\bh_{0}}+\int_{\bh_{2}} +\int_{\bh_{3}}\right)\uu{1}  && \text{by \eqref{tauM}}\\
\quad&=z_{1}(\rho^{2}+\rho)+Z_{1}(\rho+\rho^{2}) &\Rightarrow &\qquad Z_{1}=-(\bar{w}_{1}+z_{1})=\rho\bar{z}_{1}-z_{1}.
\end{align*}

Note that the expressions in \eqref{ZWtau} automatically satisfy \eqref{phiida}, \eqref{phiidb}. Also, the Riemann bilinear relations are satisfied as well.\\
Moreover, they can be rearranged with the identities above to get some new relations
\begin{equation}
 \bar{Z}_1=-\rho^{2}Z_{1}, \quad W_1=\bar{Z}_1.
\end{equation}

Finally, the Riemann period matrix takes the following form
\begin{equation}
\hat{\tau}=\left( \begin{array}{cccc}
a & b&b&b\\
b&c&d&d\\
b&d&c&d\\
b&d&d&c
\end{array}\right),
\end{equation}
in accordance with Fay-Accola Theorem \ref{fayaccolathm2}; here we are also able to find explicitly  the quantities $a$,$b$, $c$, $d$  to be
\begin{align}%
a&=\dfrac{x_1Y_0-X_0y_1}{x_1y_0-x_0y_1}, &\quad b&=\dfrac{x_1Y_1-X_1y_1}{x_1y_0-x_0y_1},\\
c&=\dfrac{2}{3}\frac{Z_1}{z_1}-\frac13\frac{x_0Y_1-X_1y_0}{x_1y_0-x_0y_1},
&\quad d&=-\dfrac{1}{3}\frac{Z_1}{z_1}-\frac13\frac{x_0Y_1-X_1y_0}{x_1y_0-x_0y_1}.%
\end{align}
\section{An alternative form for the period matrix when $\alpha=0$}\label{altpm}
We know the expression of Fay's cyclic basis in the case $\alpha=0$ (see the upper parts of Figures \ref{cyc123}, \ref{cyc0} ); in fact, this particular case was indeed our starting point in section \ref{hombasis}. To this homology basis corresponds a period matrix with a particularly simple form, different than those found in the literature thus far (see \cite{BE06} and references therein). In this section we give this alternative simple form.\\

Firstly, we remark that in this case several simplifications occur: in particular, as pointed out in section \ref{sectesbe06}\footnote{Due to a different choice of the ordering for the differentials, the formulae presented here agree with those in \cite{BE06} up to a permutation of the subscripts of $\mathcal{I},\mathcal{J}$.}, all the integrals reduce to just eight, namely
\begin{align*}
\int\limits_0^{\tilde{\alpha}}{d}u_i=\mathcal{I}_i,\quad
\int\limits_0^{\tilde{\beta}}{d}u_i=\mathcal{J}_i,\quad
i=1,\ldots,4.
\end{align*}
where $\lambda_{1}=(\tilde{\alpha}, 0), \lambda_{4}=(\tilde{\beta}, 0)$ (we recall that the branchpoints $\lambda_{i}$ are defined in \eqref{lambda}) and $\tilde{\beta}=\frac{1}{\tilde{\alpha}}$, and it is understood that these integrals are computed on the first sheet.\\
This is because, making use of the symmetries of the curve, we see that
\begin{align*}
\int\limits_{\gamma_{i}(0,\rot^{k}B_{1})}\uzero{1}&=\rho^{i+k}\mathcal{I}_{1}(\tilde{\alpha}),\quad\quad&
\int\limits_{\gamma_{i}(0,\rot^{k}B_{4})}\uzero{1}&=\rho^{i+k}\mathcal{J}_{1} (\tilde{\alpha}),\\
\int\limits_{\gamma_{i}(0,\rot^{k}B_{1})}\uzero{2}&=\rho^{i+2k}\mathcal{I}_{2}(\tilde{\alpha}),\quad\quad&
\int\limits_{\gamma_{i}(0,\rot^{k}B_{4})}\uzero{2}&=\rho^{i+2k}\mathcal{J}_{2} (\tilde{\alpha}),\\
\int\limits_{\gamma_{i}(0,\rot^{k}B_{1})}\uzero{3}&=\rho^{i}\mathcal{I}_{3}(\tilde{\alpha}),\quad\quad&
\int\limits_{\gamma_{i}(0,\rot^{k}B_{4})}\uzero{3}&=\rho^{i}\mathcal{J}_{3} (\tilde{\alpha}),\\
\int\limits_{\gamma_{i}(0,\rot^{k}B_{1})}\uzero{4}&=\rho^{i+k}\mathcal{I}_{4}(\tilde{\alpha}),\quad\quad&
\int\limits_{\gamma_{i}(0,\rot^{k}B_{4})}\uzero{4}&=\rho^{i+k}\mathcal{J}_{4} (\tilde{\alpha}),
\end{align*}
where again the integrals $\mathcal{I}_{i},\mathcal{J}_{i}$ are computed on the first sheet.\\
These integrals can be expressed in terms of hypergeometric functions as follows (cf. \cite{BE06}, p. 50\footnote{See the caveat in the previous footnote.}):
\begin{align*}
\mathcal{I}_1(\tilde{\alpha})&= \frac{4\pi^2}{9\Gamma\left(
\frac23\right)^3}\frac{\tilde{\alpha}}{(1+\tilde{\alpha}^6)^{\frac13}},\\
\mathcal{J}_1(\tilde{\alpha})&=-\frac{4\pi^2}{9\Gamma\left(
\frac23\right)^3}\frac{\tilde{\alpha}}{(1+\tilde{\alpha}^6)^{\frac13}},\\
\mathcal{I}_2(\tilde{\alpha})&=
\frac{2\pi\tilde{\alpha}^2}{3\sqrt{3}}\,{_2F_1}\left(\frac23,\frac23;1;-\tilde{\alpha}^6\right)
=\frac{2\pi}{3\sqrt{3}}\,\frac{\tilde{\alpha}^2}{(1+\tilde{\alpha}^6)\sp\frac23}
\ {_2F_1}(\frac{1}{3}, \frac{2}{3}; 1,t),
\end{align*}
\begin{align*}
\mathcal{J}_2(\tilde{\alpha})&=
\frac{2\pi}{3\sqrt{3}\tilde{\alpha}^2}{_2F_1}\left(\frac23,\frac23;1;-\frac{1}{\tilde{\alpha}^{6}}
\right)=\frac{2\pi}{3\sqrt{3}}\,\frac{\tilde{\alpha}^2}{(1+\tilde{\alpha}^6)\sp\frac23}
\ {_2F_1}(\frac{1}{3}, \frac{2}{3}; 1,1-t),\\
\mathcal{I}_3(\tilde{\alpha})&= \tilde{\alpha}^3\, {_2F_1}\left(\frac23,1;\frac43;
-\tilde{\alpha}^6\right),\\
\mathcal{J}_3(\tilde{\alpha})&=-\frac{1}{\tilde{\alpha}^3}\,
{_2F_1}\left(\frac23,1;\frac43;-\frac{1}{\tilde{\alpha}^{6}} \right),\\
\mathcal{I}_4(\tilde{\alpha})&=
-\frac{2\pi\tilde{\alpha}}{3\sqrt{3}}\,{_2F_1}\left(\frac13,\frac13;1;-\tilde{\alpha}^6\right)
=-\frac{2\pi}{3\sqrt{3}}\,\frac{\tilde{\alpha}}{(1+\tilde{\alpha}^6)\sp\frac13}
\ {_2F_1}(\frac{1}{3}, \frac{2}{3}; 1,t),\\
\mathcal{J}_4(\tilde{\alpha})&=
\frac{2\pi}{3\sqrt{3}\tilde{\alpha}}\,{_2F_1}\left(\frac13,\frac13;1;-\frac{1}{\tilde{\alpha}^{6}}
\right)=\frac{2\pi}{3\sqrt{3}}\,\frac{\tilde{\alpha}}{(1+\tilde{\alpha}^6)\sp\frac13}
\ {_2F_1}(\frac{1}{3}, \frac{2}{3}; 1,1-t),
\end{align*}
with $t=\tilde{\alpha}^6/(1+\tilde{\alpha}^6)$.

One can check that the following relations hold
\begin{equation}\label{rel}
\mathcal{R} \equiv\frac{\mathcal{I}_1(\tilde{\alpha})}{\mathcal{J}_1(\tilde{\alpha}) }
=-\frac{\mathcal{I}_3(\tilde{\alpha})}{\mathcal{J}_3(\tilde{\alpha}) }, \qquad
\mathcal{I}_2(\tilde{\alpha})+\mathcal{J}_2(\tilde{\alpha})=0, \qquad
\mathcal{I}_4(\tilde{\alpha})-\mathcal{J}_4(\tilde{\alpha})=\mathcal{I}_2(\tilde{\alpha}).
\end{equation}

Using the above expressions  for the integrals, and relations \eqref{rel}, in the arc expansion \eqref{arcexpzero} of the elements of the cyclic basis in the case $\alpha=0$, the matrices of periods simplify as follows
\begin{align}\begin{split}
\mathcal{A}_{0}&=\left( \begin {array}{cccc}  
 0&   \left( -3\,R-3 \right) \rho-3\,R   &0&   \left( -3\,R+3 \right) \rho+3  \\\noalign{\medskip}
 1+2\,\rho &   \left( 2-R \right) \rho+1+R  &  -1+\rho   &  \left( -2-R \right) \rho-1-2\,R  \\\noalign{\medskip}
   1-\rho   &  \left( 2-R \right) \rho+1+R   & -1-2\,\rho  &  \left( -2-R \right) \rho-1-2\,R  \\\noalign{\medskip} 
   -\rho-2     &  \left( 2-R \right) \rho+1+R   &  \rho+2  &   \left( -2-R \right) \rho-1-2\,R 
  \end {array} \right)
\left(\begin{array}{cccc} \mathcal{I}_1&&&\\
                           &\mathcal{J}_2\\
                           &&\mathcal{I}_1\\
                           &&&\mathcal{J}_4
\end{array}\right). \\
\mathcal{B}_{0}&= \left( \begin {array}{cccc} 
0&  \left( -6\,R-3 \right) \rho-3 \,R+3 & 0 &   \left( -6\,R+3 \right) \rho-3\,R+6 \\\noalign{\medskip}
1-\rho  & \left( -1-R \right) \rho+1-2\,R  & \rho+2 &  \left( 1-R \right) \rho+2+R \\\noalign{\medskip} 
-\rho-2  &\left( -1-R \right) \rho+1-2\,R  &  -1+\rho &   \left( 1-R \right) \rho+2+R  \\\noalign{\medskip} 
 1+2\,\rho  &  \left( -1-R \right) \rho+1-2\,R  &  -1-2\,\rho  & \left( 1-R \right) \rho+2+R 
  \end {array} \right)
 \left(\begin{array}{cccc} \mathcal{I}_1&&&\\
                           &\mathcal{J}_2\\
                           &&\mathcal{I}_1\\
                           &&&\mathcal{J}_4
\end{array}\right). 
 \end{split}
\end{align}

We remark that transforming the period matrices above to the symmetric basis of Figure \ref{be06fig}  using the change of basis matrix \eqref{Mchangeb}, the period matrices obtained agree with those found in \cite{BE06}, eq. (7.36).

\section{The Ercolani-Sinha conditions}\label{EScyclic}

As shown in \cite{BE06}, the curve $X$, with $\alpha=0$, satisfies the Ercolani-Sinha constraints for certain values of $\beta$ and $\gamma$, and hence describes indeed a monopole.\\
In this section we express the Ercolani-Sinha constraints in the different cyclic basis, and hence manage to reduce them as relations among periods on the genus 2 quotient curve $X$: this proves crucial for the solution, as will be shown in Chapter \ref{AGMchapter}.\\

The Ercolani-Sinha conditions here read
$$\mathbf{n}\sp{T}\hat{\mathcal{A}}+\mathbf{m}\sp{T}\hat{\mathcal{B}}=-2\nu\,(0,0,0,1).$$
In solving the first and third of these conditions, using \eqref{pihatsigma1} one  deduces that $n_i=n_j$ and $m_i=m_j$. Hence the remaining equations can be expressed as
\begin{equation}\label{es1}
(n_0,3n)\begin{pmatrix}x_0&y_0\\x_1&y_1\end{pmatrix}+
(m_0,3m)\begin{pmatrix}X_0&Y_0\\X_1&Y_1\end{pmatrix}=-2\nu(0,1),
\end{equation}
which are the Ercolani-Sinha equations on the reduced curve with
$n=n_{i}$, $m=m_{i}$. Alternatively, in terms of the periods on the genus 2 curve $X$, one has
\begin{align}\label{esg2}
(n_0,3n)\mathcal{A}+
(3m_0,3m)\mathcal{B}=6\nu(0,1),
\end{align}
Note that this also implies that
\begin{equation}\label{UUhat}
\hat{U}=\pi^{*}U.
\end{equation}
Also, for the Ercolani-Sinha vector of Theorem \ref{hmrthm}, 
one has
\begin{equation}\label{esvectq}
\hat{\mathfrak{c}}=\pi^{*}(\mathfrak{c}),\qquad \mathrm{where}\;
\mathfrak{c}=n_{0}\aq_{0}+3n\aq_{1}+3m_{0}\bq_{0}+3m\bq_{1}
\end{equation}

One can see that condition \eqref{es1} is satisfied at $\alpha=0$, $\gamma=\pm5 \sqrt{2}$ and
\begin{align*}
R&=-{\frac {2\,n-m}{m+n}},& { n_{0}}&=5\,n-m,&{ m_{0}}&=-3\,n.
\end{align*}
We remark that these are indeed the same values obtained in \cite{BE06}, as one can easily check transforming the Ecolani-Sinha vector of Proposition \ref{propbe09} from one basis to the other using the matrix \eqref{Mchangeb} (see Maple file \texttt{change\_basis\_alpha0.mw}, final part).\\

\numberwithin{equation}{section}
\section{The vector of Riemann constants}
Below we  calculate the vector of Riemann constants for the genus 2 curve $X$
\begin{equation}
y^{2}=(x^{3}+\alpha x +\gamma)^{2}+4\beta^{2}.
\end{equation}

For applications we need the basepoint for the Abel map to be one of the points at infinity, say $P_{+}$; but, in fact, it is easier for our calculations to choose one of the branchpoints, say $B_{1}$; then, to obtain back the vector of Riemann constants with respect to $P_{+}$ we use the following relation
\begin{equation}\label{b1infty}
\mathcal{K}_{P_{+}}=\abelmap_{P_{+}}(B_{1})+\mathcal{K}_{B_{1}},
\end{equation}
where $\abelmap_{P_{+}}$ is the Abel map with basepoint ${P_{+}}$.
\subsection{The vector \texorpdfstring{$\mathcal{K}_{B_{1}}$}{KB1}}
In the following we use the basis for the curve $X$ as in Figure 2.
The arc expansions for the cycles of eq. \eqref{basisqexp}  can be equivalently expressed as follows
\begin{align*}
\aq_{1}&=\gamma_{1}(2,1)+\gamma_{1}(1,6)+\gamma_{2}(6,1)+\gamma_{2}(1,2),\\
\bq_{1}&=\gamma_{1}(6,5)+\gamma_{1}(5,4)+\gamma_{2}(4,5)+\gamma_{2}(5,6),\\
\aq_{0}&=\gamma_{1}(3,4)+\gamma_{2}(4,5)+\gamma_{2}(5,6)+\gamma_{1}(6,1)+\gamma_{2}(1,6)+\gamma_{1}(6,5)+\gamma_{1}(5,4)+\gamma_{2}(4,3),\\
\bq_{0}&=\gamma_{1}(3,4)+\gamma_{2}(4,5)+\gamma_{1}(5,4)+\gamma_{2}(4,3).
\end{align*}
To simplify the above expressions, we proceed as in Farkas and Kra (\cite{farkaskra} VII.1.2). Using that under   the hyperelliptic involution $J:(x,y)\rightarrow(x,-y)$ one has
\begin{equation*}
\gamma_{k+1}(2j+1,2j)=-J\gamma_{k}(2j,2j+1),
\end{equation*}
which implies that
\begin{align*}
\int_{\gamma_{k}(6,5)}\boldsymbol{\omega}=\left(\int_{\gamma_{k}(1,2)}+\int_{\gamma_{k}(3,4)}\right)\boldsymbol{\omega},\qquad \qquad
\int_{\gamma_{k}(1,6)}\boldsymbol{\omega}=\left(\int_{\gamma_{k}(2,3)}+\int_{\gamma_{k}(4,5)}\right)\boldsymbol{\omega}.
\end{align*}
Using these relations, one finds
\begin{align*}
\int_{\ah_{1}}\boldsymbol{\omega}&=
2\left(\int_{\gamma_{1}(2,1)}\boldsymbol{\omega}+\int_{\gamma_{1}(2,3)}\boldsymbol{\omega}+\int_{\gamma_{1}(4,5)}\boldsymbol{\omega}\right),\\
\int_{\bh_{1}}\boldsymbol{\omega}&=
2\left(\int_{\gamma_{1}(1,2)}\boldsymbol{\omega}+\int_{\gamma_{1}(3,4)}\boldsymbol{\omega}+\int_{\gamma_{1}(5,4)}\boldsymbol{\omega}\right),\\
\int_{\ah_{0}}\boldsymbol{\omega}&=
2\left(\int_{\gamma_{1}(3,4)}\boldsymbol{\omega}+\int_{\gamma_{1}(1,2)}\boldsymbol{\omega}+\int_{\gamma_{1}(3,4)}\boldsymbol{\omega}+\int_{\gamma_{1}(5,4)}\boldsymbol{\omega}
\int_{\gamma_{1}(3,2)}\boldsymbol{\omega}+\int_{\gamma_{1}(5,4)}\boldsymbol{\omega}\right),\\
\int_{\bh_{0}}\boldsymbol{\omega}&=
2\left(\int_{\gamma_{1}(3,4)}\boldsymbol{\omega}+\int_{\gamma_{1}(5,4)}\boldsymbol{\omega}\right).
\end{align*}
From these one can solve to obtain the integrals between branchpoints as follows:
\begin{align*}
\int_{\gamma_{1}(1,2)}\boldsymbol{\omega}&=\frac{1}{2}(\pi^{(0)}+\pi^{(1)}),\\
\int_{\gamma_{1}(2,3)}\boldsymbol{\omega}&=\frac{1}{2}(e^{(0)}+\pi^{(0)}+\pi^{(1)}),\\
\int_{\gamma_{1}(3,4)}\boldsymbol{\omega}&=\frac{1}{2}(e^{(0)}+e^{(1)}+\pi^{(0)}),\\
\int_{\gamma_{1}(4,5)}\boldsymbol{\omega}&=\frac{1}{2}(e^{(0)}+e^{(1)}),\\
\int_{\gamma_{1}(5,6)}\boldsymbol{\omega}&=\frac{1}{2}(e^{(0)}+e^{(1)}+\pi^{(1)}).
\end{align*}
Hence one can easily deduce the image under the Abel map, with basepoint $B_{1}$, of each branchpoints (their characteristics are also given)
\begin{align*}
\abelmap_{B_{1}}(B_{1})&=0 
&\rightarrow&\;\;\,\left[ \begin{array}{cc}
0&0\\
0&0
\end{array}\right],\\
\abelmap_{B_{1}}(B_{2})&=\frac{1}{2}(\pi^{(0)}+\pi^{(1)})
&\rightarrow&\frac{1}{2}\left[ \begin{array}{cc}
1&1\\
0&0
\end{array}\right],\\
\abelmap_{B_{1}}(B_{3})&=\frac{1}{2}e^{(0)}
&\rightarrow&\frac{1}{2}\left[ \begin{array}{cc}
0&0\\
1&0
\end{array}\right],\\
\abelmap_{B_{1}}(B_{4})&=\frac{1}{2}(e^{(1)}+\pi^{(0)})
&\rightarrow&\frac{1}{2}\left[ \begin{array}{cc}
1&0\\
0&1
\end{array}\right],\\
\abelmap_{B_{1}}(B_{5})&=\frac{1}{2}(e^{(0)}+\pi^{(0)})
&\rightarrow&\frac{1}{2}\left[ \begin{array}{cc}
1&0\\
1&0
\end{array}\right],\\
\abelmap_{B_{1}}(B_{6})&=\frac{1}{2}(e^{(1)}+\pi^{(0)}+\pi^{(1)})
&\rightarrow&\frac{1}{2}\left[ \begin{array}{cc}
1&1\\
0&1
\end{array}\right].
\end{align*}
Note that there are only $2\,(=g)$ odd characteristics, corresponding to $B_{2}$ and $B_{6}$,  and that the rest are even.\\
Following an argument of Farkas and Kra (\cite{farkaskra} VII.1.2), we recall that $\theta\left[\begin{smallmatrix}\boldsymbol{\varepsilon\;}
\\ \boldsymbol{\varepsilon '}\end{smallmatrix}\right]$ is even/odd if the parity of the characteristic $\left[\begin{smallmatrix}\boldsymbol{\varepsilon\;}
\\ \boldsymbol{\varepsilon '}\end{smallmatrix}\right]$ is even/odd. Hence, by definition of theta with characteristic \eqref{thetachar}, the theta function vanishes at odd half periods.  Since only $g$ branchpoints have odd characteristic, we can conclude that the vector of Riemann constants takes the form
\begin{align}
\mathcal{K}_{B_{1}}&=-(\abelmap_{B_{1}}(B_{2})+\abelmap_{B_{1}}(B_{6}))\nonumber\\
&=\frac{1}{2}(e^{(0)}+e^{(1)}+\pi^{(1)})\quad\longrightarrow\quad\frac{1}{2}\left[ \begin{array}{cc}
0&1\\
1&1
\end{array}\right].
\end{align}
\subsection{The vector $\vrc_{P_{+}} $ }
Once we have calculated the vector of Riemann constants with $B_{1}$ as basepoint, making use of equation \eqref{b1infty} we can change its basepoint; to simplify future calculations, we choose one of the point at infinity, say $P_{+}$, as basepoint.\\
Numerically, using the command \texttt{AbelMap} in Maple (developed by Deconinck and Patterson\footnote{Thanks to Matt Patterson for providing additional documentation, and help in several occasions.}, see \cite{abelmap}), one can compute the Abel map of $B_{1}$ based at $P_{+}$, obtaining
\begin{equation}
\abelmap_{P_{+}}(B_{1})=\frac{2}{3}e^{(1)}+\frac{1}{2}\pi^{(1)}\quad\longrightarrow\quad\left[ \begin{array}{cc}
\frac{1}{2}&0\\
\frac{2}{3}&0
\end{array}\right].
\end{equation}
Hence, the vector of Riemann constants with basepoint $P_{+}$ is
\begin{equation}\label{vrc}
\mathcal{K}_{\infty}=\frac{1}{6}e^{(1)}+\frac{1}{2}e^{(2)}+\frac{1}{2}\pi^{(1)}+\frac{1}{2}\pi^{(2)}\quad\longrightarrow\quad\left[ \begin{array}{cc}
\frac{1}{2}&\frac{1}{2}\\
\frac{1}{6}&\frac{1}{2}
\end{array}\right].
\end{equation}
\subsection{The case $\alpha=0$}
We recall that according to the results by Fay and Accola presented in section \ref{fayaccolavrc} (see \cite{fay,accola}),  the following relation between the vectors of Riemann constants on the two curves  holds (cf. \eqref{vrcaccola})
\begin{equation}\label{vrcaccola3}
\pi^{*}\vrc_{Q}=\hat{\vrc}_{\hat{Q}}+(g-1)\sum\limits_{n=0}^{2} \abelmap( \sigma^{n}(\hat{Q}))+ \hat{\halfp}.
\end{equation}
In the case where $\alpha=0$, we have that $3\int\limits_{\BH_{i}}^{\BH_{j}}\in \Lambda$ for any branchpoint, and moreover they are sent one to the other by $\sigma$. Hence, Fay-Accola relation becomes in this case
\begin{equation}\label{vrcaccola4}
\pi^{*}\vrc_{Q}^{0}=\hat{\vrc}^{0}_{\hat{Q}}+ \hat{\halfp}.
\end{equation}
Using Fay's expression (\ref{halfpfay}) for $\halfp$, we get
\begin{equation*}
\hat{\halfp}=\pi^{*}\left ( \frac{3-1}{2\cdot3}, 0 \right) =\left ( 1, 0, 0, 0\right)\equiv \mathbf{0}.
\end{equation*}
In \cite{BE06} (section 5.3) the vector of Riemann constants $\hat{\mathcal{K}}^{0}_{\hat{\infty}}$ for the curve $\Xzero$ is computed to be a half period, namely
$$
\hat{\mathcal{K}}^{0}_{\hat{\infty}}=\dfrac12\left[\begin{matrix}1&1&1&1\\1&1&1&1\end{matrix}\right].
$$
In view of the Fay-Accola relation \eqref{vrcaccola3}, the expression \eqref{vrc} is indeed consistent with this result.
\section{Theta functions}
For completeness we observe also that applying  Fay-Accola Theorem \ref{fayaccolathm2} we obtain a simple expression for the theta function on $\Xhat$ associated to the cyclic basis of section \ref{hombasis}.
In particular we have
\begin{equation}\label{thetafay}
\frac{\hat{\theta}(3\,z_0,z_1,z_1,z_1;\hat{\tau})}
{\prod_{k=0}^{2}\theta\left[\begin{matrix}0&0
\\ \frac{k}{3}&0 \end{matrix}\right]\left(z_0,z_1;\tau%
\right)}
=c_0(\hat{\tau}%
) 
\end{equation}
where as in Fay-Accola theorem $c_0(\hat{\tau})$ is a constant.\\

Hence, the genus 4 theta function $\hat{\theta}$ can be expressed as a product of  hyperelliptic ones. This  is particularly important when attempting to solve Hitchin constraints \enf{H3} in the Ercolani-Sinha formulation \eqref{h3es}: despite this semplification, this proves to be a nontrivial task, and will be examined elsewhere. Here we only remark that, provided that the Ercolani-Sinha constraints \eqref{EScond} and \eqref{h3es} are satisfied, eq.~\eqref{thetafay} means that the monopole solutions can be expressed in terms of hyperelliptic functions.

%
%
%
%
%
\chapter{An interlude: elliptic subcovers}
\introcap{Quotienting the monopole curve $\Xhat$ by the action of the symmetry group $C_{3}$ we have obtained a genus 2 curve $X$, and are hence able to express several objects on $\Xhat$ in terms of objects on $X$, as shown in Chapter \ref{curve}. This provides a remarkable simplification of our task of imposing the Ercolani-Sinha constraints, allowing us to work in genus 2 rather than 4. Therefore, before tackling this task, we can investigate whether a further simplification is possible, namely, whether the curve $X$ also covers elliptic curve(s): if this is the case, obviously the problem would simplify much more. This is examined   in this short chapter, making use of invariant theory.}

\section{Invariants and elliptic subcovers}\label{ellsubc}
We want to investigate whether the genus 2 curve $X$ admits a \enf{maximal covering} $\psi: X\to E$ to an elliptic curve $E$, or, in other words, whether $X$ admits \enf{(degree} $\mathbf{n}$) \enf{elliptic subcovers}. Here \emph{maximal} means that the map  $\psi: X\to E$ does not factor over an unramified cover of $E$.\\
We remark here that such degree $n$ maximal elliptic subcovers always occur in pairs,  $(E,E')$; there is an isogeny of degree $n^{2}$ between the Jacobian of $X$, $\Jac(X)$ and the product $E\times E'$: one says that \enf{$X$ has $(\mathbf{n},\mathbf{n})$ split Jacobian}.\\

Here we only consider the case of a \enf{degree 2} elliptic subcover, as this is indeed the case for $\alpha=0$, as shown in \cite{BE06}; for continuity, one expects then that also in the case $\alpha\neq 0$, if there is an elliptic cover at all, this should be of degree 2.\\

Before analysing this case, let us introduce the invariants in some generality.

\subsection{Igusa invariants for a genus 2 curve}
The $j$-invariants were introduced first by Clebsh and Bolza \cite{clebinv,bolzainv}, and generalised to characteristic different from 2 by Igusa \cite{igusa}; many results used here are due to Shaska (see \cite{shaska} for a review). 
These invariants are used to study the properties of moduli of genus $2$ curves in terms of the associated sextic. \\
Recall that every genus 2 curve can be represented as a sextic\footnote{
This is due to the fact that every genus 2 curve has a representation as a double cover of $\puno$ with six ramification points $x_{i}$. This establishes a bijection between isomorphism classes of such curves and binary sextics, as every unordered 6-tuple of points can be described by such a sextic.
}, given in general form as:
\begin{align*}
\;&a_0x^6+a_1x^5+a_2x^4+a_3x^3+a_4x^2+a_5x+a_6=\\
\;&a_0(x-\alpha_1)(x-\alpha_2)(x-\alpha_3)(x-\alpha_4)(x-\alpha_5)
(x-\alpha_6)
\end{align*}
One can define an \enf{integral (or relative) invariant $J$}, as a continuous function on the space of sextics if it is invariant under the natural action of the group $SL(2)$. One can explicitly build these relative invariants (\cite{igusa}, section 2); they are homogeneous polynomials of weights 2, 4, 6 and 10, respectively given as
\begin{align*}
j_2=a_0^2\sum_{i< j< k< l< m< n}&
(\alpha_{i}-\alpha_{j})^2(\alpha_{k}-\alpha_{l})^2(\alpha_{m}-\alpha_{n})^2,\\
j_4=a_0^4\sum_{i< j< k< l< m< n}&
(\alpha_{i}-\alpha_{j})^2(\alpha_{j}-\alpha_{k})^2(\alpha_{k}-\alpha_{i})^2
(\alpha_{l}-\alpha_{m})^2(\alpha_{m}-\alpha_{n})^2(\alpha_{n}-\alpha_{l})^2,\\
j_6=a_0^6\sum_{i< j< k< l< m< n}&
(\alpha_{i}-\alpha_{j})^2(\alpha_{j}-\alpha_{k})^2(\alpha_{k}-\alpha_{i})^2
(\alpha_{l}-\alpha_{m})^2(\alpha_{m}-\alpha_{n})^2(\alpha_{n}-\alpha_{l})^2\\
\qquad &
(\alpha_{i}-\alpha_{l})^2(\alpha_{j}-\alpha_{m})^2(\alpha_{k}-\alpha_{n})^2,\\
j_{10}=a_0^{10}\prod_{j<k}(\alpha_{j}-\alpha_{k}&)^2,
\end{align*}
with $\alpha_{i}$ roots of the sextic. Here the indices run from $1$ to $6$, so that $j_{2}$ is the sum of $15$ elements, $j_{4}$ of $10$ and $j_{6}$ of $60$.\\

One can then introduce \enf{absolute invariants}, \ie continuous functions on the set of hyperelliptic curves, assuming the same value on birationally equivalent curves (see \cite{igusa}, section 6). These absolute invariants can be expressed in terms of relative invariants as follows
\begin{equation}\label{absinv}
i_1=144 \;\frac{j_2}{j_2^2},\quad i_2=-1728\; \frac{j_2j_4-3j_6}{j_2^3},\quad i_3=486\;\frac{j_{10}}{j_2^5}.
\end{equation}

It is a necessary and sufficient condition for two genus 2 curves (with\footnote{If $j_{2}=0$ one can define new absolute invariants. } $j_{2}\neq 0$) to be isomorphic that their absolute invariants are the same.\\

\subsection{Elliptic subcovers of degree $2$}
We are now in a position to give conditions on the parameters of the curve $X$ under which it is a degree 2 cover of elliptic curves, making use of invariant theory. One can show (see \cite{shaska}, section 5) the following characterisation.
\begin{proposition}\label{vanchi30}
A (non-singular) genus 2 curve has a degree 2 elliptic subcover if and only if the automorphic form $\chi_{30}$ vanishes, where
\begin{align*}
\begin{split}\label{chi30}
\chi_{30}&=-19245600\,{i_{{10}}}^{2}i_{{4}}{i_{{2}}}^{3}+
507384000\,{i_{{10}}}^{2}{i_{{4}}}^{2}i_{{2}}+
77436\,i_{{10}}{i_{{4}}}^{3}{i_{{2}}}^{4}-81\,{i_{{2}}}^{3}{i_{{6}}}^{4}-{i_{{2}}}^{7}{i_{
{4}}}^{4}+384\,{i_{{4}}}^{6}i_{{6}}\\&-78\,{i_{{2}}}
^{5}{i_{{4}}}^{5}+31104\,{i_{{6}}}^{5}-3499200\,i
_{{10}}i_{{2}}{i_{{6}}}^{3}+104976000\,{i_{{10}}}
^{2}{i_{{2}}}^{2}i_{{6}}+972\,i_{{10}}{i_{
{2}}}^{6}{i_{{4}}}^{2}+8748\,i_{{10}}{i_{{2}}}^{4
}{i_{{6}}}^{2}\\&+1332\,{i_{{2}}}^{4}{i_{{4}}}^{4}i_{{6}}-8910\,{i_{{2}}}^{3}{i_{{4}}}^{3}{i_
{{6}}}^{2}-1728\,{i_{{4}}}^{5}{i_{{2}}}^{2}i_{{6}
}+6048\,{i_{{4}}}^{4}i_{{2}}{i_{{6}}}^{2}+29376\,
{i_{{2}}}^{2}{i_{{4}}}^{2}{i_{{6}}}^{3}-47952\,i_{{2}}i_{{4}}{i_{{6}}}^{4}\\&+41472\,i_{{10}}
{i_{{4}}}^{5}-236196\,{i_{{10}}}^{2}{i_{{2}}}^{5}
+159\,{i_{{4}}}^{6}{i_{{2}}}^{3}-80\,{i_{{4}}}^{7
}i_{{2}}-125971200000\,{i_{{10}}}^{3}-592272\,i_{
{10}}{i_{{4}}}^{4}{i_{{2}}}^{2}\\&-870912\,i_{{10}}{
i_{{2}}}^{3}{i_{{4}}}^{2}i_{{6}}+3090960\,i_{{10}}{i_{{2}}}^{2}i_{{4}}{i_{{6}}}^{2}-5832\,i_{{10}}{i_{{2}}}^{5}i_{{4}}i_{{6}}+4743360
\,i_{{10}}{i_{{4}}}^{3}i_{{2}}i_{{6}}-6912
\,{i_{{4}}}^{3}{i_{{6}}}^{3}\\&+12\,{i_{{2}}}^{6}{i_{{4}}}^{3}i_{{6}}-54\,{i_{{2}}}^{5}{i_{{4
}}}^{2}{i_{{6}}}^{2}+108\,{i_{{2}}}^{4}i_{{4}}{{
\it jj}_{{6}}}^{3}-2099520000\,{i_{{10}}}^{2}i_{{4}}{i}_{{6}}-9331200\,i_{{10}}{i_{{4}}}^{2}{i_{{
6}}}^{2}
\end{split}\end{align*}
\end{proposition}
\begin{proof}
If a  non-singular genus 2 curve has a degree two elliptic subcover, then it can be put in the form
\[ Y^2=X^6-s_1X^4+s_2X^2-1, \]
with $27-18s_1s_2-s_1^2s_2^2+4s_1^3+4s_2^3\neq 0$ (see  \cite{shaska04} for a proof). Note that this determines $X$  up to a coordinate change  in the subgroup of the automorphism group of $\Sigma$ generated by $(\tau_{1},\tau_{2})$, where $\tau_{1}:x \to 1/x$ and  $\tau_{2}: x\to\rho_{6}x$, where $\rho_{6}$ is a $6$-th root of unity. Invariants for this action are: 
\[ u=s_1s_2,\qquad v=s_1^3+s_2^3. \]
In terms of these new variables, the relative invariants can be written as
\begin{align*}
j_2&=240+16 u,\\
j_4&=48v+4u^2+1620-504u,\\
j_6&=-2066 u+96 v -424 u^2 +24 u^3 +160 uv +119880,\\
j_{10}&=64(27-18u-u^2+4v)^2.
\end{align*}
Using these expressions, the absolute invariants $i_1,i_2,i_3$ can be expressed as rational functions of $u$ and $v$,  for $j_2\neq0$. Eliminating these variables  and using eq. (\ref{absinv})  give the required expression for $\chi_{30}$.\\
For more details about these calculations we refer to the Maple file \texttt{invariants4.mw}.
\end{proof}
\section{Igusa invariants for the curve $X$}
Let us now apply the previous considerations to the curve $X$ of equation \eqref{quotientcurve}, where we can rescale to set $\beta=1$ (see section \ref{AGMessect}):
\[ y^{2} = (x^3+\alpha x + \gamma)^2+4. \]
The Igusa invariants can be expressed in terms of the parameters $\alpha,\gamma$ in the following way (see Maple file \texttt{invariants4.mw}):
\begin{align*}
i_{{2}}=&-32\,{\alpha}^{3}-216\,{\gamma}^{2}-960,\\
i_{{4}}=& \,64\,{\alpha}^{6}+864\,{\alpha}^{3}{\gamma}^{2}-2496\,{\alpha}^{3}+2916\,{\gamma}^{4}+18144\,{\gamma}^{2}+
25920,\\
i_{{6}}=&-157464\,{\gamma}^{6}-1749600\,{\gamma}^{4}-6397056\,{\gamma}^{2}-7672320-69984\,{\alpha}^{3}{\gamma}^{4}\\ \qquad&-285120\,{\alpha}^{3}{\gamma}^{2}-10368\,{\alpha}^{6}{\gamma}^{2}+648960\,{\alpha}^{3}-3840\,{\alpha}^{6}-512\,{\alpha}^{9},\\
i_{{10}}=&-47775744-23887872\,{\gamma}^{2}-884736\,{\alpha}^{3}{\gamma}^{2}+3538944\,{\alpha}^{3}- 2985984\,{\gamma}^{4}-65536\,{\alpha}^{6}.
\end{align*}
Substituting these in the expression above for $\chi_{30}$, the condition of vanishing of Proposition \ref{vanchi30} reads
\begin{align*}
{\gamma}^{2}{\alpha}^{6} &\left( -8000000+1146000\,{\alpha}^{
3}-53088\,{\alpha}^{6}+784\,{\alpha}^{9}-6480000\,{\gamma}^{2}+327240\,{\alpha}^{3}{\gamma}^{2}+7128\,{\gamma}^{2}{\alpha}^{6}\right)\\&\left.-1749600\,{\gamma}^{4}-10935\,{\gamma}^{4}{\alpha}^{3}-157464\,{\gamma}^{6} \right) ^{2}=0.
\end{align*}
From this expression it is evident that $\chi_{30}$ vanishes identically when $\alpha=0$, hence the curve $X$ with $\alpha=0$ always  has an elliptic subcover; this is in agreement with the results from \cite{BE06} (lemma 7.2, also summarised earlier in section \ref{coversbe06}).\\
As for the case $\alpha\neq 0$, analysing the above equation one sees that there are real solutions for some values of $\alpha$ and $\gamma$, as plotted in Figure \ref{plot_chi}, where $\gamma$ is given in function of $\alpha$. In contrast with the case $\alpha=0$, then, the curve $X$ does not admit an elliptic subcover for all values of the parameters, but just for some special ones.\\

We come back to this graph in the last chapter, when we analyse solutions to the Ercolani-Sinha constraints \eqref{esg2}. In particular, in section \ref{secagmes} we find a curve of solutions to  \eqref{esg2} in the $\alpha, \gamma$ plane. We can hence examine the intersection of this curve with that of Figure \ref{plot_chi}, in order to find those Riemann surfaces solutions of the Ercolani-Sinha that also admit an elliptic subcover: these are particularly interesting as they correspond to monopoles solutions expressible in terms of elliptic functions.

\begin{figure}[t!]
\begin{center}
\includegraphics[height=350pt]{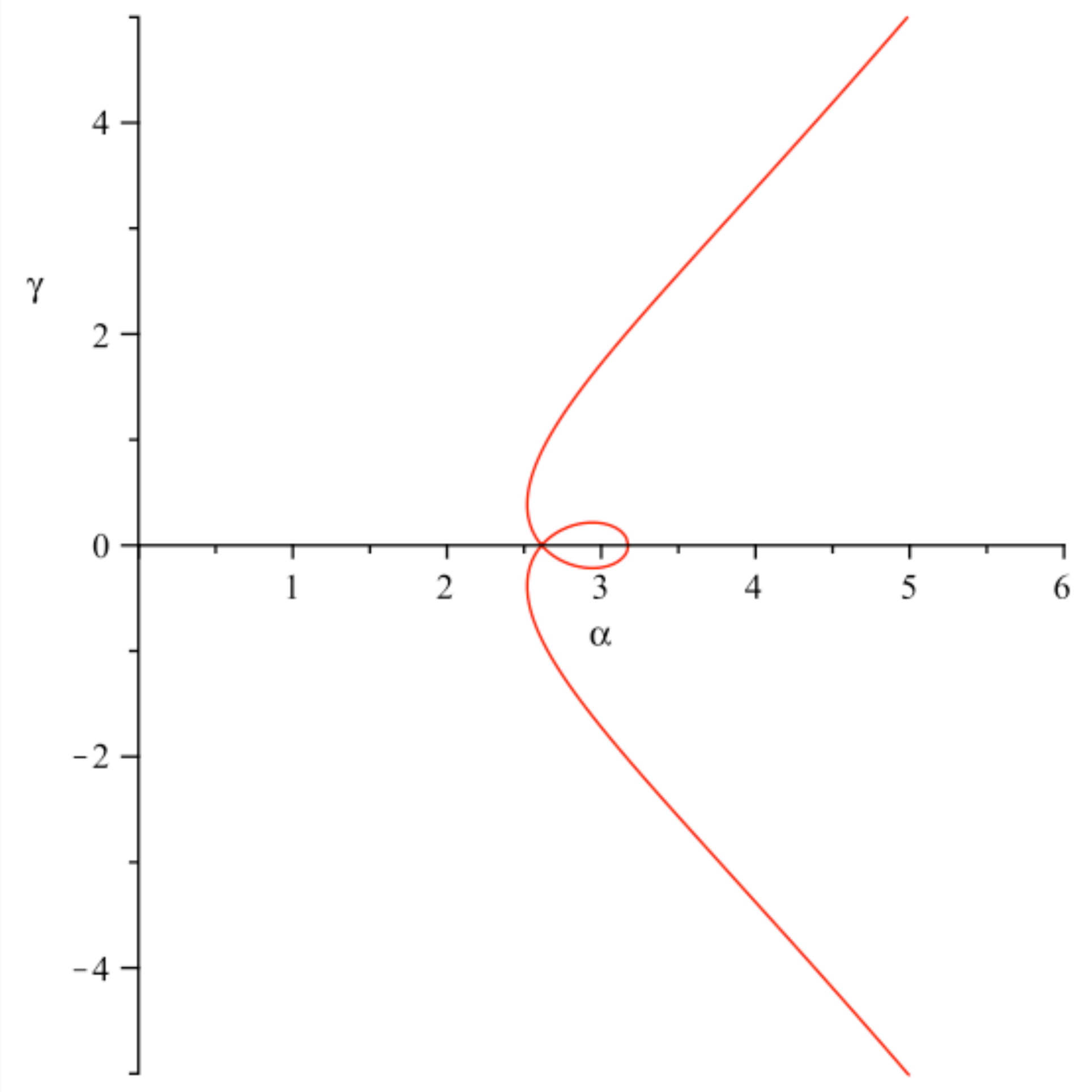}
\caption{Solution for elliptic subcovers of $X$}\label{plot_chi}
\end{center}
\end{figure}

\chapter{The AGM method}\label{AGMchapter}
\introcap{Gauss developed a rather efficient method for computing elliptic integrals, namely the Arithmetic-Geometric Mean, or AGM, first introduced by Lagrange, which was then generalised to a class of genus 2 curves by Richelot and Humbert. Here we use this method for the curve under consideration, of eq. \eqref{quotientcurve}: the AGM method cannot be applied as it is to  this class of curves, but needs to be modified.  This is done in this chapter, after a short introduction to the AGM method in general; the findings are then applied to solve the Ercolani-Sinha constraints numerically.   }

We are now in a position to solve explicitly the Ercolani-Sinha constraints, in the form given in eq. \eqref{esg2}, namely in terms of the genus 2 periods:
\begin{align*}
(n_0,3n)\mathcal{A}+
(3m_0,3m)\mathcal{B}=6\nu(0,1),
\end{align*}
In order to do this, we make use of the so called Arithmetic Geometric Mean (AGM in short) method to calculate the periods. The AGM method permits one to calculate, in a rather efficient way, all the integrals between branchpoints,
and hence, from eqs. \eqref{basisqexp}, the periods: in particular, it is possible to check the Ercolani-Sinha constraints for given values of the parameters\footnote{We can rescale so as to have $\beta=1$, see section \ref{AGMessect}.}} $\alpha$ and $\gamma$. As we know that the Ercolani-Sinha constraints have a solution for $\alpha=0$ and $\gamma=\pm5\sqrt{2}$ (tetrahedral monopole, see \cite{symmmon,BE06} and Theorem \ref{mainthmBE}), it is reasonable to look for a solution in the $(\alpha,\gamma)$ plane in a neighbourhood of these points.  This is done numerically:  starting from $\alpha=0$,$\gamma=5\sqrt{2}$  we let $\alpha$ and $\gamma$ vary infinitesimally, calculate the Ercolani-Sinha constraints for every such pair and look for a $\gamma$ such that these constraints are still satisfied, with $\alpha,\gamma\in\mathbb{R}$. \\
The results obtained are described in section \ref{secagmes}: in view of this, we give some details about the AGM method.
\section{The AGM method}
The origin of the AGM method dates back to Lagrange, but it was Gauss who truly initiated its investigation, and in fact a large part of what is known today seems to be due (or at least known) to him. In particular, Gauss examined the elliptic case in detail, while the generalisation to the hyperelliptic case was due to Richelot, even though it seems that it was known to Gauss already (for historical notes see \eg \cite{cox}).\\
Here we describe briefly the elliptic case, and then, in some more detail, the hyperelliptic case; for a more extensive treatment, we refer to \cite{cox,bm} and references therein.
\subsection{AGM: the elliptic case}\label{agmell}
Let $a,b$ be positive real numbers: their \enf{arithmetic-geometric mean}, denoted $M(a,b)$, is the common limit of the sequences defined as follows:
\begin{align}
a_{0}&=a, & b_{0}&=b, \nonumber\\
a_{n+1}&=\frac{a_{n}+b_{n}}{2}, & b_{n+1}&=\sqrt{a_{n}b_{n}}.
\end{align}
These two sequences satisfy the following
property
\begin{equation*}
a\geq a_1\geq\ldots\geq a_n\geq a_{n+1}
\geq\ldots\geq b_{n+1}\geq b_n\geq\ldots\geq b_1\geq b_0 ,
\end{equation*}
which can be proven by induction, and ensures the existence of a limit. From the observation that $b_{n+1}>b_{n}$, we have
\begin{equation*}
 a_{n+1}-b_{n+1}\leq a_{n+1}-b_n=\frac12(a_n-b_n).
\end{equation*}
So, the sequences $a_{n}$ and $b_{n}$ converge to a common limit, their arithmetic geometric mean
\begin{equation*}
\lim_{n\to\infty}a_{n}=\lim_{n\to\infty}b_{n}=M(a,b).
\end{equation*}
Moreover, one has that
\[ 0\leq a_n-b_n \leq 2^{-n}(a-b),  \]
which ensures a fast convergence (which is relevant in the present work).\\

Lagrange and Gauss introduced the Arithmetic-Geometric Mean in connection with elliptic integrals, obtaining the following result:
\begin{theorem}\enf{AGM \cite{gauss99}}
Let $a,b\in\mathbb{R}_{+}$, and $M(a,b)$ their arithmetic geometric mean; then:
\begin{equation*}
\int_{0}^{\pi/2}\frac{\mathrm{d}\phi}{\sqrt{a^2\cos^2\phi+b^2\sin^2\phi}
}=\frac{\pi}{2M(a,b)}
\end{equation*}
\end{theorem}
For the proof\footnote{The proof is based on the following change of variables
\begin{equation}
\sin\,\phi=\frac{2a\sin\,\phi'}{a+b +(a-b) \sin^2\,\phi'
}\label{jacsub}
\end{equation} and makes use of the identities $
(a^2\cos^2\,\phi+b^2\sin^2\,\phi)^{-1/2}\mathrm{d}\phi
=(a_1^2\cos^2\,\phi'+b_1^2\sin^2\,\phi')^{-1/2}\mathrm{d}\phi'$, proven by Jacobi.
} see \eg \cite{bm}.\\
Using this result, all elliptic integrals of the form $\int_{a}^{b}\frac{\de x}{\sqrt{P(x)}}$ can be expressed in terms of an appropriate arithmetic geometric mean (see \cite{bm}), after various changes of variables.
\subsection{The genus 2 case: Richelot and Humbert}
The above theory of Gauss' Arithmetic Geometric Mean was extended to the hyperelliptic case by Richelot, in \cite{richelot1,richelot2}.
This section follows the modern exposition of the work by Richelot given by Bost and Mestre in \cite{bm}. These authors remark, how Richelot's ``changes of coordinates'' is in fact a correspondence (cf. definition \ref{defcorr} below); they follow the construction given by Humbert, who  in \cite{humbert}  re-interprets Richelot's findings in terms of duplications formulae of 2-variable theta functions, \ie isogenies (of type (2,2) ) on Abelian surfaces, hence generalising  and simplifying Richelot's algorithm.\\

We remark that the Richelot-Humbert construction is only given for the case where the genus 2 curve, represented as a two-sheeted cover of $\puno$ with six branchpoints, has all \enf{real} branchpoints: in fact, it does not apply in a straightforward fashion to the case of a genus 2 curve with complex roots. In the next  section we argue a possible generalisation in this direction.\\

Consider the genus 2 curve $C'$
\begin{align}\begin{split}\label{eqcondc}
&\hskip4.5cm  y^2+P(x)Q(x)R(x)=0,\\
&P(x)=(x-a)(x-a'),\quad
Q(x)=(x-b)(x-b'),\quad
R(x)=(x-c)(x-c'),
\end{split}
\end{align}
where the real roots $a,a',b,b',c,c'$ are  ordered as
\[ a<a'<b<b'<c<c' .\] 
Associate then to the triple of (real) polynomials $(P,Q,R)$, another triple $(U,V,W)$, defined by
\begin{align}\label{eqcondc1}
U(x)&=[Q(x),R(x)], &
V(x)&=[R(x),P(x)], &
W(x)&=[P(x),Q(x)].
\end{align}
where $[f,g]:=\cfrac{\mathrm{d}f(x) }{\mathrm{d} x}g(x)-\cfrac{\mathrm{d}g(x) }{\mathrm{d} x}f(x)$.\\

In general, given two non-proportional degree 2 polynomials $P_{1},P_{2}$, the roots $r,r'$ of $[P_{1},P_{2}]$ can be obtained geometrically from the roots $p_{i},p_{i}'$ of $P_{i}$. If $\mathcal{C}$ is a conic, and $g:\puno\to \mathcal{C}$  a unicursal parametrization, then $g(r),g(r')$ are those points where the tangent to $\mathcal{C}$  passing through the intersection of the lines $g(p_{1})g(p'_{1})$ and $g(p_{2})g(p'_{2})$ meets the conic $\mathcal{C}$ itself.\\
As for the roots of the polynomials $U,V,W$, they are all real; %
denoting $u<u'$, $v<v'$, $w<w'$ the roots of $U,V,W$ respectively, the following relation between all the roots holds
\begin{equation}\label{relroots}
a \leq v \leq w \leq a'\leq  b \leq w' \leq  u \leq  b' \leq  c \leq  u' \leq  v' \leq  c' .
\end{equation}

\begin{figure}[b!]
\begin{center} 
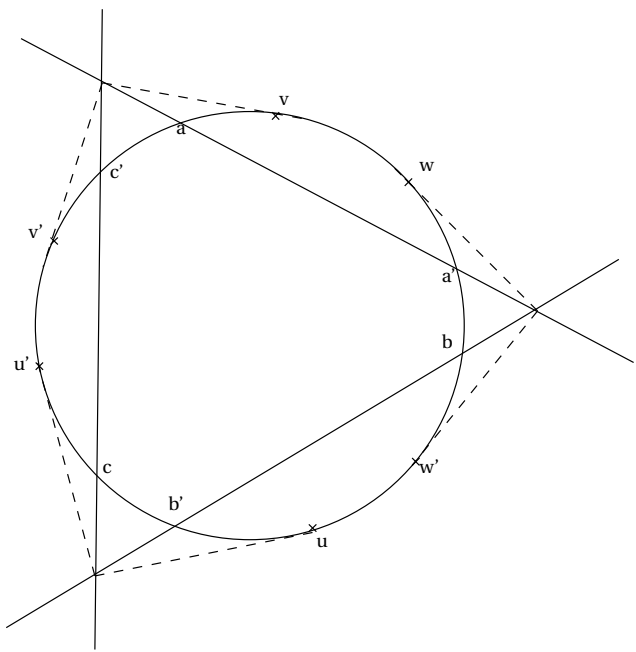
\caption{Roots of $P,Q,R$ and $U,V,W$}\label{roots}
\end{center}
\end{figure}

These inequalities can be proven algebraically (see section \ref{agmg2th}, where we give explicit expressions for the roots of $U,V,W$); in view of the above interpretation of the roots of the bracket $[P_{1},P_{2}]$, a geometric description is given in Figure \ref{roots}.\\

Therefore, this is a situation similar to Gauss' AGM case: to each pair of branchpoints one can associate another pair of points which are closer than the initial ones, and we expect a relation between the integrals of corresponding pairs. Iterating this process, we can repeat the considerations of section \ref{agmell} for every pair  to argue the existence of a limit, and hence obtain an expression for the integrals in terms of these limits. The relation between integrals  suggested and proven by  Humbert is
\begin{equation}\label{corr}
\int_{a}^{a'}\frac{S(x)}{\sqrt{-P(x)Q(x)R(x)}}\de x =
2\sqrt{\Delta} \int_{v}^{w}\frac{S(x)}{\sqrt{-U(x)V(x)W(x)}}\de x, 
\end{equation}
and similarly for the integrals between the other pairs of branchpoints,
where\footnote{The reason for the presence of this constant  is clarified in the proof given in the next subsection.} $\Delta$ is the determinant of the matrix whose entries are the coefficients of $P,Q,R$ in the basis $(1,x,x^{2})$, and $S(x)$ is a polynomial of degree at most $1$.\\

 There is, however, an important element of difference between the elliptic and the hyperelliptic case: while in the standard Gauss construction the change of variables can be interpreted as a map from one curve to the other, in this case it is given not by a map, but by a \enf{correspondence}. \\\vspace{-5mm}
\begin{definition}{\enf{\cite{gh}}}\label{defcorr} A correspondence $T:\Sigma\to \Sigma'$ of degree $d$ associates to every point $P\in \Sigma$ a divisor $T(P)$ of degree $d$ in $\Sigma'$, varying holomorphically with $P$.
\end{definition}
A correspondence can be presented both as a holomorphic  map $\Sigma\to \Sigma'^{(d)}$ from $\Sigma$ to the $d$-th symmetric product of $\Sigma'$, or equivalently by its ``curve of correspondence'':
\begin{equation*}
D=\{ (P,P'): P'\in T(P)\} \subset C\times \tilde{C}'.
\end{equation*}

In the case of the Humbert construction, the two curves are  $C$, with equation $y^{2}+P(x)Q(x)R(x)=0$, and $\tilde{C}'$, with equation $\Delta y'^{2}+U(x')V(x')W(x')=0$.  The correspondence  between $C$ and $\tilde{C}'$, which Humbert considers in \cite{humbert}, is of degree 2, and is given by the curve $Z$ in $C\times \tilde{C}'$ of equations
\begin{equation}\label{corrcurve}
\begin{cases}
P(x)U(x')+Q(x)V(x')=0, \\
yy'=P(x)U(x')(x-x').
\end{cases}
\end{equation}

In analogy with the pull-back of a map, one can introduce also for a correspondence a linear map $\delta_{Z}: \Omega^{1}(\tilde{C}')\to \Omega^{1}(C)$. 
Then eq. \eqref{corr} can be interpreted as the relation between differentials
\begin{equation}\label{corr1}
\delta_{Z}\left(  \frac{S(x')}{y'} \de x' \right)= \frac{S(x)}{y} \de x,
\end{equation}
together with an analysis of the image of the path joining $a,a'$ (resp. $b,b'$ or $c,c'$) under the correspondence $Z$.\\

Furthermore, eq. \eqref{corr} can be interpreted as an identity in $\Jac(\tilde{C}')$, the Jacobian of $\tilde{C}'$: it states that a half period in $C$ is sent by the correspondence $Z$ to a period in $\tilde{C}'$, or alternatively, the image of a half period is zero in the Jacobian. This is a slightly weaker formulation, which is explored further in Remark \ref{remjaccorr}.\\

The proof of the identity \eqref{corr}  is given in section \ref{proofid}: we state first a version of the Arithmetic-Geometric Mean for genus 2 curves, bases on the remarks above, and on eq.  \eqref{corr}.

\subsection{The AGM method for genus 2 curves}\label{agmg2th}
Consider a genus 2 curve as in eqs. \eqref{eqcondc}, \eqref{eqcondc1}. Define  6 sequences  $(a_{n})$, $(a'_{n})$, $(b_{n})$, $(b'_{n})$, $(c_{n})$, $(c'_{n})$ recursively by the conditions:
\begin{itemize}
\item $a_{0}=a$,\; $a'_{0}=a'$,\; $b_{0}=b$,\; $b'_{0}=b'$,\; $c_{0}=c$,\; $c'_{0}=c'$;
\item $a_{n+1},a'_{n+1},b_{n+1},b'_{n+1},c_{n+1},c'_{n+1}$ \;are roots of $U_{n}V_{n}W_{n}$, ordered as follows
\begin{align}\label{orderrroots}
a_{n+1}<a'_{n+1}<b_{n+1}<b'_{n+1}<c_{n+1}<c'_{n+1},
\end{align}
where, for every $n$,
\begin{align*}
P_n(x)=(x-a_n)(x-a'_n), \;\;
Q_n(x)=(x-b_n)(x-b'_n),\;\;
R_n(x)=(x-c_n)(x-c'_n),\\
U_n(x)=[Q_n(x),R_n(x)],\;\;
V_n(x)=[R_n(x),P_n(x)], \;\;
W_n(x)=[P_n(x),Q_n(x)].\qquad
\end{align*}

\end{itemize}

Bost and Mestre give in \cite{bm} an explicit expression for these sequences:
\begin{align}\begin{split}\label{sequences}
a_{n+1}&=\frac{c_{n}c_{n}'-a_{n}a_{n}'-B_n  }{c_n+c_n'-a_n-a_n'},\quad
a_{n+1}'=\frac{b_{n}b_{n}'-a_{n}a_{n}'-C_n  }{b_n+b_n'-a_n-a_n'},\\
b_{n+1}&=\frac{b_{n}b_{n}'-a_{n}a_{n}'+C_n  }{b_n+b_n'-a_n-a_n'},\quad
b_{n+1}'=\frac{c_{n}c_{n}'-b_{n}b_{n}'-A_n  }{c_n+c_n'-b_n-b_n'},\\
c_{n+1}&=\frac{c_{n}c_{n}'-b_{n}b_{n}'+A_n  }{c_n+c_n'-b_n-b_n'},\quad
c_{n+1}'=\frac{c_{n}c_{n}'-a_{n}a_{n}'+B_n  }{c_n+c_n'-a_n-a_n'},
\end{split} 
\end{align}
where
\begin{align*}
A_n&=\sqrt{ (b_n-c_n)(b_n-c_n')(b_n'-c_n)(b_n'-c_n')  },\\
B_n&=\sqrt{ (c_n-a_n)(c_n-a_n')(c_n'-a_n)(c_n'-a_n')  },\\
C_n&=\sqrt{ (a_n-b_n)(a_n-b_n')(a_n'-b_n)(a_n'-b_n')  }.
\end{align*}
These can be derived finding the roots for $U_{n},V_{n}, W_{n}$  as follows\footnote{To derive this we only use eqs. \eqref{corr} for the correspondence. For $c_{n},c'_{n}$ eq. \eqref{equiv1} is also needed. }
\begin{align}
\begin{split}\label{expruvw}
u_{n},u'_{n}&=\frac{c_{n}c_{n}'-b_{n}b_{n}'\mp A_n  }{c_n+c_n'-b_n-b_n'},\\
v_{n},v_{n}'&=\frac{c_{n}c_{n}'-a_{n}a_{n}' \mp B_n  }{c_n+c_n'-a_n-a_n'},\\
w_{n},w_{n}'&=\frac{b_{n}b_{n}'-a_{n}a_{n}' \mp C_n  }{b_n+b_n'-a_n-a_n'},
\end{split}
\end{align}
and ordering them as in eq. \eqref{orderrroots}. One can see directly from the expressions above that  $v_{n} \leq w_{n}  \leq w'_{n} \leq  u_{n} \leq  u_{n}' \leq  v_{n}'$ (cf. also eq. \eqref{relroots}); hence we set
\begin{equation}\label{rootsintermsofuvw}
a_{n+1}=v_{n}, \quad a_{n+1}'= w_{n},  \quad b_{n+1}=w'_{n}, \quad b_{n+1}'= u_{n}, \quad c_{n+1} =u_{n}', \quad   c'_{n+1}=v_{n}';
\end{equation}
the expressions in \eqref{sequences} then follow. \\

\begin{theorem}[\enf{Richelot \cite{richelot2}, Bost and Mestre \cite{bm}}]\label{richthm}
With the  above definitions, the sequences  $(a_{n})$, $(a'_{n})$, $(b_{n})$, $(b'_{n})$, $(c_{n})$, $(c'_{n})$ converge pairwise to common limits
\begin{align*}
&\lim_{n\to\infty} a_n=\lim_{n\to\infty} a_n'=\alpha\equiv M(a,a'),\\
&\lim_{n\to\infty} b_n=\lim_{n\to\infty} b_n'=\beta\equiv M(b,b'),\\
&\lim_{n\to\infty} c_n=\lim_{n\to\infty} c_n'=\gamma\equiv M(c,c').
\end{align*}
Furthermore, for any polynomial $S(x)$ of degree at most one, the following relations hold:
\begin{align}\begin{split}\label{intthm}
I(a, a')\equiv\int_{a}^{a'}\frac{S(x)\mathrm{d}x}{\sqrt{-P(x)Q(x)R(x)  }}=\pi T \frac{S(\alpha)}{(\alpha-\beta)(\alpha-\gamma)},\\
I(b, b')\equiv\int_{b}^{b'}\frac{S(x)\mathrm{d}x}{\sqrt{-P(x)Q(x)R(x)  }}=\pi T \frac{S(\beta)}{(\beta-\alpha)(\beta-\gamma)},\\
I(c, c')\equiv\int_{c}^{c'}\frac{S(x)\mathrm{d}x}{\sqrt{-P(x)Q(x)R(x)  }}=\pi T \frac{S(\gamma)}{(\gamma-\alpha)(\gamma-\beta)},
\end{split}
\end{align}
where 
\begin{equation}\label{tnt}
T=\prod_{n=0}^{\infty}t_n,\quad  t_n=\dfrac{2\sqrt{\Delta_n}}{\sqrt{(b_n+b_n'-a_n-a_n')(c_n+c_n'-b_n-b_n')(c_n+c_n'-a_n-a_n')}}.
\end{equation}
\end{theorem}\vspace*{4mm}

 The proof of convergence for the sequences $(a_{n})$, $(a'_{n})$, $(b_{n})$, $(b'_{n})$, $(c_{n})$, $(c'_{n})$   is similar, for every pair, to that for the elliptic case given in section \ref{agmell}.\\

It follows from a direct calculation that
\begin{align*}
P_{1}(x)&=[Q(x),R(x)](-a-a'+c+c')=U(x)(a+a'-c-c'),\\
Q_{1}(x)&=[P(x),Q(x)](a+a'-b-b')=W(x)(a+a'-b-b'),\\
R_{1}(x)&=[R(x),P(x)](b+b'-c-c')=V(x)(b+b'-c-c').
\end{align*}
Let us focus on the fist pair $a$, $a'$: the other cases can be dealt with in the same fashion.\\
Using \eqref{corr}, the following holds:
\begin{align}\begin{split}\label{eqintn}
\int_{a_{n}}^{a_{n}'}\frac{S(x)}{ \sqrt{ -P_{n}Q_{n}R_{n} } }\de x =& \;
2\sqrt{\Delta_{n}}\int_{a_{n+1}}^{a_{n+1}'}\frac{S(x)}{ \sqrt{ -[P_n,Q_n][Q_n,R_n][R_n,P_n] } }\de x\\
=& \;t_{n}\int_{a_{n+1}}^{a_{n+1}'}\frac{S(x)}{ \sqrt{ -P_{n+1}Q_{n+1}R_{n+1} } }\de x,
\end{split}\end{align}
It follows that
\begin{equation*}
\int_{a}^{a'}\frac{S(x)}{ \sqrt{ -PQR} }\de x=\;\left(\prod_{k=0}^{n+1}t_{k}\right)\int_{a_{n+1}}^{a_{n+1}'}\frac{S(x)}{ \sqrt{ -P_{n+1}Q_{n+1}R_{n+1} } }\de x,
\end{equation*}
Taking the limit for $n\to\infty$
\begin{align*}
\lim_{n\to\infty} \left(\prod_{k=0}^{n+1}t_{k}\right)\int_{a_{n+1}}^{a_{n+1}'}\frac{S(x)}{ \sqrt{ -P_{n+1}Q_{n+1}R_{n+1} } }\de x&=
\lim_{n\to\infty} \left(\prod_{k=0}^{n+1}t_{k}\right)\frac{1}{2}\oint_{\mathfrak{a}_{1}}\frac{S(x)}{ \sqrt{ -P_{n+1}Q_{n+1}R_{n+1} } }\de x\\
&=T \pi i \; \mathrm{Res}_{x=\alpha}\frac{S(x)}{ \sqrt{ -P_{n+1}Q_{n+1}R_{n+1} } },
\end{align*}
with $T$ as in eq. (\ref{tnt}), yields eqs. (\ref{intthm}).
\\

\begin{remark} The expressions for the integrals between branchpoints we have given are slightly different from those in \cite{bm}: in particular, the integral between $b$ and $b'$ has opposite sign. This depends on a different choice of conventions. Indeed, Bost and Mestre, in the note 2, p. 51 of \cite{bm}, claim that they want to recover the ``classical identity'' $I_{a}-I_{b}+I_{c}=0$: with our choice of convention for sheets, the relation between integrals become $I_{a}+I_{b}+I_{c}=0$, which follows from the fact that the integral around a cycle encircling all the cuts, oriented so that the upper arc goes from negative to positive real values, is zero.
\end{remark}
\subsection{The proof of a fundamental identity }\label{proofid}

For the proof of eq. \eqref{corr} we follow \cite{bm}: in particular, the authors follow the lines suggested within a problem presented at the entrance examination  to the faculty of Physics in the \'Ecole Normale Sup\'erieure, Paris, in 1988. The object of the problem was indeed to prove Richelot theorem: a number of hints were given and the proof was divided into several steps; %
here we follow these steps, providing some additional details.\\

For every pair $(x,z)$ the following fundamental equality is valid
\begin{equation}\label{equiv1}
P(x)U(z)+Q(x)V(z)+R(x)W(z)+(x-z)^2\Delta(P,Q,R)=0; 
\end{equation}
this follows from evaluating the determinant
\begin{equation*}
\det \left(\begin{array}{ccc}
P(x) & P'(x)  &P(z)\\
Q(x) & Q'(x) &Q(z)\\
R(x) & R'(x) &R(z)
\end{array}\right).
\end{equation*}
Recalling eq. \eqref{corrcurve}, we introduce the following polynomial of 2 variables $(x,z)$,
\begin{equation}\label{polf}
F(x,z)=P(x)U(z)+Q(x)V(z).
\end{equation}
Its zeros  define two functions, $z_1(x)$ and $z_2(x)$, which are the ($z$-coordinates of the) images of the correspondence $Z$ of eq. \eqref{corrcurve} in $C'$
\begin{equation} \label{equiv2} 
 F(x,z_i(x))=0, \quad i=1,2\quad \forall \;x. 
 \end{equation}
Following\footnote{We point out that Bost and Mestre use an absolute value under the square root for $y(x)$.} \cite{bm}, define, for $x\in (a,a')$,
\begin{align}
\begin{split}
y(x)&=\sqrt{-P(x)Q(x)R(x) },\\
y_1(x)&=\frac{P(x)U(z_1(x))(x-z_1(x))}{y(x)},\\
y_2(x)&=\frac{P(x)U(z_2(x))(x-z_2(x))}{y(x)}.
\end{split}
\end{align}
One can compute $y_{1,2}(x)^2$, to get\footnote{Using 
the above expression for $y(x)$ and \eqref{equiv1}
\begin{align*}
y_{1}(x)^2&=\frac{P(x)U^2(z_1(x))(x-z_1(x))^2}{Q(x)R(x)}\\
&=\frac{1}{\Delta}\frac{P(x)U^2(z_1(x))[P(x)U(z_1(x))+Q(x)V(z_1(x))
+R(x)W(z_1(x)))  ]}{Q(x)R(x)}.
\end{align*}
Recalling that, from (\ref{equiv2}), $ P(x)U(z_1(x))+Q(x)V(z_1(x))= 0$, the result follows.
}
\begin{equation}\label{yy}
 y_{1,2}(x)^2=\frac{1}{\Delta} \frac{P(x)U(z_{1,2}(x))^2 W(z_{1,2}(x))}{Q(x)} . \end{equation}
A preliminary lemma can be proven.
\begin{lemma} The following relations are valid
\begin{align}
\frac{z_1'(x)}{y_1(x)}+\frac{z_2'(x)}{y_2(x)}&=\frac{1}{y(x)},
\label{rel1}\\
\frac{z_1(x)z_1'(x)}{y_1(x)}+\frac{z_2(x) z_2'(x)}{y_2(x)}&=\frac{x}{y(x)}.
\label{rel2}
\end{align}
\end{lemma}
\begin{proof}
Firstly, using \[U(z)=\chi_1(z-c_1)(z-c_1'),\]
with $\chi_1=b+b'-c-c'$, one has
\begin{align*}
&\frac{z_1'(x)}{y_1(x)}+\frac{z_2'(x)}{y_2(x)}\\
&=\frac{y(x)}{P(x)}\left\{\frac{z_1'(x)}{U(z_1(x)(x-z_1(x)))}
+\frac{z_2'(x)}{U(z_2(x)(x-z_2(x)))}\right\}\\
&=\frac{y(x)}{P(x)\chi_1}\left\{ \frac{z_1'(x)}{(z_1(x)-c_1)(z_1(x)-c_1')(x-z_1(x))}
+\frac{z_2'(x)}{(z_2(x)-c_1)(z_2(x)-c_1')(x-z_2(x))} \right\}.
\end{align*}
This expression can be further reduced to symmetric functions of $z_1(x)$ and   $z_1(x)$,
\begin{align*}
&\frac{z_1'(x)}{y_1(x)}+\frac{z_2'(x)}{y_2(x)}\\
&=\frac{y(x)}{P(x)\chi_1}\frac{1}{c-c'}\left\{-\frac{1}{x-c'}\left[ \frac{z'_1(x)}{z_1(x)-c'}+\frac{z'_2(x)}{z_2(x)-c'}\right]
+\frac{1}{x-c}\left[ \frac{z'_1(x)}{z_1(x)-c}+\frac{z'_2(x)}{z_2(x)-c} \right]  \right\}\\
&-\frac{y(x)}{P(x)\chi_1}\frac{1}{(x-c')(x-c)}\left[ \frac{z'_1(x)}{z_1(x)-x}+\frac{z'_2(x)}{z_2(x)-x} \right]\\
&=\frac{y(x)}{P(x)\chi_1}\left\{-\frac{1}{x-c'}\left[
\frac{\mathrm{d}}{\mathrm{d} x}\ln[ (z_1(x)-c')(z_2(x)-c')]\right]
+\frac{1}{x-c}\left[\frac{\mathrm{d}}{\mathrm{d} x}\ln[ (z_1(x)-c)(z_2(x)-c)] \right]  \right\}\\
&-\frac{y(x)}{P(x)\chi_1}\frac{1}{(x-c')(x-c)}\left[ \frac{1}{z_1(x)-x}+\frac{1}{z_2(x)-x} + \frac{\mathrm{d}}{\mathrm{d} x}\ln[ (z_1(x)-x)(z_2(x)-x)] \right].
\end{align*}
The symmetric functions
\[ s_1(x)=z_1(x)+z_2(x),\quad s_2(x)=z_1(x)z_2(x) \]
are computed from the equality
\[ F(x,z)=\phi_0(x)(z-z_1(x))(z-z_2(x))\]
via Maple. We do not report the expressions for $s_{1}$ and $s_{2}$ here as they are not very enlightening, but using them the direct derivation of \eqref{rel1} is immediate. \\

Once \eqref{rel1} is proven, \eqref{rel2} follows. Indeed, we have
\begin{align*}
&\frac{z_1(x)z_1'(x)}{y_1(x)}+\frac{z_2(x) z_2'(x)}{y_2(x)}\\
&=x\frac{y(x)}{P(x)}\left\{\frac{z_1'(x)}{U(z_1(x)(x-z_1(x)))}
+\frac{z_2'(x)}{U(z_2(x)(x-z_2(x)))}\right\}+\frac{y(x)}{P(x)}\left\{\frac{z_1'(x)}{U(z_1(x))}
+\frac{z_2'(x)}{U(z_2(x))}\right\}.
\end{align*}
Using eq. \eqref{rel1} the first term in the right hand side simplifies to $ x/y(x)$. The second term vanishes, as one has
\begin{align}
\frac{z_1'(x)}{U(z_1(x))}
+\frac{z_2'(x)}{U(z_2(x))}=\frac{\mathrm{d}}{\mathrm{d} x}\ln \left( \frac{ (z_1(x)-c_1)(z_2(x)-c_1) }{ (z_1(x)-c_2)(z_2(x)-c_2) }\right),
\end{align}
but the argument of the logarithm  is a constant, namely
\[\frac{b'c'-c^2-c'^2-2bb'+bc+bc' +2\sqrt{(b'-c')(b'-c)(b-c')(b-c)}}{-b'c'+c^2+c'^2+2bb'-bc-bc' +2\sqrt{(b'-c')(b'-c)(b-c')(b-c)}}   \]
Therefore eq. \eqref{rel2} holds.
\end{proof}

Adding \eqref{rel1} and \eqref{rel2}, multiplied by arbitrary coefficients, one obtains the equality
\begin{align}\label{relsgen}
&\frac{S(z_1(x))z_1'(x)}{y_1(x)}+\frac{S(z_2(x))z_2'(x)}{y_2(x)}
=\frac{S(x)}{y(x)},
\end{align}
where we remark that $S$ is a polynomial of degree at most one\footnote{In fact, one cannot find an analogue of  eqs. \eqref{rel1} and \eqref{rel2} for higher degrees.}.\\

Now we are in a position to prove (\ref{corr1}). To do this we note that the first term on the left hand side of (\ref{relsgen}) can be transformed, by using (\ref{yy}) and then (\ref{equiv2}), as follows:
\begin{align}
\begin{split}\label{relfin}
S(z_1(x))\frac{z_1'(x)}{y_1(x)}
&=S(z_1(x))\frac{z_1'(x) \sqrt{\Delta}\sqrt{Q(x)}}{U(z_1(x)) \sqrt{P(x)}\sqrt{W(z_1(x))}}\\
&=S(z_1(x))\frac{z_1'(x) \sqrt{\Delta}}{\sqrt{-U(z_1(x))V(z_1(x))W(z_1(x))}},
\end{split}\end{align}
and similarly for the second term. Thus, combining eqs. (\ref{relsgen}) and (\ref{relfin}), we obtain eq. (\ref{corr1}).\\

\subsection{Concluding remarks}
\begin{remark}\phantomsection\label{remjaccorr}
We point out that thus far we have only proven eq. \eqref{corr1}; in order to obtain eq. \eqref{corr}, we need to integrate eq. \eqref{corr1} on the appropriate contour: on the left hand side, this is just a straight line connecting the two branchpoints, on the right hand side one needs to examine the image of the previous contour via the correspondence \eqref{corrcurve}.\\
 Note that this is not examined in detail by Bost and Mestre in \cite{bm}, although they do mention this issue.
We are not able to give an abstract description of this in general, nevertheless we propose some progress in this direction.\\

Consider the lift of the correspondence $Z$ in eq. \eqref{corrcurve} as a map between Jacobians, namely $\zeta:\Jac(C)\to\Jac(C')$, and denote $B_{a}$ the branchpoint with $z$-coordinate $a$, so $B_{a}=(a,0)$, and similarly for the others. Then the kernel of this map is
\begin{equation*}
\ker \zeta = \{  B_{a'}-B_{a},\; B_{b'}-B_{b},\; B_{c'}-B_{c}\;, 0 \}.
\end{equation*}
This can be understood analysing the images of these branchpoints under the correspondence.\\ For the first and second pair, \ie $a,a'$ and $b,b'$, using the notation of eq. \eqref{equiv2} for the $z$-coordinates of the images via the correspondence, one has
\begin{equation}\label{eqdelta12}
z_1 (a)=z_1 (a'), \qquad z_2 (a)=z_2 (a'),
\end{equation}
and similarly for the second pair. It can be seen immediately from the first of \eqref{corrcurve} that roots of $P$ correspond to roots of $V$;  writing explicitly the 2 solutions\footnote{Note that we cannot get $Y$ from the second line of eq. \eqref{corrcurve} in this case, as we have $0$, the $y$-coordinates of the branchpoints, multiplying $Y$, the unknown that we need to determine: in this case, though, the two images of a branchpoint in the first two pairs are also branchpoints for the curve $C'$, so their $Y$-coordinates are uniquely determined.} of \eqref{corrcurve}  for $X$ shows that we are indeed in the case of eq. (\ref{eqdelta12}). \\
Viewing the Jacobian as the lattice of periods, the image of the divisor  $B_{a'}-B_{a}$ (resp. $B_{b'}-B_{b}$) is a point in the lattice.\\

As for the third pair of branchpoints,  $B_{c},B_{c'}$, recalling relation (\ref{equiv1}), then the first of eqs. \eqref{corrcurve} corresponds to $R(x)W(z)+(x-z)^2\Delta(P,Q,R)=0$, hence,  when $x_{c}$ (resp. ${c'}$), $R$ vanishes, and \footnote{Note that we still cannot determine the $Y$-coordinates of the images, as for the other 2 pairs, but this time these images are not branchpoints of $C'$, so solving in the equation of $C'$ we get 2 solutions, which we can take to be $Y_{1}$ and $Y_{2}$ (resp.   $\tilde{Y}_{1}$ and $\tilde{Y}_{2}$).}
$$z_{1,2}(c)=(c,Y_{1,2})=:P_{1,2},\qquad z_{1,2}(c')=({c'},\tilde{Y}_{1,2})=:\tilde{P}_{1,2}. $$
The fact that $B_{c'}-B_{c}$ belongs to the kernel follows from the fact that
\begin{equation*}
\zeta(B_{c'}-B_{c})=\tilde{P}_{1}+\tilde{P}_{2}-{P}_{1}-{P}_{2}=\divs\left( \frac{w-c'}{w-c} \right)\equiv 0
\end{equation*}

Note that the above considerations do not provide any kind of insight about the periods
 corresponding to the images of the above divisors: indeed, they involve objects in the Jacobians, so by definition modulo periods. However, as the kernel has a group structure, there is a relation among those images (for instance, it suffices to find the images of $B_{a'}-B_{a}$ and $ B_{b'}-B_{b}$ to obtain immediately that of $ B_{c'}-B_{c}$).\\
\end{remark}
\begin{remark}\phantomsection\label{intoth}The integrals between other pairs of branchpoints, \eg $a'$ and $b$, can still be calculated using the extended AGM algorithm. This is achieved using an appropriate transformation: for example, the map 
\begin{equation}\label{chancoord}
f: x\to\frac{1}{2x-a-a'}
\end{equation}
sends the roots $a$, $a'$, $b$, $b'$, $c$, $c'$ to 6 real numbers $f(a)<f(c')<f(c)<f(b')<f(b)<f(a')$, for which the above algorithm can be used.
\end{remark}
\section{Generalisation to the genus 2 case with complex conjugate roots}
We study here the possibility to generalise the AGM method to the case where the branchpoints do not all lie on the real axis, as was the case in AGM analysis, but the polynomials $P,Q,R$ are still real: this corresponds to the three pairs of branchpoints being %
\begin{align*}
a'=\bar{a},\quad b'=\bar{b},\quad c'=\bar{c},
\end{align*}
ordered as follows
$$\Re(a)=\Re(a') < \Re(b)=\Re(b') < \Re(c)=\Re(c');$$
in the following we also take  $\Im(a)<0$, $\Im(b)<0$, $\Im(c)<0$, for definiteness.\\
This is  the case for the quotient monopole curve of eq. \eqref{quotientcurve}: recalling eq. \eqref{bpquotient} for the branchpoints, we see that indeed they can be split in complex conjugate pairs as above (cf. eq. \eqref{bpquotientcc}).\\

We remark that all the polynomial relations found in the previous sections can still be recovered in the case of complex conjugate branchpoints: hence, the relation (\ref{corr1}) between the differentials on $C$ and $C'$ still holds true for this case. The difference lies, in fact, in the images, under the correspondence $Z$ of eq. \eqref{corrcurve}, of the branchpoints and the paths connecting the branchpoints.\\

In this case, since the initial branchpoints are complex, a relation analogous to \eqref{relroots}  cannot be written; nevertheless, the roots of $U_{0},V_{0},W_{0}$ are real, as can be seen by considering their explicit expressions in eqs. \eqref{expruvw}, and can hence be ordered. In contrast with the real case, though, this ordering is not unique: in the real case the ordering of  $u$, $u'$, $v$, $v'$, $w$, $w'$ depends only on the relative ordering of $a$, $a'$, $b$, $b'$, $c$, $c'$ on the real line; in the complex case, instead, it also depends on their imaginary parts, as can be seen again from eqs. \eqref{expruvw}.\\
Depending on the ordering of  $u$, $u'$, $v$, $v'$, $w$, $w'$, equation  \eqref{eqintn} relating the integrals between  the  three pairs of branchpoints on $C$ and $C'$ needs to be modified appropriately. We have considered in detail the case of the monopole quotient curve $X$: other cases can be studied using the same method.

\subsection{The AGM method for the quotient monopole curve}
Consider the quotient monopole curve $X$ of eq. (\ref{quotientcurve}), namely
\begin{equation*}
y^{2}=(x^{3}+\alpha x +\gamma)^{2}+4\beta^{2}
\end{equation*}

If we order the branchpoints as in section \ref{properties},  we have:
\begin{align}\label{bpcc}
a=B_{4},\; a'=B_{3}; \quad b=B_{5},\; b'=B_{2}; \quad c=B_{6},\; c'=B_{1}
\end{align}
We calculate   $u$, $u'$, $v$, $v'$, $w$, $w'$ via eqs. \eqref{expruvw},  and examining their relative ordering we find the following two cases:
\begin{align}
\mathrm{\enf{case 1}:}\; \alpha>0, &\qquad   v \leq w \leq w' \leq  u \leq   u' \leq  v' ;\label{case1}\\
\mathrm{\enf{case 2}:} \;\alpha<0, & \qquad  u\leq v \leq w \leq u' \leq  v'  \leq  w'.\label{case2}
\end{align}
\enf{Case 1} is exactly the same as the situation considered in \cite{bm}: in particular, eq. \eqref{rootsintermsofuvw} still holds, and hence the sequences $a_{n}$, $a'_{n}$, $b_{n}$, $b'_{n}$, $c_{n}$, $c'_{n}$ are still given by eqs. \eqref{sequences}. Therefore, the first equality in eq. \eqref{eqintn} holds, in view of eq. \eqref{case1}, so the integrals between complex conjugate pairs of branchpoints are still expressed by eqs. \eqref{intthm}.\\
Note that here, though, it is not possible to find the other integrals between other pairs of branchpoints with a change of coordinates \eqref{chancoord} as in Remark \ref{intoth}: the images of the branchpoints after such a change of coordinates would not be complex conjugate pairs any longer. Later in this section we propose a solution to this.\\

 As for \enf{case 2}, eq. \eqref{rootsintermsofuvw} does not hold for the first step of the recurrence, due to the different ordering of eq. \eqref{case2};  it needs to be modified as follows, for $n=1$, in view of eq. \eqref{case2}:
 \begin{equation*}
a_{1}= u, \quad a_{1}'= v, \quad b_{1}=w \quad b'_{1}=u', \quad  c_{1}=v', \quad c_{1}'= w',
\end{equation*}
 Note that this change only occurs in the first step: after that, the curve $C'$ has all real branchpoints, and hence the Richelot-Humbert iteration can be applied as in \cite{bm}. In particular, let us
 denote by $I(p,q)$ the integral on $C$ between $p,q$ on the first sheet, and attach a superscript $I^{(i)}$ to denote integrals on the curve $C^{(i)}$ of equation $y^{2}+P_{i}(x)Q_{i}(x)R_{i}(x)=0$.  Then the integrals $I^{(i)}$ can be expressed by equations \eqref{intthm}, using the AGM method for the curve $C'$. \\
 We obtain an expression  for the integrals $I(p,q)$ on $C$ as follows.  Equation \eqref{corr} suggests that eq. \eqref{eqintn} for $n=0$ is modified as follows (for the three pairs):
\begin{align}\begin{split}\label{intcorr2}
I(a,a')=& \; t_{0}I^{'}(a_{1}',b_{1}),\\
I(b,b')=&\; t_{0}I^{'}(c'_{1},a_{1})= \;t_{0}(-I^{'}(a_{1}',b_{1}) +I^{'}(b_{1}',c_{1})),\\
I(c,c')=&-t_{0} I^{'}(b_{1}',c_{1}).
\end{split}
\end{align}
These relations are suggested by analogy with the Bost and Mestre case in view of the different ordering of eq. \eqref{case2}, and have been checked numerically in a variety of examples. In particular, in  the Maple file \texttt{images\_paths\_correspondence.mw} we have considered\footnote{More specifically, we have studied the images on the curve $\tilde{C}'$ of equation $\Delta y^{2}+U(x)V(x)W(x)=0$, in order to understand the first equality of \eqref{eqintn}  (as the second follows immediately from the first). We hence obtain, for instance, that the straight line between $a$ and $a'$ on $C$, call it $\gamma_{1}(a,a')$, is sent to a closed cycle encircling $a'_{1}$ and $b_{1}$ on $\tilde{C}'$: noticing that in the second equality of \eqref{eqintn} there is a factor of $1/2$, absorbed in the definition of $t_{i}$, we obtain that the image of $\gamma(a,a')$ on the curve $y^{2}+P_{1}(x)Q_{1}(x)R_{1}(x)=0$, \ie $C'$, is precisely the path from $a'_{1}$ and $b_{1}$.} the images  on $C'$ of the contours of integration on $C$, \ie  straight lines between the branchpoints,  via the correspondence \eqref{corrcurve}, to obtain the contours of integration on the right hand side of eqs. \eqref{intcorr2}.\\

Moreover, from a direct analysis of the images of the paths between every branchpoints, again performed numerically,  relations analogous to those above for every pair of branchpoint emerge:
\begin{align}
\begin{split}\label{otherint}
I(a,b)=& \,\frac{1}{2}t_{0}\,(I^{'}(a_{1},a'_{1}) + I'(b_{1}',c_{1})),\\
I(a',b')=&\,\frac{1}{2}t_{0}\,(I^{'}(a_{1},a'_{1}) - I'(b_{1}',c_{1})),\\
I(b,c)=& \,\frac{1}{2}t_{0}\,(-I^{'}(a_{1},a'_{1}) +I'(a_{1}',b_{1})-I'(b_{1},b'_{1})),\\
I(b',c')=& \,\frac{1}{2}t_{0}\,(-I^{'}(a_{1},a'_{1}) -I'(a_{1}',b_{1})-I'(b_{1},b'_{1})).
\end{split}\end{align}
Recalling that we are able to express the integrals $I'(p,q)$ applying the AGM method to the curve $C'$ (with real branchpoints) as in Theorem \ref{richthm} (combined with remark \ref{intoth}), eqs. \eqref{intcorr2} and \eqref{otherint} can be considered an extension to the AGM method, which gives  all the integrals between branchpoints on $C$.\\

Finally, we can apply considerations similar to the above also in \enf{case 1} of eq. \eqref{case1}, to find the integrals between non complex conjugate pairs of branchpoints. As before, we can express integrals between branchpoints on $C$ in terms of those on $C'$:
\begin{align}
\begin{split}\label{intalphaneg}
I(a,a')=& \; t_{0}I^{'}(a_{1},a_{1}'),\\
I(b,b')=&\; t_{0}I^{'}(b_{1},b'_{1}),\\
I(c,c')=&\;t_{0} I^{'}(c_{1},c'_{1}),\\
I(a,b)=& \frac{1}{2}\,t_{0}\,(\; I'(a_{1}',b_{1})-I'(c_{1},c'_{1})),\\
I(a',b')=&\frac{1}{2} \,t_{0}\,( I'(a_{1}',b_{1})+I'(c_{1},c'_{1})),\\
I(b,c)=& \frac{1}{2}\,t_{0}\,(\;-I^{'}(b_{1},b'_{1}) -I'(b_{1}',c_{1})-I'(c_{1},c'_{1})),\\
I(b',c')=&\frac{1}{2} \,t_{0}\,(I^{'}(b_{1},b'_{1}) +I'(b_{1}',c_{1})+I'(c_{1},c'_{1}))).
\end{split}\end{align}

\subsection{General remarks}\label{genrem}
We point out that the results obtained above hold true not simply for the monopole quotient curve, but for every curve satisfying either \eqref{case1} or \eqref{case2}. We summarise the findings of this section as follows:
\begin{agm}
The AGM method can be extended to the case of a curve $C$ with complex conjugate branchpoints, as follows:
\begin{enumerate}
\item perform the first Richelot transformation to obtain a curve with real branchpoints, $C'$;
\item  use the AGM method of eqs. \ref{intthm} to find the integrals between branchpoints on $C'$;
\item  use eqs. \eqref{expruvw} to calculate   $u$, $u'$, $v$, $v'$, $w$, $w'$ ,  and examine their relative ordering: there are several possibilities
\begin{enumerate}
\item \enf{case1}    $ v \leq w \leq w' \leq  u \leq   u' \leq  v' $
(same ordering as in Bost-Mestre case \cite{bm}): all the integrals between branchpoints on $C$ are given by eq. \eqref{intalphaneg};
\item \enf{case2} $u  \leq v \leq w \leq u' \leq  v'  \leq  w'$: the integrals between branchpoints on $C$ are given  by eqs. (\ref{intcorr2}) and (\ref{otherint}).%
\end{enumerate}
The other cases can be obtained in a similar fashion.
\end{enumerate}
\end{agm}
We remark that here we have only examined two possible orderings of  $u$, $u'$, $v$, $v'$, $w$, $w'$. To obtain the integrals between branchpoints on $C$ also for other orderings, one can perform a similar analysis of the images of the contours of integration, \ie straight lines between branchpoints, via the correspondence  \eqref{corrcurve}, to obtain the contours of integration  on $C'$.
\section[The AGM method for the quotient monopole curve]{The AGM method for the quotient monopole curve: solving the Ercolani-Sinha constraints}\label{AGMessect}
We apply the results of the previous section to the quotient monopole curve $X$ of eq. (\ref{quotientcurve}), where we scale $\beta$ to be $1$, to get
\begin{equation}\label{agmX}
Y^{2}=(X^{3}+a X +g)^{2}+4.
\end{equation}

If we split the branchpoints in complex conjugate pairs as in \eqref{bpcc}, to find the integrals between branchpoints on $X$ we work in steps, as explained in the previous section \ref{genrem}:  we perform the first Richelot transformation to go back to a curve with real branchpoints, $X'$;
 we can use eqs. \ref{intthm} to find the integrals between branchpoints on $X'$;
 finally,  to get all the integrals between branchpoints on $X$,  substitute these  in  eqs. \eqref{intcorr2}, \eqref{otherint} for $\alpha>0$, or in eqs. \eqref{intalphaneg} for $\alpha<0$.\\

Hence, for both cases we are able to use the AGM method to find all the integrals between branchpoints, and therefore, upon using the arc expansion \eqref{basisqexp}, to find the periods on $X$. This has been implemented in Maple, for this specific curve: we developed an algorithm that, for fixed values of $\alpha$ and $\gamma$, performs the AGM method as described above for case 1 and 2, giving as final output the periods on the basis of section \ref{secquotbasis}. These have been checked against the exact results obtained using Northover's code (see Appendix \ref{appmaple}) for a range of values of the parameters, and agree within $10^{-4}$. The advantage of the AGM method is that is much faster, as it deals only with polynomial manipulations; moreover, the convergence of the sequences  $(a_{n})$, $(a'_{n})$, $(b_{n})$, $(b'_{n})$, $(c_{n})$ only needs very few steps, usually 6 or 7, for the precision we require. These considerations allow us to use this method successfully in the iteration, described in the next section, to solve the Ercolani-Sinha constraints.

\subsubsection{Change of coordinates}
For convenience, we rescale $\beta$ to be 1. This is obtained using the following change of coordinates for the curve $\Xhat$
\begin{equation}\label{ccbeta}
w \to W=\beta^{-1/3} w, \quad \alpha \to a= \beta^{-2/3} \alpha, \qquad \gamma \to   g=\beta^{-1} \gamma,
\end{equation}
and for the curve $X$
\begin{equation}\label{ccbeta1}
x \to X=\beta^{-1/3} x, \quad y \to Y=\beta^{-1} y.
\end{equation}
For our purposes the only effect of this is on the constant $\nu$ in the Ercolani-Sinha constraints.\\
Indeed, recall that Theorem \ref{HMREScond} requires that
$-2\delta_{1k} = \oint_{\boldsymbol{n}\cdot\ah+ \boldsymbol{m}\cdot\bh} \Omega^{(k)}$ 
for the differentials 
$ \Omega^{(1)} = \dfrac{w^{n-2} \de z} {\frac{\partial P}{\partial w}} $,  $\Omega^{(2)} = \dfrac{w^{n-3} \de z} {\frac{\partial P} {\partial w}} ,\dots $.
Using the parametrisation \eqref{ccbeta} we then have
$$ \oint _{\gamma}\Omega^{(1)}=\oint _{\gamma} \frac{w}{3w^{2}+\alpha z^{2}}\de z= 
 \beta^{-\frac{1}{3}} \oint _{\gamma} \frac{W}{3W^{2}+\tilde{\alpha} Z^{2}}\de Z $$
Imposing 
    $-2 = \oint_{\boldsymbol{n}\cdot\ah+
\boldsymbol{m}\cdot\bh} \Omega^{(1)}  $,
we have that $$\nu = \beta^{\frac{1}{3}}.$$
This means that the genus 2 Ercolani-Sinha constraints \eqref{esg2} read
\begin{align}\begin{split}\label{escrescaled}
\oint_{\mathfrak{c}}\frac{1}{Y}\de X&=0\\
\oint_{\mathfrak{c}}\frac{X}{Y}\de X&=6\beta^{1/3}
\end{split}\end{align}
with $\mathfrak{c}$ as in eq. \eqref{esvectq}. The first of these does indeed impose a condition on $a$ and $g$, which we will solve in the next section; the second will be used to find $\beta$, and hence rescale back to $\alpha$ and $\gamma$.
\subsection{Solving the Ercolani-Sinha constraints}\label{secagmes}
We are now in a position to use the AGM method to solve the Ercolani-Sinha constraints, namely to find those values of the parameters $\alpha$ and $\gamma $ in the equation \ref{agmX} for $X$ that satisfy the constraints in the form \eqref{escrescaled}. We focus on the first of these, namely
\begin{align}\label{ESagm}
\oint_{\mathfrak{c}}\frac{1}{Y}\de X=
\left(n_{0}\int_{\aq_{0}}+3n\int_{\aq_{1}}+3m_{0}\int_{\bq_{0}}+3m\int_{\bq_{1}}\right)\UUU_{1}=0,
\end{align}
and proceed as follows.\\
We know already from the early work in \cite{symmmon} that there is a particularly symmetric monopole whose spectral curve belongs to the class \ref{agmX}: this is the \enf{tetrahedral monopole} \eqref{thetr}, corresponding to $a=0$, $g=\pm 5\sqrt{2}$ (in rescaled quantities).  We then seek a curve of solutions containing this monopole, \ie passing through the point $a=0, g=5\sqrt{2},$ in the space of parameters $a, \;g$.
The work of Braden and Enolskii in \cite{BE06}  (see also section \ref{sectesbe06}) gives a solution to the Ercolani-Sinha constraints for the case of the tetrahedral monopole, corresponding to the following values for the integers in \eqref{ESagm}, expressed in the cyclic basis of Figure  \ref{cycg2} (see section \ref{EScyclic})
\begin{align}\label{integerses}
m&=1, &n&=1, & { n_{0}}&=5\,n-m,&{ m_{0}}&=-3\,n.
\end{align}
As we seek a curve of solutions passing through the point $a=0, g=5\sqrt{2}$, by continuity, we expect that on this curve the constraints \eqref{ESagm} are satisfied with the same set of integers \eqref{integerses}.\\

Note that for the tetrahedral monopole with $a=0$ and $g=-5\sqrt{2}$ the Ercolani-Sinha constraints are satisfied for different values of integers, namely
\begin{align}\label{integersesm}
m&=0, &n&=1, & { n_{0}}&=5\,n-m,&{ m_{0}}&=-3\,n.
\end{align}
We hence expect another branch of the solution curve passing through the point $a=0, g=-5\sqrt{2}$, with constraints satisfied by the integers \eqref{integersesm}.\\

\begin{figure*} 
\begin{center}
\includegraphics[height=250pt]{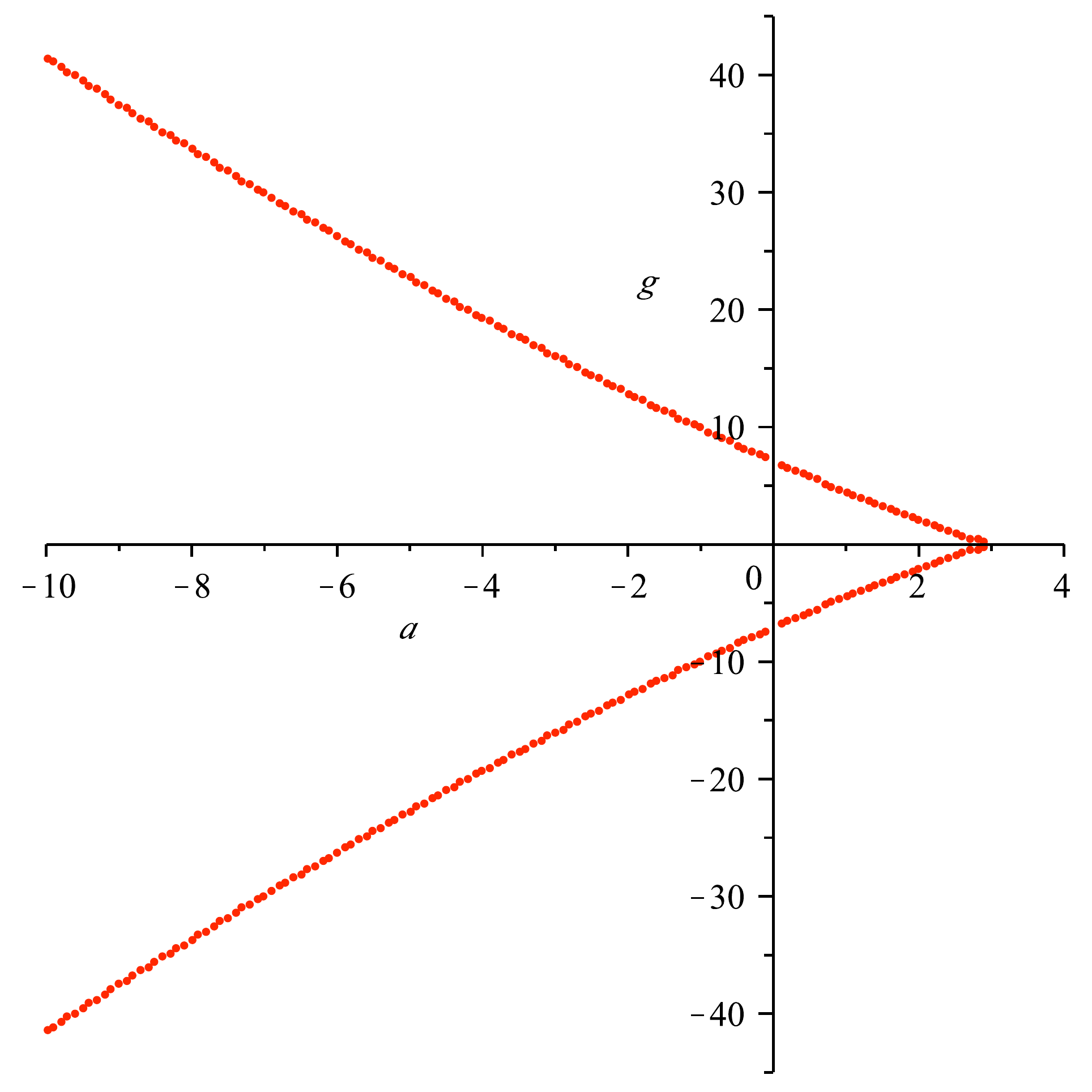}
\caption{Solutions to the Ercolani-Sinha constraints I}\label{ESsoln}
\end{center}
\end{figure*}

To begin, we wish to find solutions to \eqref{ESagm} starting from the point  $a=0$ and $g=5\sqrt{2}$, with integers   \eqref{integerses}:  we proceed iteratively as follows. We start varying $a$ by a small $\epsilon$, namely $a_{i}=i\cdot\epsilon\;$; we then vary $g$ in smaller steps, $g_{i,k}=5\sqrt{2}+k\cdot\epsilon^{2}$. For every such pair $(a_{i}, g_{i,k})$ we calculate the periods using the AGM method, and hence compute the first constraint in \eqref{ESagm} , with integers given by eq. \eqref{integerses}: for every $a_{i}$ we take the $g_{i,k}$ for which this constraint vanishes. We repeat this a sufficiently large number of times,  obtaining the curve in Figure \ref{ESsoln}. We now comment on this graph, referring to the Maple file  \texttt{AGM\_ES\_monopole.mw} for more detail.\\

 We have used a step of $10^{-1}$ for $a$, and of $10^{-2}$ for $a\in (2.8,3.0)$ to obtain a greater detail in this interval. \\ 
 For values of $a\in\{0,3\}$, the outcome is that  we have a curve of solutions in the space of parameters passing through the points $(0,5\sqrt{2})$ up to the point $(3,0)$, which, however, does not belong to this solution curve.\\

We remark that the point  $(3,0)$ is in fact a singular point, as 4 of the branchpoints collide pairwise, giving two singular points at $\pm i$: this results in a rational curve, with equation
\begin{equation}\label{ratmon}
y^{2}=(x^{3}+3x)^{2}+4=(x^{2}+4)(x^{2}+1)^{2}.
\end{equation}
Note that the curves of equation \eqref {agmX} with $g=0$ and $a\neq 0$, namely $y^{2}=(x^{3}+a x)^{2}+4$, are all hyperelliptic: the only exception is precisely the case $a=3$ above.\\
As all the cycles on the curve \eqref{ratmon} are homologous to zero, the first of eqs. \eqref{escrescaled} is trivially satisfied, but the second is zero as well: hence the Ercolani-Sinha constraints are not satisfied for this curve. Thus the values $g=0$ and $a = 3$ do not correspond to a monopole. \\

We are also able to extend this curve for negative values of $a$, on the left of the point $(0,5\sqrt{2})$: in Figure \ref{ESsoln} we plot 100 points corresponding to $a<0$, $g<0$.\\

When trying to extend the solution curve to $g<0$, we do not observe any values of the parameters satisfying  the first of eqs. \eqref{ESagm} with integers \eqref{integerses}. But we notice that, since the point $(3,0)$ does not belong to the solution curve, continuity arguments do not prevent us to use the different set of integers of eq. \eqref{integerses}: with these, we manage to extend the solution curve through the point $(0,-5\sqrt{2})$, which corresponds again to a tetrahedral monopole. We point out  that this arc of curve for $g<0$ is precisely the reflection with respect to the $a$-axis of the arc obtained for $g>0$.\\
\begin{figure}[h!]
\begin{center}
\includegraphics[height=250pt]{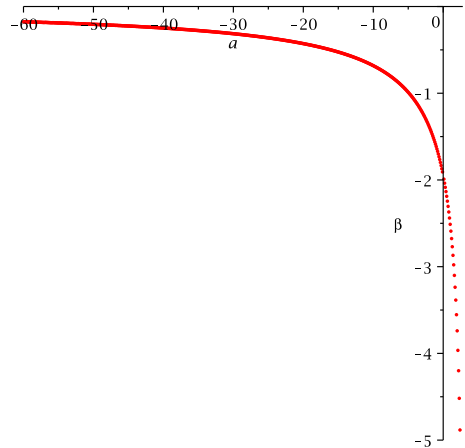}
\caption{$\beta$ as a function of $a$}\label{betaa}
\end{center}
\end{figure}

Finally, using the second equation in \eqref{escrescaled}, we can obtain $\beta$ for each pair $(a,\;g)$, again using the AGM algorithm; we prsent $\beta$ as a function of $a$ in Figure \ref{betaa}. This allows us to find the original quantities $\alpha$ and $\gamma$ in view of \eqref{ccbeta,ccbeta1}.

\subsection{Elliptic monopoles}
\begin{figure} 
\begin{center}
\includegraphics[height=250pt]{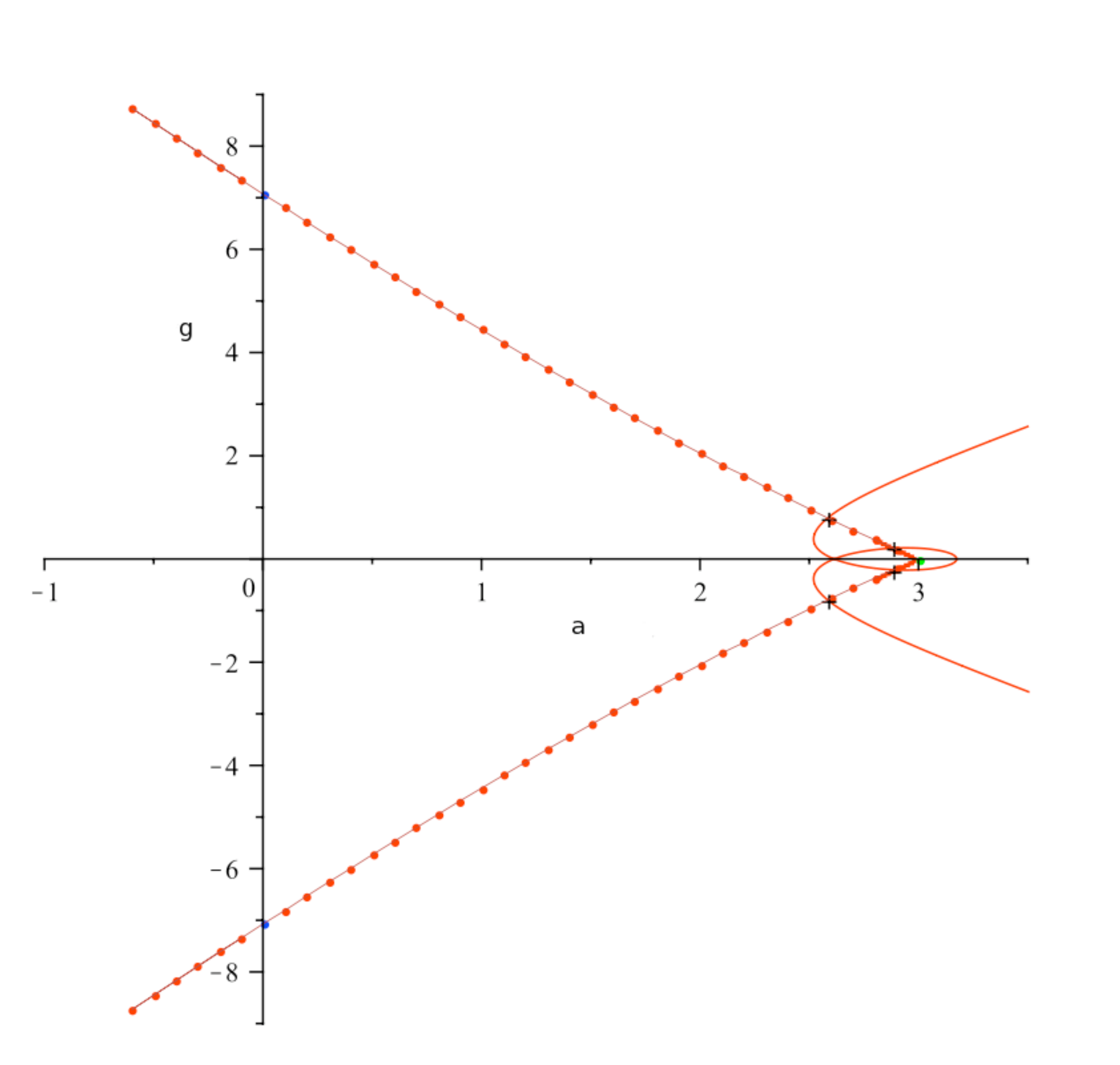}
\caption{Solutions to the Ercolani-Sinha constraints II}\label{ESchi30}
\end{center}
\end{figure}

In section \ref{ellsubc} we have examined the conditions under which the quotient monopole curve $X$ admits a degree 2 elliptic subcover: the vanishing of the invariant $\chi_{30}$ gives a curve of real solution in the parameter space $a,g$, plotted in  Figure \ref{plot_chi}. We point out here that the solutions to the Ercolani-Sinha constraints found above do intersect this curve. This is plotted in Figure \ref{ESchi30}, where we see clearly that there are four points that satisfy both the Ercolani-Sinha constraints and the $\chi_{30}$ vanishing conditions; their coordinates have been found numerically to be
\begin{align*}
a_{1}&=2.584590, g_{1}=0.795087,\\
a_{2}&=2.884209, g_{2}=0.209470,\\
a_{3}&=2.884478, g_{3}= -0.209525,\\
a_{4}&=2.584985,  g_{4}=-0.796463.
\end{align*}
This means that for the above values of the parameters the curve $X$ covers 2 elliptic curves, and hence the explicit solutions corresponding to these two monopoles can be expressed in terms of elliptic functions. This will be object of further study in the near future.

\section{Hitchin constraint \enf{H3}}
Until now we have dealt only with one Ercolani-Sinha constraint, namely that of Theorem \ref{escthm} (or equivalently Theorem \ref{hmrthm}), corresponding to Hitchin constraint \enf{H2}, which ensures that the boundary conditions \eqref{bc} are satisfied. Nothing has been said about the constraint \enf{H3}, which is equivalent to the nonsingularity of the monopole: we discuss this in the present section.\\

Recall that Braden and Enolskii showed in \cite{BE06} that \enf{H3} is equivalent to determine whether the line $sU-\tvrc$ intersects the theta divisor $\Theta$ (see Theorem \ref{beh3thm}). This condition has indeed been checked in the case of the tetrahedral monopole in \cite{BE09}: they use the same methods employed in this work of thesis to express the theta functions in terms of lower genus ones, which allows to check the above condition. Their conclusion is that the only monopole in the class \eqref{BEcurv} (corresponding to $\alpha=0$ in our case) is the tetrahedral monopole (cf. Theorem \ref{mainthmBE}).\\

We use this result in conjunction with an analysis of the moduli space of cyclically symmetric monopoles performed by Hitchin, Manton and Murray in \cite{symmmon}. They show that cyclically symmetric monopoles form $4$-dimensional totally geodesic submanifolds $\mathcal{M}_n\sp{l}$ (where $0\le l<n$) of the full moduli space of charge $n$ monopoles. In our case, $n=3$ and ignoring the rotational degrees of freedom we have a one dimensional submanifold. \\
In the previous section we have established the existence of a one parameter family of 
curves satisfying the constraints \enf{H1} and \enf{H2}, containing the tetrahedral monopole curves. In particular, the curve of Figure \ref{ESsoln} were built in such a way that they describe the only possible curves satisfying the Ercolani-Sinha constrains passing through tetrahedral monopoles.\\ Combining these two remarks, we expect that this conditions is satisfied on the whole curve of Figure \ref{ESsoln}.\\

We point out that the constraint \enf{H3}, namely the nonvanishing of  $\hat{\theta}(s\hat{U}-\tvrc)$ for $s\in(0,2)$ translate into the vanishing of genus 2 theta functions. Once can show (see \cite{B10}) that 
$$
\tilde{ \vrc} = \pi^{*}(\vrc_{\infty^{+}}-\halfp).
$$
Combining this with \eqref{UUhat} we obtain that
$$s\hat{U}-\tilde{ \vrc} = \pi^{*}(sU-\vrc_{\infty^{+}}-\halfp).$$
 Using Theorem \ref{fayaccolathm2} (see also eq. \eqref{thetafay}) we have that the nonvanishing condition above can be expressed as the nonvanishing of the 3 genus 2 theta functions
 $\theta\left[\begin{matrix}0&0
\\ \frac{k}{3}&0 \end{matrix}\right]\left(\boldsymbol{ z};\tau
\right)$.
This has indeed been checked numerically for several values of $s$, see \cite{BDE10}, confirming the prediction above.

\section{Conclusions and discussion}
Before commenting in more detail on the results obtained  so far, let us summarise the arguments presented in this work of thesis.\\
The aim was to investigate the existence of charge $3$ cyclically symmetric monopoles; this translates into verifying whether certain constraints on the coefficients of the spectral curve $\Xhat$ are satisfied. The first step consists in using the $C_{3}$ symmetry of the system to quotient the genus $4$ curve $\Xhat$ to get a genus $2$ curve $X$. This quotient is an unbranched cover, which allows us to use Fay-Accola theory to express several quantities on $\Xhat$ in terms of analogous quantities on $X$: to do this,  the use of a particularly symmetric homology basis is crucial. \\
A fundamental upshot of this is that the Ercolani-Sinha constraints reduce to constraints among the periods of the genus $2$ curve $X$. We focus in particular on the constraint corresponding to Hitchin condition \enf{H2}: we show that this basically reduces to check the vanishing of $\oint\limits_{\boldsymbol{\mathfrak{c}}}\frac{\de X}{Y}$. Solving this entails solving explicitly several hyperelliptic integrals: this is done by adapting the so called AGM method. This method is well known in the case of real branchpoints, we manage to extend it to the case of complex conjugate branchpoints. \\
Using the AGM method iteratively, we are able to find a curve of solutions for \enf{H2} in the moduli space, and we argue that for these \enf{H3} is satisfied as well. Hence we have the main result of this work of thesis, namely \enf{we establish the existence of a one parameter family of cyclically symmetric charge 3 monopoles; we do so by finding explicit values for the parameters $a$, $g$, given in Figure \ref{ESsoln} for which the curve \eqref{eq:xhat} does describe a monopole}.\\

\begin{figure}
\begin{center}
\includegraphics[height=280pt]{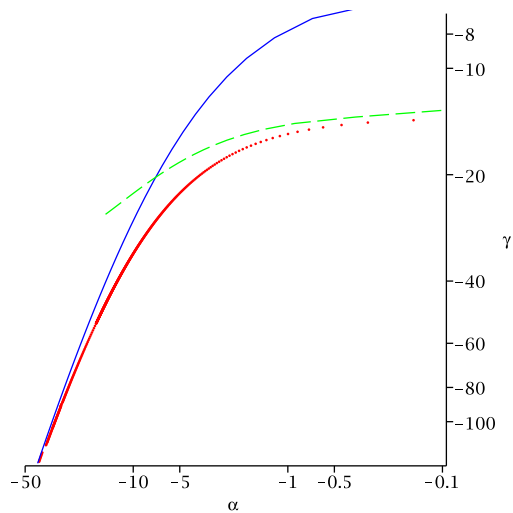}
\caption{Log-log plot of the asymptotic behaviour of $\alpha$ versus
$\gamma$ according to Hitchin, Manton and Murray (solid), Sutcliffe (dash) and
here (dots).}\label{asb}
\end{center}
\end{figure}

We conclude comparing our results with those of other authors.\\
We remark that the behaviour of the solution curve in Figure \ref{ESsoln} is consistent with the results of \cite{sutcyclic}, where Sutcliffe predicts that the curve \eqref{ratmon}, describing a configurations of three unit-charge monopoles with dihedral $D_{3}$ symmetry, constitutes an asymptotic state for a 3-monopole configuration (cf. eq. (4.16) in \cite{sutcyclic}). Moreover, both Sutcliffe \cite{sutcyclic} and Hitchin, Manton and Murray \cite{symmmon} give the following picture of the scattering of three monopoles, corresponding to  geodesic motion along one of these loci. Three unit charge monopoles  come in at the vertices of an equilateral triangle, moving towards its 
centre, in the $x_{1}-x_{2}$ plane. They then coalesce instantaneously into a tetrahedron.  Finally the tetrahedron breaks up in a unit charge monopole moving along the positive $x_{3}$-axis and an axisymmetric charge 2 monopole, moving along the negative $x_{3}$-axis. The asymptotic behaviour at this end of the scattering is the following:
\begin{equation}\label{asy}
\alpha \sim(\pi^2/4 -3b^2),\qquad \gamma \sim
2b(b^2+\pi^2/4),\qquad \beta \sim 0.
\end{equation}
We plot this relation together with the set of points we obtain from solving the Ercolani-Sinha constraints, on a log-log plot; note that here we use the values of $a$ and $g$ of Figure \ref{ESsoln}, and the corresponding values of $\beta$ to revert to $\alpha$ and $\gamma$. Figure \ref{asb} suggests that our results are indeed consistent with the asymptotic behaviour \eqref{asy}. Our results then confirm quantitatively the approximate predictions of  \cite{symmmon,sutcyclic}.

\newpage
\quad
\newpage

%
%
%
%
%
%
%
%
%
\appendix
\chapter{Riemann Surfaces}\label{apprs}
In this appendix we recall some basics about Riemann Surfaces, essentially to set notation and fix out conventions. For proofs of the theorems stated here, and a thorough exposition of these topics, we refer to \cite{farkaskra, fay, miranda}, amongst the vast literature on the subject.

\section{Cycles and period matrix}
Given a Riemann surface $\Sigma$ of genus $g$, denote by $\mathfrak{a}_{1}, \ldots , \mathfrak{a}_{g}$ , $\mathfrak{b}_{1}, \ldots , \mathfrak{b}_{g}$  a canonical basis for $H_{1}(\Sigma,\mathbb{Z})$, namely a basis such that its intersection matrix is $\pm J$, where
\begin{equation*}
J=\left(
\begin{array}{cc}
 \mathbb{O}_{g}  &   \mathbb{I}_{g}   \\
- \mathbb{I}_{g}  &   \mathbb{O}_{g} 
\end{array}
\right)
\end{equation*}
Throughout this thesis we choose the minus sign, to be consistent with the conventions of \cite{BE06}, as we aim to generalise the results therein.\\

Denote by $(\mathbf{u}_{1},\ldots,\mathbf{u}_{g})$ a basis of holomorphic differentials; we then define the matrices of periods $\mathcal{A}$ and $\mathcal{B}$ to be
\begin{align*}
\mathcal{A}&=\left(\mathcal{A}_{ki}\right)=\left(
  \oint\limits_{\mathfrak{a}_k} \mathbf{u}_{i}\right), \qquad
 \mathcal{B}=\left(\mathcal{B}_{ki}\right)=\left(
  \oint\limits_{\mathfrak{b}_k} \mathbf{u}_{i}\right), \quad i,k=1,\ldots,g
\end{align*}

Given $\mathcal{A}$ and $\mathcal{B}$ we now construct the Riemann
period matrix. We work with canonically
$\mathfrak{a}$-normalised differentials,  the period matrix is $$\tau_\mathfrak{a}=\mathcal{B}\mathcal{A}^{-1};$$
while for canonically $\mathfrak{b}$-normalised differentials it
is $\tau_{\mathfrak{b}}=\mathcal{A}\mathcal{B}^{-1}$; clearly
$\tau_{\mathfrak{b}}=\tau_\mathfrak{a}\sp{-1}$. We sometimes
denote the period matrix by $\tau$ when neither normalisation is
necessary.

\section{Theta Functions}
The canonical Riemann $\theta$-function is given by
\begin{equation}
\theta(\boldsymbol{z};\tau) =\sum_{\boldsymbol{n}\in \mathbb{Z}^r}
\exp(i\pi \boldsymbol{n}^T \tau\boldsymbol{n}+2i\pi \boldsymbol{z}^T \boldsymbol{n}),
\end{equation}
where $r\in \mathbb{N}$. The Riemann $\theta$-function is holomorphic on
$\mathbb{C}^r\times\mathbb{S}^r$ and satisfies the following periodicity conditions
\begin{equation}\theta(\boldsymbol{z}+\boldsymbol{p}\, ;\tau)=
\theta(\boldsymbol{z};\tau),\quad
\theta(\boldsymbol{z}+\boldsymbol{p}\tau;\tau)=
\mathrm{exp}\{-\imath\pi(\boldsymbol{p}^T\tau \boldsymbol{p}
+2\boldsymbol{z}^T\boldsymbol{p})\}\, \theta(\boldsymbol{z};\tau),
\label{transformation}
\end{equation}
where $\boldsymbol{p}\in\mathbb{Z}^r$.

The Riemann $\theta$-function
$\theta_{\boldsymbol{a},\boldsymbol{b}}(\boldsymbol{z};\tau)$ with
characteristics $\boldsymbol{a},\boldsymbol{b}\in\mathbb{Q}$ is
defined by
\begin{align}\label{thetachar}
\theta_{\boldsymbol{a},\boldsymbol{b}}(\boldsymbol{z};\tau)
&=\mathrm{exp} \left\{ \imath\pi
(\boldsymbol{a}^T\tau\boldsymbol{a}
+2\boldsymbol{a}^T(\boldsymbol{z}+\boldsymbol{b})))\right\}
\theta(\boldsymbol{z}+\tau\boldsymbol{a}+\boldsymbol{b};\tau)\\
& =\sum_{\boldsymbol{n}\in\mathbb{Z}^r}\mathrm{exp}
\left\{\imath\pi(\boldsymbol{n}+\boldsymbol{a})^T\tau
            (\boldsymbol{n}+\boldsymbol{a})
+2\imath\pi
(\boldsymbol{n}+\boldsymbol{a})^T(\boldsymbol{z}+\boldsymbol{b})
\right\},\nonumber
\end{align}
where $\boldsymbol{a},\boldsymbol{b}\in \mathbb{Q}^r$. An equivalent notation is
$$\theta_{\boldsymbol{a},\boldsymbol{b}}(\boldsymbol{z};\tau)=
\theta\left[\begin{matrix}\boldsymbol{a}
\\ \boldsymbol{b}\end{matrix}\right](\boldsymbol{z};\tau).
$$
The function $\theta_{\boldsymbol{a},\boldsymbol{b}}(\tau)=
\theta_{\boldsymbol{a},\boldsymbol{b}}(\boldsymbol{0};\tau) $ is
called the $\theta$-constant with characteristic
$\boldsymbol{a},\boldsymbol{b}$.\\

\begin{theorem}[\enf{Riemann theorem}]
If $\abelmap$ is the Abel map based at a point $Q_{0}$,
and $d$ is  a generic point in $\Jac(\Sigma)$,
the function $\theta(\abelmap(P)+d)$ either vanishes
identically, or has $g$ zeros $P_{1},\ldots ,P_{g}$, such that%
\begin{equation}
  \abelmap(P_{1})+\ldots \abelmap(P_{g})=d-\vrc.
\end{equation}
\end{theorem}

Here $\vrc$ is the vector of Riemann constants, independent
of $d$ but dependent on $Q_{0}$ and on the homology basis.
With this choice of sign convention for $\vrc$ one finds
 \begin{align*}
 (\vrc)_j&=\frac12
\tau_{jj}
-\sum_{k}\oint_{\mathfrak{a}_k}\boldsymbol{\omega}_k(P)\int_Q\sp{P}\boldsymbol{\omega}_j.
 \end{align*}

\subsection{Weierstrass reduction}

When a curve covers one of lower genus, the period matrix admits a reduction, \ie can be expressed in terms of a lower genus period matrix, and similarly, the associated theta functions can be expressed in terms of lower dimensional theta functions. This is known as Weierstrass reduction; a description purely in terms of the matrix of periods is given in \cite{ma92a} and more recently \cite{be02a},\cite{be02b}.\\

A $g\times 2g$ period matrix
$\boldsymbol{\Pi}$ is said to admit reduction if there exist a
maximal rank $g_{1}\times g$ matrix ${\lambda}$ of
complex numbers, a $g_{1}\times g_{1}$ matrix of complex numbers
$\boldsymbol{\Pi}_{1}$ and a maximal rank $2g_{1}\times 2g$ matrix
of integers $M$ such that:%
\begin{equation}\label{red}
  {\lambda}\boldsymbol{\Pi}=\boldsymbol{\Pi}_{1}M.
\end{equation}
where  $1\leq g_1 < g$.\\

When a period matrix admits reduction, it can be shown
(Weiestrass' theorem) that there exists an element
$\boldsymbol{\sigma}\in\mathrm{Sp}(2g,\mathbb{Z})$
such that %
\begin{equation}
  \boldsymbol{\sigma}\cdot\tau=\left(%
\begin{array}{cc}
  \tau_{1} & Q \\
Q^{T} & \tau^{\#}\\
\end{array}%
\right),
\end{equation}
where $\tau'$ and $\tau_{1}$ have the
properties of a Riemann period matrix and $Q$ is a
$g_{1}\times (g-g_{1})$ matrix with rational entries.\\
Because $Q$ here has rational entries, there exists a diagonal $(g-g_1)\times(g-g_1)$ matrix
$D=\mathrm{Diag}(d_1,\ldots,d_{g-g_1})$ with positive integer entries for
which $(QD)_{jk}\in\mathbb{Z}$. With
$(z,w)=(z_1,\ldots,z_{g_1},w_1,\ldots,w_{g-g_1})$ the theta
function associated with $ \tau$ may then be expressed in terms of
lower dimensional theta functions as
\begin{equation}\label{redthetagen}
\theta((z,w); \tau)=\sum_{\substack{
\mathbf{m}=(m_1,\ldots,m_{g-g_1})\\
0\le m_i\le d_i-1}}
\theta(z+Q\mathbf{m}; \tau_1)\,\theta\left[%
\begin{array}{c}
  D\sp{-1}\mathbf{m} \\
  0 \\
\end{array}%
\right](Dw;D \tau\sp{\#}D).
\end{equation}

\chapter{A sketch of the Tretkoff and Tretkoff algorithm}\label{apptt}
In this appendix we describe the construction by Tretkoff and Tretkoff, as in \cite{TT}, which is an algorithmic way to find a \emph{canonical} basis for the first homology group of a Riemann surface.\\
Its key elements are that the Riemann surface $X$ is seen a  branched cover of the sphere, and its Hurwitz system is used to find a cellular decomposition of $X$ (in terms of CW complexes). From this cellular decomposition one can find a system of generators for $H_{1}(X,\mathbb{Z})$, and, once calculated their intersection matrix, obtain out of it a canonical basis.\\
We now examine these elements in detail.

\section{Details of the algorithm}
Let $X$ be a Riemann surface of genus $g$, and let $p:X\rightarrow \puno$ be its branched cover of the Riemann sphere. The Hurwitz system consists of the following data:
\begin{itemize}
\item[-] $n$ sheets;
\item[-] $t$ branchpoints $b_{1},\ldots ,b_{t}$ in  $\sferar$;
\item[-] for every branchpoint, the \enf{monodromy group} associated to it (cf. section \ref{remarks}).
\end{itemize}
Also, we will use the following notation
\begin{equation*}
Z:= \sferar -\{b_{1},\ldots ,b_{t}\} ,\qquad \tilde{Z}:=p^{-1}(Z).
\end{equation*}
\subsubsection{Decomposition in CW-complexes}
One can choose on $\sferar$ an arbitrary point $a$ not coincident with any of the branchpoints $b_{1},\ldots ,b_{t}$. Joining $a$ with all the $b_{i}$ one obtains $t$ segments $l_{i}$. The 1-dimensional CW complex, or 1 skeleton, $\Gamma$ of $X$ is thus%
\begin{equation}
\Gamma = \bigcup p^{-1}(l_{i}).
\end{equation}
We observe that the Riemann sphere minus the union of the $l_{i}$, \ie  $D:=\sferar - \cup l_{i}$ is homeomorphic to the interior of a disc, namely a 2-cell: this is the prototype of the cells to be attached to the 1-skeleton $\Gamma$ to obtain a cellular decomposition of $X$.\\
Recall that
\begin{equation}\label{eq:quot}
H_{1}(X,\mathbb{Z})=
H_{1}(\Gamma,\mathbb{Z}) \diagup (\partial \bar{D}_{1},\ldots ,\partial \bar{D}_{n}),
\end{equation}
namely the quotient of $H_{1}(\Gamma,\mathbb{Z})$ with the group generated by the $\partial \bar{D}_{n}$ with the relation $\partial \bar{D}_{1}+\ldots + \partial \bar{D}_{n}=0$.\\
Using property (\ref{eq:quot}) one can find an homology basis for $X$ out of one for $\Gamma$: this is useful as it is indeed quite immediate to find a basis for  $H_{1}(\Gamma,\mathbb{Z})$, as we  discuss in the following.
\subsubsection{A basis for $H_{1}(\Gamma,\mathbb{Z})$}
Let $T\in \Gamma$ be a \emph{maximal tree}, \ie a contractible subgroup containing all the vertices in $\Gamma$, and denote $e_{1},\ldots ,e_{r}$ the edges of $\Gamma$ \emph{not} in $T$.\\ 
One can build a nontrivial cycle in the following way. For every two vertices of $\Gamma$, $P_{i},P_{j}$, there is a unique path in $T$ connecting them, denote it $\gamma_{i,j}$; so a cycle in $\Gamma$ passing through $P_{i},P_{j}$ is the union of $c_{i,j}$ and the edge  $e_{i}$ connecting  $P_{i},P_{j}$. In this way we obtain $r$ cycles $c_{1},\ldots, c_{r}$, one for every edge $e_{i}$ in $\Gamma$ not in $T$.\\
It is immediate to see that the above is a basis for  $H_{1}(\Gamma,\mathbb{Z})$, since, contracting $T$ to a single point, one gets a bouquet of $R$ circles.
\subsubsection{Intersection numbers}
We briefly discuss the way the authors propose to calculate the intersection numbers for the cycles in $H_{1}(\Gamma,\mathbb{Z})$, which is much simpler than counting ``by hand'', and has the advantage of being algorithmic.\\

Consider an open set $U\in X$ such that:
\begin{itemize}\label{open}
\item[A1.] $T\in U$
\item[A2.] it can be mapped homeomorphically and preserving orientation onto the unit disc,
\item[A3.] $\partial\bar{U}$ meets the $e_{i}$ in 2 distinct points, $P_{i}, Q_{i}$.
\end{itemize} 
The idea is to replace  the  subarc of $c_{i}$ in  the more complicated tree $\Gamma$  with the chord connecting $P_{i}, Q_{i}$ in this disc. After this replacement, the element in $H_{1}(\Gamma,\mathbb{Z})$ represented by $c_{i}$ is unchanged. Therefore the intersection between two cycles $c_{i},c_{j}$ is the same as the intersection of the 2 chords  $P_{i}Q_{i}$ and $P_{j} Q_{j}$. The latter is indeed much easier to calculate, as it is shown in section \ref{ttxhat} for the case of the curve $\Xhat$ of equation (\ref{eq:xhat}): the outcome of this is an $r \times r$ intersection matrix.
\subsubsection{A canonical basis for  $H_{1}(X,\mathbb{Z})$}
In order to get a basis for  $H_{1}(X,\mathbb{Z})$  out of that  for  $H_{1}(\Gamma,\mathbb{Z})$ found thus far, one would have to quotient by the group generated by $\partial\bar{D}_{j}$, and then consider the issue of canonicity. Tretkoff and Tretkoff remark that this can be achieved in a different, more algorithmic, easier way. Viewing the cycles $c_{i}$ as elements of $H_{1}(X,\mathbb{Z})$ , their intersection matrix is the same as the one calculated above. But an intersection matrix $K$ for a genus $g$ surface between an arbitrary number of cycles  can always been put in standard form 
\begin{equation}
A.K.A^{T}=
\left(
\begin{array}{ccc}
 \mathbb{O}_{g} & \mathbb{I}_{g}  &\mathbb{O}   \\
 -\mathbb{I}_{g} &  \mathbb{O}_{g} & \mathbb{O}  \\
   \mathbb{O}&  \mathbb{O}  &    \mathbb{O}
\end{array}
\right)
\end{equation}
Hence the matrix $A$ selects $2g$ cycles among the set of generators above, such that their intersection matrix is that required; in detail:
\begin{equation}
a_{i}=\sum_{1}^{r}A_{i,j}c_{j}, \qquad 
b_{i}=\sum_{1}^{r}A_{i+g,j}c_{j}
\end{equation}
 We remark that the fact that the intersection matrix among these cycles is non-degenerate also ensures that they are independent.
%
%
\subsubsection{The algorithm}
The previous method can also be implemented algorithmically: we  describe this, and apply in detail this algorithm to our case.\\
The idea is to give a way to construct, algorithmically, a ``planar version'' $\Delta$ of the 1-skeleton $\Gamma$, \ie a tree from which one will obtain $\Gamma$ after a quotient; then, since $\Delta$ is planar, it is easier to find the intersection matrix.\\
The algorithm works as follows.
\begin{itemize}
\item[1.] Consider as basepoint a point on the Riemann sphere, call it $a$, and call $a_{k}$ its preimages under $p$ on sheet $k$; choose one of them, say  $a_{1}$ on sheet 1. Then draw a family of circles $C_{i}$ with center in $a_{1}$, oriented clockwise. To construct the planar version of $\Gamma$, start from $a_{1}$, select points on $C_{i}$ and join them to previously constructed vertex on $C_{i-1}$. Starting from $C_{1}$, draw $t$ points on this circle, and label them $b_{1},\ldots ,b_{t}$: the labeling suggests that  these will be identified with the branchpoints. Then join the point $a_{1}$ to them.
\item[2.] On $C_{2}$, for every $b_{i}$ on $C_{1}$  put $d$ vertices, as many as the elements of the monodromy group at that branchpoint;  label them with the number of the sheet, in the same order as they are reached applying a permutation, \ie $k_{1}=\pi_{1}(1),k_{2}+\pi_{1}^{2}(1),\ldots , k_{d}=\pi_{1}^{d}(1)$ (see example in section \ref{ttxhat}). This means that for every ramification point on sheet $1$ one is applying a permutation to go to a different sheet, that is to a different $a_{k}$. Then connect each $b_{i}$ with its corresponding vertices obtained in this way.
\item[3.] On $C_{3}$, for every point $k$  on $C_{2}$ coming from a $b_{i}$ build vertices labeled  again\\ $b_{i+1},\ldots,b_{t},b_{1},\ldots ,b_{i-1} $, in this order, and connect the vertex $k$ to these points. This means that one is following a path from each $a_{i}$ to a ramification point that is different from that one is coming from.
\item[4.] Repeat this construction to get a sequence of base points on each $C_{2k}$, a sequence of ramification points on $C_{2k+1}$, connecting them as explained above, and stopping  at a point with a label that has already appeared.\\
\end{itemize}

The tree $\Delta$ built in this way is finite, and the 1-skeleton $\Gamma$ can be obtained as a quotient of $\Delta$, identifying the edges between vertices with the same label. This is done in detail in the next section, where an explicit construction is given for a special case, and also   the existence of an open neighborhood $U$ suitable for the computation of the intersection numbers is apparent, as explained above.
\section{The algorithm applied to the curve $\Xhat$}\label{ttxhat}
\begin{figure}
\begin{center}
\includegraphics*[height=600pt, trim= 80 200 0 80]{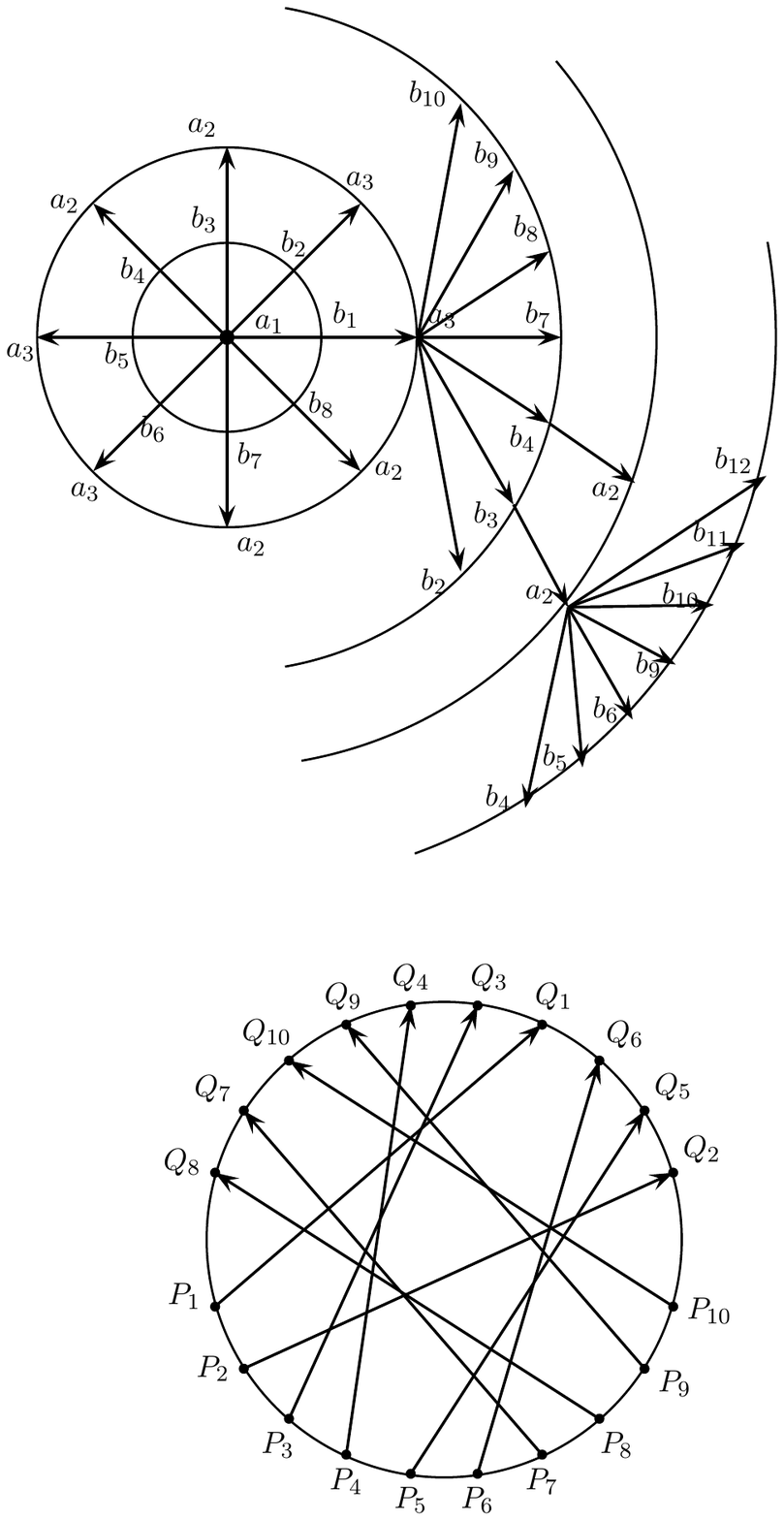}
\caption{Tretkoff and Tretkoff algorithm for the curve $\Xhat$}\label{tt1}
\end{center}
\end{figure}

We apply in detail the previous algorithm to our curve $\Xhat$ of eq. (\ref{eq:xhat}): the result obtained following the steps above is given in Figure \ref{tt1}.\\

\newpage
The cycles obtained after quotienting are clearly labeled in the figure. Their intersection numbers are computed as explained previously: we consider an open neighborhood $U'$, indicated in the previous figure, such that its image under the quotient satisfies  \ref{open}; we map $U'$ onto a disc, consider the points $P_{i},Q_{i}$ on its boundary, image of the intersection of each cycle $c_{i}$ with $\partial U'$ and compute the intersections of the chords $P_{i}Q_{i}$, $P_{j}Q_{j}$, which is the same as the intersection of $c_{i}$with $c_{j}$, as explained above. This is all represented in Figure \ref{tt1}.\\
From the figure, the intersection matrix is
\begin{equation}\label{eq:intTT}
I= \left[ \begin {array}{cccccccccc} 0&0&1&1&0&0&1&1&1&1
\\\noalign{\medskip}0&0&1&1&1&1&1&1&1&1\\\noalign{\medskip}-1&-1&0&1&0
&0&1&1&1&1\\\noalign{\medskip}-1&-1&-1&0&0&0&1&1&1&1
\\\noalign{\medskip}0&-1&0&0&0&1&1&1&1&1\\\noalign{\medskip}0&-1&0&0&-
1&0&1&1&1&1\\\noalign{\medskip}-1&-1&-1&-1&-1&-1&0&1&0&0
\\\noalign{\medskip}-1&-1&-1&-1&-1&-1&-1&0&0&0\\\noalign{\medskip}-1&-
1&-1&-1&-1&-1&0&0&0&1\\\noalign{\medskip}-1&-1&-1&-1&-1&-1&0&0&-1&0
\end {array} \right] .
\end{equation}
To have a canonical basis, this matrix needs to be put in the canonical skew-symmetric form. There is a well known algorithm that does so, and also gives the matrix of change of basis (see \eg \cite{siegel}, p.65): this has been implemented with Maple, and the result is that the intersection matrix above can be put in the form
\begin{equation}
M.I.M^{T}=
\left(
\begin{array}{ccc}
 \mathbb{O}_{4} & \mathbb{I}_{4}  &\mathbb{O}_{4,2}   \\
 -\mathbb{I}_{g} &  \mathbb{O}_{g} & \mathbb{O} _{4,2} \\
   \mathbb{O}_{2,4}&  \mathbb{O}_{2,4}  &    \mathbb{O}_{2,2}
\end{array}\right),
\end{equation}
where $M$ is found to be
\begin{equation}
M=
 \left[ \begin {array}{cccccccccc} 1&0&0&0&0&0&0&0&0&0
\\\noalign{\medskip}0&1&0&0&0&0&0&0&0&0\\\noalign{\medskip}0&0&1&0&0&0
&-1&0&0&0\\\noalign{\medskip}0&0&0&1&0&0&-2&1&0&0\\\noalign{\medskip}0
&0&0&0&0&-1&0&0&1&0\\\noalign{\medskip}-1&0&0&0&0&1&-1&1&0&0
\\\noalign{\medskip}0&-1&0&0&0&0&2&-1&-1&0\\\noalign{\medskip}0&1&-1&0
&0&0&0&0&1&0\\\noalign{\medskip}1&-1&0&0&1&-1&0&0&0&0
\\\noalign{\medskip}1&0&-1&1&0&0&-1&1&-1&1\end {array} \right] .
\end{equation}
Thus, in term of the cycles defined earlier, the required basis is 
\begin{align*}
 ( c_{{1}},c_{{2}},\;c_{{3}}-c_{{7}},\;c_{{4}}-2\,c_{{7}}+c_{{8}},\;-c_{{6}}+c_{{9}},\;
-c_{{1}}+c_{{6}}-c_{{7}}+c_{{8}},\;-c_{{2}}+2\,c_{{7}}-c_{{8}}-c_{{9}},\;c_{{2}}-c_{{3}}+c_{{9}}).
\end{align*}
\chapter{Maple tools: algcurves and extcurves}\label{appmaple}
Various results presented  in this thesis have been obtained with, or rely on work done with Maple. In particular, the \texttt{algcurves} package for Maple provides a way to compute various objects on a Riemann surface; the \texttt{extcurve} code by Northover, together with its visual aid \texttt{cyclepainter}, relies on this for implementing new tools which have been used throughout this work of thesis. Here we explain in short, referring to the relevant documentation for more details. \texttt{algcurves}, developed in collaboration by Deconinck, von Hoeij and Patterson, has been part of Maple since Maple 6: an explanation of all the commands is included in Maple's help, and a more thorough description is given in \cite{algc1}. \texttt{Extcurves} and \texttt{cyclepainter}, developed by Timothy Northover, can be downloaded at  \url{http://gitorious.org/riemanncycles}, where  the relevant documentation is also to be found.

\section{\texttt{algcurves}}
The Maple package \texttt{algcurves} allows to calculate several objects on a Riemann surface. The Riemann surface is input as a polynomial in two complex variables, and is viewed as a branched cover of the Riemann sphere; singularities are admitted but the program works with the desingularised curve. From this, using standard results from Riemann surface theory, various quantities can be computed: the genus, a basis for the differentials, the j-invariant, {etc}. . Here we focus in particular on the basis for the first homology group, and the corresponding period matrix, as these are more relevant for this thesis.

\subsubsection{Monodromy}
\begin{figure}
\begin{center}
\includegraphics*[height=300pt, trim= 0 0 0 30]{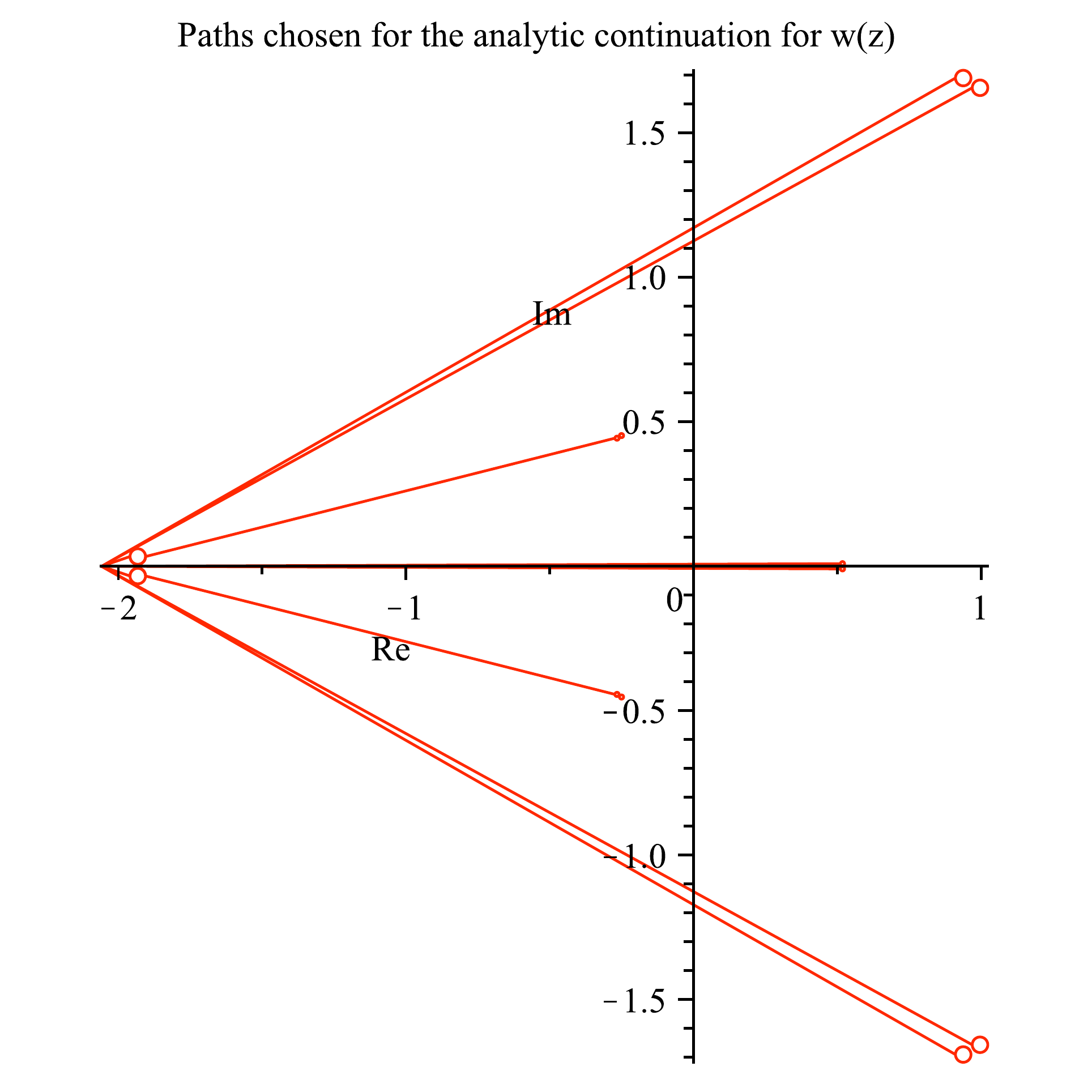}
\caption{Paths chosen for the analytic continuation for $\Xhat$}\label{cc1}
\end{center}
\end{figure}

Here a Riemann surface is viewed as a branched cover of the Riemann sphere, thus the first ingredient needed is the monodromy around each branchpoint. Recalling section \ref{remarks}, one starts by choosing a basepoint  $x^{ P}$ on the base $\puno$, and ordering its preimages on $\Sigma$, $ P= (x^{ P }, y^{P }_{1}),. . ., P= (x^{ P} , y^{P }_{n})$. The basepoint is not allowed to be a branchpoint or a singular point on $\Sigma$, and the program chooses it so that it is far away from these ``problem points'' (see section 3 in \cite{algc1} for a precise characterisation), taking a point with a sufficiently negative real part.\\
The next problem is the choice of the cut for each branchpoint: in this algorithmic approach, using simply straight lines is not feasible, as one needs a way to avoid it passing through ``problem points''. This is done by choosing a combination of segments and arcs as in Figure \ref{cc1} (again, for a precise explanation  see \cite{algc1}).\\

Now, for every branchpoint $B_{i}$ one can build a path $\gamma_{i}$ on $\puno$ that starts and end at $x^{P}$, going along the cuts previously constructed and encircling the branchpoint anticlockwise.  The n-tuple $(y^{P}_{1},y^{P}_{2},\ldots , y^{P}_{n})$ is analytically continued  along $\gamma_{i}$,  returning to the same set of $w$-values, but permuted $(y^{P}_{\sigma_{i}(1)},y^{P}_{\sigma_{i}(2)},\ldots , y^{P}_{\sigma_{i}(n)})$. This defines a permutation $\sigma_{i}$, the monodromy data for the branchpoint $B_{i}$;  the permutations $\sigma_{i}$  generate the monodromy group based at $x^{P}$. The analytic continuation is performed using the command \texttt{Acontinuation}: we describe it shortly as it was used in this work of thesis.\\ 

Consider two points on $\puno$, sufficiently close, $x^{Q}$ and $x^{R}$, and a path $\gamma$ connecting them; corresponding to $x^{Q}$ there is an ordered $n$-tuple $(y^{Q}_{1},y^{Q}_{2},\ldots , y^{Q}_{n})=:\mathbf{y}(x^{Q})$, \ie the $y$-coordinates of the $n$ ordered points $Q_{i}$  on $\Sigma$. Following the path $\gamma$ on $\puno$ from $x^{Q}$ to $x^{R}$, the entries of $\mathbf{y}(x^{Q})$ also follow the lifts $\tilde{\gamma}_{i}$ of $\gamma $ on $\Sigma$. If $x^{Q}$ to $x^{R}$ are close enough, these lifts can be approximated as straight lines:
\begin{equation*}
\mathbf{y}(x^{R})=\mathbf{y}(x^{Q})+\mathbf{y}'(x^{Q})(x^{R}-x^{Q})+{O}(|x^{R}-x^{Q}|^{2}),
\end{equation*}
where the entries of $\mathbf{y}'(x^{Q})$ are the tangents to $\Sigma$ at $Q_{i}=(x^{Q},y_{i}^{Q})$, obtained by implicit differentiation. If the two points are close enough, and the path $\gamma$ does not deviate much from a straight line\footnote{This can be made precise, \eg bounding the second derivative of $\mathbf{y}(x^{Q})$ to bound $\mathrm{O}(|x^{R}-x^{Q}|^{2}),$ }, the entries of $\mathbf{y}(x^{R})$ above give an accurate approximation for the $n$-tuple above $x^{R}$, appropriately permuted: $(y^{R}_{\sigma(1)},y^{R}_{\sigma(2)},\ldots , y^{R}_{\sigma(n)})=:\mathbf{y}(x^{R})$.\\
Given arbitrary points $x^{Q}$ and $x^{R}$ and a parametrised path $\gamma(t)$, the command \texttt{Acontinuation} iterates the above procedure along small portions of the given path, until the necessary conditions above are satisfied. It gives as output the ordered lists $\mathbf{y}(x^{Q})$ and $\mathbf{y}(x^{P})$, which means, for example, that the lift $\tilde{\gamma}$ of $\gamma$ with initial point the first entry of $\mathbf{y}(x^{Q})$, corresponding to $Q_{1}$, has final point the first entry of $\mathbf{y}(x^{R})$, corresponding to $R_{\sigma(1)}$. \\

The above results in the command \texttt{monodromy(f,x,y)}, whose output is a list with three entries: the first is the basepoint $x^{P}$; the second is the ordered n-tuple  $(x^{P}_{\sigma_{i}(1)},x^{P}_{\sigma_{i}(2)},\ldots $, $ x^{P}_{\sigma_{i}(n)})$; the third is a list of the branchpoints $B_{i}$ with the corresponding monodromy, given as a permutation $\sigma_{B_{i}}$. We give an example of how this work for the case of the curve $\Xhat$; the results are given to 4 digits for readability, but setting \texttt{Digits} in Maple yields the desired precision.

\begin{verbatim}
> f:=w^3 + w*z^2 + z^6 + 5*sqrt(2)*z^3 - 1;    # define the curve
> m:=monodromy(f,z,w,showpaths):      
       #computes the monodromy (the option showpaths produces Figure B.1)
> m[1];                                 # basepoint
                               -2.0584
> m[2];                                  # sheets
             [-1.7952, .8976-2.5795*I, .8976+2.5795*I]
> m[3];                                # branchpoints and their monodromy
  [[[.9370-1.6904*I, [[2, 3]]], [.9954-1.6567*I, [[1, 2]]], 
  [-1.9324-0.337e-1*I, [[1, 2]]], [-.2508-.4525*I, [[1, 3]]],
  [-.2665-.4435*I, [[1, 2]]],  [.5173-0.90e-2*I, [[1, 3]]],   
  [.5173+0.90e-2*I, [[1, 2]]], [-.2665+.4435*I, [[1, 3]]],   
  [-.2508+.4525*I, [[1, 2]]], [-1.9324+0.337e-1*I, [[1, 3]]],   
  [.9954+1.6567*I, [[1, 3]]], [.9370+1.6904*I, [[2, 3]]]]
\end{verbatim}

\subsubsection{A homology basis}
The construction described above gives a way to view the Riemann surface as  $n$ sheets cut and glued along the cuts $\gamma_{i}$ passing through the branchpoints $B_{i}$, each with monodromy $\sigma_{B_{i}}$: these are exactly the ingredients needed for the Tretkoff and Tretkoff algorithm of \cite{TT} (see Appendix \ref{apptt}). Given the above data, the command \texttt{homology(x,y)} follows  the  Tretkoff and Tretkoff algorithm step by step; it gives as  output both the cycles $c_{1},\ldots, c_{r}$, \ie the homology basis for the 1-skeleton $\Gamma$, and the canonical basis $\mathfrak{a}_{1} ,\ldots, \mathfrak{a}_{g}, \mathfrak{b}_{1},\ldots,\mathfrak{b}_{g}$ for  $H_{1}(\Sigma,\mathbb{Z})$, together with the matrix of change of basis (in $H_{1}(\Gamma,\mathbb{Z})$) between the two. \\
The cycle $c_{k}$  is given as a list: 
the first element specifies  the starting sheet $i$;
the second element is the coordinate of a branch point $B_{l}$ in $\puno$, 
together with the disjoint cycle of $\sigma$, which contains $i$; the third element is a sheet  $j $, and so on. Essentially, this reads: ``from sheet $i$ go to sheet $j$ , by encircling the branchpoint $B_{l}$''; the list is cyclical, giving a closed cycle. This is \texttt{h[cycles]} below.\\
The canonical cycles are also given as lists, but in a slightly different way:  the first element specifies the starting    sheet numbers; the second element is a branchpoint, together with a number indicating how many times one has to circle around it in the counter-clockwise
    direction (clockwise direction if this number is negative); the third entry is again a sheet number, etc. .  This is \texttt{h[canonicalcycles]} below.\

\begin{verbatim}
> h:=homology(f,z,w):
> h[basepoint];                                  # basepoint
                               -2.0584
> h[sheets];                                     # sheets
            [-1.7952, .8976-2.5795*I, .8976+2.5795*I]
> eval(h[cycles]);                               # cycles
table([1 = [1, [.9954-1.6567*I, [1, 2]], 2, [-1.9324-0.337e-1*I, [1, 2]]], 
     2 = [1, [.9954-1.6567*I, [1, 2]], 2, [-.2665-.4435*I, [1, 2]]], 
     3 = [1, [-.2508-.4525*I, [1, 3]], 3, [.5173-0.90e-2*I, [1, 3]]], 
     5 = [1, [-.2508-.4525*I, [1, 3]], 3, [-.2665+.4435*I, [1, 3]]], 
     4 = [1, [.9954-1.6567*I, [1, 2]], 2, [.5173+0.90e-2*I, [1, 2]]], 
     7 = [1, [-.2508-.4525*I, [1, 3]], 3, [-1.9324+0.337e-1*I, [1, 3]]], 
     6 = [1, [.9954-1.6567*I, [1, 2]], 2, [-.2508+.4525*I, [1, 2]]], 
     10 = [1, [.9954-1.6567*I, [1, 2]], 2, [.9370-1.6904*I, [2, 3]], 
           3, [-.2508-.4525*I, [1, 3]]], 
     8 = [1, [-.2508-.4525*I, [1, 3]], 3, [.9954+1.6567*I, [1, 3]]], 
     9 = [1, [.9954-1.6567*I, [1, 2]], 2, [.9370+1.6904*I, [2, 3]], 
          3, [-.2508-.4525*I, [1, 3]]]])
\end{verbatim}
 \begin{verbatim}
 > eval(h[linearcombination]);
\end{verbatim}
$$                            \left[ \begin {array}{cccccccccc} 1&0&0&0&0&0&0&0&0&0
\\0&1&0&0&0&-1&0&0&0&0\\0&0&1&0&-1
&0&0&0&0&0\\0&0&0&1&-1&-1&0&0&0&0
\\0&0&0&0&0&0&0&1&1&0\\0&0&0&0&0&1
&0&0&-1&0\\1&-1&0&0&1&0&0&1&1&0\\-
1&0&-1&0&1&0&0&-1&0&0\end {array} \right] 
$$
\begin{verbatim}
> eval(h[canonicalcycles]);                             # canonical cycles
table([b[3] = [[1, [.995401-1.656656*I, 1], 2, [-1.932407-0.33715e-1*I, -1], 
1, [-.266483-.443510*I, 1], 2, [.995401-1.656656*I, -1], 
1, [-.250849-.452536*I, 1], 3, [-.266483+.443510*I, -1], 
1, [-.250849-.452536*I, 1], 3, [.995401+1.656656*I, -1], 
1, [.995401-1.656656*I, 1], 2, [.937006+1.690371*I, 1], 
3, [-.250849-.452536*I, -1]]], a[3] = [[1, [-.266483+.443510*I, 1], 
3, [.517332-0.9026e-2*I, -1]]], b[4] = [[1, [-1.932407-0.33715e-1*I, 1], 
2, [.995401-1.656656*I, -1], 1, [.517332-0.9026e-2*I, 1], 
3, [-.266483+.443510*I, -1], 1, [.995401+1.656656*I, 1], 
3, [-.250849-.452536*I, -1]]], a[2] = [[1, [-.250849+.452536*I, 1], 
2, [-.266483-.443510*I, -1]]], b[1] = [[1, [.995401-1.656656*I, 1], 
2, [.937006+1.690371*I, 1], 3, [.995401+1.656656*I, -1]]], 
b[2] = [[1, [-.250849-.452536*I, 1], 3, [.937006+1.690371*I, -1], 
2, [-.250849+.452536*I, -1]]], a[4] = [[1, [-.266483+.443510*I, 1], 
3, [-.250849-.452536*I, -1], 1, [-.250849+.452536*I, 1], 
2, [.517332+0.9026e-2*I, -1]]], a[1] = [[1, [.995401-1.656656*I, 1], 
2, [-1.932407-0.33715e-1*I, -1]]]])
\end{verbatim}

\subsubsection{The Riemann period matrix} 
A basis $(\UUU_{1} , . . . ,\UUU_{g})$ of holomorphic differentials on the Riemann surface is given by\footnote{We refer to \cite{algc1} for the case of a singular Riemann surface.}
$$
\UUU_{k} =  \frac{P_{k} (x , y ) }{\partial_{y} F (x , y )} \de x ,
$$
with $P_{k} (x , y ) = \sum_{i +j \leq d - 3} a_{kij} x^{i} y^{j}$, where $d$ is the degree of $F (x , y )$ as a polynomial in x and y. This is implemented in \texttt{Algcurves} by the command \texttt{differentials(f,x,y)}.\\

Now, given bases for the holomorphic differentials and for $H_{1}(\Sigma,\mathbb{Z})$, the period matrix for the Riemann surface $\Sigma$ is
\begin{align*}
\UUU&=(\mathcal{A},\mathcal{B});&  \mathcal{A}_{ij} &=\int_{\mathfrak{a}_{i} }\UUU_{j}, \quad\mathcal{B}_{ij} =\int_{\mathfrak{b}_{i}} \UUU_{j}.
\end{align*}
As the paths of integration are just straight lines and arcs of circles,  they are parametrised 
by $x = \gamma (t )$ with $0 \leq t \leq 1$. The lift of $x = \gamma (t )$, denoted by $y = \tilde{\gamma}(y_{0} , t )$, is obtained by specifying a starting value of $y$, $ y_{0}$, and by analytically continuing this value $y_{0}$ along $\gamma (t )$. These integrals are evaluated numerically using Maple's numerical integration routine.

\begin{verbatim}
> differentials(f,z,w);                           # basis of holomorphic differentials
\end{verbatim}
$$\frac{\de z}{3w^{2}+z^{2}},\frac{w\de z}{3w^{2}+z^{2}},\frac{z^{2}\de z}{3w^{2}+z^{2}},\frac{z\de z}{3w^{2}+z^{2}},
$$
\begin{verbatim}
> periodmatrix(f,z,w);                            # 2g X g period matrix
\end{verbatim}
$$\left[ \begin {array}{cccc}  0.0278- 0.0143\,i& 0.301+ 0.5269\,i&-
 0.863- 0.048\,i& 0.1206- 0.279\,i\\- 0.4624+ 0.8095
\,i& 0.3181+ 0.5522\,i& 0.011+ 0.0110\,i& 0.1475- 0.3014\,i
\\- 0.4518+ 0.8229\,i&- 0.6281+ 0.0068\,i& 0.0250+
 0.0055\,i&- 0.3266+ 0.0159\,i\\- 0.4518+ 0.7961\,i&
- 0.6281+ 1.099\,i& 0.025+ 0.0167\,i&- 0.3266- 0.6187\,i
\\ 0.0115- 0.0360\,i&- 0.3174+ 0.5274\,i&- 0.435-
 0.8631\,i&- 0.1549- 0.2676\,i\\- 0.0115- 1.065\,i&
 0.3174+ 0.5162\,i& 0.435+ 0.2279\,i& 0.1549+ 0.4722\,i
\\ 0.0041- 1.119\,i& 0.3066- 0.5726\,i& 0.017-
 0.6156\,i&- 0.4814- 0.0461\,i\\ 0.4592- 0.2580\,i&
 0.0041- 1.056\,i&- 0.4770+ 0.3601\,i& 0.6531+ 0.1608\,i\end {array}
 \right] 
$$
    \begin{verbatim}                       
> periodmatrix(f,z,w,Riemann);                     # Riemann period matrix
\end{verbatim}
$$
 \left[ \begin {array}{cccc}  0.5704+
 0.9751,i&- 0.5477- 0.2433
\,i& 0.0131+ 0.7222\,i&
 0.5027- 0.4584\,i\\-
 0.4986- 0.2320\,i&- 0.1712+
 1.2201\,i&- 0.9952- 0.1650\,
i& 0.1609- 0.4954\,i\\
 0.0439+ 0.5845\,i&- 0.9591
- 0.1642\,i&- 0.3311+ 1.5383
\,i& 0.2554- 0.7874\,i\\
 0.4113- 0.3140\,i& 0.1469-
 0.5196\,i& 0.2518- 0.79099\,
i&- 0.8857+ 1.01434\,i\end {array} \right] 
$$

\section{\texttt{Extpath} and \texttt{Cyclepainter}}\label{timscode}
In the present work on symmetric monopoles a fundamental simplification of the problem was the choice and use of a specific basis for the first homology group, namely one that exploits the symmetry of the problem. This cannot be done with Maple's \texttt{algcurves}, as this package relies on the homology basis found via the Tretkoff and Tretkoff algorithm. %
Therefore the package \texttt{extcurve} has been developed, by Timothy Northover, to take care of the cases where symmetry plays an important role, which happens quite often in integrable systems.\\

The \texttt{extcurves} routines rely  on \texttt{algcurves}, so, for instance, once again the Riemann surface is specified as the zeros of a polynomial of two complex variables. The single most important feature of \texttt{extcurves} is the possibility of specifying an \texttt{extpath}, \ie describing an arbitrary path on a Riemann surface, as a sequence of straight lines on the $x$-plane together with a sheet specification at the initial point.\\
In this framework, a homology basis is given as a list of \texttt{extpaths}.\\

Paths can be created as \texttt{extpaths} in several ways: the one more used in this thesis is via CyclePainter. CyclePainter is a Java program to visualise paths on a Riemann surfaces, that can then be used by \texttt{extcurve} in Maple. The Riemann surface is pictured as a cut plane, where, very importantly, both the basepoint for monodromy and the reference point for sheet ordering can be chosen; the cuts are then straight lines from this basepoint to the branchpoints. The paths are drawn as a sequence of straight segments by simply pointing and clicking with the mouse; one can choose the starting sheet, and the segments have a different colour depending on their sheet; crossing a cut, hence changing sheet, results in a change of colour.\\
This is very useful as it allows a graphical representation of a cycle, while also producing a file which can be directly read into Maple.\\

We point out that this basepoints and cuts representation is just a pictorial representation, when the path in exported as an \texttt{extpath} no use is made of that.\\

Using CyclePainter, then, one can effectively draw a given homology basis for a Riemann surface (for example, the symmetric basis of section \ref{hombasis}), and export it as a sequence of \texttt{extpaths} in Maple. Using \texttt{extcurve}, several functions can be made to act on extpaths.\\

An important one is \texttt{isect(curve, path1, path2) }, which computes the intersection number of \texttt{path1} and \texttt{path2}. This has been widely used in this thesis to check if a given homology basis is canonical.\\

Using this, another function is defined,  \texttt{find\_homology\_transform(curve, hom1, hom2)}: this returns the matrix that transforms the homology basis \texttt{hom1} to \texttt{hom2}. This has been relevant  when comparing two different bases, like the cyclic and symmetric bases in section \ref{secarcex}. This is particularly useful when combined with the command\\ \texttt{transform\_extpath(initCurve, path, trans, finCurve := initCurve)}, which returns the image of \texttt{path} under the transformation\footnote{A \textit{caveat} here is that this function only works with linear or M\"obius transformations.} \texttt{trans}. This last function allowed us to find the image of a given homology basis under certain involutions of the given surface, and  \texttt{find\_homology\_transform} resulted in the matrix form of these involutions on the given homology basis (see for example section \ref{symmpm}).\\

Finally, the function \texttt{periodmatrix(curve, hom)} calculates the period matrix by transforming the \texttt{algcurves} version with the appropriate matrix of change of basis between \texttt{algcurves} basis and \texttt{hom}. This has been used to check numerically several results regarding the period matrix, \eg its symmetry in section \ref{symmpm}, or the Ercolani-Sinha constraints check in section \ref{EScyclic}.\\

\begin{verbatim}
> march(open, "extcurves.mla");
> with(extcurves):
> curveA, homA, namesA := read_pic("alpha0_A_final.pic"):  
                   # reads the cyclic basis from the corresponding CyclePainter file
> curveBE, homBE, namesBE := read_pic("BE_hwb.pic"):
                   # reads the symmetric basis from the corresponding CyclePainter file
> Matrix(8, (i,j) -> isect(curveA, homA[i], homA[j]));
                    # computes the intersection matrix for homA 
\end{verbatim}
$$
 \left[ \begin {array}{cccccccc} 
 0&0&0&0&-1&0&0&0\\
 0&0&0&0&0&-1&0&0\\
 0&0&0&0&0&0&-1&0\\
 0&0&0&0&0&0&0&-1\\
1&0&0&0&0&0&0&0\\
0&1&0&0&0&0&0&0\\
0&0&1&0&0&0&0&0\\
0&0&0&1&0&0&0&0
\end {array} \right] 
$$\\
\begin{verbatim}
> M:=find_homology_transform(curveA, homA, homBE);
                          # matrix of change of basis from homA to homBE
\end{verbatim}
$$
\left[ \begin {array}{cccccccc} 
1&0&0&0&0&0&0&0\\0&
1&0&0&0&0&0&0\\0&0&0&0&0&0&1&0\\0&0
&0&-1&0&0&0&-1\\-1&0&0&0&1&0&0&0\\0
&0&0&0&0&1&0&0\\0&0&-1&0&0&0&-1&0
\\0&0&0&1&0&0&0&0\end {array} \right] 
$$\\
\begin{verbatim}
> Transpose(periodmatrix(curveA, homA));
\end{verbatim}
$$
 \left[ \begin {array}{cccc}
  - 0.0003+ 0.0369\,i&- 0.0197+ 0.0358\,i&- 0.0321+ 0.0192\,i&- 0.1153+ 0.0734\,i\\
 0.0328- 0.0194\,i&- 0.0197+ 0.0358\,i& 0.00098- 0.0367\,i&- 0.1153+ 0.0734\,i\\
 -0.0314- 0.0196\,i&- 0.0197+ 0.0358\,i& 0.0325+ 0.0194\,i&- 0.1153+ 0.0734\,i\\
 - 0.0000- 0.0000\,i&0.0404- 0.0018\,i&- 0.0000- 0.0000\,i&- 0.0057- 0.4148\,i\\
  0.0330- 0.019\,i& 0.0405- 0.0007\,i&0.0325+ 0.0187\,i&- 0.0050- 0.1388\,i\\
- 0.0337- 0.0189\,i& 0.0405- 0.0007\,i&- 0.0329+ 0.0185\,i&-0.0050- 0.1388\,i\\
- 0.0006+ 0.0381\,i& 0.0405- 0.0007\,i& 0.0002- 0.0380\,i&- 0.0050- 0.1388\,i\\
- 0.0000- 0.0000\,i& 0.0619+ 0.0322\,i& 0.0000+ 0.0000\,i&- 0.3660- 0.6182\,i\end {array} \right] 
$$

Note that here we see explicitly the symmetries described in section \ref{symmpm}.

\listoffigures
\bibliographystyle{alpha}
\cleardoublepage
\phantomsection
\addcontentsline{toc}{chapter}{Bibliography}
\bibliography{biblio} 
\end{document}